\documentclass[a4paper,12pt,titlepage,usenatbib]{book}
\usepackage[T1]{fontenc}
\usepackage{amssymb}
\usepackage{times}
\usepackage{indentfirst}
\usepackage{fancyhdr}
\usepackage{color}

\usepackage[pdftex]{graphicx}

\begin{document}

 \begin{titlepage}

 \begin{center}
 \normalsize{University of Rome $\:\:$}{''SAPIENZA''}\\
 \normalsize{Chemistry Department}\\
 \end{center}

 \begin{center}
 \huge \textbf{MOLECULAR ANIONS IN CIRCUMSTELLAR ENVELOPES, INTERSTELLAR CLOUDS AND PLANETARY ATMOSPHERES: QUANTUM DYNAMICS OF FORMATION AND EVOLUTION}
 \end{center}

 \begin{center}
 \LARGE{PhD Dissertation}\\
 \end{center}

 \begin{center}
 \normalsize{presented by:$\:\:$} \Large{\itshape{Fabio Carelli}}\\
 \end{center}

 \begin{center}
 \normalsize{in partial fulfillment of}\\
 \normalsize{the requirements for the}\\
 \normalsize{degree of}\\
 \large{Doctor of Philosophy}\\
 \end{center}

 \begin{center}
 \normalsize{Presented 23 September 2011}\\
 \normalsize{Commencement: 21 Dicember 2011}\\
 \end{center}

 \begin{center}
 \large{PhD Supervisor:\ \ \ \ \ \ \ \ \ \ \ \ \ \ \ \ \ Prof. F.A. Gianturco}\\
 \large{PhD Coordinator:\ \ \ \ \ \ \ \ \ \ \ \ \ \ \ \ \ \ \ \ \ \ Prof. C. La Mesa}\\
 \large{Reviewer:\ \ \ \ \ \ \ \ \ \ \ \ \ \ \ \ \ \ \ \ \ \ Prof. E. Herbst}\\
 \end{center}

 \end{titlepage}

\tableofcontents
\listoffigures
\listoftables

\newpage

\chapter*{Acknowledgements}
\addcontentsline{toc}{chapter}{Acknowledgements}
\chaptermark{Acknowledgements}
\label{acknowledg}

\noindent Firstly I would like to express my gratitude to Prof. Gianturco, who has been an excellent supervisor, and has supported me with patience through the last three years.
Dear Franco, I hope I've learned that ten percent is inspiration, and the remaining is transpiration.

\noindent I have also benefited from his enthusiasm and his policy of providing his students with the opportunity to attend conferences: meeting a wide variety of people in the field has definitely been a great help to me.

\noindent He encouraged me to have my German (too much short, unfortunately!) experience, giving me the chance to know and work together with an excellent scientist, a very good baritone and imaginative chef: Bernd Nestmann.

\noindent I will never forget our peripatetic lessons.

\noindent Further I would like to extend my thanks to all the members of the molecular quantum structures and quantum dynamics of Rome.

\noindent I would like to sincerely thank all my friends, far and near, and all the people who gave me help and support in this study: Marco Lauricella, who worked for a long time in Ireland, Lara D'Appollonio, my favourite photographer, Federico Fraschetti, a real astrophysicist who is currently in the USA, his wife Evgenia and his sister Caterina; many thanks to Alessandro and Rosanna.

\noindent Many thanks to David Lopez, for his patience and his incredible helpfulness.

\noindent Roberto and Serafina, I'm sure you can heard me: thank you.

\noindent I have no words to thank my brother: his joie de vivre has literally saved me.

\noindent And of course, I wish to tenderly whisper my thanks to my beloved girfriend, Graziella: you have shared most of the ups and downs of this thesis, so you know what I mean.

\noindent Lastly, and most importantly, I wish to thank my parents, Laura Nolfi and Carlo Carelli.

\noindent They bored me, stressed me, made me angry, raised me, supported me, taught me, and loved me.

\noindent To them I dedicate this thesis.

\clearpage


\chapter*{Introduction}
\addcontentsline{toc}{chapter}{Introduction}
\chaptermark{Introduction}
\label{intro} 

\noindent Nowadays, it is a well known fact that most of the matter in our Solar System, in our Galaxy and, probably, within the whole Universe, exists in the form of ionized particles.
On the other hand, in our natural environment, that is to say in 'normal' conditions as on our little planet, gaseous matter generally consists of neutral atoms and neutral molecules: only under specific conditions, in fact, like along lightnings' paths, a large amount of both atoms and molecules can be found as ionized.
For decades, from this point of view, astronomers and astrophysicists believed that only positively charged ions were worthy of relevance in drawing the networks for possible chemical reactions in the interstellar medium, as well as in modeling the physical conditions in most of astrophysical environments.
Thus, negative ions (and especially molecular negative ions) received minor attention until their possible existence was observationally confirmed (discovery of the first interstellar anion, C$_{6}$H$^{-}$), about thirty years after the first physically reasonable proposal on their actual detection was theoretically surmised by E.Herbst.

\noindent From a purely theoretical point of view, negatively charged ions play a peculiar role as they can be formed in large quantities in the gas phase by attachment of low (1.5 eV < $E$ < 10 eV) and very low-energy ($E$ < 1.5 eV) free electrons, while sometimes such a formation process can occur even close to zero eV.
In an astrophysical context, roughly speaking, their role should be then found in their involvement in the charge balance as well as in the chemical evolution of the considered environment: depending on their amount and on the global gas density, in fact, the possible evolutive scenario could be susceptible of marked variations on the estimated time needed for reaching the steady state, their presence having thus also important repercussions on the final chemical composition of a given environment.
In contrast, the formation of positive ions requires an energy amount equal to or greater than the first ionization potential of the neutral molecular species considered, i.e. an energy around 8-10 eV for most of organic molecules, a value which is considerably high for several astronomical contexts, like cold dense interstellar molecular clouds and, with less importance, within diffuse molecular gaseous regions.

\noindent Low-energy electrons, usually considered in the 0-10 eV range, can be very $reactive$ in the sense that they are effectively captured by many molecules, which can then undergo either rapid stabilization toward the undissociated species or fast unimolecular decomposition toward specific fragments, where, in the latter case, the extra electron initially captured by the neutral molecular target, can in turn remain stuck to one of the fragments as well as be re-emitted again in the surroundings with a different (usually lower) kinetic energy.
Consequently, it can play a role in heating the gaseous medium at a molecular level.
Furthermore, low-energy electrons can also be 'simply' deflected and then causing vibro-rotational excitation of the target molecules, therefore, playing once more a role in the heating processes at a molecular level.
On a much larger scale, they could even produce conductivity inhomogeneity, which in turn disturbs the radiowave propagation in such a medium, like on the edge or on the tail of an interstellar shock, or in principle also moving across a dense cloud: from this point of view, electrons can be also seen as an important and versatile source for 'decoupling', at least in the low frequency range, the energy transfer from the interstellar radiative field to molecules and nanoparticles.
Hence, in a qualitative sense, all of these nanoscopic mechanisms could play a crucial role in understanding more deeply the local interactions within different 'phases' of the ISM, ranging from the cold dense molecular gas up to the diffuse molecular component.
In fact, if we look at the stars, with high-energy thermonuclear burning reactions in their interiors, as the primary sources of almost all the energy that is released in the seemingly empty interstellar space, then we can look at the free electrons (\footnote{in the ISM essentially produced either by high-energy UV photons or by stellar winds}) as a flexible 'means' which, in competition with photons, is able to participate into the complex processes responsible for the coupling between the ionized, the dust and the neutral components of the ISM.

\noindent Generally speaking, the main reasons that originally motivated us to undertake the present work, were at least two.

\noindent First of all, we intended to demonstrate the importance of resonances in forming molecular anions in different astrophysical environments.

\noindent Secondly, we were attracted by the possibility of investigating with a reliable and suitably approximate approach, the occurrence of radiationless paths like intramolecular vibrational redistributions (IVRs) to account for the dissipation of the extra energy initially carried by the impinging electron.

\noindent The former aim, which is of course a common feature to each of the molecules investigated here, can be easily justified on the basis of the most important attempts recently dedicated to the understanding of anion formation processes in several astrophysical environments, where indeed only s-wave electron attachment processes were expressly considered.
In the framework of the electron-molecule collisions, once the physical conditions like the mean free electron density and the mean kinetic electron temperature are defined according to the specific astronomical context under investigation, taking in consideration only s-wave electron attachment processes to account for the formation of molecular anions provides, in fact, a reliable but narrow point of view.
In the qualitative sense, resonant electron attachment can be viewed as the doorway for the ensuing possible formation of the stable molecular anion, provided the neutral molecule has a positive electron affinity.
It therefore follows that the resonant contributions associated to partial waves referred to non-spherical angular momentum states, can account for the formation of molecular anions also for non-vanishing collision energies which can be still relevant in the given environment.
In this framework, once one has determined which are the astrophysically relevant resonances for a given molecule, it becomes of interest to also investigate, for each of them, which is the possible evolution of that resonant species: the radiative stabilization for the extra energy content that characterizes the resonant species, in fact, is usually a slow process, that can even exceeds the resonance's lifetime, so that in such a case the autodetachment becomes strongly competitive.
On the other hand, the radiationless IVR constitutes in general another reliable evolutive path that, conversely to the spontaneous emission of a high-energy(\footnote{of several eV: in the 0-5 eV according to most of astrophysical environmental conditions}) photon, can efficiently lead to the stabilization of the metastable anion; at the same time, according to the non-ergodic vibrational redistribution, which for a pletora of molecules is an actual possibility, such a process can also account for specific fragmentation channels, as will be shown in details in the last chapter, where all the findings shall be analised and discussed.

\noindent The present PhD thesis is focusing on electron-molecule interactions with astrophysically relevant molecules, and therefore represents a theoretical/computational work which deals with an area placed at the boundary between (molecular) astrophysics, quantum collision thery, and of course theoretical chemistry.
The three molecular species that will be computationally investigated for their behaviour under low-energy electron collisions are the ortho-benzyne (o-C$_{6}$H$_{4}$), the coronene (C$_{24}$H$_{12}$), and the carbon nitride (NC$_{2}$N), respectively.
Due to their sizes, their peculiar structures, their chemical reactivity and their physical and chemical properties as well as according to astronomical observations, when available, the above three molecules provide representative examples for different astrophysical contexts.
However, each of them is linked to a specific astrophysical aspect that currently constitutes an intriguing scientific challenge, hence the present study.

\noindent The o-C$_{6}$H$_{4}^{-}$ is surmised to play a role in larger PAHs astrohysical synthesis; despite its peculiar physical and chemical properties like its large dipole moment, the neutral parent molecule has not been yet observed in spatial regions where it could appear as a reasonable molecular product.
By admitting the presence of the neutral in some astrophysical environments, like the proto-planetary atmosphere CRL-618, the ensuing reaction of the negative undissociated molecular species with cations present in the same context (toward more complex structures), could be considered as one of the possible paths according which it is possible to justify its disappearence while, at the same time, also providing one further step within the mechanism responsible for the interstellar PAHs formation.
Therefore, on the basis of the above considerations, we have investigated its feasibility in forming negative ions under low-energy electron collisions.

\noindent The C$_{24}$H$_{12}$ constitutes another aromatic species, whose size is markedly larger than the previous aromatic molecule.
Its astrophysical importance rests first of all on its (alleged) almost ubiquitous presence in several astronomical contexts, ranging from the cold dense molecular clouds up to the diffuse interstellar regions.
Therefore, it seems reasonable to assume this molecule as one of the most representative interstellar medium-size PAHs(\footnote{it is one of the largest and most symmetric between the medium-size members of interstellar PAHs family}), at least in the sense of one possible 'founder' of a family of larger PAHs.
Furthermore, according to most of astronomers and astrophysicists, it is assumed that 'large' (\footnote{with about more than 25/30 carbon atoms}) PAHs should almost soak up free electrons, if present in the same spatial region, having thus deep repercussion on the chemical evolution of such an astrophysical environment.
In this framework, due to its positive electron affinity, the coronene molecule can be viewed as one of the key interstellar polycyclic aromatic species so that, by computing the integral, the differential and the momentum transfer cross sections for the electron scattering process within a suitably large and astrophysically meaningful energy range, and with an ab initio approach based on a rigorous quantum dynamical treatment, it was possible to investigate its capability of forming (meta)stable anions under single collision conditions as well as its role in deflecting the colliding electrons.
Therefore, on the basis of the above findings, it was also possible to characterize its global efficiency in producing the transient negative species, as reflected by the estimate for the metastable anion formation's rate coefficient.
As will be fully discussed in the last chapter of the present thesis, it is not surprising that this molecule interacts with free impinging electrons leading to a relatively large number of resonances; for the moment it is important to point out that its large static polarizability coupled with the large number of $\pi$ electrons characterizing the neutral enables one to qualitatively expect such a behaviour, although the response of this molecule to colliding electrons for vanishing energies as well as its relative 'transparency' to incident electronic projectiles in the sense of a markedly small deflection of the colliding particle, are not immediately obvious.
Moreover, from the 'pure' theoretical point of view, the C$_{24}$H$_{12}$ behaviour under low-energy single electron collisions was never computationally investigated before; in this connection, such a theoretical work enables to shed light on some of the possible mechanisms responsible for the anion formation.
Due to its molecular size, it also represents a demanding challenge to relate the present theoretical findings with the available experiments, especially when one takes in consideration the lowest collision energy range ($E$ $\sim$ 0 eV) under the light of the astrophysical role played by the C$_{24}$H$_{12}^{-}$ anion.

\noindent Last but not least, the NC$_{2}$N.
This molecule was choosen as an interesting candidate that, in principle, can participate to the formation of one of the most abundant anions observed in the Titan's upper atmosphere, the CN$^{-}$ anionic species, as well as for the production of the undissociated anion, NC$_{2}$N$^{-}$.
The former would be formed if an efficient dissociative electron attachment does take place, while the latter formation involves the stabilization of the whole molecular structure without fragmentation paths.
It therefore follows that, once the metastable species is formed at the equilibrium under the observed conditions of Titan's atmosphere, the investigation of the nuclear deformation paths is a foundamental step to establish which anion is mainly produced as well as its formation probability.
More in general, such an investigation could allow to find suitable environmental conditions that favour the formation of a specific product, thus providing useful information that can be used to draw different evolutive chemical scenarios.
In this connection, I find important to stress that the available developed chemical models of Titan's atmosphere, in fact, consider the dissociative path as dominant, according which the NC$_{2}$N should mainly provide the CN$^{-}$ species.

\clearpage


\chapter[Electron driven processes in astrophysical molecular gases]{Electron driven processes in astrophysical molecular gases}
\label{e-driven-processes}

\section[Molecular negative ions, temporary molecular ions (TNI) and their stability: the electron affinity (EA)]{Molecular negative ions, temporary molecular ions (TNI) and their stability: the electron affinity (EA)}
\label{TNI-EA}

\noindent It is well known and experimentally demonstrated that many atoms and molecules can also exist as negatively charged species in the gas phase.
Such ions can in fact be produced by electron transfer processes from neutral species (indirect ionization), like in the following examples

\begin{equation}
N + M \longrightarrow N^{+} + M^{-} 
\end{equation}

\begin{equation}
N^{-} + M \longrightarrow N + M^{-}
\end{equation}

\noindent or simply by a direct collision with an incident electron (electron attachment)

\begin{equation}
e^{-} + M \longrightarrow M^{-}
\end{equation}

\noindent In the present section, I'm going to focus on the last case, the so called free electron attachment process and the eventually subsequent decomposition of the molecular anion once it is formed as in equation 1.3.
Of course, the previous two reactions refer to other possible channels by means of which it is possible to produce the gaseous anionc species $M^{-}$; it is however important to emphasize that such channels implicitly consider the presence of another negatively charged species in the same environment, so that can be viwed as secondary channels in producing complex anions.
As a qualitative rule of thumb, moreover, since the binding energy of the extra electron to $M$ is usually less when compared with the ionization energy of the other species entering the reactive collision, the first two reactions are often endothermic so that they can only occur if $N$ and $M$ contain sufficient energy (kinetic and/or internal).
An important exception to this assertion is provided by Rydberg negative ions, where the extra electron is contained in a very diffuse orbital having thus a very small binding energy: in this connection, the Rydberg anions play an important role in allowing electron transfer reactions close to the zero energy value.
Before treating the electron attachment problem more in detail, it is necessary to briefly recall the main general aspects concerning the stability of negative ions, thus introducing the electron affinity (EA), a foundamental parameter which is necessary in order to understand the reaction kinetics of all the above processes.

\noindent The possibility of an atom to form a thermodynamically stable anion is expressed by the electron affinity; by analogy with the previous definition, the same property for a molecule $M$ is expressed by the molecular electron affinity (EA) which implicitly refers to the parent $undissociated$ $thermodynamically$ $stable$ molecular anion, $M^{-}$.
Formally the molecular electron affinity defines the energy difference between the neutral species $M$ and the associated undissociated anion $M^{-}$ when both considered in their respective ground states.
By the above definition, it is possible to relate the neutral species and the anionic but only when each of them is in its own most stable geometry; this is to say that for molecules for which the extra electron is capable to induce a considerable geometry change between the neutral and the associated anion, it is necessary therefore to make a distinction between the previously defined EA (which is better identified as $adiabatic$ electron affinity, AEA) and the vertical detachment energy (VDE), as indicated in figure 1.

\begin {figure}
\begin {center}
\includegraphics[scale=0.60]{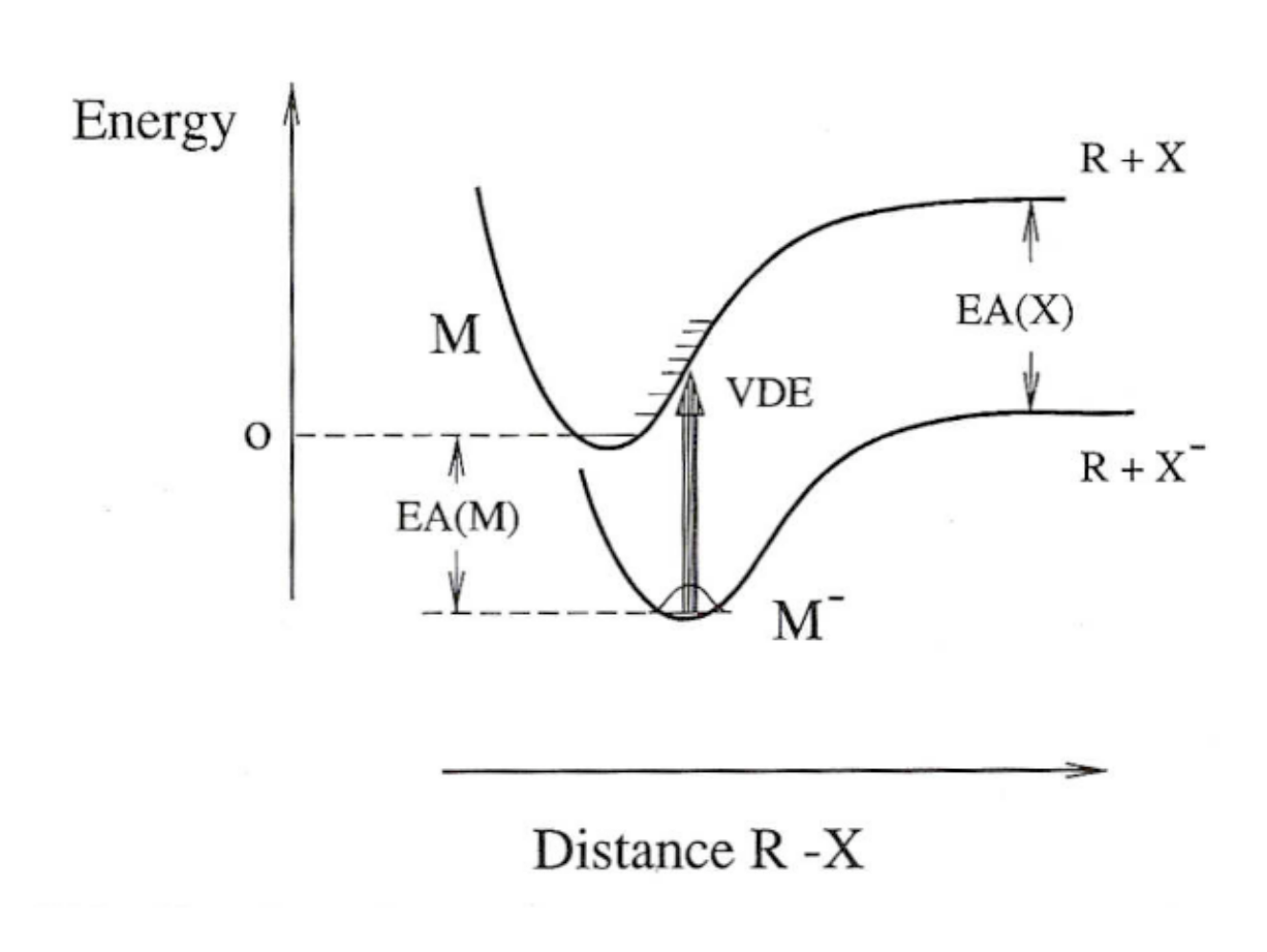}
\end {center}
\caption{\small{Hypothetical Born-Oppenheier potential energy curves for the neutral molecule $M$ and the associated anion $M^{-}$ illustrating the difference between the adiabatic electron affinity (AEA) and the vertical detachment energy (VDE). See main text for details.}}
\label{f_01_chapter_1}
\end {figure}

\noindent The importance of the VDE, when experimentally available, should be found in that such a number implicitly refers to the Franck-Condon transitions which take place in the photodetachment process from molecular anion $M^{-}$

\begin{equation}
M^{-} + h\nu \longrightarrow M + e^{-}
\end{equation}

\noindent so that, in the qualitative sense, the phisical difference between the AEA and the VDE is reminescent of the same concepts in the case of adiabatic and vertical ionization energies as occur in a photoionization process.
By convention, the electron affinity of $M$ is taken as positive when the ground state of $M^{-}$ lies below that of $M$, so that it refers to a stable molecular anion when compared to the parent neutral, and obviously it is considered negative in the opposite case, that is to say if the ground state energy of $M^{-}$ is higher than $M$.
Thus, as implicitly indicated by figure 1, a positive EA value for a molecule $M$ means that the corresponding undissociated anion can exist as thermodynamically stable and having the extra electron in a bound state.

\noindent Many molecules, when in the gas-phase undergo a direct collision with an incident free electron as in equation 1.3, are not indeed able to produce a thermodynamically stable negatively charged species; it is however a concrete possibility (the most frequent) that the impinging electron remain temporarily 'bound' to the target molecule: in such a case the complex between the electron and the molecule is indicated as transient negative ion (TNI) due to its metastable character in the sense of a limited lifetime.
The crucial point that must be emphasized at this point is that the primary electron attachment is formally a molecular electronic transition which in any case, therefore regardless the electron affinity sign, initially does create a negative species unstable with respect the ejection of the extra electron itself.
So, if the EA is positive, the electron attachment leads to an electronically excited state of the anion which can have time enough to give rise to a transition to the thermodynamically stable ground state: the last step depends in fact on how efficient the subsequent relaxation mechanisms are, since the kinetic energy carried by the colliding electron can be dissipated either into the molecular degrees of freedom or radiatively by photon emission provided the lifetime is long enough.

\begin {figure}
\begin {center}
\includegraphics[scale=0.94]{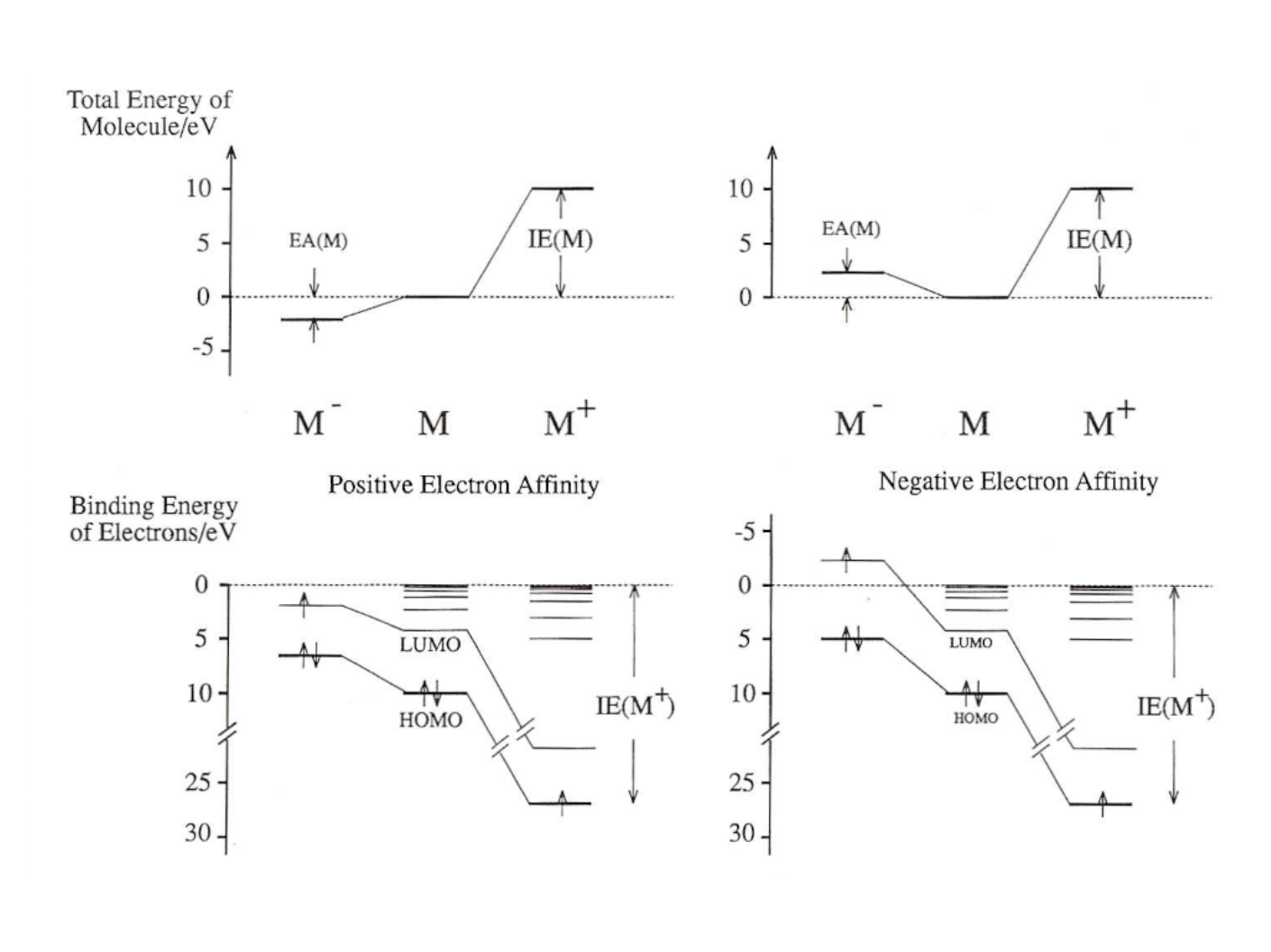}
\end {center}
\caption{\small{Total energy and molecular orbitals schematic representation for a neutral molecule $M$, the associated cation $M^{+}$ and anion $M^{-}$ when considering both a positive and a negative electron affinity.}}
\label{f_02_chapter_1}
\end {figure}

\noindent When considering a species with negative EA, on the contrary, the ground state negative ion is unstable with respect to the extra electron autodetachment, so that the stable negative ion cannot be produced.

\noindent Figure 2 pictorically summarizes the differences between a species having a positive and one having a negative EA, illustrating the situation for a neutral molecule $M$, its cation $M^{+}$ and the corresponding anion $M^{-}$ in terms of the total energy (upper half), as well as of the binding energy (lower half) of the electrons.
As such figure clearly shows, regardless the EA sign, the neutral and the cation are characterized by the presence of a large number of virtual (unoccupied) MO located below the dissociative energy limit (hence associated to bound states) and that thicken as the energy approaches the corresponding ionization limit.
Consequently, both the neutral and the cation are in principle able to capture an incident electron in a MO above the corresponding LUMO but below the dissociation limit.
In contrast to the above observations, the same figure shows a situation completely different for the anion $M^{-}$.
The latter is in fact characterized mainly by a less value for the binding energy of the extra electron, as well as by a reduced number of possible bound states, if they exist: in this connection, when the EA is negative, the figure pictorically shows in fact the extra electron within an excited MO which is not bound since it is located above the ionization limit.
Qualitatively speaking, this feature is susceptible of the following explanation: the extra electron bound in $M^{-}$ can be considered as interacting with a neutral 'core' (i.e. $M$) and therefore, increasing its distance from the neutral molecule, it 'feels' an induced-dipole potential like

\begin{equation}
V(r) = -\frac{\alpha e^{2}}{2 r^{4}}
\end{equation}

\noindent $\alpha$ being the polarizability of the neutral $M$.
An analogous (and qualitative!) approach involving both the neutral $M$ and the cation $M^{+}$ therefore leads us to recognize that the extra electron should be viewd as 'bound' to a single and a double positively charged core, respectively: in these cases the potential 'felt' by the electron is clearly Coulombic

\begin{equation}
V(r) = -\frac{e^{2}}{r}
\end{equation}

\noindent where it is sufficient to think about the hydrogen atom problem to immediately justify the larger number of possible bound states as well as their becoming thicker as the energy approaches the ionization limit.

\section[Collisions of electrons with molecules. TNI formation and resonant states]{Collisions of electrons with molecules. TNI formation and resonant states}
\label{TNI-formation}

\noindent The present section focuses on the interaction of a free electronic particle with a gas-phase molecule under the so called under collision conditions: we are thus considering here conditions which require a mean free path lenght large enough to ensure that secondary collisions (either with other atomic/molecular species or with other electrons) have very low probability to take place.
The above constraint, as we shall see also in the next chapter, turns out to be reasonably satisfied when referring to different astrophysical environments within the interstellar medium (ISM) where the density of matter is low enough to allow its application as a realistic approximation.
In this framework, the interaction of a free incident electron on a molecular target can be divided into two classes, namely $direct$ scattering and $resonant$ scattering.
As the name itself suggests, the former refers to the situation where the impinging electron with kinetic energy $+h^{2}k^{2}/2m_{e}$ collides with the target molecule and will eventually deflected from its original trajectory, as depicted in the upper part of figure 3.

\begin {figure}[here]
\begin {center}
\includegraphics[scale=0.74]{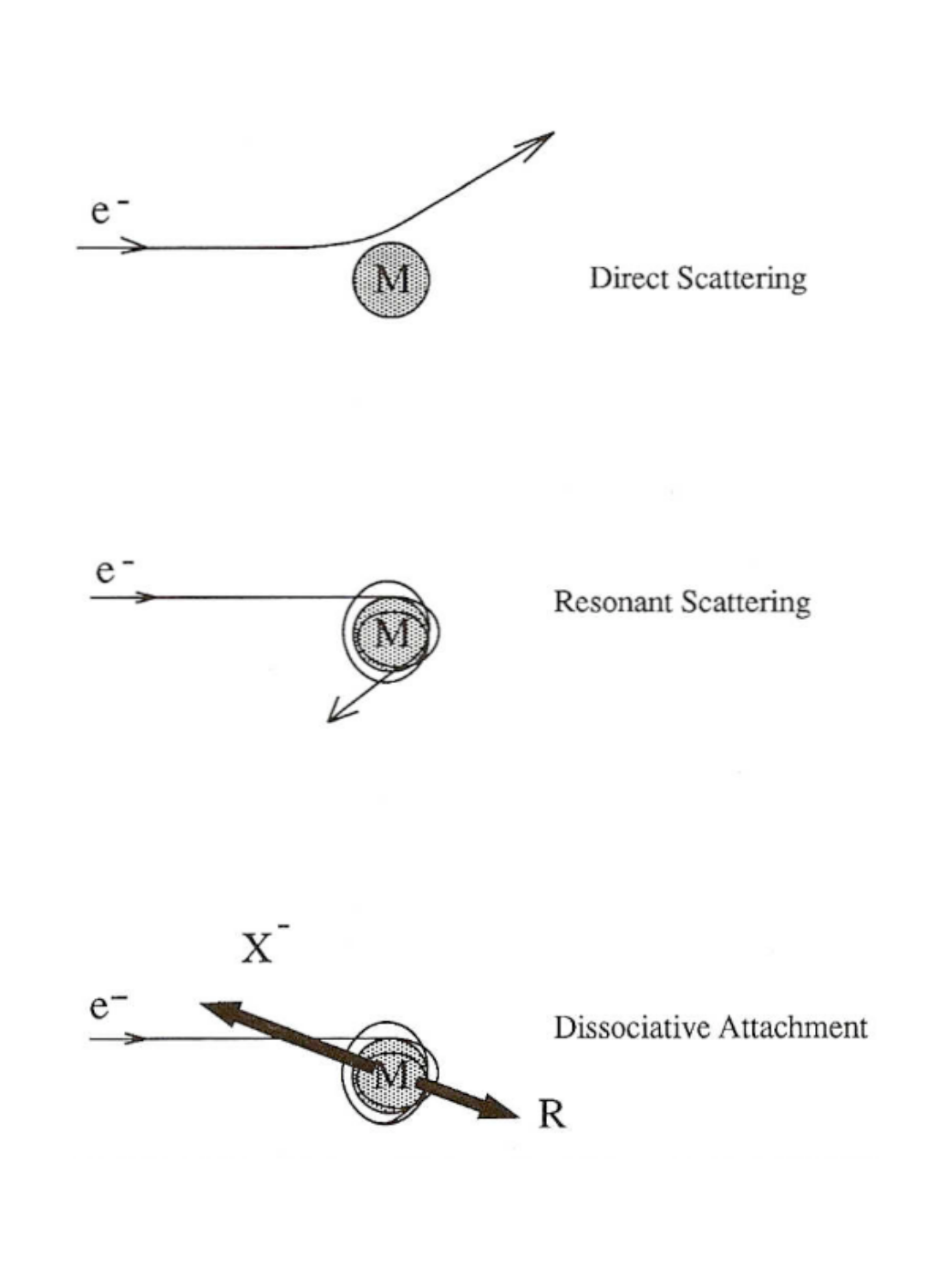}
\end {center}
\caption{\small{Pictorical representation of direct, resonant and dissociative scattering.}}
\label{f_03_chapter_1}
\end {figure}

\noindent Obviously, if the energy of the colliding particle results unaffected by the impact (the moduli of the initial and final wave vectors are equals), the scattering is also elastic.
Furthermore, the colliding electron can instead loose energy to some extent, which therefore results in the excitation of the internal degrees of freedom of the target (or in its electronic excitation) so that the scattering event is inelastic.
Keeping in mind that the mass ratio between the electronic particle and the lightest nucleus is a very small number ($m_{e}/m_{H} \sim 1/1836$), the direct excitation of the roto-vibrational state of the molecular target is not prevented, but strongly unlikely; hence the direct inelastic collisions are mainly responsible for the electronic excitation of the target which, in turn, could evolve in yielding an excited roto-vibrational state of the ground electronic state by internal conversion and, in general, depending on how the strength of the coupling between the metastable electronic state and the nuclear dynamics is.
On the other side, the resonant collision does occur when the incident electron remains 'partially' bound on the molecular target for a time longer than that needed to only cross the region of space occupied by the molecule itself, where partially should be then red in the sense of temporarily.
At this point it is appropriate to point out that an electron capture process can occur only if the kinetic energy of the colliding electron is 'suitable', i.e. only if the incident electron kinetic energy is 'in resonance' with the formation energy of the transient negative ion, so that the expressions 'transient', 'metastable' and 'resonant' are synonymous.
Having thus definied a resonant state, it is implicit that the captured particle can decay after a finite time.
Consequently, in the framework of a stationary representation of the scattering process, the time dependence of such a metastable state is reasonably expressed by the well known imaginary exponential function

\begin{equation}
\Psi \propto exp\left(-\frac{iEt}{\hbar}\right)
\end{equation}

\noindent where the total energy $E$ is complex, according to $E$ = E$_{R}$ - i$\Gamma$/2.
Taking advantage of this allows one to write 

\begin{equation}
\left| \Psi\right|^{2} = \Psi\Psi^{*} \propto exp\left( -\frac{\Gamma t}{\hbar}\right)
\end{equation}

\noindent so that it is clear that the resonant state decays after a lifetime $\tau$ given by

\begin{equation}
\tau \approx \frac{\hbar}{\Gamma},
\end{equation}

\noindent the last espression being reminiscent of the time-energy uncertainty principle ($\hbar \sim 6.6\;\cdot\;10^{16}\;eV\;sec^{-1}$). 
The imaginary component $\Gamma$/2 of the total scattering energy $E$ is the resonance halfwidth.
Generally, the resonance lifetimes $\tau$, depending on both the collisional energy of the electron (i.e. the modulus of its linear moment) and the molecular size, are susceptible to cover a large window, ranging from few vibrational periods ($N_{2}^{-} : \tau\sim 10^{-13} - 10^{-14}$sec, E$_{R}$ $\sim$ 2.3 eV \cite{intro-birtwistle, intro-berman}) up to much longer values ($SF_{6}^{-} : \tau\sim 10^{-6}$sec, \cite{intro-fenzlaff, intro-fenzlaff2}): in this connection it might be useful to emphasize that a free electron of about 1 eV needs of 5$\;\cdot\;10^{-16}\; sec$ to cover a distance of 3$\;\cdot\;10^{-10}\; m$, while a chemical bond typically has a spatial extension which is restricted between $1.2\;\cdot\;10^{-10}\; m$ and $2.8\;\cdot\;10^{-10}\; m$.

\section[Resonance classification]{Resonance classification}
\label{res-classification}

\noindent So far almost nothing has been said about the mechanisms by which a resonant state can be produced.
In this regard, there are many different ways such that an extra electron impinging on a molecule can remain temporarily bound on it, and according to them it is therefore possible to make a general distinction between a $Feshbach$ resonance and a $shape$ resonance.
So it is appropriate to emphasize that a quantitative discussion concerning a resonant scattering process is based on the formalism of quantum mechanics; without going into details, we should however keep in mind that, in the context of a quantum approach, one has to do with a wave packet responsible for the physical description of an electron (collimated) beam, incident on a molecule, which is represented in the following form \cite{intro-sakuraibook}

\begin{equation}
\Psi(z,t) = \int_{0}^{\infty}\; dk\; A(k)\; e^{ikt}\; e^{-i\frac{E_{k}}{\hbar}t}
\end{equation}

\noindent where 'z' in turn represents the incident direction of the beam while $A(k)$, the weight function for the plane waves $e^{ikz}$ continuum summation expressed by the previous integral, is supposed to have compact support centered on a certain $k_{0}$ value.
Therefore, in relation to the possibility of developing each plane wave in spherical waves

\begin{equation}
e^{ikz} = \sum_{l=0}^{\infty} i^{l} \sqrt{4\pi(2l+1)} j_{l}(kr)Y_{l}^{0}(\vartheta) = \sum_{l=0}^{\infty}i^{l}(2l+1)j_{l}(kr)P_{l}(\cos\vartheta),
\end{equation}

\noindent where '$l$' labels each partial angular component of the plane wave, it is possible to qualitatively understand the basis mechanism of a resonant scattering process: such a process is always describable as a constructive interference between the target molecule and one or more partial waves associated to the wavepacket which properly represents the dynamical evolution of the colliding beam.
Having so established a plausible definition, which has a general nature resulting effective as long as it is understood as qualitative and heuristic, let me now add somtething little more specific from the theoretical point of view, always remaining in the scope of a general description.

\subsection[Feshbach resonances]{Feshbach resonances}

\noindent It is now essential to consider the electronic configuration of a generic temporary negative molecular ion.
If the extra electron is able to remain trapped on the molecular target, so producing a resonant state, but without introducing any variation into the electronic configuration of the target molecule itself, we are speaking about a 'one' particle resonance: such a process can therefore be imagined as that one where the TNI is ideally obtained simply adding an electron in one of the virtual (thus available) molecular orbitals of the neutral species involved in the collision.
On the other side, if the extra electron does cause a variation of the target electronic configuration while remaining temporarily bound to it, we would be dealing with a so-called 'two-particle' resonance, also known as 'core-excited' resonance.
In the latter case, the energy of the metastable species (generally indicated as $M^{*-}$) can be located either above or below the energy of the corresponding electronically excited (but not negatively charged) species $M^{*}$.
Depending on whether the energy of the excited species $M^{*-}$ is greater than those of $M^{*}$ or not, the negative electronically excited state is an open or a closed-channel core-excited resonance, respectively; hence, the closed-channel resonance, also identified as electronic Feshbach resonance, cannot directly decay into the ground electronic state.
As figure 4 clearly shows, such a possibility is necessarily accompanied by a modification in the electronic configuration, so that electronic Feshbach resonances have usually long lifetimes.

\begin {figure}[here]
\begin {center}
\includegraphics[scale=0.54]{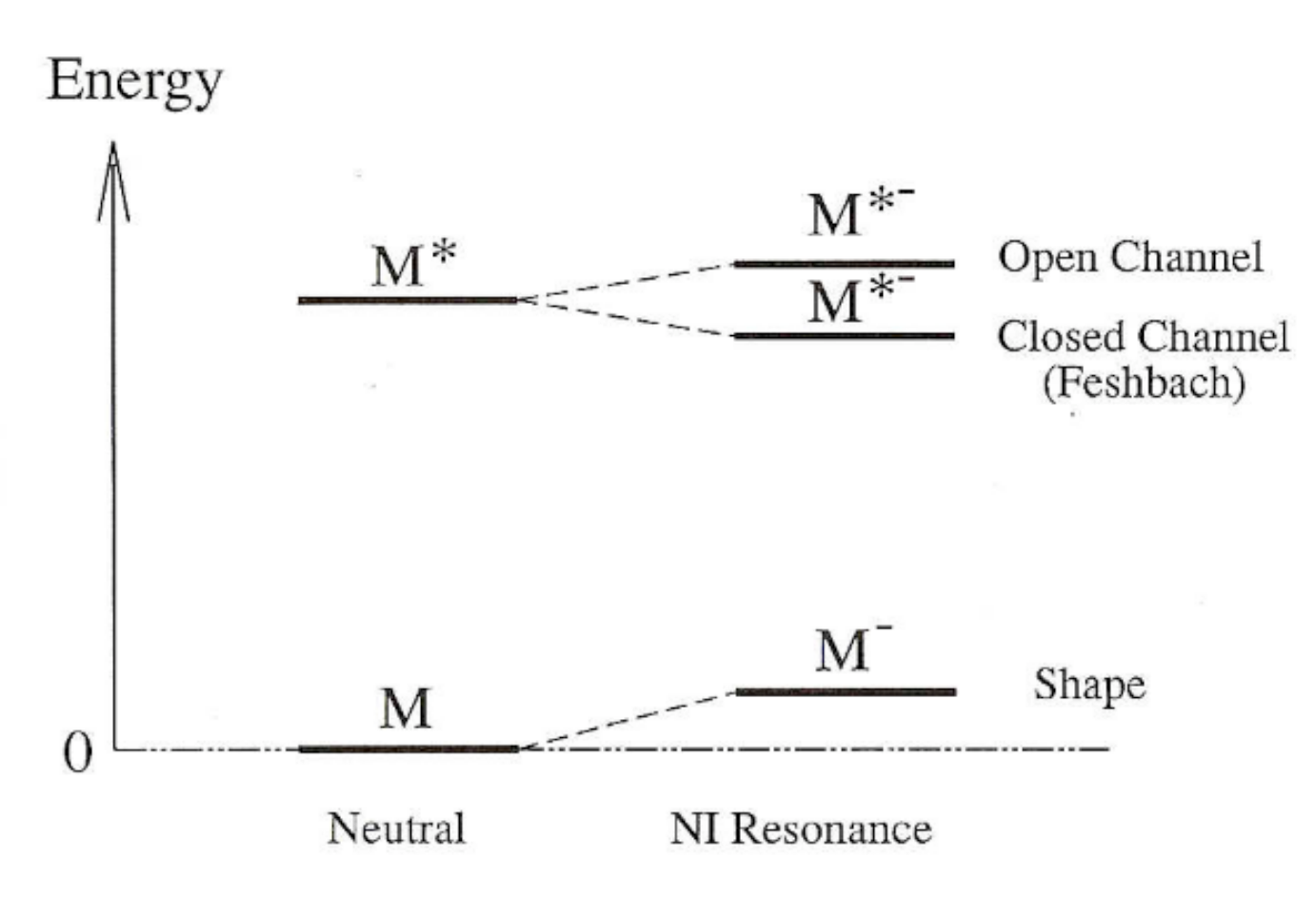}
\end {center}
\caption{\small{Representation of $M$, $M^{*}$, $M^{-}$ and $M^{*-}$ (as displayed by increasing energies) to schematically illustrate the differences between a core-excited open channel resonance and a Feshbach electronic resonance. The figure also make possible a qualitative comparison with a shape resonance. 'NI' = Negative Ion. See text for details.}}
\label{f_04_chapter_1}
\end {figure}

\noindent However, always in the qualitative sense, the colliding electron can also induce a coupling with the nuclear dynamics which in turn is sufficiently strong so that the energy transfer from the electronic to the nuclear degrees of freedom can actually participate in dissipating its excess (kinetic) energy amount, thus preventing the ejection of the extra electron and eventually leading to a nuclear Feshbach resonance.

\noindent At this point, what else about the general mechanism responsible of the electron 'capture' in a Feshbach state?
When the electron projectile induces an electronic excitation into the target molecule, providing an electronic Feshbach resonance, the extra electron is simply detained within the molecular potential field generated by the excited molecule: having loose too much energy to cause the electronic excitation, in fact, it cannot escape since now it has no sufficient energy available so that the target remains in the excited state; hence, before the electron can be emitted, it must reabsorb energy from the excited target.
It is of course possible to assume a little more technical point of view, without any pretense of completeness.
Let us therefore immagine the total resonant wavefunction $\Psi_{n}(\textbf{r},q,\textbf{R})$ developed in the following form, like in a close-coupling expansion

\begin{equation}
\Psi_{n}(\textbf{r},q,\textbf{R}) = \sum_{i} \phi_{i}(q,\textbf{R})\chi_{i}(\textbf{R})f_{n,i}(\textbf{r})
\end{equation}

\noindent where $\textbf{r},q,\textbf{R}$ indicate the electronic projectile spatial coordinates, the ensamble of molecular electronic coordinates (both spatial and spin) and the nuclear coordinates, respectively.
The symbols $\phi_{i}, \chi_{i}$ and $f_{n,i}$ refer then to the i-th electronic eigenstate for the molecule kept with fixed nuclei, the i-th nuclear state, and the wave function for the incident electron.

\noindent Based on the above premise, it is then possible to say that a 'pure' electronic Feshbach resonance does occur when the impinging electron excites one(\footnote{for the sake of simplicity, we are considering here the excitation of only one electronic state}) electronic state, for example the j-th, for which the vibrational levels are all located above the resonance energy, so that the summation reported in eq. 12 should collapse in one term, the j-th.
Consequently, is fairly straightforward to introduce a nuclear Feshbach resonance in a similar way, but now considering, as final achievement, a vibro-rotational excited level of a certain molecular electronic state which, as final result of the collision, can or cannot remain that one initially possessed by the neutral molecular target.

\noindent A conclusive observation.

\noindent If, when treating electronic Feshbach resonances, it is reasonably allowed (as moreover has been implicitly done) to consider a molecular geometry with fixed nuclei, using thus a Born-Oppenheimer description of the molecular electronic levels, moving on nuclear Feshbach resonances such an assumption is no longer valid.
For a nuclear Feshbach resonant state, in fact, the energy exchange due to the coupling between the electronic and the nuclear degrees of freedom comes to play a crucial role, thus excluding a Born-Oppenheimer dynamical description: it is just the fixed nuclei approach that prevents such a 'dialog' between the 'external' (electronic) and the 'internal' (nuclear) degrees of freedom.
In other words, if the molecular nuclei were considered kept as fixed in a certain geometry configuration, the resonant (metastable) nuclear Feshbach negative state would be stable against the electron ejection, so that the nuclear Feshbach state would therefore be meaningless.

\subsection[Shape resonances]{Shape resonances}

\noindent Another mechanism, which is applied to physically explain both one-particle resonances and open channel core-excited resonances, describes the electron trapping by means of the shape of the effective interaction potential between the incident particle itself and the molecule.
The electronic projectile, in fact, remains trapped by means of, or better, within the interaction potential because of its shape: according to this statement, in fact, one speaks about $shape$ resonances.
To get a qualitative idea of that, it is sufficient to assume that the colliding electron experiences an attractive potential in a spatial region centered in the molecular center of mass which is in turn surrounded by a region of repulsive potential whose extension is spatially limited, thus vanishing for large distances.
Depending on the electron kinetic energy, the electron can tunnel the external barrier; in such a case, then, the combined action of the latter (which is now an obstacle for the electron's escape) with the internal attactive region can prevent the electron re-ejection (autodetachment), at least for the time necessary for the negative transient species to be experimentally observed (for example, by electron energy loss spectroscopy or electron transmission spectroscopy).
At this point is legitimate, even necessary, to look for the origin of such a potential: its shape is in fact responsible of the dynamical trapping of the incident electron.
For the sake of symplicity, we can thus suppose the target molecule to have a spherical polarizability, so that we come to the following approximated expression for the effective interaction potential

\begin{equation}
V_{eff}(r) = -\frac{\alpha e^{2}}{2 r^{4}} + \frac{l(l+1)\hbar^{2}}{2mr^{2}},
\end{equation}

\noindent where the first term was already introduced (thus not accidentally; eq. 1.5) in order to qualitatively describe the attractive interaction between the external electron, in a given $l$-th state of angular momentumi, and the 'neutral, soft' internal core in a molecular anion.
Thus, following the present heuristic model, it is just the combination of the negative (attractive) dipole-induced potential with the positive (repulsive) branch, associated to the angular momentum of the electron, that globally results in producing the potential repulsive barrier.

\begin {figure}[here]
\begin {center}
\includegraphics[scale=0.54]{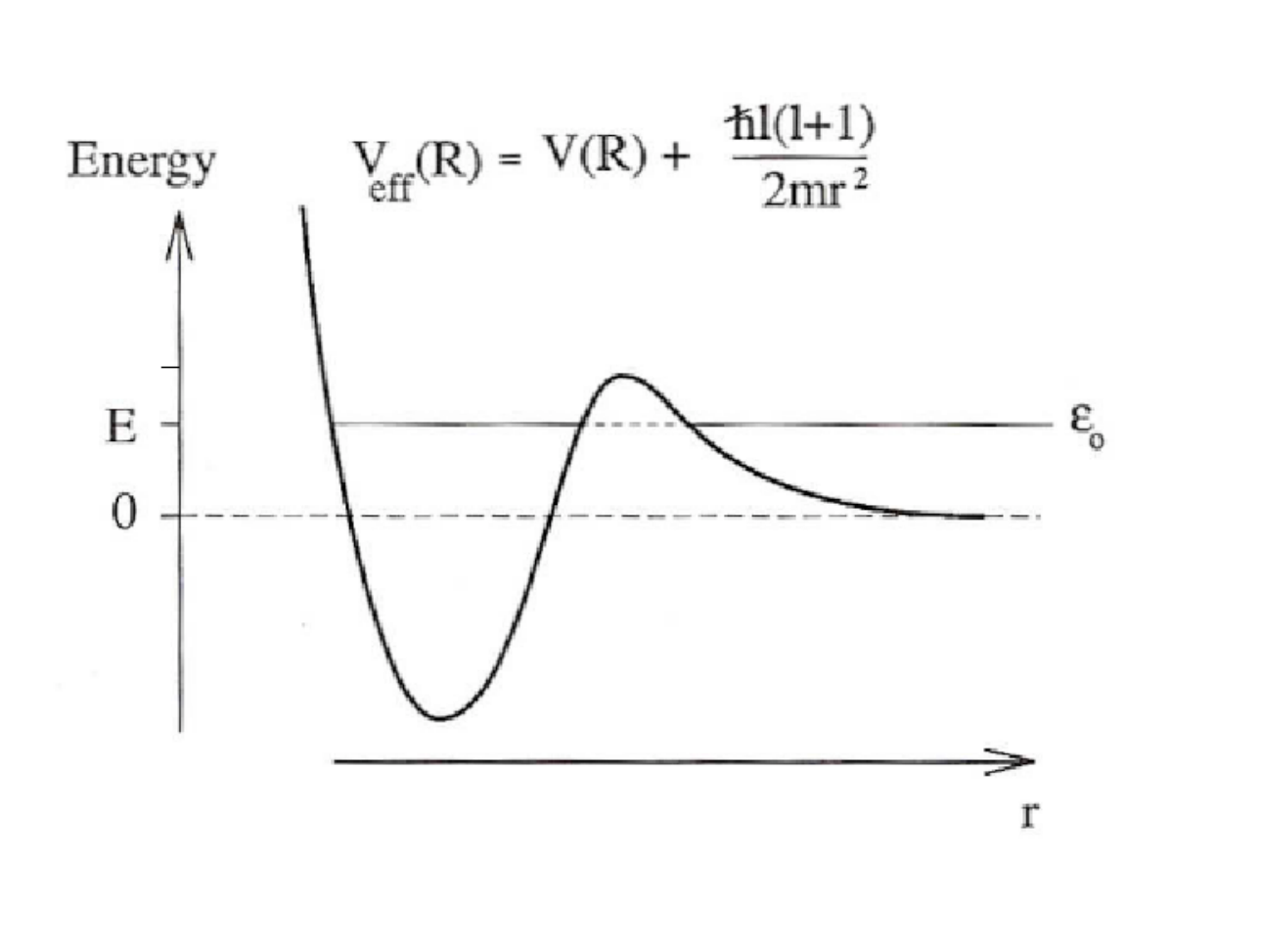}
\end {center}
\caption{\small{Representation of the radial effective interaction potential responsible for the dynamical trapping of an incoming electron in a given state of angular momentum $l\;\geq\;0$. $\epsilon$ indicates the extra electron kinetic energy.}}
\label{f_05_chapter_1}
\end {figure}

\noindent Of course, including only the polarization term in the total interaction potential $V_{eff}(r)$, provides an understandble but not physically realistic picture by which one describes the capture process.
As we will discuss more in detail in chapter 2, in fact, the dipole-induced term (which moreover usually dominates at very low-energy) should be added with the exchange interactions as well as with another term which accounts for the electron correlation; however, for the sake of simplicity, it might be sufficient here to take in consideration only the indicated term.
More in detail, as it is illustrated in figure 1.5, for short distances from the molecular center of mass it dominates the repulsive Coulombic electron-electron interaction, which gives rise to the rapidly increasing branch of repulsive potential close to the zero distance value.
Covering the remaining radial distance, then going away from the target center of mass, that figure clearly shows the presence of an attractive well (just before the centrifugal barrier) which in principle allows the metastable state (a bound state placed at positive energy) to exist for a limited time.

\noindent It is now possible to write down the following general statement: if the target molecule has a virtual MO, energetically accessible and characterized in terms of its spatial symmetry by a suitable value of electronic angular momentum $l$, then an incoming electron that is in turn characterized by the same angular component $l$ $\textbf{can}$ be temporarily captured by tunneling the above barrier, being thus located within the attractive potential region; in other words, since the above point of view is strongly reminiscent of wave mechanics, we can also say that it could take place a constructive interference involving the $l$-th partial wave of the spherical waves summation (eq. 1.11) describing the colliding electron and the above suitable virtual MO.
Of course, the energy of such a virtual MO, generally speaking, do not exactly matches the resonance energy, i.e. the electron kinetic energy for which the metastable anion is formed.
The reason should be found in the fact that eventhough such a description seems to be physically reasonable, on the other hand it is not considering the electron as a dynamical perturbation to the molecular potential field, so that such a virtual MO could or also could not exist before the capture process itself takes place; in this sense, the metastable state should be viewd as the $electronic$ $dynamical$ $molecular$ $response$ to the impact of a free electron with a suitable kinetic energy: hence the energy spectra of the virtual MOs appears to be modified, the corresponding energies being consequently shifted.

\noindent Moreover, the resulting TNI constitutes an example of $discrete$ state embedded in the continuum: according to this, such a species is also called quasi-bound state.
Obviously, resonant states like these cannot exist in principle for an $s$-wave attachment process, unless the effective interaction potential deprived from the centrifugal term is however able to give rise to a similar behaviour, i.e. in producing some barrier behind which an attractive well may exist.

\begin {figure}[here]
\begin {center}
\includegraphics[scale=0.54]{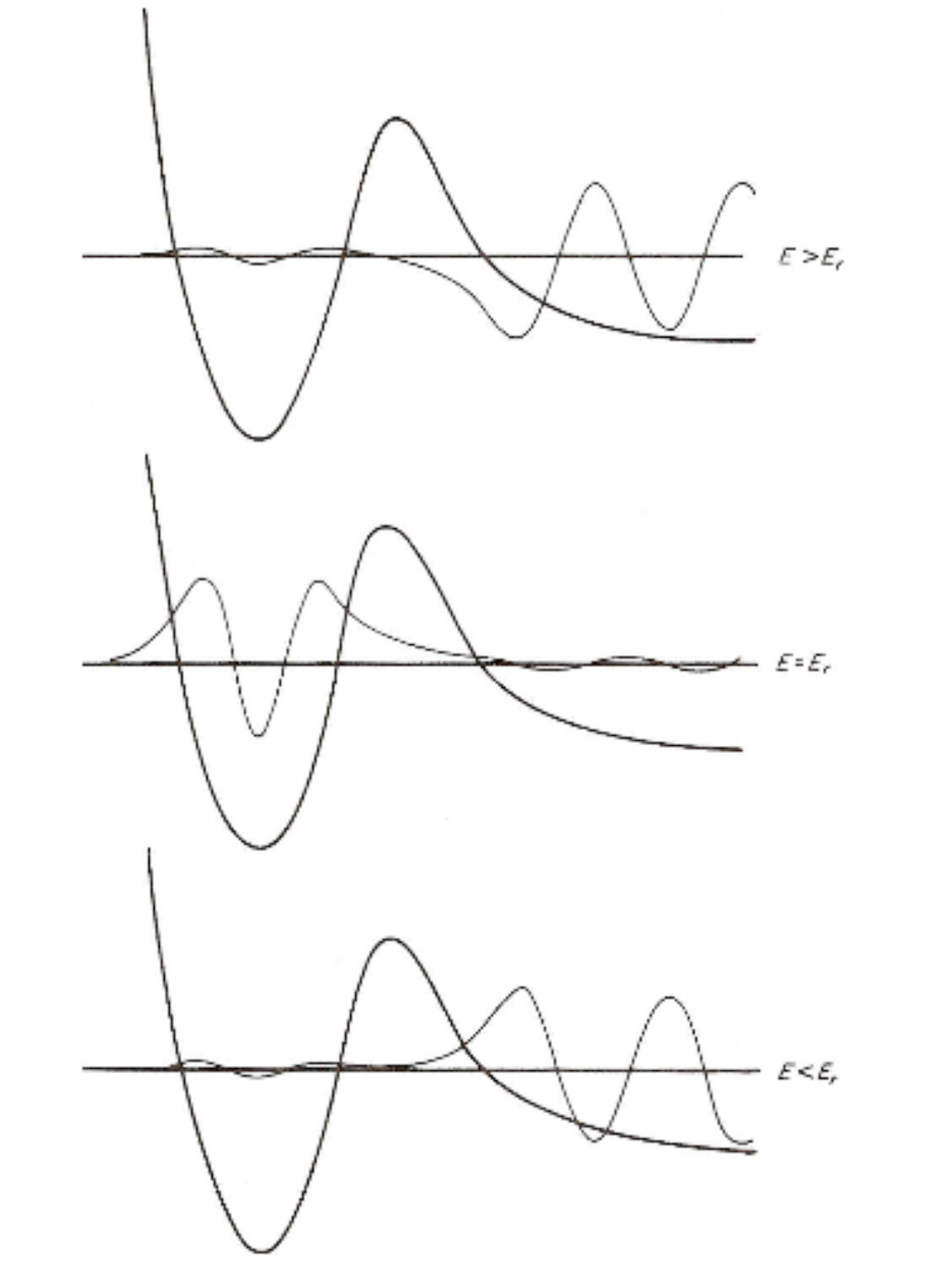}
\end {center}
\caption{\small{Behaviour of the radial scattering wavefunction as the kinetic energy crosses the resonant value ($\epsilon$ = $\epsilon_{r}$). It is implicitly assumed that the interaction potential is not affected by little energy variations around the resonant value. See text for details.}}
\label{f_06_chapter_1}
\end {figure}

\noindent As figure 1.6 pictorically shows, the radial scattering wavefunction behaviour is susceptible of substantial variations depending on the colliding kinetic energy: only when the collision energy matches the resonance value ($\epsilon$ = $\epsilon_{r}$, central inset of figure 6), the incident particle is represented by a wavefunction which has a not negligible squared modulus inside the centrifugal barrier region, then justifying its dynamical trapping and at the same time allowing to the extra electron to be released by the same tunneling mechanism, but in the reverse direction.
When such a condition is not satisfied (upper and lower insets in the same figure), the scattering wavefunction get to be 'simply' reflected by the angular barrier: in this sense the formation of a TNI provides a growth of probability density of the extra electron just within the spatial region corresponding to the attractive well.
Depending of the attractive potential branch deepness, the spatial region where the dynamical trapping does occur can be shifted closer the molecule or not, resulting then in a metastable species where the extra electron is less or more diffuse, respectively.
As we will see in the case of ortho-benzyne metastable anion (chapter 3), this feature qualitatively provides useful information about the anion stability in the sense that the less diffuse the resonant charge is, the more stable the TNI: in such a case, in fact, the subsequent dynamical evolution toward the stable anion by radiationless processes does appear to be in general helped.

\noindent To bring to a conclusion the present subsection, let us now consider the case of a finite-ranged electron-molecule interaction potential, that we can generally write as

\begin{equation}
V_{eff}(r) = V(r) + \frac{l(l+1)\hbar^{2}}{2mr^{2}}.
\end{equation}

\noindent In the realistic case of a finite barrier (i.e. not too much close to the molecular target), the electron can be trapped inside the attractive well, but as we have previously discussed it cannot be trapped forever.
From the pure theoretical point of view, a rigorous non-relativistic quantum mechanical description demostrates that to each scattered $l$-th partial wave there corresponds a phaseshift $\delta_{l}$(\footnote{which is a function of the electron collision energy or, equivalently, of the associated linear momentum $k$}) the behaviour of which is indicative of the possible resonant capture occurrence: for each partial wave involved in a resonance, in fact, as the collision energy crosses the quasi bound metastable state value, the associated phase shift rises through the $\pi/2$ $rad$ value up to about $\pi$.
At the same time, the corresponding partial wave cross sections $\sigma_{l}$, which in general provide the transition probabilities for the TNI formation per unit time, per unit target scatterer and per unit relative flux of the incident electrons with respect to the molecular target, pass though their maximum values given by $\frac{4\pi}{k^{2}}(2l+1)$, so that when a shape resonance dominates the $l$-th partial-wave cross section around the collision energy $E_{r}$, the well known Breit-Wigner resonance formula \cite{intro-taylorbook} does apply(\footnote{provided the resonance is reasonably narrow so that variation in $1/k^{2}$ can be ignored})

\begin{equation}
\sigma_{l} = \frac{4\pi}{k^{2}}(2l+1)\frac{(\frac{\Gamma}{2})^{2}}{(E - E_{r})^{2}+(\frac{\Gamma}{2})^{2}},
\end{equation}

\noindent where $\Gamma$ refers to the full width at half maximum which, according to the energy-time uncertainty as expressed in eq. 1.9, provides an estimate for the resonance lifetime.

\subsection[Low-energy scattering and bound states]{Low-energy scattering and bound states}

\noindent At very low energies, partial waves for higher $l$ are, in general, unimportant.
This point may be obvious classically because the particle cannot penetrate the centrifugal barrier; as a result the potential inside the attractive well has no effect.
As previously introduced, the effective potential for the $l$-th partial wave is given by eq. 14 so that, in terms of quantum mechanics, unless the potential attractive region is strong enough to accomodate $l$ $\ne$ 0 quasi-bound or bound states near $E$ = 0 eV, the behaviour of the radial wavefunction is largely determined by the centrifugal barrier term, which means that it must resemble the $\textbf{free}$ radial wavefunction $j_{l}(kr)$.
A little more quantitatively, in the framework of the partial waves method, it is possible to estimate the behaviour of the phase shift $\delta_{l}$ using the integral equation for the corresponding $l$-th partial wave, obtaining \cite{intro-burkebook}

\begin{equation}
\frac{e^{i\delta_{l}}\sin\delta_{l}}{k} = -\frac{2m}{\hbar^{2}}\int_{0}^{\infty}j_{l}(kr)V(r)f_{l}(k,r)r^{2}dr
\end{equation}

\noindent where $f_{l}(k,r)$ under integration on the r.h.s. is the $\textbf{scattering}$ radial wavefunction.
If $f_{l}(k,r)$ is not too much different from $j_{l}(kr)$, and $1/k$ is much larger than the potential range, it can be shown that the right hand side would vary approximately as $k^{2l}$ (\footnote{according to one of the properties of the spherical Bessel functions, for very low incident energies (so that $k\;\sim\;$0) it results $j_{l}(kr)\;\approx\;\frac{(kr)^{l}}{(2l+1)!!}$}); if one now supposes that $\delta_{l}$ is small, so that we are implicitly dealing with an interaction potential not so strong, the left hand side must consequently vary as $\delta_{l}/k$.
It therefore follows that, for low-energy incident electrons ($k$ $\sim$ 0) on a neutral molecule (the interaction being reasonably finite-ranged), the partial phase shift goes to zero roughly as $\delta_{l}$ $\sim$ $k^{2l+1}$: this means that, even excluding the partial wave contribution corresponding to $l$ $>$ 0, the s-wave ($l$ = 0) scattering $\textbf{can}$ be important.

\subsubsection[Zero energy scattering: $s$-wave capture, bound and virtual states]{Zero energy scattering: $s$-wave capture, bound and virtual states}

\noindent To gain a little deeper insight in the last statement, we can start from the radial equations describing the scattering as provided by the formalism of partial waves:

\begin{equation}
\frac{d^{2}}{dr^{2}} u_{\ell}(k,r) + \left[k^{2} - \frac{2m}{\hbar^{2}}V(r) -\frac{\ell\left(\ell+1\right)}{r^{2}} \right] u_{\ell}(k,r,) = 0.
\end{equation}

\noindent Focusing on the s-wave case, for very low-energies one obtains

\begin{equation}
\frac{d^{2}}{dr^{2}} u_{0}(k,r) - \frac{2m}{\hbar^{2}}V(r)u_{0}(k,r) = 0.
\end{equation}

\noindent In the region where $r$ >> $R_v$, $R_v$ representing the range of the potential interaction, the previous equation becomes

\begin{equation}
\frac{d^{2}}{dr^{2}}u_{0}(k,r) = 0
\end{equation}

\noindent therefore, literally admitting as solution a straight line

\begin{eqnarray}
u_{0}(k,r) = c\cdot\left(r-\alpha\right),\qquad r >> R_v
\end{eqnarray}

\noindent where

\begin{equation}
\alpha = \lim_{k\rightarrow 0^{+}} -\frac{\tan \delta_{0}(k)}{k}
\end{equation}

\noindent defines the scattering length \cite{intro-joachainbook}.
In what follows, the crucial point rests on the simple geometrical meaning for $\alpha$: the scattering length, in fact, is the intersection of the asymptote of $u_{0}(k,r)$ with the $r$ axis.
Neglecting for simplicity sake the case of a repulsive potential, it follows that the scattering length can assume either positive or negative values so that, depending on the attractive potential's depth (see figure \ref{f_06_chapter_1}), the sign of $\alpha$ is indicative of the presence($\alpha$ > 0)/absence($\alpha$ < 0) of a bound state.

\begin {figure}[here]
\begin {center}
\includegraphics[scale=0.34]{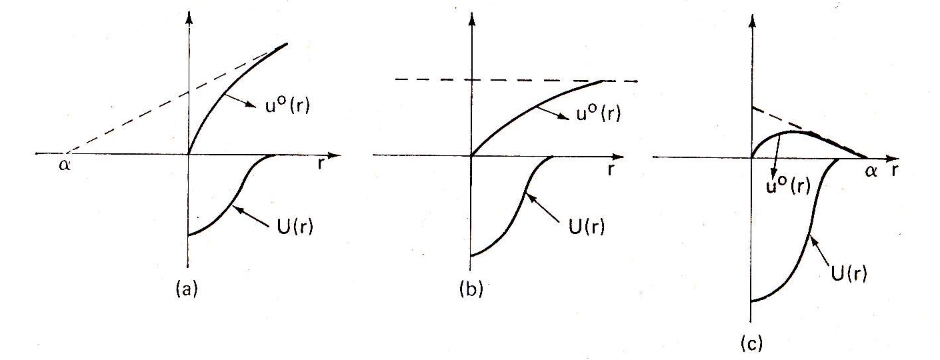}
\end {center}
\caption{\small{Behaviour of the scattering length for various attractive potentials. From left to right, as the interaction potential becomes more and more attractive, the panels illustrate three cases: (a) $\alpha$ < 0 for which a bound s-wave state cannot exist, (b) $\alpha$ = $\infty$ for which the potential is almost able to support a real bound s-wave state (virtual state) and (c) $\alpha$ > 0 correponding to an attractive interaction strong enough to support a bound s-wave state.}}
\label{f_06_chapter_1}
\end {figure}

\noindent Thus, once again, the dipole induced interaction (hence, the neutral molecular polarizability) plays a foundamental role in the sense that for vanishing energies it can yield an s-wave bound state: in other words, an s-wave attachment not necessarily produces a stable negative ion.
Keeping in mind that the conditio sine qua non in order to have a stable negative ion is represented by a positive electron affinity value, the above simple analysis also suggests that depending on the molecular polarizability, a molecule can provide not efficient s-wave attachment and, at the same time, efficient higher-order partial waves resonances that occur for energies higher than 0 eV but still astrophysically relevant.

\noindent As further shown by figure \ref{f_06_chapter_1}, it can also happen that $\alpha$ = $\infty$: this peculiar case corresponds, as we will discuss in detail about the coronene molecule, to a zero energy resonance (\footnote{also called virtual state}).
In such a case, on the basis of the geometrical meaning of $\alpha$, the interaction potential is nearly strong enough to support an s-wave bound state so that, depending on the nuclear motion, such a virtual state can subsequently evolve toward a bound state or not.

\noindent An observation to conclude the present section.

\noindent In the framework of the partial waves method, under the hypothesis according which $\delta_{0}(k)\;\sim\;$0 when $k\rightarrow 0^{+}$, it is possible to write down the following relations for the $\ell=$0 partial amplitude \cite{intro-joachainbook}:

\begin{equation}
\lim_{k\rightarrow0^+}\mathcal{F}_{\ell=0} = \lim_{k\rightarrow0^+}\frac{1}{k} e^{i\delta_{0}(k)}\sin \delta_{0}(k) = \frac{1}{k}\sin\delta_{0}(k)\sim|\alpha|,
\end{equation}

\noindent On the basis of the relation

\begin{equation}
\frac{d\sigma}{d\Omega} = \left|\mathcal{F}(\Omega)\right|^{2},
\end{equation}

\noindent which represents the link between theory and experiments, it follows that

\begin{equation}
\sigma_{tot} = \int_{0}^{4\pi}\frac{d\sigma}{d\Omega}d\Omega = 4\pi\sum_{\ell=0}^{\infty}\left|\mathcal{F}_{\ell}\right|^{2} \sim 4\pi\left|\mathcal{F}_{0}\right|^{2} \sim 4\pi\alpha^{2}
\end{equation}

\noindent Therefore, for a neutral molecular target without permanent dipole, a very large value of the integral cross section at very low energies, is indicative of the possible presence of a zero energy resonance.

\section[Negative ions in the interstellar medium]{Negative ions in the interstellar medium}

\subsection[General overview]{General overview}

\noindent With the expression 'molecular plasma', one implicitly refers to a (partially) ionized gaseous medium which contains rather equal amounts of positively and negatively charged atomic and molecular species as well as neutral molecules.
Nowadays we have learned that matter in Universe is mostly in the state of plasma, so that it is not so surprising to encounter also different kinds of molecular plasmas in the huge seemingly empty interstellar space of the known cosmos.
Between the typical examples of a molecular plasma, without going too much away from our little planet, one is provided by the ionosphere surrounding the Earth and some other planets/satellites (like Titan's upper atmosphere \cite{intro-vuitton}): in this portion of the atmosphere, both atoms and molecules are ionized mainly by the high-energy UV and X-ray photons which, in turn, are produced either from the Sun or from other intense radiation sources located outside the Solar System.
Another example of a region of the Universe which can be globally considered as a molecular plasma is then represented by a stellar atmosphere, which little more in detail constitutes a kind of high-temperature molecular plasma (thus usually characterized by a greater degree of ionization).
An interstellar cloud and a circumstellar envelope are regions of the interstellar medium (ISM) characterized by rather dense gaseous and solid matter and which usually contain a variety of different atoms and molecules.
Since the interstellar/circumstellar clouds often contain electrons and ions (ranging in different amounts and depending on the environmental astrophysical conditions), they also could be viewd as examples of immense molecular plasmas.
Thus, more in general, the whole interstellar space, seemingly almost empty, is filled with low and mainly very-low density matter, sometimes very tenuous and tipically with non-uniform mean local density, that often is in the form of a molecular plasma due to the ionization caused by the interstellar radiative field and by the cosmic rays (very high-energy electrons and protons).

\noindent For tens of years it was a common thought that negative ions and especially molecular anions were species which did not deserve so much attention due to their difficulty in surviving in an environment rich of ionizing radiation.
In other words, even though the cosmic-ray penetration as well as the ionization caused by the presence of UV photons, the latter produced by vicinal stars and acting at the outer edges of a given interstellar gaseous cloud, do produce an initial assortment of positive ions and free electrons, it had been concluded however that the subsequent formation processes of molecular negative ions through electron attachment were scarce, so that they should only constitute a negligible contribution to the ion-molecule interstellar synthetic processes.

\noindent Historically, however, the possible existence and role of negative ions in the interstellar medium had been the object of a few intriguing investigations \cite{intro-dalgarno, intro-sarre}.

\noindent The first who really suggested and revealed in advance that 'large' molecular negative ions could actually exist in dense interstellar clouds was E. Herbst \cite{intro-herbst}, which surmised that the large electron affinity of carbon chain molecules and hydrocarbon radicals with more than 4-5 atoms would have been responsible to yield rather high radiative electron attachment rate coefficients, thus resulting in anion-to-neutral ratios on the order of a few percent.
To be precise, in such a pioneering work he speculated that for specific linear carbonaceous radical species like C$_{4}$H, C$_{3}$N, C$_{5}$N, C$_{7}$N, C$_{9}$N and other oxigen-containing species \cite{intro-herbst}, with electron affinity (EA) values greater than 2 eV, the corresponding abundance ratios [$M^{-}$]/[$M$] could be high enough (0.01-0.1) so that the corresponding undissociated molecular anion $M^{-}$ might be detectable in such an astrophysical environment, provided the associated microwave spectral emission/absorption lines were determined by laboratory measurements.

\noindent The confirmation of this hypothesis was achieved many years later following the laboratory measurements of the rotational spectrum of the C$_{6}$H$^{-}$ molecular anion \cite{intro-mccarthy}.
The above measurements, in fact, from already exsisting astronomical spectra \cite{intro-kawaguchi}, allowed the verification of the presence of C$_{6}$H$^{-}$ in the molecular envelope of the C-rich evolved star IRC+10216, with an abundance of 1-5\% that of the neutral parent species.
The ensuing successfull astronomical search for the same molecular negative ion in TMC-1, the dense molecular cloud in Taurus 1, determined in this dense region an anion-to-neutral ratio of about 2.5 \% \cite{intro-mccarthy}.

\noindent Nowadays, we know that six different anions do exist in the ISM: C$_{6}$H$^{-}$, C$_{4}$H$^{-}$, C$_{8}$H$^{-}$, C$_{3}$N$^{-}$, C$_{5}$N$^{-}$ and CN$^{-}$, each of them being a carbonaceous polyyne-like linear-chain structure which includes periferic atoms like H or N.

\noindent The rotational spectra of C$_{4}$H$^{-}$, C$_{8}$H$^{-}$ and C$_{3}$N$^{-}$ have since been measured in laboratory \cite{intro-gupta, intro-thaddeus} with subsequent detections of C$_{4}$H$^{-}$ \cite{intro-cernicharo}, C$_{8}$H$^{-}$ \cite{intro-remijan} and C$_{3}$N$^{-}$ \cite{intro-thaddeus} in the molecular envelope of IRC+10216, where C$_{8}$H$^{-}$ was observed also in TMC-1 \cite{intro-brunken}.
Prompted by the detection of several carbon-chain molecules with large abundances in the protostar L1527 \cite{intro-sakai2}, the C$_{6}$H$^{-}$ anion was first searched for and then confirmed to be present also in this source \cite{intro-sakai}, in turn followed by the observation of C$_{4}$H$^{-}$ \cite{intro-agundez}.
A survey of galactic molecular sources by Gupta and coworkers \cite{intro-gupta2} detected once again C$_{6}$H$^{-}$ in two further sources, the pre-stellar cloud L1544 and the protostar L1521F, therefore suggesting the likely ubiquitousness of molecular anions in the interstellar/circumstellar medium and, in particular, of C$_{6}$H$^{-}$.
The next to last known anion, the C$_{5}$N$^{-}$, was observed by Cernicharo and coworkers \cite{intro-cernicharo2}, which have attributed a series of rotational lines observed in the envelope of IRC+10216 to this new anionic species.
Last but not least, \cite{intro-agundez2} the CN$^{-}$ ion was detected, with unexpectedly high abundances, in the molecular circumstellar envelope of IRC+10216.

\noindent On the other side, polycyclic aromatic hydrocarbon (PAH) molecules in different states of charge (including also anions) and hydrogenation, which are thought to be an important component in the ISM \cite{intro-bakes, intro-wakelam, intro-flower, intro-draine} since some of their spectroscopic features are observed in the infrared emission bands \cite{intro-leger, intro-leger2, intro-allamandola, intro-salama, intro-cesarsky, intro-smith}, are currently included in several astrophysical models even though no specific PAH responsible of the above feature has been unambiguously identified \cite{intro-mulas}.
Their role, generally speaking, is found when considering the charge balance, the chemical and the dynamical evolution (ambipolar diffusion, UV screening effect, general degree of ionization, presence of free electrons and their repercussions on both the attenuation and reflection of radio-waves) within several regions of the ISM.
In this connection, currently there are few experiments and/or theoretical investigations that focuse on their behaviour under low-energy electron collisions, and despite the fact that some consolidated hypothesis exist that point toward their actual presence in the ISM \cite{intro-tielensbook, intro-allamandolarew}, together with some good agreement of a few astrochemical models involving PAH anions \cite{intro-bakes, intro-wakelam}, they are not constrained enough to be explicitly (i.e., as individual species) considered in astrochemical models: beyond the unresolved problem about their sizes and abundances as functions of the global physical conditions of a given astrophysical region, no realistic ab-initio estimates of their electron capture rate coefficients are currently available.

\noindent As previously mentioned, Herbst \cite{intro-herbst} surmised that, subjected to specific structural constraints and specific conditions inherent their electron affinities, specific molecular negative ions could be efficiently synthesized, in dense interstellar clouds, by rapid radiative electron attachment processes

\begin{equation}
M + e^{-} \longrightarrow M^{-} + h\nu 
\end {equation}

\noindent where '$M$' refers to the neutral molecular species.
During the previous years, the importance of radiative association processes for the interstellar synthesis of molecules began to be rapidly explored with enthusiasm \cite{intro-herbst2, intro-herbst3, intro-herbst4, intro-herbst5}, and a similar approach was then followed in order to provide reasonable estimates for the formation rate coefficients of molecular anions associated with appropriate neutral species \cite{intro-herbst4, intro-herbst5, intro-herbst, intro-petrie, intro-osamura}.
The first step of the original theory, as illustrated with full particulars in \cite{intro-herbst4}, is to assume that such a process does occur in two distinct stages, where the free electron $e^{-}$ and the neutral molecular target $M$ form a long-lived metastable complex ($M^{-}$)$^{*}$ which in turn either dissociates back into the reactants themselves or is stabilized by the emission of a photon, as reported by the following reaction scheme
\begin{eqnarray}
 M + e^{-}  &\stackrel{\scriptsize{k_{f}}}{\longrightarrow}& (M^{-})^{*} \\
(M^{-})^{*} &\stackrel{\scriptsize{k_{b}}}{\longrightarrow}&  M + e^{-}  \\
\\ \nonumber
(M^{-})^{*} &\stackrel{\scriptsize{k_{rs}}}{\longrightarrow}&  M^{-} + h\nu
\end{eqnarray}

\noindent The symbols $k_{f}$ (cm$^{3}$/sec), $k_{b}$ (sec$^{-1}$) and $k_{rs}$ (sec$^{-1}$) refer to the rate coefficient for the formation of the metastable species (or transient negative ion, TNI), the rate coefficient for the backward process by means of which the TNI get destroyed by the ejection of the extra electron (autodetachment) and the rate coefficient associated with the radiative stabilization, respectively.
The overall rate coefficient for the whole radiative association process, $k_{Rad}$, is thus given by the following relation

\begin{equation}
k_{Rad} = \frac{k_{f} k_{rs}}{(k_{b} + k_{rs})}
\end{equation}

\noindent if the metastable species is assumed to be at steady state \cite{intro-herbst4, intro-herbst}.
In the customary limit according which $k_{b} >> k_{rs}$, the last expression reduces to

\begin{equation}
k_{Rad} = \frac{k_{f}}{k_{b}} k_{rs};
\end{equation}

\noindent in the unlikely event that $k_{rs} > k_{b}$, $k_{Rad}$ 'saturates' at $k_{f}$, so that in general the value of $k_{Rad}$ cannot exceed the rate coefficient for temporary electron capture $k_{f}$ \cite{intro-herbst5, intro-herbst}.
Supposing the thermal equilibrium, the capture and auto-ionization rate coefficients ($k_{f}$ and $k_{b}$, respectively) must be related by the detailed balance principle, according which the probabilities for the forward process and for its inverse are equal.
Following this assumption, and assuming that no vibrational excitation on the neutral species 'M' need to be considered, it has then been possible to achieve to an analitical approximated expression to get the rate coefficient for the whole negative ion formation process, as reported in eq. \ref{Krad-herbst}

\begin{eqnarray}
k_{Rad} = \frac{10^{3}h^{3}(KT)^{-1/2}}{(2\pi\mu_{e})^{3/2}}\;\cdot\;\frac{g_{M^{-}}}{2g_{M}}\;\cdot\;n_{v}(EA)
\label{Krad-herbst}
\end{eqnarray}

\noindent where $k_{Rad}$ is expressed in $cm^{3}\;sec^{-1}$, $h$ is the well known Planck's constant, $K$ the Boltzmann's constant, $T$ the temperature, $g_{M^{-}}$ and $g_{M}$ are the electronic degeneracy factors of the neutral and anionic species ($M$ and $M^{-}$ respectively), $\mu_{e}$ refers to the reduced collisional mass and $n_{v}(EA)$ means the vibrational density of states of the negative species $M^{-}$ at an internal energy equal to the electron affinity of the neutral.
As clearly shown by the above formula, the rate of attachment strongly depends on the density of vibrational states of the anion and more in general, even if not directly deducible from eq. \ref{Krad-herbst}, in the framework of the phase-space theory where the angular momentum is conserved, one assumes that once the metastable negative ion, the TNI, is formed by electron attachment, all the vibrational states in all accessible electronic states \cite{intro-herbst4, intro-osamura} can be rapidly formed with restrictions inherent only the electronic spin, therefore intuitively suggesting the importance of internal conversion.
Following this approach, when considering only s-wave electrons ($^{2}$S), one easily account for the production of the $^{1}\Sigma$ and $^{3}\Sigma$ states as the possible products associated to a neutral open-shell molecule $^{2}\Sigma$, the same multiplicity ripartition being valid for a $^{2}\Pi$ neutral open-shell state which can thus produce $^{1}\Pi$ and $^{3}\Pi$.

\noindent On the basis of direct detections (when available), or equivalently, following the results provided by several models where negative ions are surmised to play an important role in different regions of the ISM, one could say that including anions in an astrophysical model means to also realistically (as much as possible) account for their initial formation: in this sense, a pure s-wave attachment coupled with the only radiative stabilization path to account for the ensuing dissipation of the extra energy content carried by the colliding electron might be little reductive.
When looking at the resonances as doorways to form negative ions, in fact, our well consolidated ab-initio approach enables us to take in consideration also the non-spherical angular contribution, therefore going beyond the s-wave attachment and allowing us to provide further useful information.
In line with that, once the TNI is formed at energies that of course must agree with the environmental conditions for the astrophysical region under investigation, we find also crucial to investigate with a captivating and not so much computationally demanding approach the TNI evolution: this will be described in sec. 1.4.3.
Generally speaking, beyond the fact that the ab-initio realistic estimate that we provide for the metastable anion formation rate coefficient $k_{f}(T)$ (see sec. 1.4.2) already constitutes a very important parameter requested by astrophysicists, our results can be helpful in predicting which are the likely products of the electron collision given a specific astrophysically relevant neutral molecule; at the same time, the comparison with the available experiments enables us to gain a deeper insight in the nanoscopic mechanisms that are foundamental in additionally suggesting which are the chemical species that might be introduced in an astrochemical model.

\subsection[The metastable anion formation rate coefficient, $k_{f}(T)$]{The metastable anion formation rate coefficient, $k_{f}(T)$}

\noindent From a purely theoretical point of view, the rate coefficient for reactive processes between two colliding species is defined by $\langle \sigma_{n}(v) v \rangle$, i.e. by the thermal average of the product of the state dependent reaction cross section $\sigma_{n}(v)$, $n$ denoting the internal states of the species involved in the binary collision, times the relative collision velocity, $v$.
The above definition can be also applied in the case of metastable anion formation processes by electron resonant capture as well as for dissociative electron attachment processes, where therefore the reactant species are the free incoming electron and the neutral molecular target \cite{intro-hotop}.

\noindent By assuming that the electrons, having kinetic temperature T$_{e}$, and the target molecules described by their internal temperature T$_{M}$ are in thermal equilibrium corresponding to the temperature T = T$_{e}$ = T$_{M}$, it therefore follows that the general expression for the metastable anion formation rate coefficient $k_{f}$ is given by

\begin{eqnarray}
k_{f}(T) &=& \frac{1}{Q_{v}Q_{r}} \sum_{n_{\nu j}} e^{-\frac{\epsilon_{n_{\nu j}}}{K_{B}T}} \langle \sigma_{n}(v) v \rangle \\ \nonumber
\\ \nonumber
         &=& \frac{1}{Q_{v}Q_{r}} \sum_{n_{\nu j}} e^{-\frac{\epsilon_{n_{\nu j}}}{K_{B}T}} \int v\; \sigma_{n}(v)\; f(v,T)\; dv
\label{kf}
\end{eqnarray}

\noindent where $\sigma_{n}(v)$, as a function of the relative electron velocity, is the integral cross section (ICS) for the electron-molecule collision for a given initial vibrorotational state $n$, $\epsilon_{n}$ is the vibrorotational energy of the neutral target molecule (from the bottom of the potential), $Q_{v}Q_{r}$ is the product of the vibrational and rotational partition functions and $f(v,T)$ is the Maxwell-Boltzmann velocity distribution function

\begin{equation}
f(v,T) = 4\pi \left( \frac{\mu_{e}}{2\pi K_{B} T} \right)^{3/2}\; v^{2}\; e^{-\frac{\mu_{e}\; v^{2}}{2\; K_{B}\; T}}.
\end{equation}

\noindent Note that the electron velocity is usually much higher than the molecular traslational velocity so that the relative collision velocity $v$ is simply given by the electron velocity; in line with that, it also follows that the reduced mass could be approximated to the electronic mass, $\mu_{e}$.
However, to be precise, when the electrons and the molecules are described by two different mean kinetic temperatures, a convolution of two Maxwell-Boltzmann functions could be also use, as indicated in \cite{intro-flowerbook}.
According to our theoretical approach to calculate the elastic electron-molecule ICS (see chapter 2), we will assume the target molecule with its nuclei fixed at the equilibrium geometry and, as our integral cross sections are rotationally summed, once the variable of integration is changed to the collision energy, the above equation (eq. \ref{kf}) becomes:

\begin{equation}
k_{f}^{FN}(T) = \left( \frac{8 K_{b} T}{\pi \mu_{e}}  \right)^{1/2}\; \frac{1}{\left( K_{B}T \right)^{2}} \int E\; \sigma(E)\; e^{-\frac{E}{K_{B}T}}\; dE
\end{equation}

\noindent This is the expression which will be used for the estimate of $k_{f}(T)$ in the present work.

\noindent Moreover, since in a partially-ionized plasma like an interstellar cloud or a circumstellar envelope the electron transport is primarily governed by momentum-transfer cross sections (MTCSs) for elastic scattering of the present electrons, we find important to also provide realistic estimates of the elastic differential cross sections (DCSs) for electrons scattered off gas-phase molecules: despite the direct and useful physical information given by the DCSs, in fact, their knowledge enables us to estimate the MTCS which is defined as

\begin{equation}
\sigma_{m}(E) = 2\pi \int_{0}^{\pi} \left( 1 - \cos \theta \right) \frac{d\sigma}{d\Omega}(E) \sin \theta\; d\theta
\end{equation}

\noindent It therefore follows that eq. \ref{kf} expressing the metastable formation rate coefficient can be employed with our computed elastic (rotationally summed) ICS as well as with our computed MTCS, thus providing realistic upper and lower limits for the real $k_{f}$ value.

\subsection[The dynamical evolution of the metastable anion: pseudo-1D approach. Investigating the IVR occurrence]{The dynamical evolution of the metastable anion: pseudo-1D approach. Investigating the IVR occurrence}

\noindent Once the metastable negative species is formed, in order to follow the corresponding post-attachment evolution, one should need in principle a very detailed and computationally demanding treatment based on complex kinetic and/or statistical method \cite{intro-troe, intro-robinson}.
We instead use a much more simplified treatment, but well consolidated \cite{intro-carelli0, intro-carelli2, intro-carellicf2}, which is $qualitatively$ indicative of the possible occurrence of the intramolecular vibrational redistribution (IVR) toward the stabilization of the full-size undissociated negative ion or to its fragmentation.

\noindent The IVR, i.e. the process by which the vibrational energy initially localized in a particular mode is redistributed among all the vibrational modes of a given molecule, is an ubiquitous and foundamental process which has strong influence in the kinetics and in the final outcome of many chemical reactions, ranging from the unimolecular decomposition to protein folding.
From the theoretical point of view, since the molecular vibrational modes cannot be assumed perfectly harmonic, they in principle can interact so that the unimolecular reaction under investigation does occur when the superposition of these modes causes the reaction coordinate, on the corresponding potential energy surface, to reach the top of the reaction barrier.
In this connection, one should note that in the case of a metastable anionic state (with positive EA), the complex transient species essentially needs to reduce its internal energy content, since the required energy to reach the top of such a barrier is provided by the colliding electron.
Following the Fermi's golden rule, the IVR rate, given the initially excited state, strictly depends on the root mean squared value of the coupling strength $\langle V^{2}\rangle$ as well as on the density of vibrational states, $\rho$

\begin{equation}
\Gamma = \frac{2\pi}{\hbar}\; \langle V_{v,v'}^{2}(E_{res}) \rangle\; \rho(E_{res}),
\end{equation}

\noindent $E_{res}$ being the real resonance energy component.
While predicting the density of bath states to a sufficiently good approximation can be currently accomplished \cite{intro-romanini}, even a rough estimate of the average coupling strength requires in principle a very accurate knowledge of the involved (complex, for metastable anions) potential surface and thus extensive calculations that account for presence/absence of intermediates transient states \cite{intro-marcus}, large amplitude motion \cite{intro-bethardy} as well as extreme motion states \cite{intro-gambogi}, which can change rather unpredictably the expected size of couplings by several orders of magnitude.

\noindent When considering the electron attachment process where the excess energy brought in the neutral target by the resonant electron gets transferred into the internal energy of the molecular bonds, one is interested in gathering if the breakupof the target with the ensuing formation of stable anionic fragments takes place or, conversely, if the creation of a long-lived full-size (meta)stable anion with a little residual energy occurs, the latter being in turn a good candidate for the subsequent radiative stabilization by which the stable anion is eventually produced in the highly diluted, practically collisionless, ISM.
In this framework, we note that Herbst and coworkers \cite{intro-petrie, intro-osamura} assume an efficient attachment of s-wave electrons at zero energy, calculating the radiative electron attachment rates for linear molecules like CN$_{3}$ \cite{intro-petrie} and H$_{2}$C$_{n}$/HC$_{n}$H with n = 4, 6, 8 \cite{intro-osamura}.
However, higher energy astrophysically relevant resonances still need to efficiently dissipate their excess energy before the autodetachment occurs, so that it seems reasonable to consider as a very likely possibility the energy transfer within the molecular network as the molecular vibrational modes couple with the metastable electron.
It therefore follows that, according to a simplified pseudo monodimensional (pseudo-1D) picture for the electron attachment mechanism, one is searching for a possible $dominant$ fragmentation path which finally occur via the beakup of that particular bond and the stabilization of a molecular anion.
As will be discussed with full particulars in connection with the ortho-benzyne and dicyanogen TNIs (see sections 3.1.6 and 3.3.3, respectively), we carried out several calculations using different molecular structures for which the resonant electron density maps showed, at the initial molecular equilibrium geometry, the presence of nodal planes and of partial localization along specific bonds of the resonant electron density.
The ensuing pseudo-1D potential energy curves, as cuts over the real potential energy surface associated with both the neutral N-electron target and the metastable (N+1)-electron TNI, follow a selected stretching deformation and provide the energy balance as 

\begin{equation}
E_{N+1}^{Tot} = E_{res}(\textbf{R}) + E_{N}(\textbf{R}) - E_{N}(\textbf{R}_{eq})
\label{Etotn+1}
\end{equation}

\noindent where $E_{N}$ is the computed electronic energy (at the same level of accuracy of the scattering calculations at the equilibrium geometry) for the neutral molecule at a given deformed geometry identified by $\textbf{R}$, and $E_{res}(\textbf{R})$ is the real part of the computed resonant electron (complex) energy over the same range of molecular geometries.
The corresponding widths, $\Gamma_{res}(\textbf{R})$, of the resonances associated with the (N+1)-electrons states are also given at each computed molecular geometry via the well known relation \cite{intro-taylorbook}

\begin{equation}
E_{res}^{complex}(\textbf{R}) = E_{res}(\textbf{R}) + i\Gamma_{res}(\textbf{R}).
\label{Erescomplex}
\end{equation}

\clearpage

\chapter[The theoretical method]{The theoretical method}
\label{theo}

\section[Total and effective Hamiltonian]{Total and effective Hamiltonian}

\noindent In order to take up the theoretical treatment of any many-body system from the quantum mechanical point of view, as the corresponding Schroedinger equation is in principle responsible to describe all the system properties, the first step consists in writing down the associated interaction Hamiltonian in a suitable form so that the eigenfuctions for the states of interest can be either numerically or explicitly determined.
In the electron-molecule scattering process the particles participating to the collisional event are the incident electron, the electron of the molecule (n$_{e}$) and the molecular nuclei (N$_{n}$).
Therefore the Hamiltonian of the system in principle shall depend on the 3+3n$_{e}$+3N$_{n}$ degrees of freedom.
However, with some approximation it becomes possible to separate the motion of the molecular bound electrons from that of the other particles; in such an approach, thus, the wavefunction determination is then reduced to the determination of a wavefunction with 3N$_{n}$+3 degrees of freedom, thereby reducing the complexity of the scattering problem.
At the same time, in principle, such a procedure allows the nuclear dynamics during the collision event to be also expressly included.

\noindent The total electron-molecule interaction Hamiltonian expressed in atomic units is

\begin{equation}
H_{TOT} = - \frac{\triangle_{\textbf{r}_{e}}}{2} + H_{mol}(\textbf{r},\textbf{R}) + V_{int}^{eM}(\textbf{r}_{e},\textbf{r},\textbf{R})
\end{equation}

\noindent with

\begin{equation}
V_{int}^{eM}(\textbf{r}_{e},\textbf{r},\textbf{R}) = +\sum_{i=1}^{n_{e}} \frac{1}{|\textbf{r}_{e} - \textbf{r}_{i}|} - \sum_{j=1}^{N_{n}} \frac{Z_{j}}{|\textbf{r}_{e} - \textbf{R}_{j}|}
\end{equation}

\noindent and

\begin{eqnarray}
H_{mol} & = & -\sum_{i=1}^{n_{e}} \frac{\triangle_{\textbf{r}_{i}}}{2} -\sum_{j=1}^{N_{n}} \frac{\triangle_{\textbf{R}_{j}}}{M_{j}} -\sum_{i=1}^{n_{e}}\sum_{j=1}^{N_{n}} \frac{Z_{j}}{|\textbf{r}_{i} - \textbf{R}_{j}|} + \nonumber \\ 
\nonumber \\
        & + & \sum_{i=1}^{n_{e}}\sum_{j=i+1}^{n_{e}} \frac{1}{|\textbf{r}_{i} - \textbf{r}_{j}|} + \sum_{j=1}^{N_{n}}\sum_{k=j+1}^{N_{n}} \frac{Z_{j}Z_{k}}{|\textbf{R}_{j} - \textbf{R}_{k}|} \nonumber \\
\nonumber \\
        & = & \hat{T}_{N} + \hat{T}_{el} + V_{eN} + V_{ee} + V_{NN}
\nonumber \\
        & = & \hat{T}_{N} + \hat{T}_{el} + \mathcal{V}_{int}^{mol}
\nonumber \\
        & = & \hat{T}_{N} + \mathcal{H}_{mol}^{el}.
\end{eqnarray}

\noindent In the above equations, $\textbf{r}$ represents the positions of the molecular electrons, $\textbf{r}_{e}$ is the position of the colliding electronic particle and $\textbf{R}$ refers to the positions of the nuclei.
For the sake of simplicity, all the interaction terms are written in atomic unit.
It is also possible to consider that the previous equations are written in the molecular center of mass which, on the basis of the huge difference between the free colliding electron weight and the molecular weight, can be qualitatively seen as coincident with the total system (molecule + incident electron) center of mass: in this connection we can reasonably neglect the spatial motion of the total system, focusing instead on its internal dynamical evolution.
Moreover, in $H_{mol}$ the spin-orbit interaction has been neglected as well as other interactions involving the electronic and nuclear spin (like the spin-spin and the hyperfine structure term).

\noindent If we now denote with $\phi(\textbf{r},\sigma,\textbf{R})$ the wavefunction of the electronic and spin parts (the label $\sigma$ refers collectively to the molecular electron spins) of the molecule with eigenvalue $E_{n}(\textbf{R})$ within the Born-Oppenheimer (BO) approximation, one can write the eigenfunctions of the total electron-molecule stationary Hamiltonian (eq. 1) as

\begin{equation}
\Psi(\textbf{r}_{e},\textbf{r},\sigma,\textbf{R}) = \sum_{\phi_{n}}|\phi_{n}(\textbf{r},\sigma,\textbf{R})\rangle \psi_{n}(\textbf{r}_{e},\textbf{R})
\end{equation}

\noindent The above close-coupled expansion is exact, and also includes the continuum scattering states of the system ($\psi_{n}$).
For electronically elastic collisions, it is further possible to neglect in eq. 4 the contributions coming from the excited states ($\phi_{n}$, n$\neq$0) of the molecule, so that

\begin{equation}
\Psi(\textbf{r}_{e},\textbf{r},\sigma,\textbf{R}) \approx |\phi_{0}(\textbf{r},\sigma,\textbf{R})\rangle \psi_{0}(\textbf{r}_{e},\textbf{R}).
\end{equation}

\noindent When writing the molecular Hamiltonian $H_{mol}$ as in eq.3, then the eigenvalues equation $H_{TOT}\Psi$=$E\Psi$ with the total Hamiltonian of eq.1 applied to eq.5 will give

\begin{eqnarray}
\left[ - \frac{\triangle_{\textbf{r}_{e}}}{2} + \hat{T}_{N} + \mathcal{H}_{mol}^{el} + V_{int}^{eM}(\textbf{r}_{e},\textbf{r},\textbf{R}) \right] (|\phi_{0}(\textbf{r},\sigma,\textbf{R})\rangle \psi_{0}(\textbf{r}_{e},\textbf{R})) = E_{TOT} |\phi_{0}\rangle \psi_{0} \nonumber \\
\nonumber \\
|\phi_{0}\rangle \left(-\frac{\triangle_{\textbf{r}_{e}}}{2} \right)\psi_{0} +\hat{T}_{N}\left(|\phi_{0}\rangle \psi_{0} \right) + \mathcal{E}_{0}^{el}(\textbf{R})|\phi_{0}\rangle \psi_{0} + V_{int}^{eM}|\phi_{0}\rangle \psi_{0} - E_{TOT}|\phi_{0}\rangle \psi_{0} = 0 \nonumber \\
\nonumber 
\end{eqnarray}

\noindent where $\mathcal{E}_{0}^{el}(\textbf{R})$ refers to the ground molecular electronic state energy within the BO approximation.
Multiplying now from the left for the ground electronic molecular function $\langle \phi_{0}|$ and integrating with respect to the position variables of the molecular electrons (as indicated by the label $r$) one obtains

\begin{equation}
\left[-\frac{\triangle_{\textbf{r}_{e}}}{2} + \langle \phi_{0}|\hat{T}_{N}|\phi_{0}\rangle_{r} + \hat{T}_{N} -\sum_{i=1}^{N_{n}}\langle\phi_{0}|\nabla_{\textbf{r}}|\phi_{0}\rangle_{r} \cdot \nabla_{\textbf{r}} - \frac{k^{2}}{2} + \langle \phi_{0}|V_{int}^{eM}|\phi_{0}\rangle \right]\psi_{0} = 0
\end{equation}

\noindent having esplicitated the free electron kinetic energy as $k^{2}/2$ = ($E_{TOT}$ - $\mathcal{E}_{0}^{el}(\textbf{R})$).
This is the complete integro-differential scattering equation, where also the nuclear dynamics is susceptible to 'feel' the incoming electron: since the molecular ground electronic state BO wavefunction is given by $\phi_{0}(\textbf{r},\sigma,\textbf{R})$ = $\varphi_{0}^{el}(\textbf{r},\sigma) \chi_{J,\nu}(\textbf{R})$, $\chi_{J,\nu}(\textbf{R})$ being the nuclear wavefunction, the second and the fourth term on the left hand side implicitly contains the adiabatic corrections to the BO potential surface.
It obviously follows that using the complete close-coupled expansion as in eq.5, the same terms will contains also the non-adiabatic corrections.
\noindent One should note at this point that the BO approximation, previously introduced to treat indipendently the electronic and the nuclear molecular degrees of freedom, is the most common and natural one which is encountered in any collisional problem. It follows that, when treating the scattering problem including the nuclear dynamics inside the scattering equation, as in eq.6, the BO electronic energy $\mathcal{E}_{0}^{el}(\textbf{R})$ does indeed provide the effective potential energy for the nuclear motion.

\noindent Furthermore, when making the approximation of eq.5, for the scattering wavefunction $\Psi$ the most important terms which have been neglected are 

\begin{eqnarray}
\langle \phi_{0}| V_{int}^{eM}(\textbf{r}_{e},\textbf{r},\textbf{R})|\phi_{n}\rangle \nonumber.
\end{eqnarray}

\noindent They represent what is classically described as the polarisation interaction.
Hence the eq.6 takes in consideration the projectile interaction with the molecular electronic cloud in its unperturbed ground electronic state, and does not consider the deformation induced by the approaching free electron.
So, identifying the $\langle \phi_{0}|V_{int}^{eM}|\phi_{0}\rangle$ term with the static interaction, the neglected terms consequently represent just the polarisation interaction

\begin{eqnarray}
V_{stat}^{eM} = \langle \phi_{0}| V_{int}^{eM}(\textbf{r}_{e},\textbf{r},\textbf{R}) |\phi_{0} \rangle_{r} = V_{int}^{eM,00} \nonumber \\
V_{pol}^{eM} = \sum_{\phi_{n}}\langle \phi_{0}|V_{int}^{eM}(\textbf{r}_{e},\textbf{r},\textbf{R})|\phi_{n}\rangle = \sum_{n}V_{int}^{eM,0n}
\end{eqnarray}

\noindent As shall be discussed in detail when treating the single center expansion method, this interaction can be brought back into the eq.6 using a modellistic form; for the same reason, it is foundamental to point out now that although we are dealing with fermions (the $n_{e}$ molecular electrons and the colliding particle), we have not suitably antisymmetrized the expansions in eq.4 and/or eq.5.
Such a physical constraint cause the complete scattering equation (eq.6) to contain also non-local potential terms whose meaning is thus justified on the basis of the Pauli principle; on the other side, such constraint would cause the same scattering equation to be rather impossible to be computationally resolved: for the same reason as above, also the exchange interaction contribution will be introduced using a suitable optical potential (which will be described later).
In other words, if in the Schoredinger equation with the total Hamiltonian given by eq.1 we would expand the total wave function in terms of all possible target states (rotational, vibrational and electronic), and then the antisymmetrizing operator would be applied to ensure the validity of the Pauli principle, the ensuing coupled integro differential equations will automatically include all types of acting forces without any approximation.
However, on the other hand, such an expansion would be not entirely feasible since in this way the equations will contain non-local and often complex terms that would be either very complicated to handle numerically or too much demanding computationally.
Therefore, the way we follow to bypass this obstacle consists in resorting to semiempirical method thus using a DFT approach for the short range forces and a local (energy-dependent) model optical potential for the exchange interactions.
For the purpose of the present section, in conclusion, if one writes

\begin{equation}
\hat{V}_{int}^{eM}(\textbf{r}_{e},\textbf{R}) = V_{stat}^{eM}(\textbf{r}_{e},\textbf{R}) + V_{pol}^{eM}(\textbf{r}_{e},\textbf{R}) + \hat{W}_{ex}^{eM}(\textbf{r}_{e},\textbf{r},\textbf{R})
\end{equation}

\noindent then the complete 'exact' scattering equation including the adiabatic corrections takes the following form:

\begin{equation}
\left[-\frac{\triangle_{\textbf{r}_{e}}}{2} + \langle \phi_{0}|\hat{T}_{N}|\phi_{0}\rangle_{r} + \hat{T}_{N} -\sum_{i=1}^{N_{n}}\langle\phi_{0}|\nabla_{\textbf{r}}|\phi_{0}\rangle_{r} \cdot \nabla_{\textbf{r}} - \frac{k^{2}}{2} + \hat{V}_{int}^{eM}(\textbf{r}_{e},\textbf{R})  \right]\psi_{0} = 0
\end{equation}


\section[Fixed nuclei (FN) approximation.]{Fixed nuclei approximation.}

\noindent The most simple approximation that is possible to use for the full Hamiltonian of eq.9 is the one where all the nuclear motions of the target molecule are neglected, so that the total FN Hamiltonian appears in the following simplified form:

\begin{equation}
H_{TOT}^{FN} = \left[-\frac{\triangle_{\textbf{r}_{e}}}{2} + \hat{V}_{int}^{eM}(\textbf{r}_{e},\textbf{R})  \right]
\end{equation}

\noindent Thus the scattering equation is given by

\begin{equation}
\left[-\frac{\triangle_{\textbf{r}_{e}}}{2} - \frac{k^{2}}{2} + \hat{V}_{int}^{eM}(\textbf{r}_{e},\textbf{R})  \right]\psi_{0}^{FN} = 0
\end{equation}

\noindent However, this approximation is good only when the energy of the projectile is high enough that in the spatial region in which there is interaction, the electronic projectile spends effectively little time so that the molecule does not manage to move during that time.
In other words, such an approximation should be considered strictly valid for direct and non-resonant scattering.
In this sense, then, the FN approach is expected to be reliable, provided that the kinetic energy of the incident electron is appreciably larger than the molecular rotational energy differences of interest.
On the other side, when not excited, complex and 'heavy' polyatomic molecules are considered as target, the latter thus being characterized by relatively slow nuclear (vibrorotational) motions, due to the problem complexity, it still makes sense to describe the collision event within the framework of a FN description.
Moreover, for very low-energy incident electrons (E < 1$\sim$1.5 eV), the colliding particle usually has not the kinetic energy needed to deeply penetrate the molecular electronic cloud, so that its coupling with the molecular nuclear dynamics can be viewed as partially negligible, as we will in fact argue for the electron-coronene case (see chapter 3, section 3.2, 'The C$_{24}$H$_{12}$ molecule').
However, to be precise, for very low incident energy, the collision time $\tau_{c}$ may exceed the vibrational period $\tau_{\nu}$ but it might still be such that $\tau_{c}\sim\tau_{r}$, $\tau_{r}$ being the molecular rotational characteristic period.
Since the above condition might be considered as satisfied also at thermal energy (for low-lying rotational states in fact the energy spacing can be of the order of 10$^{-2}\;\sim\;$10$^{-3}$ eV), one should thus expect that rotationally inelastic collisions play a role in the scattering process; more in general, therefore, one should expect that vibrationally (or vibrorotationally) inelastic collisions play a role in scattering and also in resonant processes involving large molecules, where this actually takes place but expecially in the differential cross sections (DCS) behaviour: according to the previous qualitative argumentations, we thus simply surmise that for the purposes of the present investigation such inelastic contributions, even though (very) likely, are not so important, as then confirmed by our findings for the coronene molecule (see chapter 3, sec. 3.2.4 and 3.2.5).

\noindent When considering polar molecular target, the FN approximation also fails.
The detailed description for the above breakdown would deserve more complex and detailed analysis being thus beyond the purpose of the present section as well as of the whole PhD dissertation.
It will however be considered again, without entering the complex theoretical details, when the conclusions for the ortho-benzyne molecule shall be summarized (see chapter 3, sec. 3.1.9).
For the moment it is sufficient to emphasize that, once the $T$ matrix is introduced to calculate the relevant scattering cross sections, it shows a very slow convergence in the partial waves number.
It is possible to glean a reliable justification for this, using a simple and qualitative physical picture.
High $l$ partial waves are mainly 'restricted' to the long-range spatial region due to their intense centrifugal barriers so that their phase accumulation (\footnote{the phase shifts $\delta_{l}$, see chapter 1}) is essentially due to the contribution associated to the electron-permanent dipole interaction.
The latter statement, in fact, is easily confirmed when keeping in mind that polar molecules are characterized by the presence of a permanent dipole that has with the incoming electron a long-range interaction ($V_{int}^{dipole}\sim\;1/r^{2}$) stronger than those given either by the static quadrupole ($V_{int}^{quadrupole}\sim\;1/r^{3}$) or associated with both the isotropic and anisotropic polarizabilities (i.e. the induced dipole: $V_{int}^{\alpha}\sim\;1/r^{4}$).
In conclusion, when the dipole is not allowed to rotate, it is possible to show that the associated DCS diverges in the forward direction, causing then the ICS to be also divergent, even though logaritmically.
This means that, as discussed in section 3.1.3 in connection with our findings for the ortho-benzyne molecule, the larger partial waves number the more reliable the ICS values are.

\section[Single determinant expansion. The SE, SECP and SMECP approximations]{Single determinant expansion. The SE, SECP and SMECP approximations.}

\noindent The purpose of the previous sections was to derive the total hamiltonian for the electron-molecule scattering problem and consequently to introduce the FN approximation thus separating the nuclear and the electronic dynamics during the collision event.
Now we move to separate the scattered electron coordinates from the molecular bound electrons: this passage is complicated by the fact that the $n_{e}+1$ electrons involved in the present case are undistinguishable fermions so that they are affected by specific restrictions in their wavefunction which, asaccording to the Pauli principle, have to be suitably antisymmetrized with respect to the exchange of each couple of electrons.
The final aim is to write down the scattering equations when the static-exchange (SE), the static-exchange-correlation-polarization (SECP) and the static-model-exchange-correlation-polarization (SME CP) approximations are made, respectively.
This will enables us to illustrate in section 2.4.6 the meaning of the adiabatic-static-model-exchange-correlation-plarization (ASMECP) approaches which have been also used to obtain some of the findings about the molecules involved in the present work.

\noindent For simplicity sake, we now omit the esplicit dependance of the scattering wavefunction from the nuclei position (\footnote{having assumed the FN approximation to be valid, it is implicit for what follows that the scattering wavefunction depends parametrically on the nuclear geometry $\textbf{R}$}) so that, introducing the antisymmetrization operator $\hat{\mathcal{A}}$, eq.4 can be rewritten as follows

\begin{equation}
\Psi(\textbf{r}_{e},\textbf{r}) = \hat{\mathcal{A}}\sum_{n}|\phi_{n}(\textbf{r},\sigma)\rangle \psi_{n}(\textbf{r}_{e})
\end{equation}

\noindent It is now possible to make the following distinction, depending on the number of terms which in the previous summation are retained: when only the first one is considered (so that only the molecular electronic ground state is taken in consideration for the collision process) we obtain the SE approximation, otherwise the SEP.
In what follows the associated scattering equation will be derived, at the same time introducing the second important constraint that is crucial for our theoretical model: the single determinant expansion.

\noindent For electronically elastic collisions, only the first term in the expansion provided by eq. 2.12 can be considered.
When the molecular electronic ground state wavefunction $\phi_{0}^{FN,el}$ is expressed at the Hartree-Fock level as a single Slater determinant, one can write

\begin{equation}
\Psi(\textbf{r}_{e},\textbf{r}) = \hat{\mathcal{A}}||\varphi_{1}(1),\varphi_{2}(2), ... ,\varphi_{n_{e}}(n_{e})||\psi_{0}(n_{e}+1)
\end{equation}

\noindent where the symbol $||...||$ refers to the Slater determinant, having thus indicated the $i$-th one-electron spin-orbital as $\varphi_{i}(i)$.
It then follows that the esplicit form for the $\hat{\mathcal{A}}$ operator is given by

\begin{equation}
\hat{\mathcal{A}} = \frac{1}{\sqrt{(n_{e}+1)!}}\sum_{p=1}^{n_{e}+1}(-1)^{p}\hat{\mathcal{P}},
\end{equation}

\noindent $\hat{\mathcal{P}}$ being the permutation operator acting on each possible couple of electrons.
The FN-SE equation for the scattered electron wavefunction $\psi_{0}(n_{e})$ is then obtained by projecting the stationary FN scattering equation $H_{TOT}^{FN}\Psi^{FN}=E_{TOT}^{FN}\Psi^{FN}$, containing the FN Hamiltonian (eq.11) and the SE expansion (eq.13) for $\Psi^{FN}$, onto the $bra$ $\langle \;\;||\varphi_{1}(1)\varphi_{2}(2) ... \varphi_{n_{e}}(n_{e}) || \;\;|$ associated to the closed shell electronic molecular ground state, so that the successive integration on the $n_{e}$ bound molecular electronic coordinates provides (in a.u.)

\begin{eqnarray}
\left[-\frac{1}{2}\nabla^{2}_{\textbf{r}_{e}} -\frac{1}{2}k^{2}_{e} + V_{stat}(\textbf{r}_{e},\textbf{R}) \right]\psi_{0}(\textbf{r}_{e}) = \nonumber \\
= \sum_{j=1}^{n_{e}}\int\;d\textbf{r}'\varphi_{j}^{*}(\textbf{r}')\frac{1}{|\textbf{r}_{e} - \textbf{r}'|}\psi_{0}(\textbf{r}')\cdot\varphi_{j}(\textbf{r}_{e})
\end{eqnarray}

\noindent with

\begin{equation}
V_{stat}(\textbf{r}_{e},\textbf{R}) = \sum_{j=1}^{n_{e}}\int\;d\textbf{r}'\varphi_{j}^{*}(\textbf{r}')\frac{1}{|\textbf{r}_{e} - \textbf{r}'|}\varphi_{j}(\textbf{r}') - \sum_{k=1}^{N_{n}} \frac{Z_{k}}{|\textbf{r}_{e} - \textbf{R}_{k}|}
\end{equation}

\noindent and

\begin{eqnarray}
& k_{e}^{2} = \left|2\left(E_{TOT}^{FN,SE} - \epsilon_{0}(\textbf{R}) \right)\right| & \nonumber \\
& H_{mol}^{BO}||\varphi_{1}(1)\varphi_{2}(2) ... \varphi_{n_{e}}(n_{e})|| & = \epsilon_{0}(\textbf{R})||\varphi_{1}(1)\varphi_{2}(2) ... \varphi_{n_{e}}(n_{e})||
\end{eqnarray}

\noindent On the basis of the adopted approximations, it is easy to recognize that the SE scattering equation does contain only the static (eq. 2.16) and the exchange interaction (given by the sum on the r.h.s of eq. 2.15).
The SECP approximation level is then obtained from he previous one, but now, besides the static and the exchange interactions, one takes in consideration also the dynamical effects given by the molecular response to the colliding electron by introducing an optical phenomenological local potential $V_{cp}(\textbf{r}_{e},\textbf{r})$ to include the molecular polarization; consequently, the scattering equations take the following form

\begin{eqnarray}
\left[-\frac{1}{2}\nabla^{2}_{\textbf{r}_{e}} -\frac{1}{2}k^{2}_{e} + V_{stat}(\textbf{r}_{e},\textbf{R}) + V_{cp}(\textbf{r}_{e},\textbf{r}) \right] \psi_{0}(\textbf{r}_{e}) = \nonumber \\
= \sum_{j=1}^{n_{e}}\int\;d\textbf{r}'\varphi_{j}^{*}(\textbf{r}')\frac{1}{|\textbf{r}_{e} - \textbf{r}'|}\psi_{0}(\textbf{r}')\cdot\varphi_{j}(\textbf{r}_{e})
\end{eqnarray}

\noindent To conclude the present section, we note that the previous are clearly non-local equations; it is then possible to further simplify them by replacing the sum of non-local terms on the r.h.s. with another local energy-dependent potential $W_{ex}(\textbf{r};\varepsilon_{coll},$ $ I_{p})$.
The latter will be described little more in detail within the framework of the single center expansion (see sec. 2.4.4).
The resulting SMECP equations finally represent our local differential scattering equations:

\begin{eqnarray}
\left[-\frac{1}{2}\nabla^{2}_{\textbf{r}_{e}} -\frac{1}{2}k^{2}_{e} + V_{stat}(\textbf{r}_{e},\textbf{R}) + V_{cp}(\textbf{r}_{e},\textbf{r}) + W_{ex}(\textbf{r};\varepsilon_{coll}, I_{p}) \right] \psi_{0}(\textbf{r}_{e}) = 0 \nonumber \\
\end{eqnarray}

\section[The Single Center Expansion]{The Single Center Expansion}

\noindent In the present section, we shall illustrate in some detail the guidelines of our method by means of which we computationally solve the scattering equations describing the elastic resonant collision process; the method is based on the general single center expansion (SCE) approach adapted for electron- polyatomic molecules elastic scattering.
Generally speaking, however, the scattering processes which occur in the collision of electrons with polyatomic targets have been object of intensive theoretical studies in the past years, so that in order to solve the final scattering equation several methods do indeed exist.
As previously mentioned, our method is based on the coordinate representation of the interaction forces, so that a little detailed description for the local model exchange as well as for the static and both the short-range (correlation) and long-range (polarization) interaction terms will be illustrated.

\subsection[The SCE bound state wavefunction]{The SCE bound state wavefunction}

\noindent The first step in the SCE method consists in obtaining the bound state wavefunction that will be later included into the scattering equations.
One therefore needs to interface with an initial quantum chemistry code (like Gaussian, \cite{th-frisch}) that is thus employed to generate the Single Determinant description (at the Hartree-Fock level) of the target electronic wavefunction.
As mentioned above, in this way we obtain the closed shell electronic wavefunction for the neutral target as a single Slater determinant of one-electron bound spinorbitals, indicated as $\varphi_{i}(\textbf{r};\textbf{R})$ with $i=1:n_{e}$, or equivalently of $n_{e}/2$ two-electrons multicenter MOs labelled as $\hat{\varphi}_{i}(\textbf{r}_{\zeta};\textbf{R})$ with $i=1:n_{e}/2$ and $\zeta=1:2$.
Each of the above functions is in turn susceptible to be expressed as a truncated linear sovrapposition of optimized known functions $g_{k}(\textbf{r};\textbf{R})$ (Gaussian Type Orbitals, GTOs), depending on the choosen basis set, according to

\begin{equation}
\varphi_{i}(\textbf{r}_{\zeta};\textbf{R}) = \sum_{\lambda=1}^{N_{n}} \sum_{k=1}^{\hat{k}_{max}} \sum_{\nu=1}^{\hat{\nu}_{max}} d_{\nu}^{\lambda k} C_{i}^{\lambda k}(\textbf{R}_{\lambda}) g_{\nu}^{\lambda k}(\textbf{r}_{\zeta};\textbf{R}_{\lambda})
\end{equation}

\noindent where $i$ refers to the $i$-th multicenter molecular wavefunction, and the $k$-th primitive gaussian function for the $\lambda$-th atomic center is

\begin{equation}
g_{\nu}^{\lambda,k}(\textbf{r}_{\zeta};\textbf{R}_{\lambda}) = \mathcal{N}_{k}(a,b,c;\gamma)(r_{ix}-R_{\lambda x})^{a}(r_{iy}-R_{\lambda y})^{b}(r_{iz}-R_{\lambda z})^{c}e^{-\gamma\left|\textbf{r}_{i} - \textbf{R}_{\lambda} \right|^{2}},
\end{equation}

\noindent $\mathcal{N}(a,b,c;\gamma)$ being the normalization coefficient

\begin{equation}
\mathcal{N}(a,b,c;\gamma) = \left[\left( \frac{2\gamma}{\pi} \right)^{3/2} \frac{(4\gamma)^{a+b+c}}{(2a-1)!!(2b-1)!!(2c-1)!!} \right]^{1/2}
\end{equation}

\noindent The $C_{i}^{\lambda k}$ is the GTO coefficient of the $\nu$-th GTO at a given molecular geometry as concisely expressed by the argument ($\textbf{R}_{\lambda}$).
The index $\nu$ labels each primitive gaussian function within the subgroup that belongs to a given contraction coefficint $d_{\nu}^{\lambda,k}$, so that the contracted gaussian function $G^{\lambda,k}$ is

\begin{equation}
G^{\lambda k} = \sum_{\nu=1}^{\hat{\nu}_{max}} d_{\nu}^{\lambda k} g_{\nu}^{\lambda k}(\textbf{r}_{\zeta};\textbf{R}_{\lambda})
\end{equation}

\noindent The single center expansion method can be now introduced.
Since we want to generate the full electron-molecule interaction potential as a function of the electronic density $\rho(\textbf{r})$ of the target molecule in a suitable way, the crucial step consists therefore in expanding both the bound electron functions and the continuum scattered function around the center of mass 

\begin{eqnarray}
\varphi_{i}^{p\mu}(\textbf{r};\textbf{R}) & = & \frac{1}{r}\sum_{\ell h} u_{\ell h}^{p\mu, i}(r) X_{\ell h}^{p\mu, i}(\theta,\phi) \\
\psi^{p\mu}(\textbf{r}_{e};\textbf{R}) & = & \frac{1}{r}\sum_{\ell h} f_{\ell h}^{p\mu}(r) X_{\ell h}^{p\mu}(\theta,\phi)
\end{eqnarray}

\noindent In the above formulas, $i$ labels a specific mutlicenter MO, which contributes to the bound electrons density $\rho(\textbf{r})$ in the polyatomic target, while the $p\mu$ indices refer to the $\mu$-th component of the $p$-th relevant irreducible representation (IR), respectively.
The $h$ index labels a specific basis, for a given angular momentum $\ell$ value, for the $p$-th IR that is under consideration.

\noindent At this point, a crucial passage which fully deserves to be pointed out consists in the construction of the symmetry-adapted, generalized harmonics indicated as $X_{\ell h}^{p\mu}(\theta,\phi)$.
Ror a given $\ell$, the associated 2$\ell$ + 1 spherical harmonics $Y_{\ell}^{m}(\theta,\phi)$ form a basis for a (2$\ell$ + 1)-dimensional IR of the full rotational group, and at the same time they constitute the basis of a reducible representation of the molecular point group to which the considered molecule in the choosen frozen geometry belongs: the molecular point group is in fact a subgroup of the full rotational group.
Therefore, the generalized harmonics $X_{\ell h}^{p\mu}(\theta,\phi)$ can be defined as basis for IRs of the molecular point group so that they can be reasonably expanded as linear combinations of spherical harmonics

\begin{equation}
X_{\ell h}^{p\mu}(\theta,\phi) = \sum_{m} b_{\ell h m}^{p\mu} \; Y_{\ell}^{m}(\theta,\phi)
\end{equation}

\noindent where the $b_{\ell h m}^{p\mu}$ coefficients, choosen so that the $X$ functions transform as the $\mu$-th component of the $p$-th IR of the molecular point group considered, are described and tabulated in \cite{th-altman}.
Moreover, since the $X_{\ell h}^{p\mu}(\theta,\phi)$ functions satisfy the following orthonormality relations

\begin{equation}
\int X_{\ell h'}^{p'\mu' *}(\theta,\phi) X_{\ell h}^{p\mu}(\theta,\phi)\;\sin\theta\;d\theta\;d\phi = \delta_{pp'}\delta_{\mu\mu'}\delta_{hh'}
\end{equation}

\noindent then the $b_{\ell h m}^{p\mu}$ coefficients constitute a unitary transformation between the $X$s and the $Y$s, in the sense that $\forall \ell$ it can be shown

\begin{eqnarray}
\sum_{m}b_{\ell h' m}^{p'\mu' *} b_{\ell h m}^{p\mu} = \delta_{pp'}\delta_{\mu\mu'}\delta_{hh'} \\
\sum_{p,\mu,h}b_{\ell h m'}^{p\mu *} b_{\ell h m''}^{p\mu} = \delta_{m'm''}
\end{eqnarray}

\noindent Thus, writing the multicenter electronic wavefunction which describes the target neutral molecule as an antisymmetrized product of multicenter MOs $\varphi_{i}$ as in eq. 2.20, it is then possible to generate each of the radial coefficients $u_{\ell h}^{p\mu, i}(r)$ for the bound molecular electrons by the following numerical quadrature

\begin{eqnarray}
u_{\ell h}^{p\mu, i}(r) = \sum_{\lambda=1}^{N_{n}} \sum_{k=1}^{\hat{k}_{max}} \sum_{\nu=1}^{\hat{\nu}_{max}} \sum_{m=-\ell}^{+\ell} \int_{0}^{\pi} \int_{0}^{2\pi} & \sin(\theta) \; d\theta & \nonumber \\
b_{\ell hm}^{p\mu, i} \; Y_{\ell}^{m}(\theta,\phi) r & C_{i}^{\lambda k}(\textbf{R}_{\lambda}) & d_{\nu}^{\lambda k} \; g_{\nu}^{\lambda k}(\textbf{r};\textbf{R}_{\lambda}) \; d\phi 
\end{eqnarray}

\subsection[The SCE one-electron density]{The SCE one-electron density}

\noindent Having obtained the radial coefficients $u_{\ell h}^{\mu}(r)$ from eq. 2.30, each bound one-electron wavefunction 

$$\varphi_{i}(\textbf{r}) = \frac{1}{r}\sum_{\ell h}u_{\ell h}^{p\mu, i}(r) X_{\ell h}^{p\mu}(\theta,\phi)$$

\noindent is then expanded around the center of mass in terms of $X_{\ell h}^{p\mu}$ functions so that the one-electron density function $\rho(\textbf{r})$ can be written as usual in the following form

\begin{eqnarray}
\rho(\textbf{r}) & = & \int \left|\;\; det||\varphi_{1}(\textbf{r})\varphi_{2}(\textbf{x}_{2})\varphi_{3}(\textbf{x}_{3})...\varphi_{n_{e}}(\textbf{x}_{n_{e}}) ||\;\; \right|^{2} dx_{2}\;dx_{3}\;...dx_{n_{e}} \nonumber \\ 
& = & 2\cdot\sum_{i=1}^{n_{e}}\left| \varphi_{i}(\textbf{r})\right|^{2}
\end{eqnarray}

\noindent where the numerical factor 2 is due to the sum over the spin, and the sum is over each $i$-th doubly occupied orbital.
At this point it should be noticed that the $\rho(\textbf{r})$ function belongs to the total-symmetric ($A_{1}$) IR.
Therefore, once the quantity $\rho(\textbf{r})$ is obtained as above explained, it can be expanded in terms of symmetry-adapted total-symmetric general harmonics:

\begin{equation}
\rho(\textbf{r}) = \frac{1}{r} \sum_{\ell h (m)} \rho_{\ell h (m)}(r) X_{\ell h (m)}^{A_{1}}(\theta,\phi)
\end{equation}

\noindent where 

\begin{equation}
\rho_{\ell h} = 2\cdot\sum_{i=1}^{n_{e}}\int_{0}^{\pi}\;\sin(\theta)\;d\theta\;\int_{0}^{2\pi}\;d\phi\;\varphi_{i}(\textbf{r})\cdot\varphi_{i}(\textbf{r})
\end{equation}

\noindent The knowledge of these coefficients, in principle, enable us to numerically estimate all the interaction terms between the incoming electron and the target molecule.
Let us see how.

\subsection[The SCE static potential]{The SCE static potential}

\noindent The static potential exerted by the molecular electrons and nuclei on the surrounding molecular volume is given by the sum of two contributions

\begin{eqnarray}
V_{stat}(\textbf{r}, \{\textbf{R}_{j}\}) &=& \int\;d\textbf{s}\frac{\rho(\textbf{s})}{\left|\textbf{r} - \textbf{s}\right|} - \sum_{j=1}^{N_{n}}\frac{Z_{j}}{\left|\textbf{r} - \textbf{R}_{j} \right|}\\ \nonumber
                     &=& V_{el}(\textbf{r}) + V_{nuc}(\textbf{r}, \{\textbf{R}_{j}\})
\end{eqnarray}

\noindent the first of which is the repulsive electron-electron contribution, and the second, with the sum over the number of atomic centers $N_{n}$, refers to the nuclear attractive effect on the colliding electron.
The substitution of eq. 2.32 and 2.33 in the above formula (eq. 2.34), after using the following generalized form of the well known spherical harmonics sum theorem,

\begin{equation}
\frac{1}{| \textbf{r} - \textbf{s}|} = \sum_{p\mu}\sum_{\ell h} \frac{4\pi}{2\ell +1} \gamma_{\ell}(r,s) X_{\ell h}^{p\mu *}(\theta_{r}, \phi_{r}) X_{\ell h}^{p\mu}(\theta_{s}, \phi_{s})
\end{equation}

\noindent with

\begin{equation}
\gamma_{\ell}(r,s) = \frac{r_{<}^{\ell}}{r_{>}^{\ell +1}},
\end{equation}

\noindent enables us to obtain the electronic contribution to the static potential expressed as

\begin{equation}
V_{el}(\textbf{r}, \{\textbf{R}_{j}\}) = \sum_{\ell h} V_{el, \ell h}(r) X_{\ell h}^{A_{1}}(\theta_{r}, \phi_{r}).
\end{equation}

\noindent where the potential radial expansion coefficients $V_{el, \ell h}(r)$ are obtained via radial integration

\begin{equation}
V_{el, \ell h}(r) = \frac{4\pi}{2\ell +1} \int\; ds\; \rho_{\ell h}\; \gamma_{\ell}(r,s)
\end{equation}

\noindent It therefore follows that the single center expansion of the nuclear static contribution $V_{nuc}(\textbf{r})$, using once again the above expressions 2.35 and 2.36, is given by

\begin{equation}
V_{nuc}(\textbf{r}, \{\textbf{R}_{j}\}) = \sum_{\ell h} V_{nuc, \ell h}(r, \{\textbf{R}_{j}\}) X_{\ell h}^{A_{1}}(\theta_{r}, \phi_{r})
\end{equation}

\noindent with the radial coefficients assume the expression

\begin{equation}
V_{nuc, \ell h}(r, \{\textbf{R}_{j}\}) = \frac{4\pi}{2\ell +1} \sum_{j=1}^{N_{n}}Z_{j}\gamma_{\ell}(r,R_{j})X_{\ell h}^{A_{1}}(\theta_{R_{j}},\phi_{R_{j}}).
\end{equation}

\subsection[The exchange interaction: the Free Electron Gas Exchange (FEGE) Model Potential]{The exchange interaction: the Free Electron Gas Exchange (FEGE) Model Potential}

\noindent To circumvent the difficulty of having to handle and solve integro-differential radial scattering equations, the necessary exchange effects can be suitably modelled through the use of simpler, energy dependent potential forms which can be added to the exact static potential discussed in the previous section within the framework of the SCE approach.
One of the most widely used for polyatomic molecular targets, and that is also used for all the scattering calculations of the present thesis, is the Free-Electron-Gas-Exchange (FEGE) introduced many years ago by Hara \cite{th-hara}.
The actual form of this exchange potential derives from two approximations in the integral exchange terms of the exact-exchange scattering equations (see eq. 2.15, 2.18 and 2.19 in sec. 2.3).
First, the molecular electrons are treated as a free-electron gas, with a given charge density $\rho(\textbf{r}, \textbf{R})$, determined for particular electronic state of the target under investigation (initially the ground electronic state).
Second, the distorsion of the continuum function is neglected so that the colliding electron is treated as a plane undeformed wave.
The resulting FEGE potential is then given by the following form:

\begin{equation}
V_{FEGE}(\textbf{r}, \{\textbf{R}_{j}\}) = -\frac{2}{\pi} K_{F}(\textbf{r}, \{ \textbf{R}_{j} \}) \left[ \frac{1}{2} + \frac{1 - \eta^{2}}{4\eta} ln \left| \frac{1+\eta}{1-\eta} \right| \right]
\end{equation}

\noindent where the wavevector up to the top of the Fermi surface is given by the usual free-electron relation

\begin{equation}
K_{F}(\textbf{r}, \{ \textbf{R}_{j} \} ) = \left[3\pi^{2}\rho(\textbf{r}, \{\textbf{R}_{j} \}) \right]^{1/3}
\end{equation}

\noindent and $\eta$ holds the ratio between the actual wavevector $k$ for the scattered electron and the one at the top of the Fermi surface for the electron gas, $\eta(\textbf{r})$ = $k(\textbf{r})$/$K_{F}(\textbf{r}, \{\textbf{R}_{j} \})$, where the local momentum is given by

\begin{equation}
k(\textbf{r}) = \left[2\left(\varepsilon + I_{p} \right) + K_{F}^{2}(\textbf{r}, \{\textbf{R}_{j} \})  \right]^{1/2}.
\end{equation}

\noindent The collision energy $\varepsilon$ is in turn given by the initial-channel energy and $I_{p}$ refers to the first ionization potential for the neutral target molecule.
The final FEGE exchange is then obtained by means of the integration scheme already outlined and used in the above subsections for the bound state wavefunction and for the electron density:

\begin{equation}
V_{FEGE}(\textbf{r}, \{\textbf{R}_{j}\}) = \sum_{\ell h}V_{FEGE, \ell h}(r)X_{\ell h}^{A_{1}}.
\end{equation}

\subsection[The correlation-polarization (V$_{cp}$) potential]{The correlation-polarization (V$_{cp}$) potential}

\noindent Another important type of interaction between the impinging electron and the target molecule comes from the description of the response function of the target to the incoming charger projectile.
Due to the fact that this interaction depends on the distance of the incoming electron in the sense that it varys from a purely polarisation effect at large distances up to including exchange-correlation interaction with the bound electrons whithin the molecular volume, the model correlation-polarization potential $V_{cp}$ that is implemented in our SCE treatment of polyatomic molecules makes a distinction between a long-range region and a short-range of interaction.
Within the former, the perturbative polarisation effects are dominant, the potential is correctly susceptible of a local analitic form and is adiabatic on the incident electron's velocity; conversely, within the latter, both non-adiabatic and non-local effects play an important role.
It therefore follows that one can write the overall correlation-polarisation interaction as

\begin{eqnarray}
V_{cp}(\textbf{r}, \{\textbf{R}_{j}\}) &=& V_{corr}(\textbf{r}, \{\textbf{R}_{j}\}),\qquad r\leq r_c \\ \nonumber
                                       &=& V_{pol}(\textbf{r}, \{\textbf{R}_{j}\}),\qquad r > r_c
\end{eqnarray}

\noindent where the connecting spatial factors $r_c$ are usually obtained from the crossing radii of the lower coefficients in the SCE expansion

\begin{equation}
V_{cp}(\textbf{r}, \{\textbf{R}_{j}\}) = \sum_{\ell h} V_{cp, \ell h}(r) X_{\ell h}^{A_{1}}
\end{equation}

\subsubsection{The short-range correlation term}

\noindent The short-range contribution, $V_{corr}$, rest on a DFT treatment a full description of which has been given elsewhere \cite{th-fag-VcorrDFT}, so that here we only briefly outline its leading features.
As the relevant quantities reported so far, the $V_{corr}$ potential must be expanded over symmetry adapted spherical harmonics

\begin{equation}
V_{corr}(\textbf{r}, \{\textbf{R}_{j}\}) = \sum_{\ell h} V_{corr, \ell h}(r) X_{\ell h}^{p\mu}(\theta, \phi) = \frac{\delta}{\delta\rho}\left[E_{c}(\rho(\textbf{r}, \{\textbf{R}_{j}\}))\right]
\end{equation}

\noindent where $E_c$ is the correlation energy which depends on the one-electron density $\rho(\textbf{r})$.
Apart from the one-electron density, it can be shown tha the $V_{corr}(\textbf{r})$ potential also depends on the first and second functional derivatives of $\rho(\textbf{r})$: accordingly, unlike the other potentials so far described, the $V_{corr}(\textbf{r})$ requires a further single center expansion of the bound state wavefunction.

\subsubsection{The long-range polarization term}

\noindent Within the outermost asymptotic region we use a long-range potential which, depending mainly on the static electrical properties of the neutral molecular target (i.e., dipole and higher moments and associated polarizabilities) is susceptible of the following general form for a given molecular geometry $\{\textbf{R}_{j}\}$:

\begin{equation}
V_{pol}^{A_{1}}(\textbf{r}, \{\textbf{R}_{j}\}) = - \lim_{r\rightarrow +\infty} \sum_{\ell = 1} \frac{\alpha_{\ell}}{2r^{2\ell +2}}
\end{equation}

\noindent Let us first focus on the case of a permanent dipole moment and of its associated polarizability $\alpha$.
When only the symmetrical spherical polarizability is known (often referred to as the average dipole polarizability $\overline{\alpha}_{av}$ = 1/3$(\alpha_{xx} + \alpha_{yy} + \alpha_{zz})$) it is customary to assume that $\overline{\alpha}_{av}$ = $\alpha_{xx}$ = $\alpha_{yy}$ = $\alpha_{zz}$ while $\alpha_{xy}$ = $\alpha_{xz}$ = $\alpha_{yz}$ = 0.
In this case the polarizability tensor is written as

\begin{equation}
\begin{array}{ccc}
\overline{\alpha}_{av} & 0 & 0 \\
0 & \overline{\alpha}_{av} & 0 \\
0 & 0 & \overline{\alpha}_{av}
\end{array}
\end{equation}

\noindent On the other hand, if all of the six terms of $\alpha$ are known, then the polarizability tensor becomes

\begin{equation}
\begin{array}{ccc}
\alpha_{xx} & \frac{1}{2}\alpha_{xy} & \frac{1}{2}\alpha_{xz} \\
\frac{1}{2}\alpha_{xy} & \alpha_{yy} & \frac{1}{2}\alpha_{yz} \\
\frac{1}{2}\alpha_{xz} & \frac{1}{2}\alpha_{yz} & \alpha_{zz}
\end{array}
\end{equation}

\noindent A common case is when only the $\alpha_{0}$ and $\alpha_{2}$ polarisabilities are known: in this case the polarisation potential, for a linear molecule, has the simple form

\begin{equation}
V_{pol}(\textbf{r}, \{\textbf{R}_{j}\}) = - \frac{\alpha_0}{2r^4} - \frac{\alpha_2}{4r^4}[3\cos^{2}(\theta) -1]
\end{equation}

\noindent where the angular dependence is provided by the Legendre polynomial $P_{2}[\cos(\theta)]$, and $\alpha_{xx}$ = $\alpha_{yy}$ = $(\alpha_{0} - 1/2\alpha_{2})$ while $\alpha_{zz}$ = $(\alpha_{0} + \alpha_{2})$, the other off-diagonal terms of the polarizability tensor being zero.
Besides this special case, one has to take into account all the components of the dipole static polarizability so that the long range polarisation potential expressed in atomic units assumes the form

\begin{equation}
V_{pol}(x,y,z) = - \frac{1}{2r^6} \sum_{i=1}^{3} \sum_{j=1}^{3} q_{i} q_{j} \alpha_{ij}
\end{equation}

\noindent where $q_{i,j}$ = x, y, z for $i,j$ = 1, 2, 3, respectively.
The last equation is that one implemented with the options to select all the different polarizabilities reported above, as a function of the available data.
Once the $V_{pol}(x,y,z)$ is calculated and expanded over the symmetry adapted spherical harmonics belonging to the $A_{1}$ irreducible representation

\begin{equation}
V_{pol}(\textbf{r}, \{\textbf{R}_{j}\}) = \sum_{\ell h} V_{pol, \ell h}(r)X_{\ell h}^{A_{1}}(\theta, \phi)
\end{equation}

\noindent the matching point $r_c$ is found selecting the matching direction along the line connecting the molecular center of mass and a given atom or, equivalently, along the line connecting the center of mass to a generic point with cartesian coordinates x, y, z provided by the user.

\subsection[The ASMECP level of approximation]{The ASMECP level of approximation}

\noindent In order to examine the mechanism and the qualitative characteristics of a possible low-energy, one-electron resonance as well as its evolution along the lower part of the molecular potential energy curve according to our pseudo-1D model for the resonance energy redistribution during the temporary anionlifetime, we need a model which is simple enough to be computationally attractive and which at the same time includes sufficient details of the full scattering problem to reproduce the essential features of the physics involved. 
Thus, we look at the low-energy resonances also by using a simple, purely local model potential thatwe have called the adiabatic static model-exchange correlation-polarization (ASMECP) potential \cite{th-lucchese}.
We start by noting that the standard, symmetry adapted angular momentum eigenstates, $X_{\ell h}^{p\mu}$, do not form the most compact angular set for the e-molecule scattering problem: an alternative basis expansion is provided, in fact, by the angular eigenfunctions obtained by diagonalizing the angular Hamiltonian at each radius r. 
These distance-dependent angular eigenstates are referred to as the adiabatic angular functions $Z_{k}^{p\mu}(\theta, \phi, r)$: at each radial value, they are linear combinations of the symmetry-adapted 'asymptotic' harmonics discussed before

\begin{equation}
Z_{k}^{p\mu}(\theta, \phi, r) = \sum_{\ell h}X_{\ell h}^{p\mu}(\theta, \phi) C_{\ell h,k}(r)
\end{equation}

\noindent where the expansion coefficients are solutions of the matrix eigenvalue equation

\begin{equation}
\sum_{\ell h} \left( V_{\ell' h', \ell h}^{p\mu}(r) + \delta_{\ell \ell'} \delta_{hh'} \frac{\ell(\ell+1)}{2r^{2}} \right) C_{\ell h,k}(r) = V_{k}^{p\mu, \textbf{A}}(r) C_{\ell'h',k}(r)
\label{eq-asmecp}
\end{equation}

\noindent The eigenstates $V_{k}^{p\mu, \textbf{A}}(r)$ now form an adiabatic radial potential for each index k over the selected range of the electron-molecule distances (for the meaning of $V_{\ell' h', \ell h}^{p\mu}(r)$ see eq. 2.58 in the next section).
The spatial extent of the resonantwave function can be determined from the well and angular momentum barrier of such adiabatic potential terms and the physical mechanism for the resonance is that of a trapped electron tunneling through the potential barrier.
In order to avoid the nonadiabatic coupling terms between adiabatic curves, we actually employ a piecewise diabatic (PD) representation of the potential whereby the radial coordinate is divided into a number of regions so that the $i$-th sector is defined as $r_{i-1}<r<r_{i+1}$, with $r_{0}$ = 0.
In each radial region we average the coupling potential $V_{\ell'h',\ell h}(r)$ over r and the resulting averaged potential is diagonalized as in eq. \ref{eq-asmecp} to yield a set of angular functions $Z_{k,i}^{p\mu}(\theta, \phi)$.
Then, in the $i$-th region the scattering potential is transformed into the new representation in which it is nearly diagonal. 
The resulting equations are solved using the full scattering potential in each region with the further approximation of ignoring the off-diagonal couplings in that region: to solve the radial equations using the PD approach requires matching of the radial functions and their derivatives at the boundary between radial regions. 
The transformation of the radial functions from one region to the next is given by the transformation matrix $\mathbb{U}_{k,k'}^{(i+1\leftarrow i)}$ defined by

\begin{equation}
\mathbb{U}_{k,k'}^{(i+1\leftarrow i)} = \sum_{\ell h} C_{\ell h,k}^{(i+1)}C_{\ell h,k'}^{i}.
\end{equation}

\noindent When the size of the angular momentum eigenfunction basis used is larger than the size of the diabatic angular basis set, the transformation matrix $\mathbb{U}_{k,k'}$ is not in general unitary: we however accomplish the unitarization of $\mathbb{U}^{(i+1\leftarrow i)}_{k,k'}$ using simple Graham-Schmidt orthonormalization on the columns of that matrix.

\noindent It is now possible to conclude the present short but important section anticipating that the above (ASMECP) treatment, being easier in terms of computational cost, usually finds the resonances (located as maxima in the integral cross section, each associated to $\pi$-jump in the corresponding eigenphase) slightly shifted from those given by the full coupled-channel solutions \cite{th-lucchese} which will be derived and discussed with full particulars in the next section (see eq. 2.57). 
However, their nature and general features remain unchanged and of the same physical significance, while allowing for a reduction of the computational effort.

\subsection[The scattering equations; S and K matrices]{The scattering equations; S and K matrices}

\noindent To derive the equations describing the radial motion of the scattered electron, as previously introduced, we make two approximations.
Firstly, we assume that the fixed nuclei approximation, where the main symmetry molecular axis is taken to be stationary during the collision and pointed toward the incoming electron, can be made; secondly, we adopt a single centre expansion of both the bound molecular orbitals and the scattered electron orbitals about the centre of mass.
Having in the previous sections already discussed both of them, now we move to write down and to solve the relevant radial scattering equations the solutions of which enable us to numerically calculate the $\mathbb{K}$ matrix (and consequently both the $\mathbb{T}$ and $\mathbb{S}$ matrices) that in turn provides the scattering amplitudes and thus the scattering cross sections.

\noindent Once all the interaction potential terms ($V_{stat}$, $V_{FEGE}$, $V_{cp}$) as well as each of the N+1 one-electron wave functions (both the bound and the free particles) are recasted according ot the SCE method, their substitution into the eq. 2.19 followed by multiplying the whole equation by $X_{\ell' h'}^{p' \mu' *}$ and finally integrating all over the angular coordinates provides the following ensamble of coupled radial differential equations for the radial scattering function

\begin{equation}
\left[\frac{d^2}{dr^2} + k_{0(=\ell)}^{2} - \frac{\ell_i(\ell_i +1)}{r^2} -\mathcal{U}_{ii}^{p\mu} \right] f_{ii}^{p\mu}(r) = 2\sum_{j=1;j\ne i}^{n_c} \mathcal{U}_{ij}^{p\mu} f_{ij}^{p\mu}(r)
\end{equation}

\noindent where $i$ and $j$ (the same for $k$) label a couple of angular channels $\ell_i h_i, \ell_j h_j$ among the $n_c$ coupled angular channels, so that

\begin{eqnarray}
\mathcal{U}_{ij}^{p\mu}(r) = \langle X_{i}^{p\mu}\left| V_{stat}(\textbf{r}, \{\textbf{R}_{j}\}) + V_{FEGE}(\textbf{r}, \{\textbf{R}_{j}\}) + V_{cp}(\textbf{r}, \{\textbf{R}_{j}\}) \right| X_{j}^{p\mu}\rangle \nonumber
\end{eqnarray}

\begin{eqnarray}
= \int\; \sin\theta_{r}\;d\theta_{r}\;d\phi_{r}\:X_{i}^{p\mu *}(\theta_{r}, \phi_{r})\left[V_{stat} + V_{FEGE} + V_{cp} \right] X_{k}^{p\mu}(\theta_{r}, \phi_{r}) \nonumber
\end{eqnarray}

\begin{eqnarray}
= \sum_{k (=\ell_k h_k)} V_{k}^{TOT}(r) \int\;d\omega_{r} X_{i}^{p\mu *}(\theta_{r}, \phi_{r}) X_{k}^{p\mu}(\theta_{r}, \phi_{r}) X_{j}^{p\mu}(\theta_{r}, \phi_{r})
\end{eqnarray}

\noindent For $n_c$ coupled differential equations (for each state $p\mu$) there will be a set of $n_c$ independent solutions which satisfy the 'internal' boundary condition $f(0)$ = $0$: this is the reason according which we introduce the additional label $j$ by which we therefore write the solutions as $f_{ij}^{p\mu}$.
The $\mathbb{S}$ matrix is defined by looking for solutions with the asymptotic form \cite{th-joachainbook, th-taylorbook}

\begin{equation}
f_{ij}^{p\mu}(r) \approx e^{-i(kr-\frac{\pi}{2}\ell_{j})} \delta_{ij} - \mathbb{S}_{ij}^{p\mu} e^{+i(kr-\frac{\pi}{2}\ell_{j})}.
\end{equation}

\noindent We can form a general solution $F_{i}^{p\mu}$ which is a linear combination of these independent solutions

\begin{equation}
F_{i}^{p\mu} = \sum_{j=1}^{n_c} a_{j}^{p\mu}f_{ij}^{p\mu}(r)
\end{equation}

\noindent from which we can then construct the total function for the scattered electron

\begin{equation}
\mathcal{F}(\textbf{r}) = \sum_{i=1}^{n_c} \sum_{p\mu} \frac{1}{r_i} F_{i}^{p\mu}(r) X_{\ell_i h_i}^{p\mu}(\theta_{r}, \phi_{r})
\end{equation}

\noindent The $a_{i}^{p\mu}$ numerical coefficients in eq. 2.60 are choosen such that the outgoing ($r\rightarrow+\infty$) boundary condition is satisfied, namely

\begin{equation}
\mathcal{F}(\textbf{r}) \approx e^{i\textbf{k}_{0}\cdot\textbf{r}} + \tilde{f}_{FN} (\hat{\textbf{k}_{0}} \cdot \hat{\textbf{r}})   \frac{e^{ik_{0}r}}{r}
\end{equation}

\noindent where the $\tilde{f}_{FN}(\hat{\textbf{k}_{0}} \cdot \hat{\textbf{r}})$ is the FN scattering amplitude, and $\hat{\textbf{k}_{0}}$, $\hat{\textbf{r}}$ are the initial and final directions of the electronic projectile.
Taking advantage from the following expansion for the plane wave

\begin{equation}
e^{i\textbf{k} \cdot \textbf{r}} = 4\pi \sum_{\ell}\sum_{h, p\mu} i^{\ell} \frac{j_{\ell}(kr)}{kr} X_{\ell h}^{p\mu *}(\hat{\textbf{k}}) X_{\ell h}^{p\mu}(\hat{\textbf{r}}),
\end{equation}

\noindent obtained starting from the expression for a plane wave as a function of spherical Bessel functions and Legendre polynomials in which one introduce the spherical harmonics sum theorem coupled with the asymptotic form for the spherical Bessel functions $j_{l}(kr)$(\footnote{$j_{\ell}(kr) \sim \sin(kr - \ell \frac{\pi}{2})$ for $r\;\rightarrow\;+\infty$}), one can easily deduce the explicit form for the $a_{j}^{p\mu}$ coefficients.
In fact, by writing the asymptotic total scattering function $\mathcal{F}(\textbf{r})$ as

\begin{equation}
\mathcal{F}(\textbf{r}) \approx \mathcal{F}_{inc}(\textbf{r}) + \mathcal{F}_{scatt}(\textbf{r})
\end{equation}

\noindent one should first substitute into this expression the equations 2.59-2.61 as well as the asymptotic expansion 2.63 for $\mathcal{F}_{inc}(\textbf{r})$.
Now, even though $\mathcal{F}_{scatt}(\textbf{r})$ is not still known (as the scattering amplitude $\tilde{f}_{FN} (\hat{\textbf{k}_{0}} \cdot \hat{\textbf{r}})$, too), we know that $e^{-ikr}$ and $e^{ikr}$ are two linearly independent functions as well as $\mathcal{F}_{scatt}(\textbf{r})$ should not have any incoming wave component behaving like $e^{-ikr}$, so that by equating the coefficients of $e^{-ikr}$ on both sides of 2.64 it is possible to show that the expansion coefficients $_{j}^{p\mu}$ are given by

\begin{equation}
a_{j}^{p\mu} = (-1)i^{l_{j}}\frac{2\pi}{ik_{0}}X_{\ell_j h_j}^{p\mu}(\hat{\textbf{k}_{0}}).
\end{equation}

\noindent Now the scattering amplitude $\tilde{f}_{FN}$ can be obtained \cite{th-huo} by first substituting in eq. 2.62 all the equations 2.59-2.61 (\footnote{remember the factor $\frac{1}{r}$ as reported in eq. 2.25}) and then by taking the difference between the asymptotic total function $\mathcal{F}(\textbf{r})$ and the expansion 2.63 for the plane wave, so that, finally

\begin{equation}
\tilde{f}_{FN} (\hat{\textbf{k}_{0}} \cdot \hat{\textbf{r}}) = \sum_{ij} \sum_{p\mu} \frac{2\pi}{ik_{0}} i^{\ell_{i} - \ell_{j}} X_{i}^{p\mu}(\hat{\textbf{k}_{0}}) X_{j}^{p\mu} (\hat{\textbf{r}}) \left( \mathbb{S}_{ij}^{p\mu} - \delta_{ij} \right).
\end{equation}

\noindent As computations are usually done with real functions, we therefore introduce the $\mathbb{K}$ matrix which is defined from the following $real$ boundary conditions

\begin{equation}
f_{ij}^{p\mu}(r) = \sin\left(kr - \ell_i \frac{\pi}{2} \right)\delta_{ij} + \mathbb{K}_{ij}^{p\mu} \cos\left(kr - \ell_i \frac{\pi}{2}  \right),
\end{equation}

\noindent where the $\mathbb{S}$ and $\mathbb{K}$ matrices are related by the following transformation

\begin{equation}
\mathbb{S} = \frac{\mathbb{I} + i\mathbb{K}}{\mathbb{I} - i\mathbb{K}}
\end{equation}

\subsection[The cross sections in the molecular frame]{The cross sections in the molecular frame}

\noindent The integral elastic cross section, often improperly called the total elastic cross section, $\sigma$, describes the scattering from randomly oriented molecules.
It is obtained by the integration of $\left|\tilde{f}_{FN}(\hat{\textbf{k}_{0}} \cdot \hat{\textbf{r}})\right|^{2}$ over all the orientations $\hat{\textbf{r}}$ of the scattered electron and then averaging over all the orientations $\hat{\textbf{k}}_{0}$ of the impinging beam

\begin{equation}
\sigma = \frac{1}{4\pi} \int \; d\hat{\textbf{k}_{0}} \; d\hat{\textbf{r}}\;\left|\tilde{f}_{FN}(\hat{\textbf{k}_{0}} \cdot \hat{\textbf{r}})\right|^{2}
\end{equation}

\noindent which is easily shown to be equivalent to

\begin{equation}
\sigma = \frac{\pi}{k_{0}^{2}} \sum_{ij} \sum_{p\mu} \left|\mathbb{S}_{ij}^{p\mu} - \delta_{ij} \right|^{2}
\end{equation}

\noindent It is also instructive to express $\sigma$ in terms of eigenphases $\eta_{i}^{p\mu}$ which are in turn defined via the further diagonalization of each $\mathbb{S}^{p\mu}$ matrix corresponding to each contributing molecular symmetry.
We could obtain a unitary transformation $\mathbb{U}$ such that $\Lambda^{p\mu}$ is consequently susceptible of a diagonal representation:

\begin{equation}
\Lambda^{p\mu} = \mathbb{U} \mathbb{S}^{p\mu} \mathbb{U}^{\dag} = 
\left(
\begin{array}{cccc}
e^{2i\eta_{1}^{p\mu}(k_0)} & 0 & \cdots & 0 \\
0 & e^{2i\eta_{2}^{p\mu}(k_0)} & \cdots & 0 \\
\vdots & \vdots & \ddots & \vdots \\
0 & 0 & \cdots & e^{2i\eta_{n_{c}}^{p\mu}(k_0)}
\end{array}
\right)
\end{equation}

\noindent Another equivalent way to define the eigenphases $\eta_{i}^{p\mu}$ is via the eigenvalues of the $\mathbb{K}^{p\mu}$ matrix, which is hermitian and therefore has real eigenvalues.
Likewise there is an unitary transformation by which the $\mathbb{K}^{p\mu}$ is diagonalized as

\begin{equation}
\left(
\begin{array}{cccc}
\tan\eta_{1}^{p\mu}(k_0) & 0 & \cdots & 0 \\
0 & \tan\eta_{2}^{p\mu}(k_0) & \cdots & 0 \\
\vdots & \vdots & \ddots & \vdots \\
0 & 0 & \cdots & \tan\eta_{n_{c}}^{p\mu}(k_0)
\end{array}
\right)
\end{equation}

\noindent Having defined the diagonal elements of $\Lambda_{p\mu}$ to be given by exp($2i\eta_{i}^{p\mu}$), where the $\eta_{i}^{p\mu}$ are the eigenphases for each contributing symmetry $|p\mu\rangle$, it therefore follows that it is easy to show that

\begin{equation}
\sigma = \frac{4\pi}{k^2}\sum_{i} \sum_{p\mu} \sin^{2}(\eta_{i}^{p\mu})
\end{equation}

\noindent To obtain an expression for the differential cross section one needs to introduce then a coordinate system $S'$ which is space-fixed (SF), the so-called LAB system, and another coordinate system $S$ which is fixed to the molecular highest symmetry axis and that is identified as body-fixed (BF).
The latter is obtained from $S'$ by a rotation through the appropriate Euler angles ($\alpha$, $\beta$, $\gamma$).
The previously discussed expression for the FN scattering amplitude (see eq. 2.66), through the symmetry adapted functions $X_{\ell h}^{p\mu}(\hat{\textbf{k}}_{0}))$ and $X_{\ell h}^{p\mu}(\hat{\textbf{r}}))$, contains angular factors like $Y_{\ell m}(\hat{\textbf{k}}_{0})$ and $Y_{\ell m}(\hat{\textbf{r}})$ that are in turn defined with respect the BF system, $S$ ($\hat{\textbf{k}}_{0}$ and $\hat{\textbf{r}}$ are defined with respect the S reference).
The corresponding expression in the SF frame is obviously given by 

\begin{equation}
Y_{\ell m}(\hat{\textbf{r}}') = \sum_{m'} D_{m m'}^{\ell}(\alpha, \beta, \gamma) Y_{\ell m'}(\hat{\textbf{r}})
\end{equation}

\noindent where $\hat{\textbf{r}}'$ is defined with respect to $S'$ and $\hat{\textbf{r}}$ with respect to $S$ and the coefficient $D_{m m'}^{\ell}(\alpha, \beta, \gamma)$ are the Wigner functions \cite{th-sakurai}.
The differential cross section for scattering by randomly oriented molecules is then obtained by averaging the SF quantity $\left| \tilde{f}_{SF}(\hat{\textbf{k'}}_{0} \cdot \hat{\textbf{r}}';\;\alpha, \beta, \gamma) \right|^{2}$ over all the molecular orientations ($\alpha$, $\beta$, $\gamma$):

\begin{equation}
\frac{d\sigma}{d\Omega} (\hat{\textbf{k}}_{0}' \cdot \hat{\textbf{r}}') = \frac{1}{8\pi^{2}} \int\;d\alpha d\beta d\gamma \sin\beta |\tilde{f}_{SF}(\hat{\textbf{k}}_{0}' \cdot \hat{\textbf{r}}';\;\alpha, \beta, \gamma)|^{2}.
\end{equation}

\noindent This angular integration can be done analytically and lead to an expansion of differential cross section in terms of Legendre polynomials $P_{\ell}$:

\begin{equation}
\frac{d\sigma}{d\Omega}(\hat{\textbf{k}}_{0}' \cdot \hat{\textbf{r}}') = \sum_{\mathcal{L}} A_{\mathcal{L}}(\mathbb{S}_{ij}^{p\mu}) P_{\mathcal{L}}(\cos\theta)
\end{equation}

\noindent where the description with full particulars inherent the calculations of the $A_{\ell}$ coefficients can be found in the first appendix and particularly in \cite{th-fag-physrep, th-sanna}.

\noindent To conclude the present section, it is useful to emphasize the fact that a body-frame treatment like that one briefly illustrated in this chapter, is always appropriate as long as the incident electron wavefunction overlaps the molecule, that is to say in the so called 'inner region': in this spatial region, in fact, the colliding electron is moving faster than the nuclei.
A qualitative explanation is found when one notes that in general, due to the influence of molecular attraction, the projectile velocity varies during the collision in the sense that it is reasonable to say that an incident electron which is $slow$ at large distances becomes $faster$ at short distances.
It therefore follows that the whole three dimensional space is divided into two parts.
In the inner region the use of the BF frame implies the Born-Oppenheimer approximation, which in turn regards the projection of the electron orbital angular momentum along the figure axis (i.e. the main symmetry molecular axis) as a good quantum number, so that generally either the Hund's case 'a' or 'b' is applicable to couple the incident electron orbital angular momentum to the molecular electronic angular momentum and to the angular momentum of nuclear rotation.

\clearpage

\chapter[Results and discussion]{Results and discussion}
\label{resdisc}

\section[The ortho-benzyne molecule, o-C$_{6}$H$_{4}$]{The ortho-benzyne molecule, o-C$_{6}$H$_{4}$}
\label{coronene_res}

\subsection[Introduction: metastable anions of o-benzyne]{Introduction: metastable anions of o-benzyne}

\noindent In a carbon-rich red giant or in a proto-planetary nebula (PPN) atmosphere, the major carbon-bearing chemical species that is initially formed is acetylene (C$_{2}$H$_{2}$), rather than the carbon monomer, due in principle to the wide availability of hydrogen.
Generally speaking, specific chemical reactions, depending on both density and temperature conditions, could then lead to the formation of planar molecules of polycyclic aromatic hydrocarbons (PAHs), large aromatic molecules with five or six-membered carbonaceous rings.
Due to their marked electronic delocalization, they are very stable against photodissociation, thus they could also be present in circumstellar environments.
In this connection, the still unassigned infrared (IR) bands detected in different regions of the interstellar medium (ISM) are commonly related to infrared fluorescent emissions from highly excited PAHs \cite{allamandola}.
Moreover, some of these bands have been observed in planetary nebulae \cite{witterborn} and have also been detected in stars that were in transition from the asymptotic red giant branch (AGB) to the planetary nebula phase (PN) \cite{geballe}.
It seems reasonable, in fact, that PAHs are formed in the envelopes of C-rich stars and within regions spectroscopically characterized by the featureless, continuum infrared (IR) emission attributed to amorphous carbon and to the formation of C-particulates \cite{frenklach}.
A similar mechanism may also be at work in circumstellar shells where PAH molecules could be viewed as providing the building blocks of stellar dust \cite{tielens}.

\noindent In line with the above considerations, some chemical schemes assume PAHs production in the molecular envelopes of carbon-rich stars where, just assuming the acetylene as the major carbon source, thay take advantage of classical nucleation theory in order to provide estimates of the formation rates \cite{stein}.

\noindent More detailed kinetic mechanisms for gas-phase formation of PAHs were already put forward in \cite{frenklach2}, whereby the onset of their production in the ISM was predicted to occur within the temperature range between 900 and 1100 K, and require particularly dense and slow stellar winds with high initial acetylene abundances.
Because of the presence of a fair amount of photoionized electrons within that stellar wind, it is also a likely possibility that very reactive, metastable negative ions are formed under the above conditions and therefore play a significant role during the PAHs formation.

\begin {figure}[here]
\begin {center}
\includegraphics[scale=1.14]{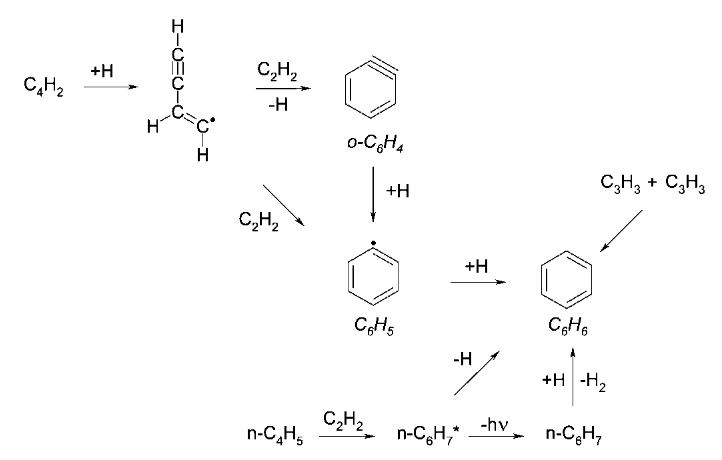}
\end {center}
\caption{\small{The radical-based reaction scheme for PPN benzene chemistry (adapted from \cite{frenklach2}) in which the o-C$_6$H$_4$ is expressly suggested as an actual competitor to benzene, thus leading to the larger PAHs.}} 
\label{fig_ch3.1_01}
\end {figure}

\begin {figure}[here]
\begin {center}
\includegraphics[scale=0.26]{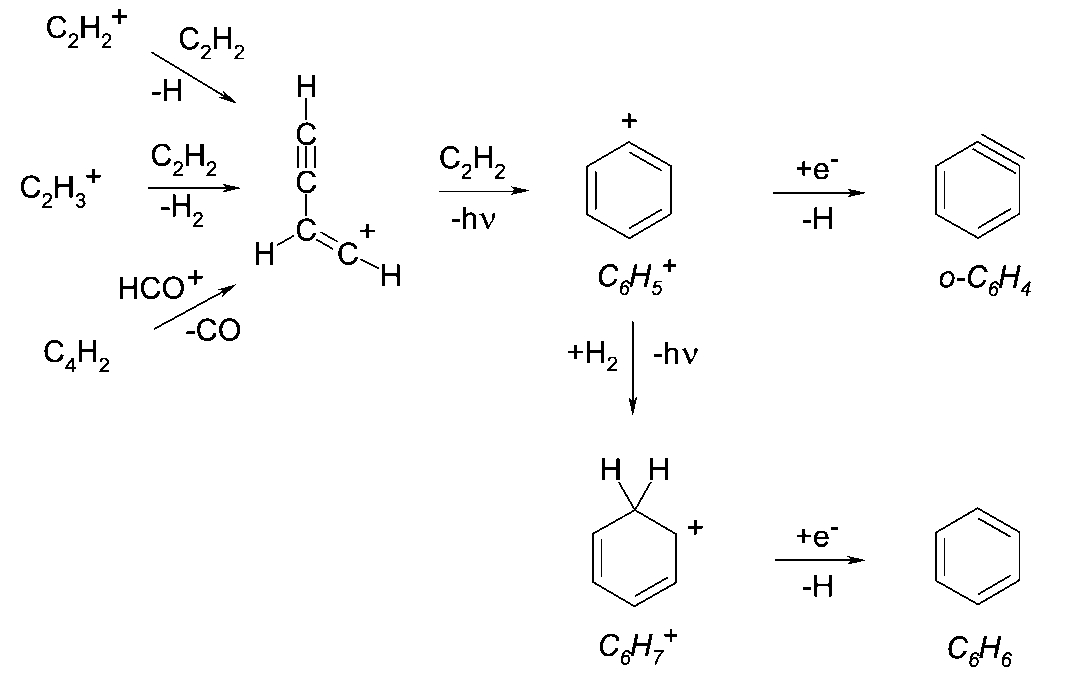}
\end {center}
\caption{\small{The ion-based reaction scheme for PPN benzene chemistry (adapted from \cite{woods2}). See text for details.}} 
\label{fig_ch3.1_02}
\end {figure}

\noindent Unfortunately, although significant laboratory spectroscopic work has been done, and dedicated to PAHs in order to support the astronomical observations, their identification has always been elusive.
Only benzene has been tentatively detected \cite{cernicharo} towards the proto-planetary nebula CRL 618, since its $\nu_{4}$ bending mode (the strongest infrared feature for C$_{6}$H$_{6}$) has been observed, and consequently the column density as well as the kinetic temperature have been determined to be respectively of  5x10$^{15}$ cm$^{-2}$ and of 200 K.
The discovery of benzene molecules \cite{cernicharo}, which had never been found outside the solar sistem, thus represents a fundamental step for our understanding of the possibile interstellar and/or circumstellar synthesis of PAHs because of its being the simplest, basic aromatic unit.
Moreover, the detection of circumstellar benzene molecules indicates that both physical and chemical conditions in PPN CRL 618 are right to activate a rich and diverse chemistry.
This PPN, in fact, is an evolved object with a central B0 star \cite{sanchez} ($T_{*}$ = 30.000 K) and an ultracompact H II region surrounded by a carbon-rich (C/O > 1) molecular envelope \cite{sanchez}.
In addition, it shows optical high-velocity bipolar outflows as well as a low velocity expanding torus \cite{sanchez2}, the latter being characterized by an intense molecular emission by means of which it has been possibile to realize a rich molecular inventory including a large variety of hydrocarbons \cite{sanchez2, pardo}.
It then follows that either the UV photons from the rapidly evolving B0 central object, or the strong shocks associated with the high-velocity stellar winds (up to 200 Km s$^{-1}$), could in principle drive the polymerization of acetylene and consequently form benzene as the starting point for PAHs' synthesis.
In this connection, therefore, the circumstellar(\footnote{for the outflows of AGB and post-AGB C-rich stars}) chemical formation of large aromatic species has been modelled not only in strong analogy with soot formation processes which take place in terrestrial combustion environments (see figure \ref{fig_ch3.1_01}, \cite{frenklach2, cherchneff}), but also surmising a complex chain of reactions driven by photoionization and dissociative electron recombination processes (see figure \ref{fig_ch3.1_02}, \cite{woods2}).

\noindent The carbon-rich nature of proto-planetary nebula CRL 618, coupled with the C$_{6}$H$_{6}$, C$_{6}$H$_{2}$, C$_{4}$H$_{2}$ detections \cite{cernicharo}, makes it an excellent target for additional benzene derivative searches, although no spectral features associated with o-C$_{6}$H$_{4}$ has yet been detected towards this object \cite{widicus}.
Keeping in mind that we do not have as yet reliable predictions of C$_{6}$H$_{6}$, C$_{6}$H$_{5}$ and o-C$_{6}$H$_{4}$ abundances in PPN (i.e. on the key species involved in the ionic and in the radical-based reaction mechanisms as those given in figures \ref{fig_ch3.1_01} and \ref{fig_ch3.1_02}), new observations and better theoretical data on both C$_{6}$H$_{5}$ and o-C$_{6}$H$_{4}$ would yield, in principle, important information as well as new insight for future modelling efforts.
We therefore present here a study directed at providing one of the possible reasons for which the ortho-benzyne molecule might not have been detected during searches over CRL 618 \cite{widicus}, despite of its large electric dipole moment of 1.68 D \cite{kraka} and  of its lack of hyperfine splitting features, both of which would make it a very good candidate for detection with infrared and microwave spectroscopy.
In fact, at least qualitatively, one could say that a molecular system such as o-C$_{6}$H$_{4}$ is likely to be present in the torus of CRL 618, where higher densities could shield it from photodissociation as well as photoionization, while at the same time allowing its reaction with low-energy thermalized electrons, present in that environment \cite{sanchez}, to form the corresponding metastable resonant anion.
The present study is further encouraged by the fact that three molecular anions such as C$_{4}$H$^{-}$, C$_{6}$H$^{-}$ and C$_{8}$H$^{-}$ have recently been detected (and in large quantities) in different astrophysical environment \cite{cernicharo3, mccarthy, sakai, remijan, brunken} like the envelope of the carbon-rich AGB star IRC+10216 \cite{cernicharo3, mccarthy, remijan}, the latter being considered the AGB star archetype.

\noindent The possibility that a not so small but marginal fraction of interstellar molecular material might be in the form of anions was first suggested by Herbst \cite{herbst} in relation to carbon chains and other radicals with electron affinities large enough to exhibit in principle large radiative attachment rate.
This characteristic is linked to the fact that the higher the electron affinity of a molecule the more states are accessible for the redistribution of the excess energy carried by the attached electron while, at the same time, auto-ionization is less likely to occur before the competing radiative stabilization of the excited anions could take place.

\noindent Thus, the question we wish to address here is the possible existence of metastable anions of o-C$_{6}$H$_{4}$ as "doorway states" for the ensuing possible stabilization of different negative species which, in turn, can lead to ionic reaction chains on the way to the formation of increasingly more complex PAH species.

\subsection[Possible chemistry and role of stabilized  o-C$_{6}$H$_{4}$ negative ions in PPN atmosphere.]{Possible chemistry and role of stabilized  o-C$_{6}$H$_{4}$ negative ions in PPN atmosphere.}

\noindent Although significant laboratory spectroscopic work has been dedicated to PAHs in order to support the astronomical observations, their identification has always been elusive: none of them has been unambiguously detected and only benzene have been tentatively revealed toward the proto-planetary nebula (PPN) CRL 618 \cite{cernicharo}.
Hence, the first step to understand the possible involvement of the ortho-isomer of benzyne (o-C$_{6}$H$_{4}$), and specifically of its negative ion, in a realistic circumstellar gas-phase PAH synthesis is to analyze the benzene-related circumstellar chemistry i.e. its currently suggested formation pathways at the physical conditions that characterize the proto-planetary nebula stellar phase: in this regard, the benzene (C$_{6}$H$_{6}$) could be considered as the corner-stone of carbonaceous polycyclic aromatic structures.

\noindent Generally speaking, as mentioned in the introduction, PAHs are considered to be produced in carbon-rich star outflows, as it seems not very likely that such large molecules could be produced by gas-phase or by grain-assisted chemistry within the interstellar medium, given the much lower pressure (and density) of that environment.
The PPN phase, which for stars of low and intermediate initial mass represents the link (the transition object) from the asymptotic giant branch (AGB) phase and that of the planetary nebula (PN), is at present both physically and chemically rather poorly understood, and even though its evolution is tipically fast as compared to the initial AGB and final PN phase lifetimes, it is however an established fact that investigating its microscopic (chemical) evolution could provide a deeper insight in predicting the nature of the molecular and atomic material returned to the ISM as the central star approaches the end of its life.
The circumstellar chemistry for a PPN, or a post-AGB atmosphere, appears to be strongly dependent on both the radiative flux from the central star and its density, because circumstellar material undergoes energetic processing.
Various observations \cite{meixner} have suggested that during the PPN phase a second, more violent, mass-loss event occurs with high velocity winds that partially destroy the remnant AGB envelope, so that over a relatively short period that envelope is depleted and gradually exposes the central star \cite{kwok2}.
As the envelope thins down, the stellar radius decreases, raising in turn the temperature of the star to T $\geq$ 30000 K.
Finally, the increasing UV output from the central star itself begins to ionize the remnant envelope, thus leading to a PN \cite{kwok2}.

\noindent In this framework PPN CRL 618 represents an extreme example of an evolved object, at present characterized by optical high-velocity bipolar outflows, where the central B0 star, having lost its dusty shroud, is producing sufficient UV radiation to create a small, compact nearby HII region \cite{sanchez} surrounded by a carbon rich molecular envelope: this slowly expanding toroidal envelope (showing a rich molecular emission \cite{sanchez2, sanchez3}) is a high-energy, high-density environment so that its chemistry could be expected to be similar to the one observed in combustions \cite{frenklach2}.
Consequently, during the PPN evolution the surrounding gaseous shell expands, the density decreases and the material in the circumstellar envelope is subjected to photoprocessing from the central star, thus very likely leading to predominant ion-molecule chemistry \cite{woods}.

\noindent It is for the above reasons that we can surmise as a formation mechanism for large molecules the likely combination of radical- and ion-molecule reactions: in fact both a radical-based mechanism \cite{frenklach2, cherchneff} and an ion-molecule mechanism \cite{woods2} have been proposed for the benzene molecule circumstellar synthesis (see figures \ref{fig_ch3.1_01} and \ref{fig_ch3.1_02}).
Moreover, it should be noted that while the former \cite{cherchneff} directly supports and accounts for the possible presence of ortho-benzyne (neutral ortho-benzyne), the original ionic chemical model \cite{woods2} does not include that formation reaction.
Such a reaction, however, could be realiably induced by free electrons and accompanied by the phenil cation (C$_{6}$H$_{5}^{+}$) dehydrogenation: in fact the ionic chemical model involves ionized hydrocarbon - neutral acetylene reactions as well as hydrogenation and/or (dissociative) electron recombination processes, thereby implicitly accounting for the presence of low-energy electrons that we argue here to play an important role.
Additionally, one of the benzene derivatives formed in the radical network, the phenil open-shell radical (C$_{6}$H$_{5}$), is a suggested precursor to PAHs in circumstellar environments \cite{frenklach, cherchneff} due to its high reactivity with acetylene.
However, at present no PPN model includes both the radical- and the ion-based benzene reaction networks, thus making difficult the interpretation of observational results.

\noindent Some current studies of high-energy environments indicate that, if circumstellar chemistry is similar to plasma or combustion chemistry, then C$_{6}$H$_{5}$ and o-C$_{6}$H$_{4}$ may coexist with C$_{6}$H$_{6}$: theoretical investigations showing that benzene unimolecular decomposition during combustion actually leads to C$_{6}$H$_{5}$ and o-C$_{6}$H$_{4}$ production do exist \cite{mebel}, as well as experimental studies indicating that gas-phase benzene electrical discharges produce C$_{6}$H$_{5}$ and o-C$_{6}$H$_{4}$ \cite{mcmahon}.
However, no spectral features associated with o-C$_{6}$H$_{4}$ were detected in the MHz frequency range during its astronomical search toward CRL618, despite of its rather large permanent electric dipole moment coupled with the lack of hyperfine splitting.
Such features should in principle yield stronger lines for the neutral molecule, making it a more likely and easier candidate for detection than C$_{6}$H$_{6}$, if their abundances were comparable.
A molecule such as o-C$_{6}$H$_{4}$ is however likely to be present in the toroidal molecular envelope of CRL618, where higher densities could shield it from both photodissociation and photoionization.
More in general, such a species could also be present in more extended regions when actually produced by ion-molecule chemistry.

\noindent We therefore follow here the suggestion of \cite{widicus} and make additional investigation on possible chemical networks involving o-C$_{6}$H$_{4}$.
As will be shown in what follows, from the results involving the o-benzyne molecule at the equilibrium geometry \cite{carelli1}, we learn that this molecule is an efficient producer of compact, fairly long-lived metastable anionic intermediates.
In line with that, we then present the logical extension of that initial investigation, in keeping also with the occurrence of fast radiationless electronic transitions to an energetically lower-lying electronic state for a similar molecule \cite{thanopulos}, and suggest a likely stabilization of the metastable negative ion via an intramolecular vibrational redistribution processes, the latter chiefly involving the triple bond within the exagonal aromatic structure.
This picture, in fact, offers not only a qualitative explanation for the surprising absence of spectroscopic features associated with neutral o-C$_{6}$H$_{4}$ toward CRL618, but enables us to surmise a direct involvement of the stabilized negative ion of ortho-benzyne in the complex chain of reactions eventually leading to circumstellar PAHs, as we shall further discuss below.
In other words, we shall show that the anionic benzyne can provide an highly reactive intermediate that is very likely to disappear fairly rapidly via the ionic reactive path which, in turn, we shall qualitatively illustrate in the following sections.

\subsection[The integral cross sections]{The integral cross sections}

\begin {figure}
\begin {center}
\includegraphics[scale=0.30]{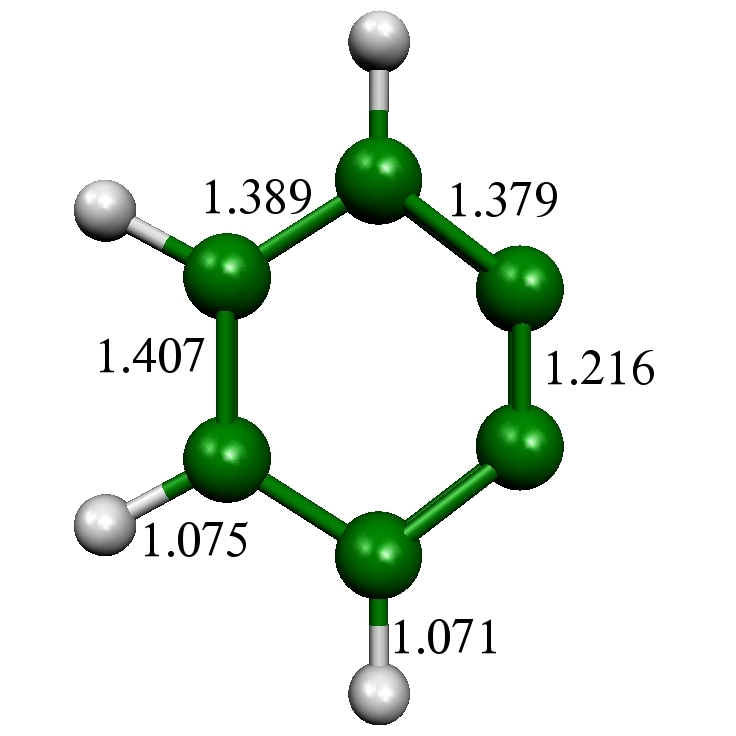}
\end {center}
\caption{\small{Computed optimized equilibrium geometry and bond values ( in \AA) for the ortho-benzyne molecule. See text for further details.}} 
\label{fig_ch3.1_03}
\end {figure}

\noindent Equilibrium structure, geometry optimizations and the potential energy of the neutral species involved in this study, as well as the starting molecular orbitals (MOs) used for the present scattering calculations, have been performed at the HF/aug-cc-pVTZ level using the Gaussian suite of programs \cite{frisch}.
These ab initio calculations yielded for the title system a total SCF energy of -229.467905 hartrees.
The molecular structure obtained with the above Gaussian basis-set shows good agreement with the earlier more accurate ab-initio calculations \cite{groner}: see figure \ref{fig_ch3.1_03} for their values.
Using the above basis set the calculated wave function for the neutral target yield a permanent dipole moment of 1.77 D, in good agreement with the theoretical value of 1.68 D \cite{kraka}.
According to the molecular symmetry point group to which the ortho-benzyne belongs at its equilibrium geometry, reported by figure \ref{fig_ch3.1_03}, the elastic cross sections and poles of the S-matrix for the scattering process

\begin{eqnarray}
o-C\mathrm{_6}H\mathrm{_4} (^1\Sigma) + e\mathrm{^-} \rightarrow ( o-C\mathrm{_6}H\mathrm{_4} \mathrm{^-}) \mathrm{^*} (^2\mathrm{A}) 
\nonumber
\end{eqnarray}

\noindent have been computed within the four irreducible representations of C$\mathrm{_{2v}}$ group using our scattering program \cite{epoly}: these calculations have been initially performed by us at the equilibrium geometry.

\noindent For the S-matrix calculations we expanded the wave function up to a maximum angular momentum value of $L_{Max}$ = 40; the corresponding partial wave expansion for the scattering potential was thus carried out up to $l\mathrm{_{MAX}}$ = 80; moreover, a maximum of 25 partial waves ($L_{MaxA}$ = 25) was used by us for the corresponding scattering electron expansion within the piecewise diabatic potential approach.
We note that to increase either values of the maximum angular momentum ($L_{Max}$ and $L_{MaxA}$ respectively up to 60 and 30) had no significant effect on the final results.
The size of the physical box that contains the diabatized partial terms, as described in the previous section, was 50 \AA, then the overall radial and angular grids ($r,\theta,\phi$) included 1760 x 80 x 80 points.
The model correlation-polarization potential $V_{cp}$ is the one described in a previous section (2.4.5) and in our earlier work \cite{lucchese, tele, telega, carelli0}; it was used in the present calculations with the total asymptotic polarizability value of 64.98 $a_{0}^{3}$ by us calculated.
We shall describe below in greater detail how this term was included in our calculations.

\begin {figure}
\begin {center}
\includegraphics[scale=0.10]{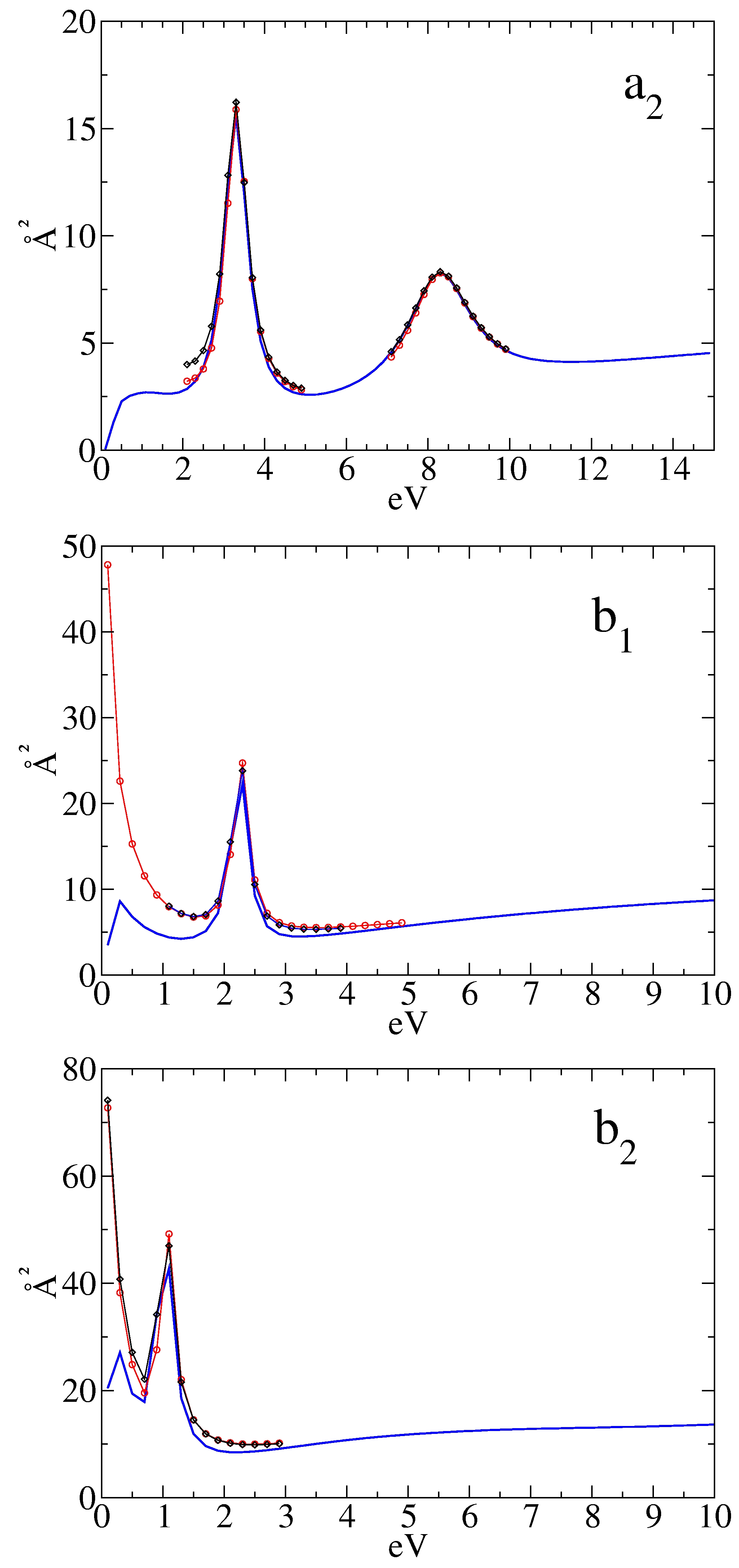}
\end {center}
\caption{\small{Computed elastic integral cross sections of the three molecular symmetry contributions. For the definition of the various curves shown see the main text.}} 
\label{fig_ch3.1_04}
\end {figure}

\noindent In what follows, we shall describe and discuss our computed elastic (rotationally summed) integral cross sections for the equilibrium geometry of the title system.
The actual energy behaviour of the dominant symmetry components which make up the total electron scattering elastic cross-sections is depicted, in the energy range of interest, within the three panels of figure \ref{fig_ch3.1_04}.

\noindent We report there the partial elastic cross sections for the $a_2$ (top panel), $b_1$ (middle panel) and $b_2$ (bottom panel) contributions; no maxima in the integral cross sections are found in the totally symmetric ($a_1$) component, the latter not being shown here.
We are initially interested in revealing the possible existence of low-energy shape resonances in the o-C$\mathrm{_6}$H$\mathrm{_4}$ ($^1\Sigma$), features which are qualitatively due to short-range effects leading to dynamical trapping of the impinging electron.
However, the ortho-benzyne molecule, due to its triple bond, has a large permanent electric dipole moment as well as a large spherical polarizability having their origin in the large number of $\pi$ electrons in the aromatic pseudo-exagonal ring.
Thus we cannot in principle discard the long-range polarizability contributions on the resonance parameters, and in the present case we indeed expect they shall play an important role in the scattering process.

\noindent Our long-range polarization potential contribution to the $V_{cp}$ potential is obtained by using a potential form which asymptotically agrees with the value given by the static polarizability $\alpha$ of the molecule under investigation.
This can be done either by assuming a single polarization center or by partitioning the static polarizability $\alpha$ on different centers where the individual atomic polarizabilities are subjected to the constraint that the total molecular polarizability is reproduced in the long-range region.

\noindent  We have actually performed both such calculations including the long-range part of the electron-molecule interaction for all the elastic partial cross sections in the three contributing symmetries and also increasing the physical box up to 50 \AA.
The corresponding results are reported in figure \ref{fig_ch3.1_04}: the continuous lines represent the elastic partial integral cross sections computed without using the o-C$\mathrm{_6}$H$\mathrm{_4}$ ($^1\Sigma$) polarizability and at the same time referring to a physical box with a radius of 15 \AA, while the circles and the diamonds refer respectively to a calculation which is  now including either the spherical or the multicenter polarizability (the latter weighed and scaled by the Mulliken charges on each atomic center) as well as considering the larger physical box of 50 \AA.
We can immediately recognize that both the $a_2$ maxima and the higher energy maxima of $b_1$ and $b_2$ contributions appear at the energy values for which we also found the maxima without the dipole-induced polarization corrections.
The situation is instead different for each of the very low-energy features visible in the partial cross sections of $b_1$ and $b_2$ symmetries: when the polarizability is introduced in the calculation, they both disappear below threshold thus suggesting that they are actually artefacts created by the limited size box employed initially.
All further calculations were therefore done employing the larger integration "box" mentioned before (50 \AA).

\subsection[The $\sigma^{*}$ and $\pi^{*}$ resonances]{The $\sigma^{*}$ and $\pi^{*}$ resonances}

\begin {figure}
\begin {center}
\includegraphics[scale=0.08]{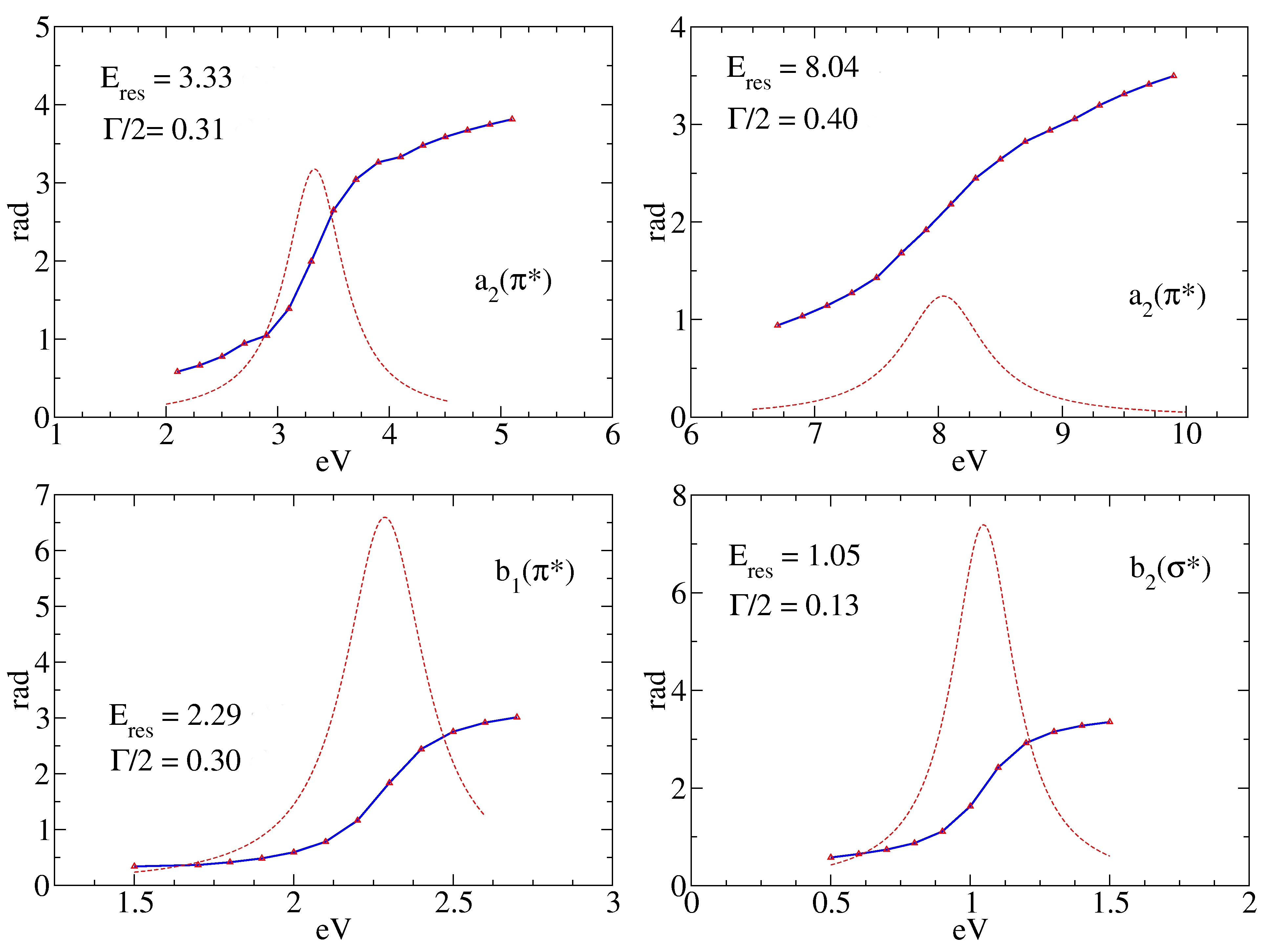}
\end {center}
\caption{\small{Computed eigenphase sums within each of the scattering contributions for the three symmetries discussed in the main text. All values are in eV.}} 
\label{fig_ch3.1_05}
\end {figure}

\begin {figure}
\begin {center}
\includegraphics[scale=0.25]{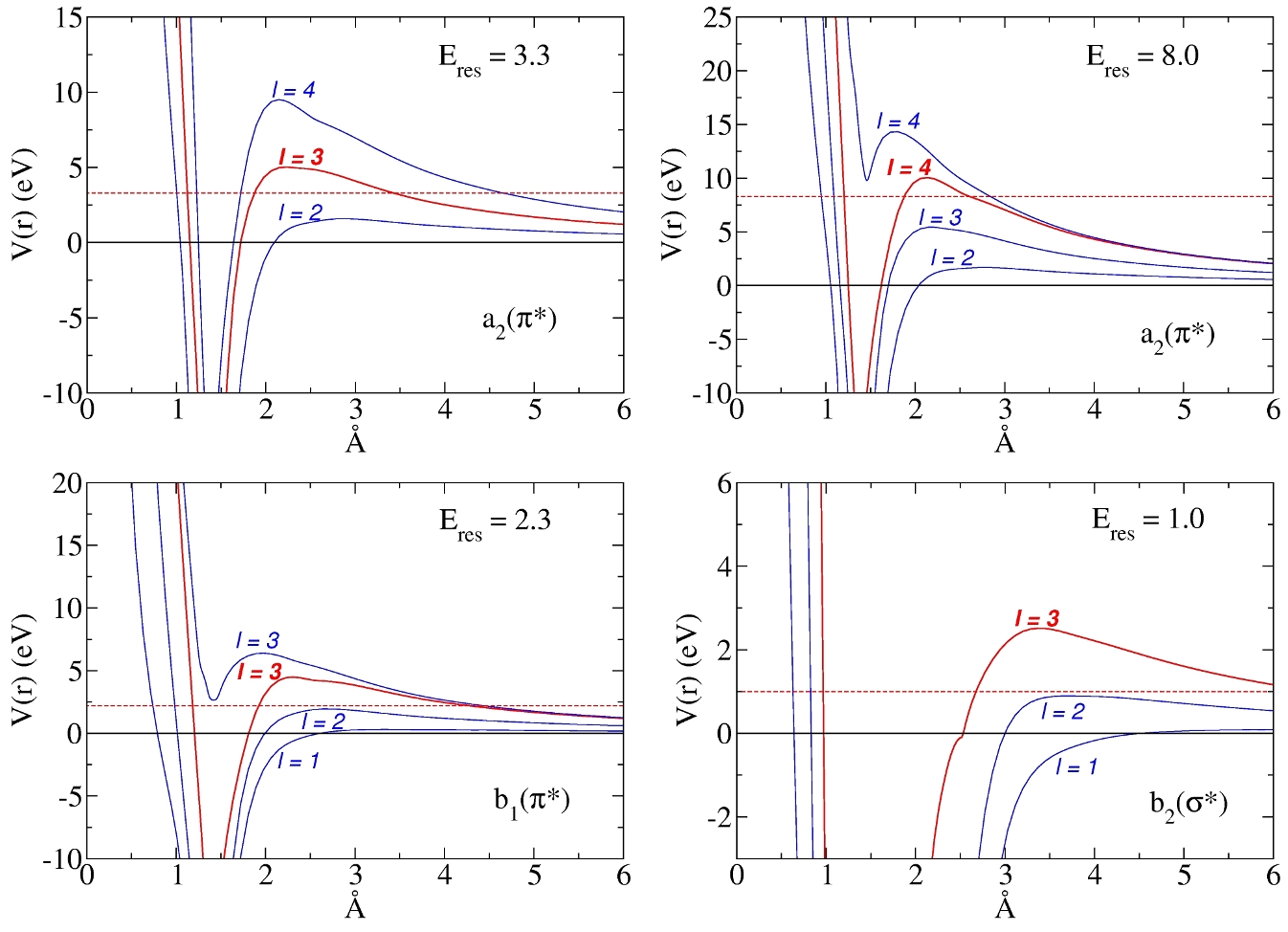}
\end {center}
\caption{\small{Computed diabatic potential energy curves employed for the calculation of each partial elastic integral cross section. See main text for details. All energies are in eV.}} 
\label{fig_ch3.1_06}
\end {figure}

\noindent We discuss below our results on the computed resonance parameters, i.e. the eigenphase sums and the diabatic potential curves, reported respectively in figures \ref{fig_ch3.1_05} and \ref{fig_ch3.1_06}.
We are in fact interested in characterizing the metastable anions obtained from resonant trapping of the impinging low-energy electron because of their possible role as reaction intermediates in the circumstellar environment.

\noindent In figure 3.5 we show the behaviour of the eigenphase sums together with their first derivatives for each symmetry in which we find a maximum in the corresponding partial cross-section.
The four eigenphase sums were each fitted to a Breit-Wigner formula, using the familiar functional form \cite{taylorbook}

\begin{equation}
\delta(E)=a+b(E-E_{res})+c(E-E_{res})^{2}+arctan\left[\frac{\Gamma}{2(E-E_{res})}\right] 
\nonumber
\end{equation}

\noindent in order to extract the pure resonant component from the background and to obtain the resonance width from the lorentzian fit of the corresponding delay matrix.
It is clearly seen that the four isolated resonant contributions in the three different symmetries $a_2$, $b_1$ and $b_2$, show a sharp $\pi$ jump when the collision energy of the charged projectile crosses the value for which each of the corresponding cross sections shows a maximum: these findings confirm that the four maxima in the elastic partial cross sections correspond to pure shape-resonances.
The energy dependence of the lower diabatic potential components for each resonance are shown in the four panels of figure \ref{fig_ch3.1_06}, where angular momenta are reported up to $l=3$ ($b_1$ and $b_2$ contributions) as well up to $l=4$ (for the two $a_2$ resonances).
We find it important to point out that the diabatic potential curves of the $b_1$ and $b_2$ symmetries are such that the corresponding impinging electron energies are in both cases degenerate with the minimum value needed to overcome their barriers; this particular feature is of importance in controlling the width of a resonance.
The two resonances of $a_2$ symmetry, as clearly shown in the two upper panels of figure \ref{fig_ch3.1_06}, are instead due to the tunnelling across a barrier generated by the $l=3$ ($f$ wave) and $l=4$ ($g$ wave) angular momenta, features that therefore explain their being located at higher energies.

\noindent By examining the contour plots of the resonant wavefunctions of the three irreducible representations (IRs) in which we find shape-resonances (which we do not report here for the sake of brevity), we have discovered that three of the four resonances have $\pi^*$ nature (respectively the two of $a_2$ symmetry and the one of $b_1$ symmetry), while the remaining $b_2$ resonance has $\sigma^*$ nature.

\begin {figure}
\begin {center}
\includegraphics[scale=0.17]{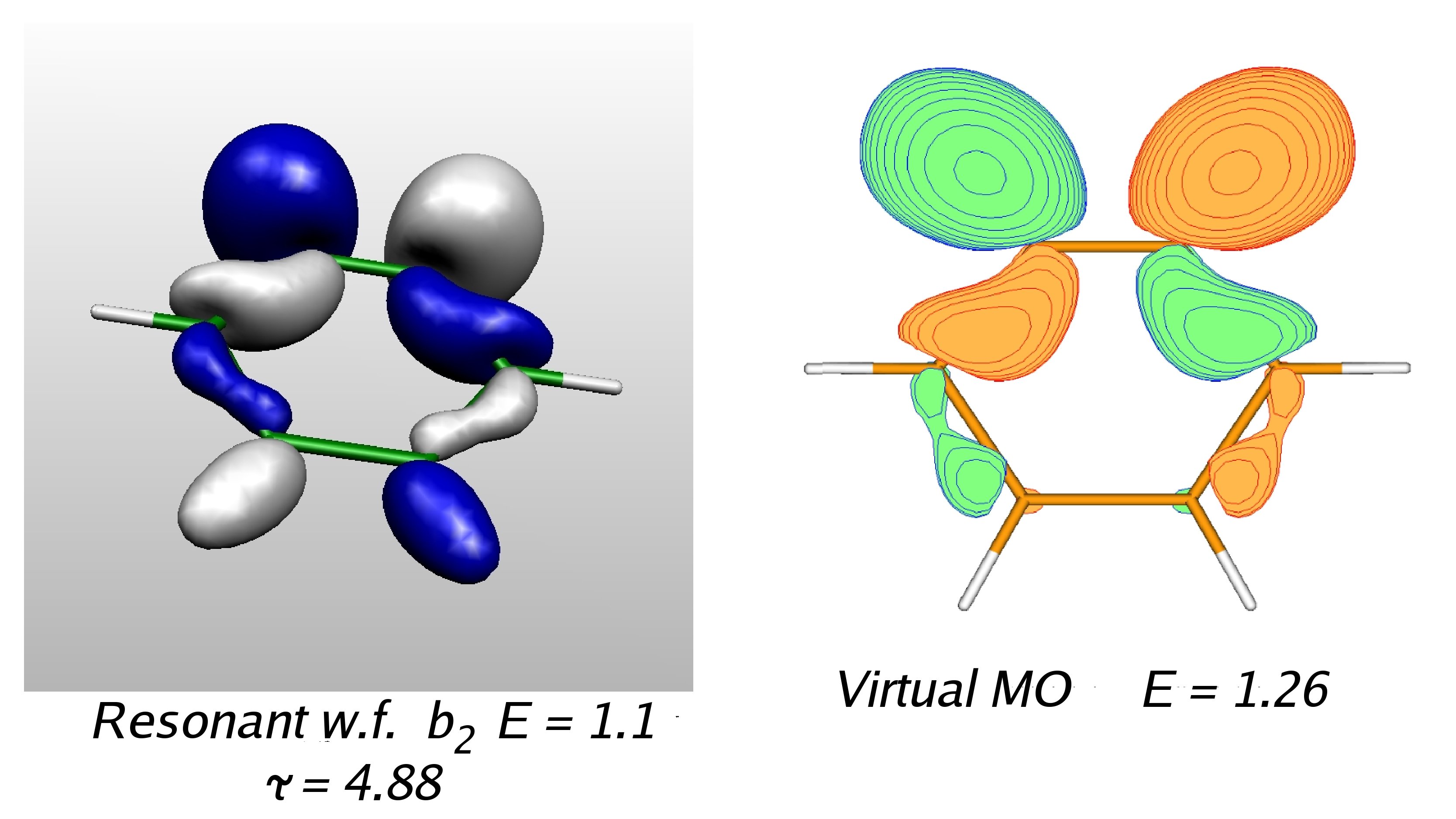}
\end {center}
\caption{\small{Left-hand panel: shape of the real part of the scattering wave function (w.f.) for the $\sigma^*$ resonance computed using the piecewise diabatic model interaction (color online). Right-hand panel: HF/aug-cc-pVTZ  computed virtual MO of the ground electronic state of the o-C$\mathrm{_6}$H$\mathrm{_4}$ molecule for the $b_2$ symmetry where the $\sigma^*$ resonance is located. All energy values in eV. The lifetime $\tau$ in units of $10^{-15}$ s.}} 
\label{fig_ch3.1_07}
\end {figure}

\noindent The $b_2$($\sigma^*$) is now explicitely shown in figure \ref{fig_ch3.1_07} and compared there with the nearest virtual orbital (MO) obtained from the HF calculations on the target molecule.
It shows that the distribution of electronic charge associated with the impinging electron is displaced asymmetrically, mainly on the carbon atoms of the triple bond where a nodal plane clearly exists perpendicular to that bond: this strongly suggests its antibonding behaviour across this CC bond.
Thus, according to these findings, the $b_2$($\sigma^*$) resonance at the equilibrium geometry could be reasonably considered a good candidate for adding antibonding behaviour to the triple bond in the (N+1) electron system and therefore for being responsible of a possible ring-breaking process guided by the non-adiabatic coupling between this $\sigma^*$ resonant electron and the nuclear dynamics which then would chiefly involve the triple bond.
The following section shall further elaborate on this point.

\clearpage

\subsection[o-C$_{6}$H$_{4}$ versus C$_{6}$H$_{6}$: a comparison between their equilibrium metastable anions]{o-C$_{6}$H$_{4}$ versus C$_{6}$H$_{6}$: a comparison between their equilibrium metastable anions}

\noindent From a structural point of view one of the main differences between benzene and ortho-benzyne, both considered in their equilibrium geometries, lies in the lack for the latter of a vicinal couple of hydrogen atoms to which corresponds the presence of the triple bond.
The ensuing appearance of a large permanent electric dipole moment (1.68 D) in the ortho-benzyne, in spite of the o-C$_{6}$H$_{4}$ and C$_{6}$H$_{6}$ having aromatic rings, turns out to be responsible, at least qualitatively, for many of the differences that we shall find below between the behaviour of the resonant states of these molecules.

\noindent Let us briefly summarize the properties of the two systems:

\noindent i) the benzene molecule in its equilibrium geometry belongs to the D$\mathrm{_{6h}}$ symmetry group, while the ortho-benzyne belongs to the C$\mathrm{_{2v}}$group;

\noindent ii) benzene is characterized by resonances in seven distinct irreducible representations which correspond to symmetries for which there are no occupied MOs ($a\mathrm{_{2g}}$, $b\mathrm{_{2g}}$, $e\mathrm{_{2u}}$) as well as to symmetries where occupied MOs of the target ground electronic state exist ($a\mathrm{_{1g}}$, $e\mathrm{_{2g}}$, $b\mathrm{_{1u}}$, $e\mathrm{_{1u}}$) \cite{gianturco3};

\noindent iii) according to the diabatic potential curves and to the eigenphase sums behaviour as functions of the impinging electron energy, the first group of computed resonances \cite{gianturco3} can be viewed as actual metastable negative ions of C$\mathrm{_{6}}$H$\mathrm{_{6}}$;

\noindent iv) for the resonances which belong to an IR that also contains occupied MOs (with the exception of $a\mathrm{_{1g}}$ contribution), the corresponding diabatic potential curves indeed exhibit their lowest angular components with an $l$ value greater than zero and located at rather low energy, thus having a barrier too low to support a shape resonance.
In fact, each of their cross sections also shows one low-energy structure which could be qualitatively related to the top of the centrifugal barriers for the lowest contributing angular momenta \cite{gianturco3}.

\noindent In order to make a more efficient comparison between the benzene and the ortho-benzyne resonances, we can use the correlation table C$\mathrm{_{2v}}$ $\leftarrow$ D$\mathrm{_{6h}}$ shown in table 3.1.
A first important difference between these molecules can be found in the fact that no resonance features appear in the totally symmetric contribution ($a\mathrm{_{1}}$) for the ortho-benzyne, while for benzene two one-electron resonances, both with symmetry $a\mathrm{_{1g}}$ (for which also bound MOs exist) can be identified.

\noindent Each of the partial cross sections in the $b\mathrm{_{1}}$ and $b\mathrm{_{2}}$ symmetries for o-C$\mathrm{_{6}}$H$\mathrm{_{4}}$ shows, as discussed before, that high energy maxima do not undergo variation of their energy location once polarization effects are included.
Moreover, their eigenphase sums as energy functions show a sharp $\pi$ jump, thus indicating the dynamical trapping of the incident electron due to a nonadiabatic coupling between the $l=2$ and $l=3$ angular momentum contributions (see Table 3.2).

\begin{table}[here]
\begin{center}
\begin{tabular}{||c|c||}
\hline
 D$\mathrm{_{6h}}$  & C$\mathrm{_{2v}}$                    \\
\hline
\hline
 A$\mathrm{_{1g}}$  & A$\mathrm{_{1}}$                     \\
\hline
 A$\mathrm{_{2g}}$  & A$\mathrm{_{2}}$                     \\
\hline
 B$\mathrm{_{1g}}$  & B$\mathrm{_{1}}$                     \\
\hline
 B$\mathrm{_{2g}}$  & B$\mathrm{_{2}}$                     \\
\hline
 E$\mathrm{_{1g}}$  & B$\mathrm{_{1}}$ + B$\mathrm{_{2}}$  \\
\hline
 E$\mathrm{_{2g}}$  & A$\mathrm{_{1}}$ + A$\mathrm{_{2}}$  \\
\hline
 A$\mathrm{_{1u}}$  & A$\mathrm{_{2}}$                     \\
\hline
 A$\mathrm{_{2u}}$  & A$\mathrm{_{1}}$                     \\
\hline
 B$\mathrm{_{1u}}$  & B$\mathrm{_{2}}$                     \\
\hline
 B$\mathrm{_{2u}}$  & B$\mathrm{_{1}}$                     \\
\hline
 E$\mathrm{_{1u}}$  & B$\mathrm{_{2}}$ + B$\mathrm{_{1}}$  \\
\hline
 E$\mathrm{_{2u}}$  & A$\mathrm{_{2}}$ + A$\mathrm{_{1}}$  \\
\hline

\end{tabular}
\end{center}
\caption{\small{Correlation table C$\mathrm{_{2v}}$ $\leftarrow$ D$\mathrm{_{6h}}$ between the symmetry groups associated with benzene and benzyne.}}
\end{table}

\noindent Through the above correlation table is also possible to connect each of the four ortho-benzyne resonant states in table 3.2 with those found for the C$\mathrm{_{6}}$H$\mathrm{_{6}}$ molecule (reported in table 3.3).
In general, however, with the exception of the $a_2$ (o-C$\mathrm{_{6}}$H$\mathrm{_{4}}$; E$_{res}$ = 8.0 eV) $\rightarrow$ $a_{2g}$ (C$\mathrm{_{6}}$H$\mathrm{_{6}}$; E$_{res}$ = 21 eV) connection, which is geometrically rather simple to do, the remaining three metastable states for the ortho-benzyne molecule do not appear to be comparable with any of those found for the benzene molecule in the calculations of ref \cite{gianturco3}.
In other words, the removal of a vicinal couple of hydrogen atoms from the aromatic ring and the consequent presence of the large permanent electric dipole moment due to the triple bond, make the resonant metastable states for the ortho-benzyne very distinctive and it is then difficult to compare them with those found for the benzene, as they are in general different not only in the geometrical shape of the real parts of the resonant wave functions but are also associated with different dynamical mechanisms, as we shall further see below.

\begin{table}
\begin{center}
\begin{tabular}{||c|c|c|c||}
\hline
         IR         &      $E_{res}(eV)$     &      $\Gamma(eV)$           &     $l_{dominant}$         \\
\hline
\hline
b$\mathrm{_{2}}$    &      1.05              &      0.27                   &               2/3          \\
\hline
b$\mathrm{_{1}}$    &      2.29              &      0.61                   &               2/3          \\
\hline
a$\mathrm{_{2}}$    &      3.33              &      0.63                   &                3           \\
\hline
a$\mathrm{_{2}}$    &      8.04              &      0.81                   &                4           \\
\hline

\end{tabular}
\end{center}
\caption{\small{Computed resonances for the o-C$\mathrm{_{6}}$H$\mathrm{_{4}}$ target. See text for further details.}}
\end{table}

\begin{table}[here]
\begin{center}
\begin{tabular}{||c|c|c|c||}
\hline
          IR         &      $E_{res}(eV)$     &      $\Gamma(eV)$      &     $l_{dominant}$  \\
\hline
\hline
e$\mathrm{_{2u}}$    &      4.66              &      1.50              &       3           \\
\hline
b$\mathrm{_{2g}}$    &      9.02              &      2.26              &       4           \\
\hline
e$\mathrm{_{1u}}$    &      12.25             &      4.75              &       5           \\
\hline
a$\mathrm{_{1g}}$    &      12.32             &      6.78              &       4           \\
\hline
e$\mathrm{_{2g}}$    &      15.46             &      7.04              &       4           \\
\hline
b$\mathrm{_{1u}}$    &      16.52             &     12.81              &       5           \\
\hline
a$\mathrm{_{2g}}$    &      21.49             &      4.82              &       6           \\
\hline
e$\mathrm{_{1u}}$    &      21.96             &     14.63              &       5           \\
\hline
b$\mathrm{_{1u}}$    &      24.48             &      5.31              &       5           \\
\hline
a$\mathrm{_{1g}}$    &      25.73             &     10.09              &       6           \\
\hline

\end{tabular}
\end{center}
\caption{\small{Computed resonance parameters for the benzene molecule as reported in \cite{gianturco3}.}}
\end{table}

\noindent There are also additional differences that have to do with the global behaviour of the metastable anions o-C$\mathrm{_{6}}$H$\mathrm{_{4}}^-$ and C$\mathrm{_{6}}$H$\mathrm{_{6}}^-$, both in their ground electronic states, as summarized in the tables 3.2 and 3.3.
With the exception of the $e\mathrm{_{2u}}$ contribution, the benzene molecule exhibits resonances located at markedly higher energies than its ionization potential (9.24 eV, \cite{nist}).
Furthermore, the corresponding widths allow one to obtain for each benzene resonance a lifetime which is shorter than that associated with their nuclear vibration dynamics.
The ortho-benzyne, on the other hand, appears to be somewhat more "reactive" in forming metastable anions: the ones calculated and discussed here correspond to energies which are well below its ionization potential (9.02 eV, \cite{zhang}) and, as reported in the third column of table 3.3, the ortho-benzyne resonance widths are smaller, corresponding to greater lifetimes which then become comparable with the times for vibrational motions within the bound molecule.

\noindent On the basis of the above findings we can therefore suggest that in the energy range of a few eV, that correspond to kinetic temperatures up to about 10.000 K for the circumstellar atmosphere of CRL 618 \cite{sanchez}, the ortho-benzyne molecule could actually enter the complex chain of interstellar reactions by forming metastable anionic states more efficiently and less transiently than benzene.

\subsection[Pathways to ortho-benzyne stabilization: computational findings]{Pathways to ortho-benzyne stabilization: computational findings}

\noindent In the previous sections \cite{carelli1}, we have investigated the quantum dynamics of low-energy electron scattering from gaseous o-C$_{6}$H$_{4}$, at its equilibrium geometry.
We have shown there that electron-benzyne single collision events give rise to resonant scattering states, since o-C$_{6}$H$_{4}$ shows four resonances under 10 eV one of which is around 1 eV, a fairly long-lived $\sigma^{*}$ resonance with clear antibonding character across the triple bond.
Such a metastable state, because of its longer lifetime and of its relatively small excess energy, has been considered by us a likely candidate which could play a crucial role in ring-opening paths, in specific bond cleavage (each of the CH bonds or specific CC bonds), or also in radiationless stabilization by intramolecular vibrational energy redistribution (IVR).
By assuming the formation of neutral ortho-benzyne in an astrophysical environment like a circumstellar atmosphere of a carbon star \cite{fangtong}, where there are also free electrons as suggested in \cite{woods2}, a ring-opening path would provide a negatively charged intermediate which could then initiate condensation reactions with the acetylene molecules of the same environment; furthermore, on the other hand, a stabilization of the whole negative ion would lead in principle to a highly reactive anionic species that can in turn react with carbonaceous cations (like C$_{2}$H$_{2}^{+}$ and C$_{4}$H$_{3}^{+}$), participating in a second aromatic ring closure toward the formation of the naphthalene structure.

\noindent In this section we thus want to focus on the investigation of such evolutionary channels of the metastable o-C$_{6}$H$_{4}^{-}$ by analysing its behaviour when the nuclear structure is deformed: in other words, we want to consider the possible elementary process where the excess resonant electron energy could couple with, and transfer into, the internal energy of the network of molecular bonds.
Whenever one specific bond can be reasonably considered as more directly responsible for energy transfers to the vibrational degrees of freedom, by selecting different one-dimensional (1D) pictures for the electron attachment process one should observe a corresponding shift in both the resonance energy position (the real scattering energy) and width (the imaginary energy contribution).
Since for a polyatomic molecule like the o-C$_{6}$H$_{4}$ system the dynamical coupling of the resonant electron with its vibrational degrees of freedom is a multidimensional process, to account for it as realistically as possible we carried out several scattering calculations using partially optimized molecular (nuclear) structures in which various specific molecular bonds were selected (and deformed) as cuts of a larger multidimensional molecular deformation path.

\begin {figure}
\begin {center}
\includegraphics[scale=0.23]{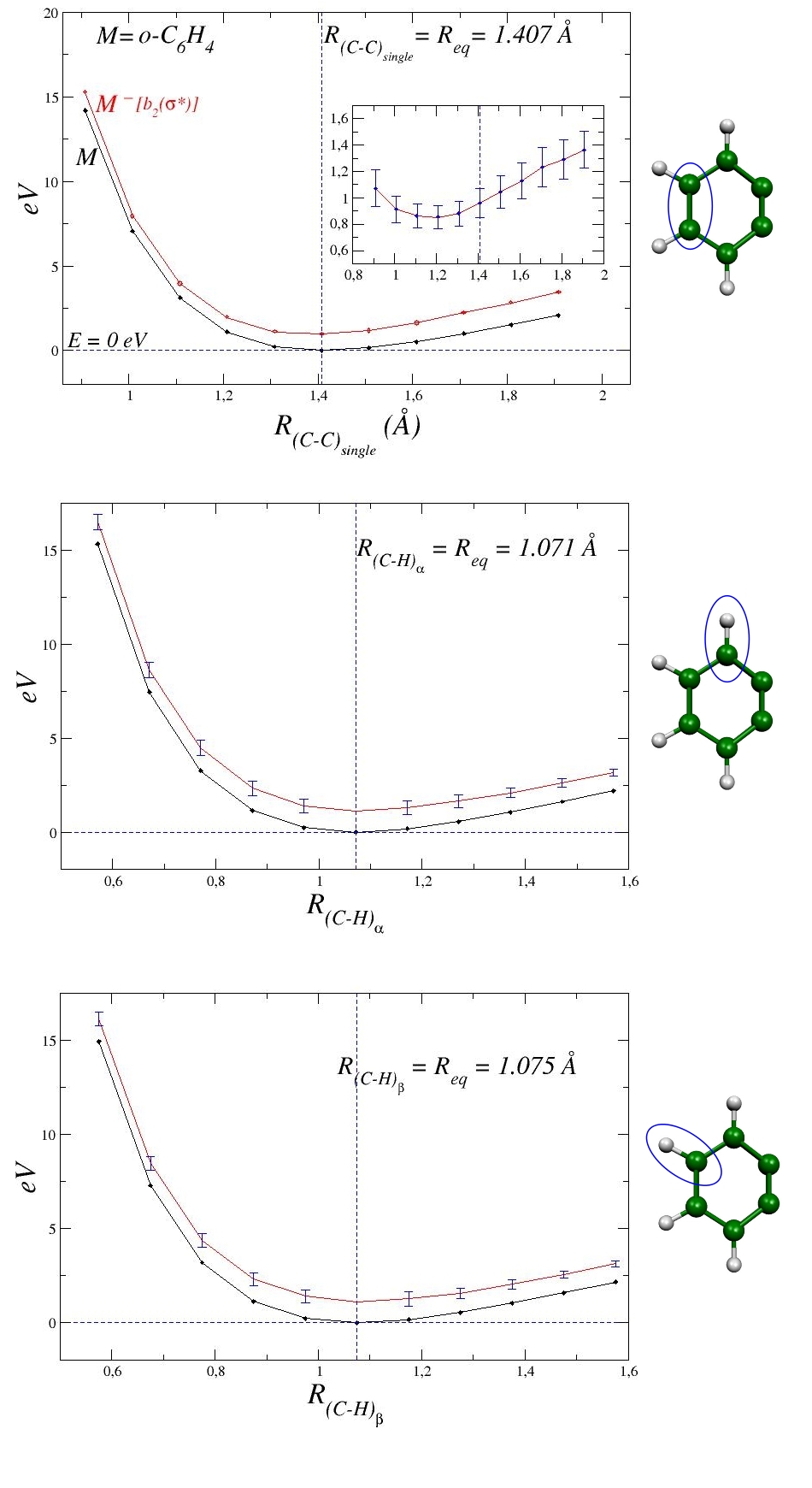}
\end {center}
\caption{\small{Upper panel: behaviour of the real part of potential energy curve for the resonant $\sigma^{*}$($b_{2}$) anion (o-C$_{6}$H$_{4}$)$^{-}$ (upper red curve) and for the neutral species o-C$_{6}$H$_{4}$ as a function of the single CC bond deformations (lower black curve). The small inset shows the computed changes of the resonance features ($E_{Res}$ and $\Gamma_{Res}$) as a function of the same single CC bond lenght. Central panel: the same as the upper panel, but for the CH bond located in $\alpha$ position respect the triple CC bond. Lower panel: the same as the upper panel, for the $\beta$ CH molecular bond. See main text for details.}}
\label{fig_ch3.1_08}
\end {figure}

\begin {figure}
\begin {center}
\includegraphics[scale=0.35]{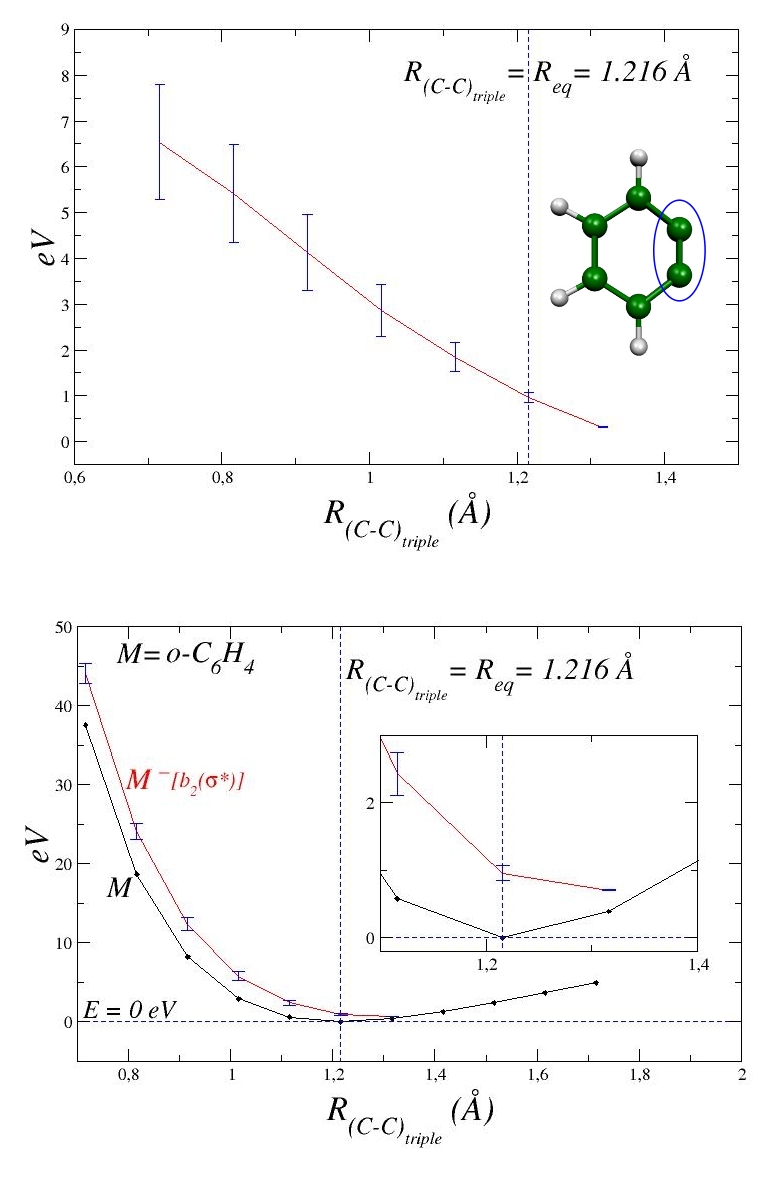}
\end {center}
\caption{\small{Upper panel: computed variations in the resonant features ($E_{Res}$ and $\Gamma$$_{Res}$) as the triple CC bond is deformed. The inset shows an enlargement around the equilibrium value for the triple CC bond. Lower panel: shape of our computed potential energy curves for the resonant anionic species (upper red curve) and for the corresponding neutral parent molecule (lower black curve) as functions of the triple CC bond deformations. See main text for details.}}
\label{fig_ch3.1_09}
\end {figure}

\noindent The calculations previously presented \cite{carelli1} have shown a partial localization of the extra electron as well as a clear presence of nodal planes across the triple and the single bond respectively.
Thus, one possible TNI evolution is given by the ring-breaking pathways which, in a simpler 1D model, can be seen as lengthening of either the triple CC bond or the single CC bond on the opposite side of the molecule.
A different pathway to stabilization is represented by the cleavage of each CH bond located respectively in the $\alpha$ and $\beta$ positions with respect to the triple bond, a process that can cause the loss of neutral hydrogen atom (forming a highly reactive closed-shell radical anion C$_{6}$H$_{3}^{-}$) or the detachment of a negative atomic ion H$^{-}$ (leading then to a highly reactive radical C$_{6}$H$_{3}$).
We have therefore computed the corresponding pseudo-1D potential energy curves associated with both the neutral N-electron target and the metastable ($N+1$)-electron negative ion (figures \ref{fig_ch3.1_08} and \ref{fig_ch3.1_09}), setting the energy balance by the following formula:

\begin{equation}
\label{1dpot}
E_{\bf Tot}(\bf R) = E_{Res}(\bf R) + E_{N}(\bf R) - E_{N}(\bf R_{eq}).
\end{equation}

\noindent In this expression $E_{N}(\bf R_{eq})$ is the electronic energy of the neutral parent molecule at its equilibrium geometry, computed at HF level with the $aug$-$c$-$pVTZ$ quality of the basis set, $E_{N}(\bf R)$ is the computed electronic energy (at the same level of accuracy as for the equilibrium geometry) for the neutral deformed molecule at a set of geometries identified by the involved single-bond coordinate $\bf R$, and $E_{Res}(\bf R)$ is the real part of the computed resonant electron energy over the same range of molecular geometries.
The corresponding widths $\Gamma(\bf R)$ of each resonance, associated with the ($N+1$)-electron states, are related to the complex scattering energy via the well known relation \cite{taylorbook}

\begin{equation}
\label{gamma}
E_{\bf Res}^{\bf complex}(\bf R) = E_{Res}(\bf R) + \mathrm{i}\Gamma_{Res}(\bf R).
\end{equation}

\noindent According to the energy-time uncertainty, the above widths are then indicative of the TNI lifetime at each computed molecular geometry.

\noindent We report in each panel of figure \ref{fig_ch3.1_08} the real part of our computed 1D potential energy curves when the single CC bond and each of the CH molecular bonds are deformed.
The black solid lines identify the Born-Oppenheimer potential energy curves for the neutral parent molecule, while the red ones are the real branches of potential energy for the metastable anion; the vertical bars refer to the imaginary resonant energy component ($\Gamma_{Res}(\bf R)$) for the deformed nuclear structures.
For each of the pathways reported in figure \ref{fig_ch3.1_08}, we can easily recognize a non-fragmenting behaviour for our metastable state: generally speaking, in fact, depending on the shape of the potential energy curve for the metastable anion the TNI itself can evolve either toward bond fragmentation or anion stabilization.

\noindent From the inset in the upper panel, containing the computed real and imaginary (vertical bars) resonance energies as a function of CC single bond deformation, we see that the extra electron is relatively weakly coupled with this nuclear motion and that it remains attached to the target molecule since the lifetime increases as the CC bond is lengthened.
A very similar situation characterizes the CH bond deformations located in $\alpha$ and $\beta$ positions, for which the extra electron also appears to be uncoupled with these 1D nuclear motions.
In sum, our analysis of each of the above deformations suggests that these evolutions of the nuclear dynamics yield a bound metastable negative ion where the extra electron seems to remain attached to the target molecule for at least one vibration.

\noindent In the case of the triple CC bond stretching (see figure \ref{fig_ch3.1_09}), the resonant electron is seen instead to be strongly coupled with this nuclear motion, as suggested by the steep downward behaviour of the real scattering energies when the bond is deformed (upper panel) moving from a rather compressed geometry to a slightly stretched bond with respect the equilibrium value for the neutral.

\begin {figure}
\begin {center}
\includegraphics[scale=0.30]{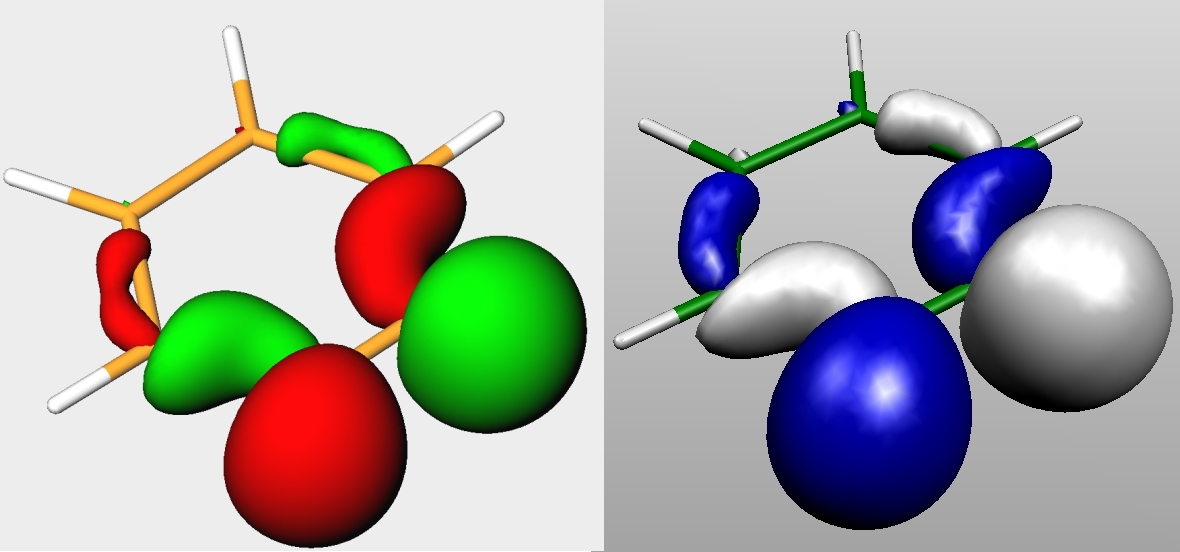}
\end {center}
\caption{\small{Right-hand panel: shape of the real part of the resonant wave function for the $\sigma^{*}$($b$$_{2}$) resonance associated with the +0.1\AA\ lengthening for the triple bond. Left-hand panel: DFT/cc-pVTZ computed HOMO of the ground doublet electronic state of the $stable$ (o-C$_{6}$H$_{4}$)$^{-}$ anion at its equilibrium. See main text for details.}}
\label{fig_ch3.1_10}
\end {figure}

\noindent Furthermore, the resonance disappears into a bound state for a deformation of $\sim$ 0.1 \AA\ beyond its equilibrium value: in particular we are able to locate the last resonance as a pole in the S-matrix.
Our findings, summarized in the lower panel of the same figure, indicate that a relatively small lengthening of the triple bond causes the potential energy curve associated with the metastable negative ion to cross the one of the neutral, and to move toward the stable anion.
To support this occurrence, accurate ab-initio calculations on the $stable$ negative ion (o-C$_{6}$H$_{4}^{-}$) indicate that the triple bond has to be stretched with respect to that of the neutral \cite{nash}.
We therefore report in figure \ref{fig_ch3.1_10} a comparison between the shape of the resonant wave function (on the right) associated with the last TNI geometry we found during C$\equiv$C bond lengthening, and the highest occupied molecular orbital (HOMO, on the left) that corresponds to the stable negative ion minimum energy structure: the figure clearly shows there is no difference between the two electron density maps.

\noindent We are thus encouraged to surmise that the residual amount of energy nedeed to form the stable anion (o-C$_{6}$H$_{4}$ has a positive electron affinity (EA) of about 0.56 eV, \cite{wenthold, leopold}) could actually be dissipated via fast radiationless IVR.
Such a stabilization mechanism is also indirectly supported by the calculations of such an internal conversion time involving the S$_{2}$$\rightarrow$S$_{1}$($\sim$20 fs) singlet electronic states for the pyrazine \cite{thanopulos}: the pyrazine molecule has in fact the same number of atoms as in ortho benzyne, thus a very similar density of vibrational modes likely to be coupled with the excess electron during an IVR process.

\subsection[Estimating the rate coefficients to electron attachment]{Estimating the rate coefficients to electron attachment}

\noindent The intense radiative flux from the central star is chiefly responsible for the ionization of the inner edge of the dense molecular envelope in CRL618 and therefore the energy distribution of the electrons created by photoionization is linked to the energy of the stellar photons.

\noindent In principle, low-energy electrons are more likely to recombine with protons (if available) and with molecular cations and thus could be selectively removed from the free electron pool.
However, electrons thermalize very quickly because of the very large electron-electron collision cross sections, so finally their recombination rates with molecules or atoms could be considered slow in comparison with the interaction among electrons, thereby allowing a local thermodyamic equilibrium (LTE) condition as a good macroscopic approximation for the free electron ensemble.
One therefore requires a single parameter (the free electron kinetic temperature, T$_{e}$), to characterize the free electron energy distribution, reasonably described by a Maxwell-Boltzmann distribution.
In this framework, the rate coefficient for the formation reaction of the metastable ortho-benzyne anion by free electron capture

\begin{equation}
\label{reaction} C_{6}H_{4} + e^{-} \rightarrow C_{6}H_{4}^{-}
\end{equation}

\noindent could be estimated by the convolution \cite{flowerbook} between our computed elastic integral cross sections and two Maxwell-Boltzmann energy distribution functions, one referred to T$_{e}$ $\leq$ 10000 K \cite{sanchez} and the other describing the molecular kinetic temperature, T$_{M}$ = 200 K \cite{cernicharo}.
Such calculation, for the above values of T$_{e}$ and T$_{M}$, yields a value of $k_{f}$ = 4.2 x 10$^{-9}$ cm$^{3}$ sec$^{-1}$.
At this point, we find crucial to emphasize the fact that recently the elementary gas-phase reaction of D1-ethynyl radical (C$_{2}$D) with vinylacetylene (C$_{4}$H$_{4}$) was studied under single collision conditions via crossed molecular beam experiments and electronic structure calculations \cite{fangtong}.
The corresponding findings point toward the formation of the neutral o-C$_{6}$H$_{4}$ which is in fact identified between all the products and for which the computations predict branching-ratios of up to 10 per cent \cite{fangtong}; the authors consequently argue that since the reaction path has no barrier, it might be accessed even in a cold molecular cloud like TMC-1, where the environmental conditions are deeply different than those of a circumstellar molecular envelope.
In this connection, we find intriguing to also estimate the metastable o-C$_{6}$H$_{4}$ anion formation rate coefficient at those environmental conditions, therefore assuming a much lower kinetic electron temperature, T$_{e}$ = 30 K; in this case the rate coefficient appears to be increased of one order of magnitude, being $k_{f}$ = 1.62 x 10$^{-8}$ cm$^{3}$ sec$^{-1}$.

\noindent Since we use here the elastic integral cross sections for the resonant scattering event, these findings represent an upper limit for the actual rate coefficient $k$; it is however interesting to see that such calculated estimates suggest fairly large rates originating from electron-molecule interactions \cite{herbst}.
Hence the combination of the above values for the metastable o-C$_{6}$H$_{4}^{-}$ formation rate coefficient with the possible fast radiationless stabilization of such a resonant anion, enables us to suggest that the o-benzyne (meta)stable anion might actually play a role in the circumstellar chemistry of CRL618 and more in general in the chemistry of either PPN atmospheres or cold molecular clouds, if the neutral parent molecule is actually produced in those environments.

\subsection[The stable o-benzyne anion: a quantum chemistry view]{The stable o-benzyne anion: a quantum chemistry view}

\noindent To further strengthen the computational exploration on the possible effects of ring-deformations and bond-deformations on the fate of the benzyne anion, we have carried our quantum structural calculations on the stable negative ion of o-benzyne.
The quantum calculations employed a Density-Functional-Theory (DFT) approach \cite{valiev} using a basis-set expansion defined by the well known acronym B3LYP/cc-pVDZ \cite{frisch}, generating all anionic wavefunctions as spin-polarized doublets; all "passive geometries" were optimized at each point of the deformation curves described below.
Energy convergence was tested within 10$^{-6}$ hartrees and the optimization convergences were based on the maximum and root mean square gradients, with accuracies of about 10$^{-4}$ hartrees \cite{valiev}.

\begin {figure}
\begin {center}
\includegraphics[scale=0.37]{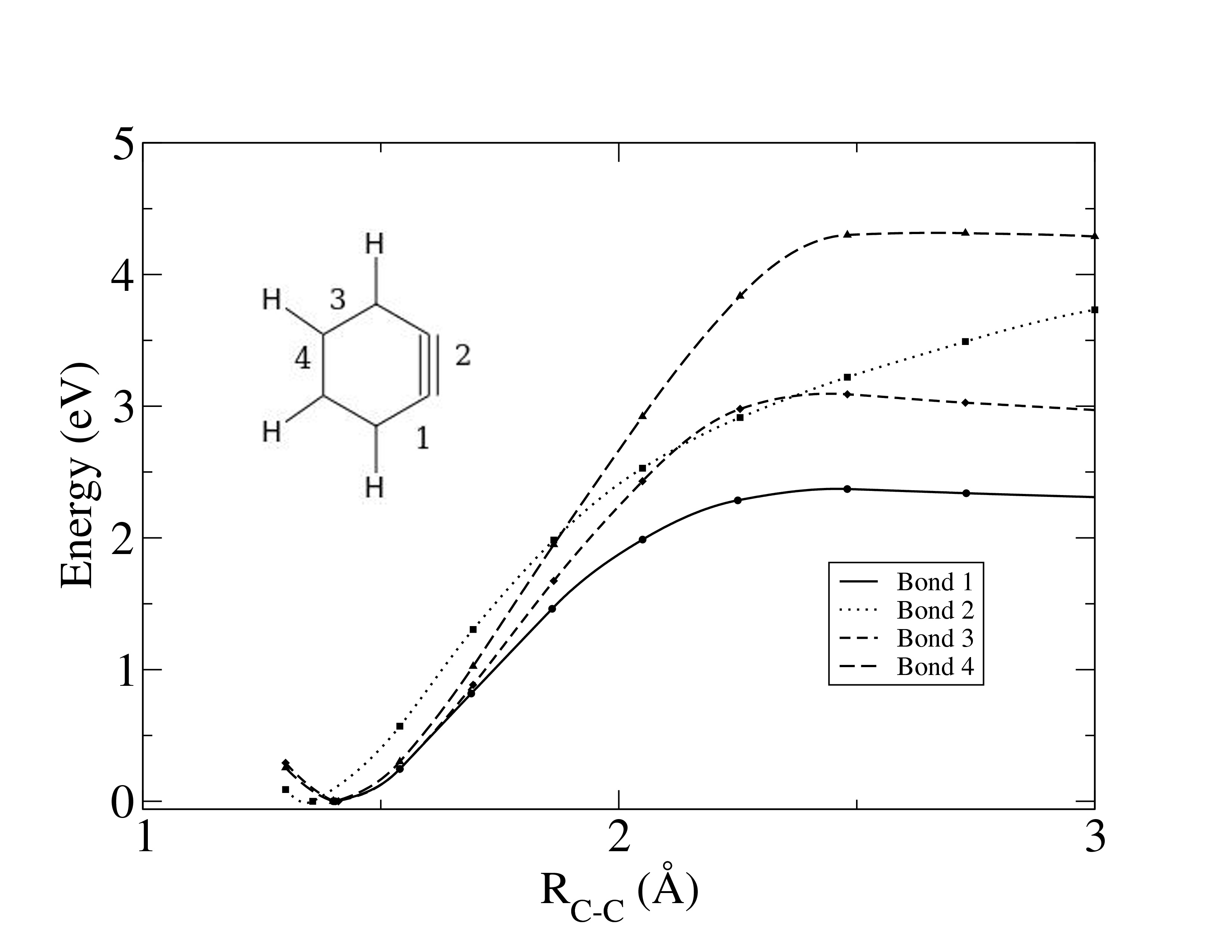}
\end {center}
\caption{\small{Computed reative energies changes of o-benzyne anion by stretching different CC bonds of the molecular ring. See main text for details.}}
\label{fig_ch3.1_11}
\end {figure}

\noindent The data of figure \ref{fig_ch3.1_11} report the energy changes upon the stretching, from the minimum geometry of the anion, of four different CC bonds, labelled according to the figure.
One clearly sees there that all stretching deformations present rather substantial energy barrier when (o-C$_{6}$H$_{4}$)$^{-}$ is deformed away its equilibrium geometry.
Hence, it is reasonable to argue that electron attachment stabilization would not be able to cause ring-opening processes as direct consequences of anion formation: the stable molecular anion, in fact, would need a substancial amount of excess energy in order to undergo such ring-breaking processes.
In conclusion, our suggestion that electron attachment stabilization paths would still leave unbroken the molecular structure is further supported by the model calculations of figure \ref{fig_ch3.1_11}.

\noindent It is also interesting to note here that the stable anion of the minimum structures of figure \ref{fig_ch3.1_11} is also an even more markedly polar molecule: the permanent dipole of o-C$_{6}$H$_{4}$$^{-}$, in fact, is computed here to be around 4.25 D, a value which is nearly three times that of the stable neutral molecule of o-benzyne.

\subsection[Present conclusions]{Present conclusions}

\noindent The present theoretical investigation has been directed at the computational analysis of the possible role which could be played by the ortho-benzyne molecule as a reactive intermediate in proto-planetary atmospheres.
In particular, we have studied the quantum dynamics of the scattering of low-energy (< 10 eV) electrons off gaseous o-C$\mathrm{_{6}}$H$\mathrm{_{4}}$ in order to verify its feasibility for forming long-lived negative ions as "doorway states" which, after dissipating the extra energy content toward the thermodynamically stable negative ion, can evolve through chemical reactions participating to the complex reactions chain responsible of the formation of PAHs species.

\noindent We find it important to emphasize that the present theoretical investigation on this particular aromatic system was encouraged by the fact that o-C$_{6}$H$_{4}$ molecule was already conjectured in previous experimental work \cite{frenklach2, cherchneff, mcmahon} to be a possible intermediate during the benzene formation in several astrophysical environments. 
In this connection, therefore, in the present work we have endevoured in order to tentatively analyse the possibility of an actual negative ion chemistry involvement in the astrophysical PAHs synthesis, where the anionic o-benzyne, due to the positive EA \cite{wenthold, leopold} that characterizes the parent neutral species, could be consequently viewed as a reasonable important aromatic precursor of larger species belonging to this family.

\noindent All the calculations have been carried out by treating the scattering events within a time-independent formulation of the problem, solving the ensuing multichannel equations within a SCE description of the scattering states and via a model, local potential for the interaction between the low-energy electrons and the benzyne molecules, as described with sufficient detail in chapter 2.

\noindent When considering the o-C$\mathrm{_{6}}$H$\mathrm{_{4}}$ as frozen at its equilibrium geometry, the results indicate that this molecule indeed exhibits four fairly narrow resonances below 10 eV, three of which can be classified as $\pi^*$ resonances while one of them is a low-energy $\sigma^*$ resonance exhibiting antibonding character across the triple bond.
Hence, we single out this resonance as the most likely dynamical mediator state which could in principle lead to ring-opening processes.
A comparison with similar scattering resonances which involve the benzene species (C$\mathrm{_{6}}$H$\mathrm{_{6}}$), indicates instead that the latter shows only fairly broad metastable states which start to appear above about 4.66 eV.
Our conclusions, therefore, rest also on having observed marked differences of behaviour between the more stable resonant states of o-benzyne and those appearing at higher energy in benzene, a species considered to be the corner-stone step for the chain of reactions leading to PAH formation.
As discussed in the previous sections, however, this comparison should be grasped only in the qualitative sense.
In this connection, in fact, one should note that computational parameters like the number of partial wave for the incoming electron description and the number of interaction potential terms employed for the benzene and o-benzyne investigations are little different; due to the fact that the latter has been studied with the same approach but several years later, nowadays a larger partial wave expansion like that used for the o-benzyne can be easily used since it is not so much demanding from the computational point of view.
Moreover, we find it important to emphasize that the present calculations as those on the benzene molecule employ a realistic, but approximate, form of interaction, and therefore possible experimental resonances would very likely be shifted in energy with respect to our estimates.
However, since for both the above aromatic molecules we have used the same theoretical approximated approach, according to the last consideration, we expect that the use in both cases of more realistic, improved model potentials would not modify our comparative conclusions.
On the other hand, we know that the benzene molecule has a large negative electron affinity (-1.12$\pm$0.003 eV, \cite{burrow}), so we think that even surmising the formation of metastable negative ions for this molecule by low-energy electron resonant attachment, it seems unlikely their actual involvement in such an astrochemical reaction network.

\noindent In the framework of the above qualitative comparison, we can therefore collect the following useful information.
Since the excess energy carried by the extra electron has to come from the environmental medium in the atmosphere, the attachment processes to benzene do require much higher electron temperatures than those needed for the formation of metastable anions of benzyne.
Furthermore, the (o-C$\mathrm{_{6}}$H$\mathrm{_{4}}^{-}$)$^{*}$ temporary negative ions are seen to have markedly longer lifetimes, thereby allowing more easily the dynamical coupling of the extra electron with the nuclear degrees of freedom which play a crucial role in the ensuing possible molecular break-up or also in possible non-dissociative stabilization paths.
It therefore follows that our present calculations for the equilibrium structure (sections 3.1.3, 3.1.4, 3.1.5) support the picture of an intermediate anionic formation for the o-benzyne species that could then rapidly decay either by ring-opening reactions, a feature which therefore could prevent the actual detection of the parent neutral species by observational spectroscopy \cite{widicus}, or by non-dissociative stabilization mainly due to intramolecular vibrational redistribution processes.
Hence, we surmise that the possible presence of neutral o-C$\mathrm{_{6}}$H$\mathrm{_{4}}$ in proto-planetary atmospheres like CRL-618, where free low-energy thermalized electrons might be also present, could help in producing the negative, metastable species that in turn could then be likely to contribute to the chain of reactions relevant for PAHs formation.
Admitting in fact the above hypothesis, we find reasonable that such reactions could thus be easily triggered also by the anionic o-benzyne molecule than by benzene itself, a feature that could then explain the observation of the latter neutral species \cite{cernicharo} with respect to the absence of the former neutral one \cite{widicus}.
At this point it is useful to emphasize that, currently, two different mechanisms for the benzene synthesis in proto-planetary stellar atmospheres were developed, one involving neutral and cations \cite{woods} and the other being radical-based \cite{frenklach2}, where only the latter directly accounts for the possible presence and production of neutral o-benzyne; in connection with this, as already illustrated in the introduction (sections 3.1.1 and 3.1.2), since the above mechanisms are associated with rather different physical conditions, the formation of anionic o-benzyne could be viewd as a first step in drawing a new and more complex reaction network for benzene and more in general for PAHs formation in proto-planetary atmospheres: no chemical models, in fact, include both the radical- and the ion-based chemical networks, so that the lack of a single model makes the observational results quite difficult.\\

\noindent After having shown that metastable anionic formation for o-C$\mathrm{_{6}}$H$\mathrm{_{4}}$ is a likely option in the circumstellar environment, a feature that would therefore suggest its involvement in PAHs formation in C-rich proto-planetary atmospheres like CRL-618, we consequently started to investigate further aspects of this problem as discussed in sections 3.1.6, 3.1.7 3.1.8, where our findings about the possible dynamical evolution once the metastable anion is formed are illustrated and discussed with full particulars.

\noindent The pseudo-1D quantum dynamics which we have employed to study the above process, as discussed in section 3.1.6, has indeed shown: 

\noindent (i) the formation of a long lived metastable anion at fairly low electron energies, 

\noindent (ii) the stabilization of the initial molecule by small ring deformations which cannot lead to ring-breaking effects, and 

\noindent (iii) the formation of a bound anion by small direct stretching of the triple bond.

\begin {figure}
\begin {center}
\includegraphics[scale=0.09]{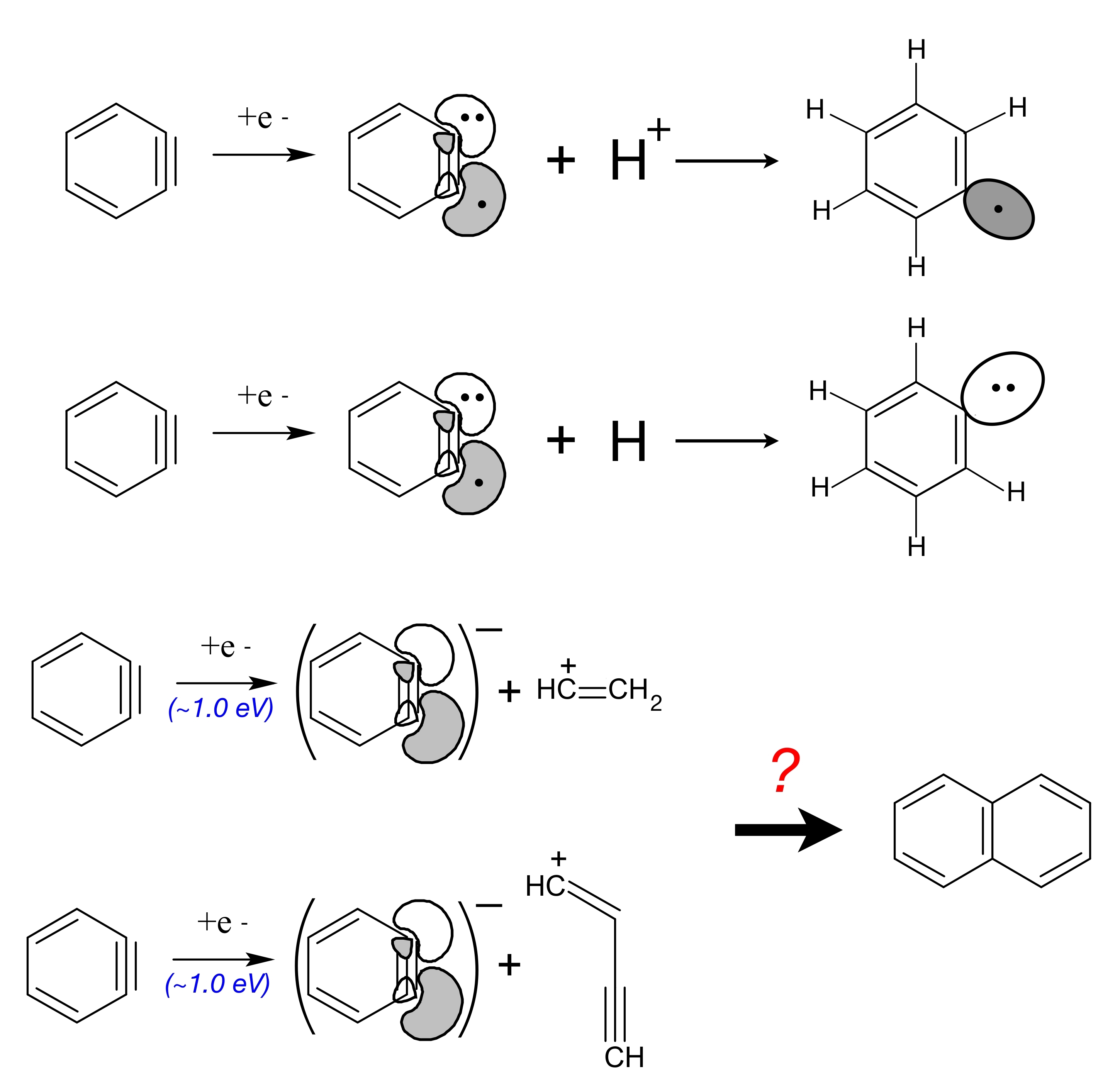}
\end {center}
\caption{\small{Surmised schematics of the gas-phase ionic reactions involving the benzyne anionic intermediate and either ionized/neutral hydrogen or polyynic cations.}}
\label{fig_ch3.1_12}
\end {figure}

\noindent The last step of the quantum dynamics futher indicates the possibility of reaching the bound (o-C$_{6}$H$_{4}$)$^{-}$ ground state by fast internal rearrangement of vibrational states of the benzyne anion.
The above conjecture, proven to be realistic by our quantum scattering calculations, could then allow the very reactive, anionic radical to participate in reactions with $H$ and/or $H^{+}$, and also in possible condensation reactions with acetylene and acetylene-like cationic species (see figure \ref{fig_ch3.1_12}) or even with the corresponding neutrals of the same molecules, hence explaining its negligible presence within column density measurements.

\noindent The cationic molecular partner surmised in the coarse outline of figure \ref{fig_ch3.1_12} are two of the most likely abundant carbonaceous species in the proto-planetary atmosphere of CRL-618 \cite{woods2}.
One should note, in fact, that A$^{+}$ + B$^{-}$ neutralization reactions have been estimated to have very large reaction rates at the conditions of interest ($k$ $\sim$ 10$^{-7}$ - 10$^{-8}$ cm$^{3}$ sec$^{-1}$ \cite{herbst}, and therefore could be very efficient initiators of ring-addition reactions as those outlined above.

\noindent An important observation to bring to a conclusion the present section.

\noindent As introduced and qualitatively discussed in chapter 2 (sec. 2.2), the FN nuclei approximation for polar molecule like the o-benzyne molecule, is not the best one.
In the framework of this approximation, in fact, it can be shown that the ICS diverges logarithmically due to the slow convergence in the $T_{\ell,\ell'}$ matrix elements which dramatically affects the large $\ell$,$\ell'$ range (where the static dipole interaction dominates).
More precisely, for polar molecules the divergence of the elastic integral cross section for vanishing energies comes from the divergence of the differential cross section in the the forward direction, which thus in turn causes the ICS to become infinite\cite{altshuler}.

\noindent From a practical point of view, this means that one should be aware of the fact that the larger $l$ value, the more trustworthy the computed ICSs are.

\noindent In the present investigation we have used $l$ = 40 as the maximum partial waves number, which corresponds to 441 coupled channels considered for the ICS calculations, where the $l$ = 60 value used for the necessary convergence tests had no significant effect on locating the resonances (see sec. 3.1.3).
Therefore, since the rate coefficient for the TNI (o-C$_{6}$H$_{4}^{-}$)$^{*}$ formation is numerically estimated as the integral of our computed elastic ICSs weighted by the Maxwell-Boltzmann distribution function

\begin{equation}
k(T_{e}) = \sqrt{\frac{8\:K\:T}{\pi\:\mu}} \frac{1}{(K\:T)^{2}} \int_{0}^{\infty} \sigma(E)\:E\:e^{\frac{-E}{K\:T}}\:dE,
\end{equation}

\noindent our rate coefficient estimates appear to be reasonably large and physically meaningful already when referring to ICSs computed with an $l$ = 40 partial wave expansion.
Furthermore, although the logarithmic divergent behaviour in the $a_{1}$ partial contribution to the elastic ICS, due to the Boltzmann exponential, at most we expect that using a greater partial waves number for the ICSs calculation could then give rise to little larger rate constant values, therefore again supporting our conclusions.

\noindent In any case, we have planned for the near future to calculate the ICS (and also the momentum-transfer cross sections, MTCS) for the present molecule using an improved approach which was already used in the past by our group \cite{sanna}.

\clearpage

\section{The coronene molecule, C$_{24}$H$_{12}$}
\label{coronene_res}

\subsection[Introduction: PAH anions in the interstellar medium. Role of C$_{24}$H$_{12}$ anion]{Introduction: PAH anions in the interstellar medium. Role of C$_{24}$H$_{12}$ anion}

\noindent Polycyclic aromatic hydrocarbons (PAHs) are a family of planar molecules, consisting of carbon atoms arranged in a typical honeycombed lattice structure of fused six-membered rings, typically characterized at the external edges by C-H sigma bonds.

\noindent Due to their four valence electrons, each carbon atom in polycyclic aromatic structure forms three covalent $\sigma$ bonds with C (and H) atoms, thus usually leading to a planar structure; the remaining electron, located perpendicularly to the molecular plane in an hydrogen-like p-orbital, overlaps with the other adiacent electrons (one per C atom) to form $\pi$ molecular orbitals, which can be pictorically represented as diffuse delocalized electronic clouds above and below the molecular plane, respectively.
This peculiar electronic structure of PAHs enables to introduce the general rule according which, qualitatively speaking, the more extensive the delocalized electronic cloud, the more stable the molecule.
Moreover, because of the typically large binding energy of aromatic carbon atoms in their planar exagonal lattice, neutral and singly charged cationic PAHs are known to be particularly stable among the family of organic compounds: they need, in fact, large amount of energy to be photodissociated (\cite{tielensbook}), so that they are currently considered to be potential carriers for at least a subset of the absorption diffuse interstellar bands (DIBs, \cite{zwet, crawford}), as well as reliable candidates to efficiently convert the absorbed energy in infrared emission (the so-called unidentified infrared bands, UIBs \cite{giard, allamandola2, hudgins, bakes}).

\noindent Besides the fact that the history of PAHs in the Universe has been a real fascinating saga, nowadays these organic molecules are considered to be almost ubiquitous in the cosmos: free gas-phase PAH molecules in different charge and hydrogenation states, in fact, are commonly considered to be an important component of the interstellar medium (ISM) \cite{leger, allamandola, henning}, of diffuse molecular clouds \cite{bakes2}, of dense molecular cores \cite{wakelam}, as well as of circumstellar environments (like the planetary nebula NGC 7027, \cite{peeters}).
Unfortunately, although significant and intensive astronomical searches have been dedicated to PAHs, their detection has always been elusive.
Only tentative identifications of neutral benzene \cite{cernicharo}, pyrene and anthracene \cite{vijh} were proposed in the last years; the first was detected toward the protoplanetary nebula CRL 618 since its $\nu_{4}$ bending mode was observed, while the last two species (containing respectively three and four fused aromatic exagonal C-rings) are considered to be likely responsible for the blue fluorescence observed in the Red Rectangle Nebula.

\noindent Through astronomical observations, laboratory spectroscopy, computational characterization as well as increasingly detailed modeling of several and different astrophysical contexts, during the last 25 years it was possible to unfold their multiple role in the chemistry of the interstellar medium (ISM), and more interestingly, to surmise that most of interstellar processes involving PAHs requires a detailed knowledge of their charge distribution.

\noindent One of the first important attempts to model the chemical evolution of PAHs and the modifications to interstellar chemistry that consequently would result from the presence of such 'large' molecules in the interstellar gas, was provided by Lepp \& Dalgarno (\cite{lepp}).
In such a study, the authors concluded that for PAHs consisting of molecules with 30-50 carbon atoms(\footnote{thus little larger than coronene}), with fractional abundances greater that 10$^{-8}$, the corresponding negatively charged species, PAH$^{-}$, would be the main source(\footnote{in the sense of 'carriers'}) of negative charge.

\noindent It has then been tentatively shown that, both in dense \cite{wakelam, draine3, weingartner} and in diffuse \cite{bakes2} clouds, electron attachment to neutral PAHs and to nanoscopic grains can have significant effects not only on the free electron density: the formation and the possible presence of PAH negative ions in those environments have been in fact determined to be particularly important in relation to the ensuing recombinations involving positive ions, both atomic and molecular.
Conversely to what usually occurs for 'direct' recombinations between free electrons and positive ions, one should in fact note that, in general, the recombination processes involving negatively charged PAHs are less exothermic, due to the PAH's electron affinity \cite{batesbook}.
In this connection, when considering recombinations between $molecular$ cations and negatively charged PAHs, the latter could be viewed as actual 'energetic shock absorbers' in the sense that some of the recombination exothermicity can be efficiently stored in the internal energy of the anionic PAH species itself (due to their typically large number of vibrational degrees of freedom), so that less is available to dissociate the positive molecular ion upon neutralization.
On the other hand, specifically in relation to positive atomic ions(\footnote{as distinct from the molecular cations}), negatively charged PAHs could instead act like 'catalysts' in the neutralization reactions: beside the dielectronic recombination, the recombination processes of $atomic$ cations with free electrons present in the surrounding gaseous medium, in fact, are mediated by the emission of a photon, so that they are intrinsically very slow processes.

\noindent Accordingly, the presence of PAH anions in photodissociation regions (PDR) influences the C$^{+}\;\rightarrow\;$C$\;\rightarrow\;$CO transitions \cite{lepp2}, as investigated by \cite{bakes}.

\noindent The presence of PAH negative ions has also repercussions on the ionic (atomic and molecular) abundances for other important and widely diffused interstellar species like H$^{+}$, H$_{3}^{+}$ and HCO$^{+}$: the importance rest on the fact that the abundances of these ions directly affect the ortho:para H$_{2}$ ratio, which in turn has very important repercussions on the ensuing global chemical evolution (\cite{flower3}), since for example it plays a crucial role either in the kinetics of specific excitation processes (like the reaction between ortho/para molecular hydrogen and sulfur dioxide, \cite{cernicharo2}), or in establishing the ortho:para ratios for other species, including the deuterated forms of H$_{3}^{+}$ (\cite{flower3}).

\noindent As furthermore suggested by recent theoretical and laboratory studies (\cite{cecchi, steglich}), aromatic species like PAHs show strong UV absorption in two broad bands centered at 200 nm ($\pi^{*}$$\rightarrow$$\pi$ excitation, $\sim$6 eV) and 70 nm ($\sigma^{*}$$\rightarrow$$\sigma$ excitation, $\sim$17.7 eV) respectively, thus they are considered to be good candidates to explain the 217 nm (5.7 eV) bump as well as the far-UV rise of the interstellar extinction curve (ISEC, \cite{mulas2}).
Moreover, since PAHs absorb efficiently dissociating photons ($\lambda\:\leq$ 100 nm, $E$ $\geq$ 12.4 eV), they are considered to be responsible (at least partially) in controlling the atomic fraction obtainable by photodissociations and, in this sense, the global degree of excitation of interstellar/circumstellar clouds.
This UV screen could be also increased if PAHs are more abundant at cloud edges (limb brightening effect, \cite{boulanger, bernard, rapacioli, berne, velusamy}).

\noindent In this framework, the role of the charge state of PAHs in ultraviolet extinction has been the object of some recent theoretical investigations \cite{cecchi}.
More in detail, the authors of \cite{cecchi} use a specific extinction model \cite{cecchi2, iati} where the PAH contribution to extinction \cite{iati} is represented by two Lorentz profiles, one for the $\pi$$\rightarrow$$\pi^*$ electronic transition accounting for the bump feature and the other for the $\sigma$$\rightarrow$$\sigma^*$ plasmon resonance whose low-energy tail contributes to the far-UV rising feature in the interstellar extinction curve (ISEC).

\noindent The crucial point of that work is that it shows how a combination of classical dust particles and mixtures of real PAHs satisfactorily matches the observed ISECs, where variations of the spectral properties of PAHs in different charge states (including the anions), produce changes consistent with the varying relative strengths of both the bump and the non-linear far-UV rise features \cite{cecchi, iati}.

\noindent Recently, the role of neutral and anionic PAHs has been also the object of some investigations focused on the fractional ionization in dark molecular clouds \cite{flower2}.

\noindent In this connection, even in absence of intense external sources of radiation and of the background interstellar field, thus admitting a very efficient screening by the surrounding PAHs(\footnote{and, of course, by H$_{2}$ and by the solid phase represented by the dust particles}), a not negligible amount of high-energy free electrons and UV ionizing radiation can be 'locally' produced due to the interactions of high-energy cosmic rays (CR) mainly with H$_{2}$ \cite{prasad}; it was in fact demonstrated as a very likely possibility that the following ionization reactions

\begin{eqnarray}
H_{2} + CR \rightarrow H_{2}^{+} + e^{-} \\
H_{2} + CR \rightarrow H^{+} + H + e^{-}
\end{eqnarray}

\noindent do produce highly energetic secondary electrons (\cite{cravens}, with mean kinetic energy of about 30 eV), which in turn are able to cause the excitation of the Lyman, Werner and the other bands of the H$_{2}$ molecules present in the same medium, so that in the ensuing energetic cascade toward the X$^{1}\Sigma_{g}^{+}$ ground electronic state for the molecular hydrogen, the produced UV photons can be finally able to either ionize or dissociate other molecular species.
Despite the apparent complexity, which depends on the abundance ratios between molecular and atomic (neutral and ionized) hydrogen, on PAH abundances and sizes, on the dust phase represented by PAH clusters and carbonaceous nanoparticles, as well as on the other molecular species (both neutral or ionized) present in the same environment, the degree of ionization for a dark molecular region like TMC-1 has been tentatively estimated.
Such a complex mechanism, in fact, has been studied in connection with the determination 
of the C$_{n}$H$^{-}$/C$_{n}$H ratios, with n = 6, 4, (8), because all of these anions have been recently discovered (C$_{6}$H$^{-}$ and C$_{8}$H$^{-}$ in TMC-1 and in the C-rich dense envelope of IRC+10216 \cite{mccarthy, brunken, remijan}, while C$_{4}$H$^{-}$ only in IRC+10216, \cite{cernicharo3}): even if currently there are no direct observations of negatively charged PAHs in the interstellar space, since the electron affinities of both C$_{4}$H and C$_{6}$H \cite{taylort} are larger than those of some medium-size PAHs \cite{modelli} including the C$_{24}$H$_{12}$ whose EA is $\sim$ 0.5 eV \cite{chen, duncan}, the above chemical model \cite{flower2} has been recently investigated in order to take in consideration also the possible electron transfer from PAH$^{-}$ to the above neutral linear polyynes for (n = 6, 4).
The results have led the authors to surmise that, for such an astrophysical environment, the key factors in determining the C$_{n}$H$^{-}$/C$_{n}$H ratios (n = 6, 4) are the free electron density and the density of atomic hydrogen: a high fractional abundance of PAHs, i.e. a low fractional abundance of free electrons, yields in fact a qualitatively rather good agreement with the observations.

\noindent When specifically considering dense interstellar environments like dense molecular clouds, several observations suggest that the abundances of smaller PAHs (N$_{C}$ < 30) are decreased \cite{boulanger, bernard, rapacioli}.
According to the absence of UV photons, which in fact are thought to be not able to efficiently penetrate those regions, Rapacioli and his coworkers \cite{rapacioli, rapacioli2} suggested that small-size gas-phase PAHs could form larger PAHs just in such dense regions; in line with that, therefore, smaller free-flying PAHs observed at the border of illuminated clouds may be formed by photodissociation of these larger species, provided an intense UV source exists in the surroundings.
Despite the fact that many spectroscopic observational indicators about the PAHs presence almost all over the ISM do exist, no specific types of PAH responsible of the previously mentioned features (UIBs, DIBs) has been unambiguously identified. 
The form and intensity of the spectra, in fact, strongly vary from one region to another \cite{draine4}, hence suggesting that differences in sizes and abundances are present; moreover, one should also note that the actual existence of PAHs in dark molecular clouds is difficult to be established observationally, since their fractional abundance is surmised to depend on the extent of their accretion onto the nanoscopic solid grain particles \cite{flower2}.
As a consequence, the aggregation process responsible for cluster and nanoscopic carbonaceous grains formation in the above regions is still poorly understood, and the abundance of smaller gas-phase PAHs remaining screened by UV radiation in such dense regions, as well as the time needed for the aggregation process to occur, remain currently undetermined.
Furthermore, since the general constraints on the PAH size and abundance, both as functions of the physical conditions typical of dense interstellar regions, are currently weak, it therefore follows that most of the detailed gas-phase chemical models of dense molecular clouds do not $expressly$ include PAHs in the sense that PAHs are not treated as individual species.
In this connection, the recent work of Wakelam and Herbst \cite{wakelam} constitutes an exception: in such an interesting work, in fact, in order to analyse the role of PAHs in dense cloud chemistry (like TMC-1 or L134N), the authors explicitly include free medium-size PAHs with a mean radius of 4\AA\ (N$_{C}$ = 30) and a fractional abundance with respect $n_{H}$ of 3 x 10$^{-7}$; due to the fact that PAHs may not be free in dense cold environments, they also include PAH clusters of radius 10 \AA\ (N$_{C}$ $\sim$ 470) and 100 \AA\ (N$_{C}$ $\sim$ 4.7 x 10$^{5}$) having proportionally lower fractional abundances (2 x 10$^{-8}$ and 2 x 10$^{-11}$, respectively).
In order to investigate the possible chemical scenario at a fixed temperature and density, the authors of the above work include electron attachment processes to neutral PAHs, recombinations of PAH anions with smaller positive ions and photodetachment processes of negative PAH ions triggered by UV radiation.
The PAH anions formation is treated as a pure $radiative$ electron attachment

\begin{eqnarray}
PAH + e^{-} \rightarrow PAH^{-} + h\nu \nonumber
\end{eqnarray}

\noindent whose rate coefficient, according to \cite{lepage}, is surmised to be $k_{e}$ = 1.3 x 10$^{-6}$ cm$^{3}$/s for the attachment of electrons to neutral gas-phase PAHs containing 30 carbon atoms.
We find it important to emphasize that the authors themselves, originally, pointed out that their results depend on the above value of $k_{e}$ only if it is decreased by at least two order of magnitude.\\

\noindent So far, I have summarized some between the recent and most important models concerning neutral and negatively charged PAHs in order to provide a short survey(\footnote{absolutely not comprehensive!}) about their physical and chemical role in different astrophysical environments.

\noindent Currently, in fact, it is a very demanding task to realistically include PAHs in the chemical modeling of different regions of ISM, including dense molecular clouds and diffuse clouds.
Due to the fact that PAH molecules occupy a middle ground in size between small gas-phase species and dust particles, they are consequently seen as an 'intermediate' phase among the molecular gas components and the solid dust phase of the interstellar matter.
It therefore follows that in order to include PAHs properly, as suggested in \cite{wakelam}, it should be necessary to look at them as individual species, but the single aromatic species that are surmised to make up the PAH content of interstellar clouds currently are not constrained enough from spectral observations and laboratory measurements.
Consequently, as emphasized by the authors of \cite{wakelam}, the inclusion of PAHs in a realistic chemical model at present means that one is obliged to treat them as small grains and thus follow only their charge state.

\noindent The present theoretical investigation on metastable C$_{24}$H$_{12}$ anions formation by electron resonant collisions should therefore be taken into account just in connection with the above general comments: the coronene molecule, in fact, with an equilibrium structure of six aromatic rings surrounding a central exagonal C-ring and with a positive electron affinity ($\sim$ 0.47 eV, \cite{duncan}) whose value is comparable with that for two neutral tentatively observed PAHs(\footnote{pyrene and anthracene, EA = 0.41 eV \cite{malloci} and 0.53 eV \cite{song}, respectively}), could be viewed as an important and representative medium-size PAH among the surmised interstellar aromatic molecules. 
Keeping in mind that the electron affinity is essentially indicative of the anion $thermodynamical$ stability, looking for the possible presence of long-lived resonances in an energy range which agrees with the global environmental astrophysical conditions, might be helpful in characterize the feasibility of this PAH to efficiently form stable anions by electron attachment and, more in general, its feasibility to react with cold electrons.
Furthermore, when considering the possible presence of C$_{24}$H$_{12}^{-}$ in an interstellar/circumstellar region as mainly produced by resonant electron collisions, some experiments \cite{abouaf, denifl, khakoo} show that the undissociated anion production takes place for colliding particles having energy very close to threshold while the vibrational excitation by electron impact should occur well below 10 eV.
Since this energy range appears to be reasonable also for the general ISM conditions, (ranging from dense/dark clouds to diffuse clouds and hot circumstellar envelopes), we find useful to make a comparison between the experimental results and our findings in order to better characterize the behaviour of this molecule in an astrophysical environment where also free electrons are.

\noindent In the interstellar space, in fact, free electrons represent an alternative way to transport energy; when compared to photons in the non-relativistic energy range, they have a much lower energy/momentum ratio, hence they can correspondingly carry much more momentum.
In other words, although energy-exchange between the ionized gas, the neutral gas and the dust component might be mainly radiative (although still depending on the environmental constraints), free electrons can also play a significant role because the neutral molecular gas can be heated by energy-transfer electron collisions.
Moreover, in a partially-ionized plasma, the electron transport is primarily governed by the momentum-transfer cross-section $\sigma_{m}$ for elastic scattering of the present electrons which is defined as

\begin{equation}
\label{sigmamt}
\sigma_{m} = 2\pi \int_{0}^{\pi} (1-cos\theta) \: \frac{d\sigma}{d\Omega} \: sin\theta \: d\theta,
\end{equation}

\noindent $d\sigma$/$d\Omega$ being the elastic differential cross section (DCS). 
Since experiments show the coronene molecule to be very ''reactive'' under low- and very low-energy electron collisions [20,21], here we want to investigate the angular redistributions that the impinging particles undergo after the collisional event with the title molecule. 

\noindent Therefore, in the framework of the multiple roles played by PAH molecules in interstellar or circumstellar environments, we find important to provide realistic estimates of the elastic DCS for electrons scattered off gas-phase coronene molecules.
The angular redistributions of scattered electronic particles can in fact affect other physical processes: under the application of an external electric field (originated by radiowaves propagating through the plasma), electrons start to move away from the field direction where the degree of deviation of the electron trajectory is determined, at least qualitatively, by the momentum-transfer which has occurred during the collisional event.
This means that the propagation (both reflection and attenuation) of radiowaves in a plasma can be considered as controlled by $\sigma_m$ [22] in the sense that, generally speaking, the larger $\sigma_m$ the larger will be the deviation of the scattered electrons from the field direction, whereby the global electron motion would contribute less to the conductivity of the partially ionized gases.

\clearpage

\subsection{The partial and total elastic cross sections at low energies}

\begin {figure}[here]
\begin {center}
\includegraphics[scale=0.4]{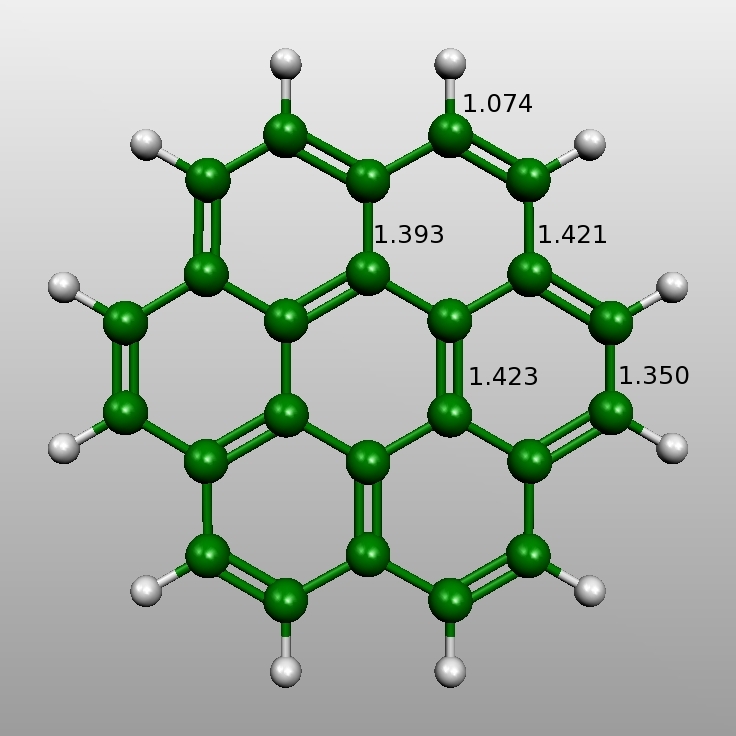}
\end {center}
\caption{\small{Coronene otpimized equilibrium structure at 6-31++G** level}} 
\label{fig_ch3.2_01}
\end {figure}

\noindent We start our analysis of the electron scattering on gas-phase coronene by assuming that the C$_{24}$H$_{12}$ molecule is in its ground electronic state as well as in its equilibrium geometry.
The neutral closed shell target wave function was obtained using first an expansion with Gaussian functions centered on the nuclei of the molecule using a 6-31++G** basis set.
The total SCF energy provided by calculations was $-915.992787$ hartrees.
The equilibrium geometry for the coronene molecule obtained with the above Gaussian basis set is reported in figure \ref{fig_ch3.2_01}.

\noindent According to the molecular symmetry point group to which the coronene molecule belongs in its equilibrium structure, the full nonspherical nature of the interaction potential with the incoming free electron was included in the present scattering calculations using the $D_{6h}$ symmetry point group.
In order to include in the interaction potential the long-range polarizability contributions we used the model correlation-polarization potential $V_{cp}$ with the total asymptotic tensorial polarizability of 105.928, 343.560 and 343.560 $a_{0}^{3}$ ($\alpha_{xx}$, $\alpha_{yy}$ and $\alpha_{zz}$, respectively) calculated by us.
To provide the relevant scattering K-matrix elements, the parameter-free calculations at the adiabatic-static-model-exchange-correlation-polarization (ASMECP) level were carried out in the molecular frame of reference by solving the close-coupled equations where we have employed a partial wave expansion up to $l_{max}$ = 60 both for the bound molecular orbitals and for the scattering wave functions, so that the corresponding expansion of the static contribution included 2$l_{max}$ terms for both the electron-electron and electron-nuclei Coulombic interactions.
For the S-matrix poles as well as the ICS and DCS calculations the size of the physical box that contains the diabatized partial wave terms was taken by us to be 50 \AA.

\begin {figure}[here]
\begin {center}
\includegraphics[width=\textwidth]{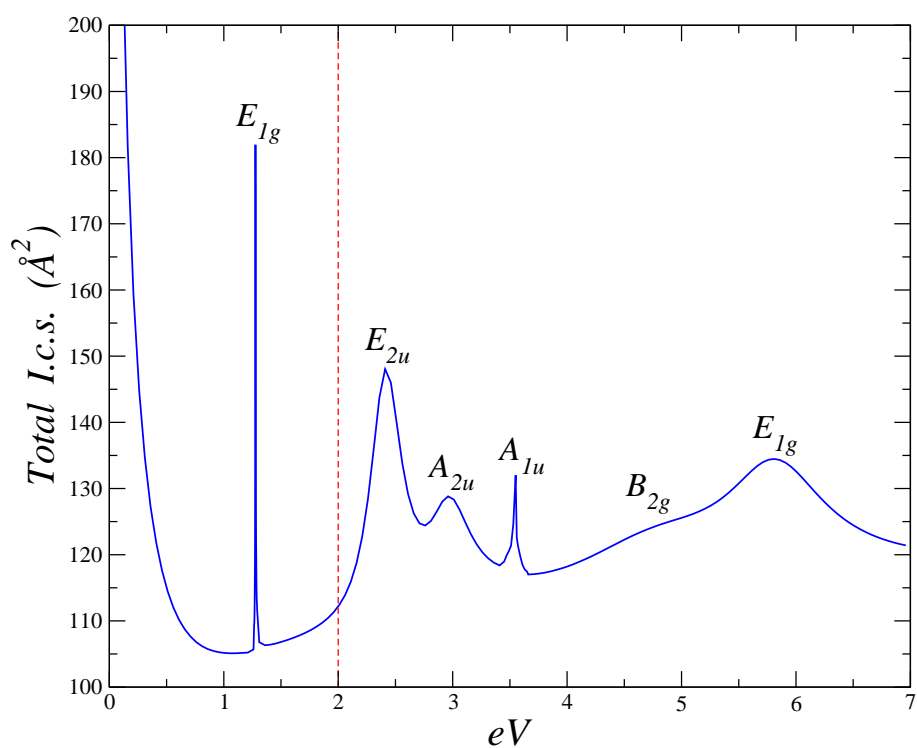}
\end {center}
\caption{\small{Computed elastic (rotationally summed) integral cross sections. }} 
\label{fig_ch3.2_02}
\end {figure}

\noindent We show in figure \ref{fig_ch3.2_02} our computed total elastic integral (rotationally summed) cross sections in the 0-7 eV energy range.
Keeping in mind that the realistic, albeit approximate, form of interaction with the free incoming electron, as well as the necessary truncation of the $l$ expansion, could reasonably provide the main reasons according which the resonance energies could be located at somewhat different position than what expected by experiments, the perusal of the results shown in figure \ref{fig_ch3.2_02} already provide the following general information on the expected features of the low-energy dynamic response of the coronene molecule to its interaction with environmental electrons:

\noindent i) in the 0-7eV energy window, the computed cross sections show rather marked structural peaks which can be identified as one particle shape resonances (i.e. formation of metastable C$_{24}$H$_{12}$ anions) in the E$_{1g}$ (E = 1.277 and 5.86 eV), E$_{2u}$ (E = 2.40 eV), A$_{2u}$ (E = 2.98 eV) A$_{1u}$ (E = 3.56 eV) and B$_{2g}$ (E = 4.80 eV) symmetry components of the D$_{6h}$ point group, respectively.
The resonant parameters are summarized in Table 3.4, where we report the symmetry, the energy at which the metastable anion is formed, the imaginary component width (both in eV) and the corresponding lifetime; within the last column on the right we report the way we have assessed them ('BW' refers to the occurrence of a clear $\pi$ jump in the eigenphase sum, hence the ICS lorenzian peak is fitted at the Breit-Wigner level; 'S' means identification via the associated S-matrix pole on the unphysical sheet of the complex energy plane).

\begin{table}
\begin{center}
\begin{tabular}{|l|l|l|l|l|}
\hline
$ Symmetry $ & $ E_{res} \:\: (eV) $ & $ \frac{\Gamma}{2} \:\:\ (eV) $ & $ \tau \:\: (sec) $ & $ \: $ \\ \hline
\hline
\hline
$ ^{2} E_{1g} $ & 1.277 & 0.0046 & $\sim$ 715 $\cdot 10^{-16}$ & BW \\
\hline
$ ^{2} E_{2u} $ & 2.26  & 0.13   & $\sim$ 25 $\cdot 10^{-16}$ & S \\
\hline
$ ^{2} A_{2u} $ & 2.97  & 0.197  & $\sim$ 17 $\cdot 10^{-16}$ & S \\
\hline
$ ^{2} A_{1u} $ & 3.62  & 0.013  & $\sim$ 253 $\cdot 10^{-16}$ & S \\
\hline
$ ^{2} B_{2g} $ & 4.61  & 0.758  & $\sim$ 4 $\cdot 10^{-16}$ & S \\
\hline
$ ^{2} E_{1g} $ & 5.79  & 0.471  & $\sim$ 7 $\cdot 10^{-16}$ & S \\
\hline

\end{tabular}
\caption{\small{Computed one particle shape resonance real and imaginary energy components.}}
\end{center}
\end{table}

\noindent Furthermore, we find it important to point out now that the likely presence of vibrational coupling effects, which are not considered by the present fixed nuclei calculations, could cause the above electron attachment cross sections to be both broader and smoother.
It is also interesting to note that in the energy range considered in the present investigation, which contains the main contributions from electrons having a mean kinetic temperature $T_{e}\;\leq\;10.000\;K$, the computed ICS are not particularly large: apart from the sharp resonance features (thus neglecting for the moment the possible resonant capture of the incoming charged particle) and the rapidly increasing branch close to 0 eV, the background cross sections are in fact a little greater than 100\AA$^2$, a value which is not much different from the rigid sphere limit given by the geometrical size of the neutral system with a diameter of roughly 10 \AA.
This feature is evident between 0.5 eV and 2 eV, a range corresponding to electronic kinetic temperatures of $T_{e}\sim\:$5000 K and $\sim\:$ 20000 K, respectively;

\noindent ii) above 2 eV we find five relatively crowded and rather marked maxima corresponding to five resonances which, with the exception of the $A_{1u}$ symmetry contribution, are relatively broad.
Their resonant character was confirmed by both the $\pi$ jump in each eigenphase sum and by locating the corresponding S-matrix poles on the unphysical sheet;

\noindent iii) below the collision energy of 2 eV, we locate instead only one very narrow peak of $E_{1g}$ symmetry at 1.277 eV.
The resonant character of this feature is provided by the clear and very sharp $\pi$ jump in the corresponding eigenphase sum, which we have fitted to a Breit-Wigner formula, yielding a resonance width of 0.0046 eV;

\noindent iv) although the molecule has not a permanent electric dipole moment, due to the highly symmetric structure of its equilibrium geometry, our calculations show a markedly increasing behaviour of the elastic ICS close to threshold.
One can surmise here that this feature indicates that the slowest (the 'coldest') component of the free-electron gas present in the ISM can strongly interact with the gaseous neutral C$_{24}$H$_{12}$ molecule, if present, as already suggested by laboratory experiments \cite{abouaf,denifl} and as we shall further discuss below.

\begin {figure}[here]
\begin {center}
\includegraphics[scale=0.15]{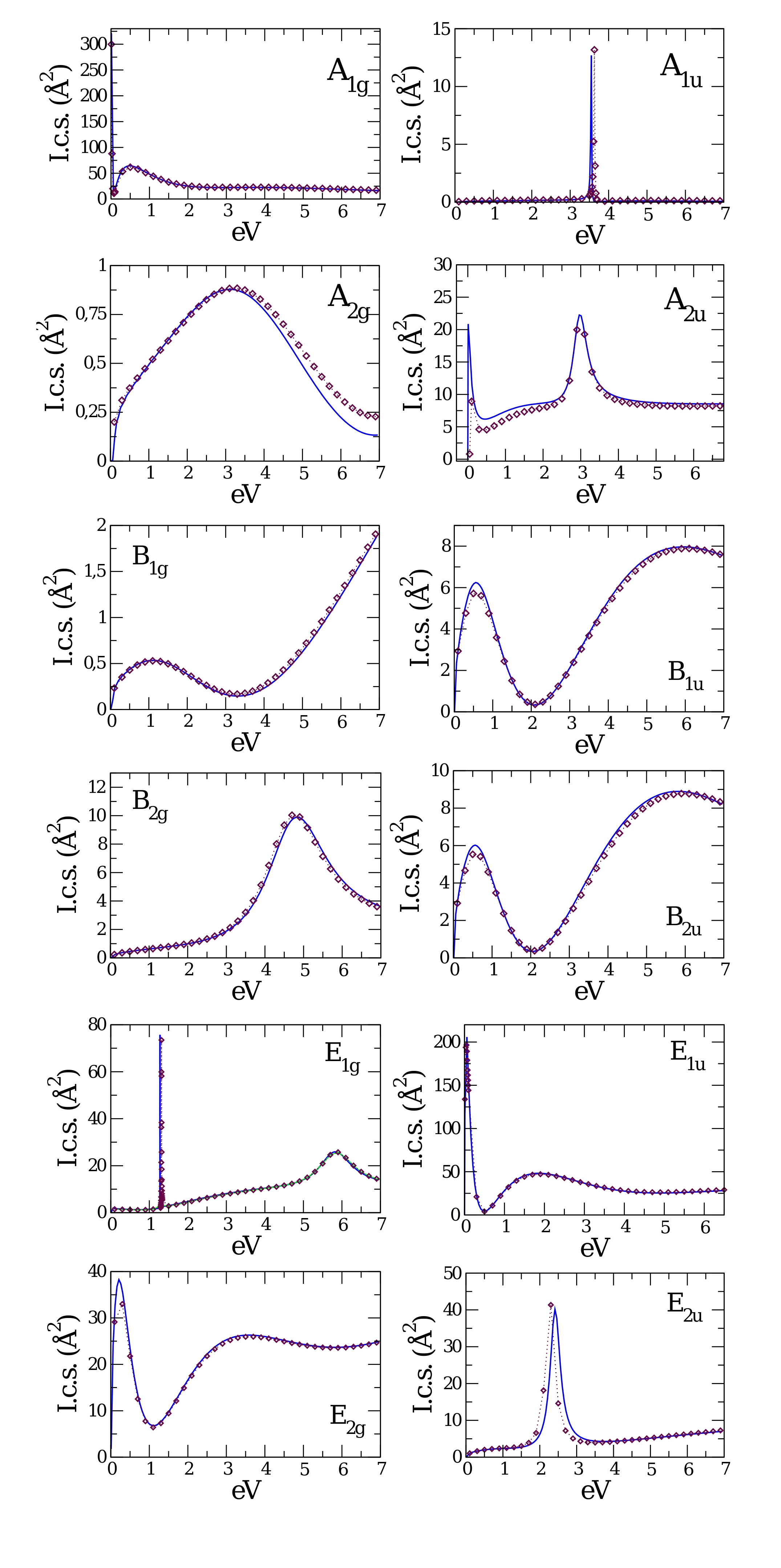}
\end {center}
\caption{\small{Computed partial ICS for the tewlve irreducible representations of D$_{6h}$ point group.}} 
\label{fig_ch3.2_03}
\end {figure}

\noindent In order to gain a deeper insigth in the different contributions which make the total ICS, we report all the partial ICS coming from the twelve irrreducible representations of the $D_{6h}$ point group in the twelve panels of figure \ref{fig_ch3.2_03}.
Within our present modeling of the scattering processes it is obviously important to ensure that all the necessary indicators of numerical convergence (e.g. size of radial integration box, size of the multipolar expansion of the interaction potential, size of the partial wave expansion for the scattered electron, etc.) be tested to provide final, converged cross sections.
To this purpose, we thus report in each panel of figure \ref{fig_ch3.2_03} the partial ICSs computed using different values for the interaction box size (50 \AA (blue straight line) and 80 \AA (maroon diamonds)), as well as two different values of $l_{max}$ ($l_{max}$ = 60 (blue straight line) and $l_{max}$ = 80 (maroon diamonds)) for the corresponding partial wave expansion.
As showed in that picture, the numerical convergence is already reached when using a box of 50 \AA and assuming $l_{max}$ = 60\ in the sense that, in the present case, the final cross sections can be reasonably considered as converged within a few percent of their reported values.

\noindent From a purely physical point of view, we can now add a few additional comments on the general and global behaviour of the ICSs for the coronene molecule under low-energy electron collisions.
First of all, in connection with the A$_{1g}$ partial contribution (first upper panel on the left in figure \ref{fig_ch3.2_03}), it is important to stress the ICS behaviour close to zero energy: such a threshold behaviour, which is distinctive of the totally symmetric component (the one including the s-wave scattering contribution), together with the large and negative scattering length ($\alpha$ = -17.63) obtained at the smallest energy value examined by the present calculations (E = 10$^{-4}$eV), suggests the formation of a zero energy resonance \cite{joachain} due to the 'strong' long-range attractive potential provided by the large polarizability of this molecule.
However, as also shown by the other panels of the same figure, we further see that the E$_{1u}$ and E$_{2g}$ symmetry components also contribute to the strong rise of the total ICS in the low-energy range, since each of them has a well resolved maximum when the collision energy downs to 0.06 and 0.21 eV, respectively.
We can then surmise the likely formation of a loosely quasi-bound state \cite{joachain} with components in all these symmetries.
Such a feature will however discussed more in detail in the next sections.
To conclude the present section, we underline that our calculations indicate markedly enhanced interaction effects between electrons and coronene close to the vanishing energy range of the surrounding electrons.
The implications for its possible behaviour, if present, in interstellar dense clouds (low T$_{e}$) and protoplanetary atmospheres (high T$_{e}$) are therefore that most of the efficient probing of this molecule by the surrounding electron swarms will be occurring via the very low-energy tail of their distributions and chiefly through the lower partial waves describing the interacting electrons.
Before to analyse the angular redistributions of electrons scattered off coronene molecule, in the next section we will focus on the mechanisms responsible for the electron capture process, discussing our findings first of all for very low-energy impinging electrons (0-2 eV) and then for little more energetic (2-7 eV) colliding particles.

\clearpage

\subsection{Threshold behaviour and resonances in electron scattering from gas-phase coronene}

\noindent In the present section we focus more in detail on the mechanisms which could be active and then responsible for the dynamical trapping of the electronic impinging particle in each of the previuosly located shape resonances and in particular when the colliding energy is moving down to its threshold. 
We shall also discuss in what follows some non-resonant features associated with the maxima in the A$_{1g}$, E$_{1u}$ and E$_{2g}$ symmetry contributions.
We shall in fact speculate that such features, even if not properly ascribed to resonances, could indeed play a role in trapping the incident electron, and therefore for the production of the stable bound anion: the lack of a permanent electric dipole moment for this molecule could qualitatively justify their non-resonant character in the framework of our fixed nuclei approximation.

\noindent Generally speaking, at impact energies below 10 eV, the interaction of an electron with a molecule is mainly dominated by the existence of short lived anions (called 'resonances') due to the temporary capture of the incident electron by the target molecule.
From the experimental point of view, when the lifetimes are 'short' (e.g. 10$^{-12}$ sec $\sim$ 10$^{-16}$ sec) their direct observation is generally rather challenging: they are thus observed in many cases through their decay, either by autodetachment or by dissociation into different fragments, one of which carries the extra electron.
If the lifetime of the resonance under investigation lengthens, the ejection of the extra electron can also occur at an internuclear distance different from those of the ground state, hence giving rise to vibrational excitation.
This channel is however in competition with the channel which yields a stable negative fragment plus one or several neutral ones via the dissociative electron attachment process.

\noindent The most recent among the available experiments shows that the major resonant interaction takes place for low and especially vanishing energies \cite{abouaf}, where for higher collision energy values (1.5, 2.0, 2.5, 3.5, 4.5 7.0 and 9.0 eV) five groups of excited vibrational modes are observed with electronic energy losses ranging from 0.065 eV to 0.375 eV confirming that several more energetic resonant states appear to contribute to the global interaction.
These experiments also confirm that the formation of the corresponding not dehydrogenated negative ion occurs at zero energy, and at the same time they shed new light on the real behaviour of the C$_{24}$H$_{12}$ molecule under very low-energy electron collisions: if on one side the presence of one intense peak centered at zero energy in fact confirms the findings of \cite{denifl}, the better electron energy resolution allows to demonstrate the existence of another peak at 0.3 eV, in sharp contrast with the previous data. 
The nature of this second, less intense, peak is not obvious: the authors argue that it can therefore correspond either to a particular long-lived shape resonance or to resonant state having a core excited character, the latter reasonably not directly coupled to the ground state of the anion.
Moreover, since the spacing between this peak and the larger one at zero energy is reminiscent of a specific C-H stretch mode (0.375 eV), a vibrational Feshbach resonance associated with this mode could also be a possibility.

\noindent Keeping in mind that elastic scattering and electron capture cross sections can often be related since the same resonant features can in principle contribute to both processes, we intend now to analyse and discuss our findings by making a comparison (when possible) with experiments in order to provide some useful information about the possible formation channels for low-energy metastable negative ion (C$_{24}$H$_{12}^{-}$)$^{*}$ which can actually be the gateway for the production of the thermodynamically stable, long-lived parent anion, the electron affinity for the neutral species being in fact positive: +0.54 $\pm$ 0.1 eV according to the kinetic measurements in \cite{chen} and +0.47 $\pm$ 0.09 eV according to the photodetachment experiments reported in \cite{duncan}.
Since the theoretical modelling of low-energy electron scattering for large polyatomic molecules is indeed computationally very challenging, we have employed in the present study our ASMECP model of potential interaction which has been previously shown to provide good results not only for small molecular objects: it has been in fact proved that such a potential is accurate enough for examining the low-energy attachment on the  systems as large as the C$_{60}$ molecule (\cite{lucchese3}).

\subsubsection{Non resonant very-low energy features: E$_{1u}$ and E$_{2g}$ partial contributions.}

\begin {figure}[here]
\begin {center}
\includegraphics[width=\textwidth]{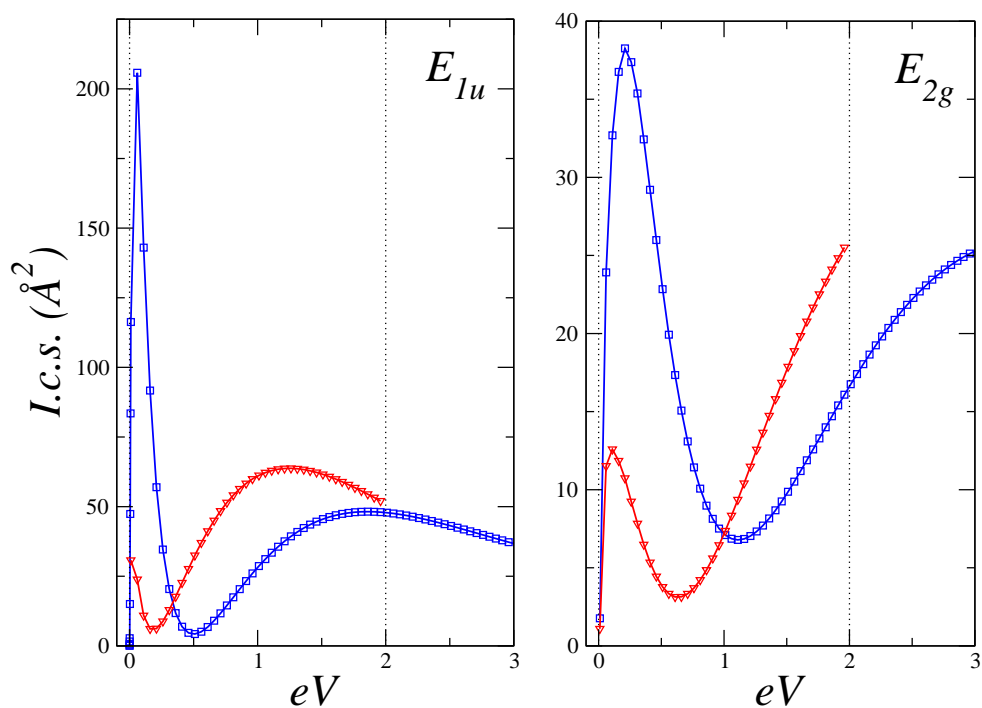}
\end {center}
\caption{\small{Low-energy (0-2 eV) ICS behaviour for the E$_{1u}$ and E$_{2g}$ symmetry. See text for details.}} 
\label{fig_ch3.2_04}
\end {figure}

\noindent The two panels of figure \ref{fig_ch3.2_04} report the behaviour of our computed ICS in the 0-2 eV energy range for the partial contributions where we find a pronounced maximum, clearly below 0.5 eV.

\noindent The blue squares in each panel refer to ICS calculations including all four interaction terms (Coulombic, exchange, correlation and polarization) as previously illustrated (see chapter 2), while the triangles refer to the computed ICS in the same symmetry contribution but without including the correlation-polarization potential, V$_{cp}$.
We can easily recognize that for the E$_{1u}$ and E$_{2g}$ symmetries the low-energy maxima, located at 0.06 eV and 0.22 eV respectively, are not numerical artifacts: due to the large polarizability of the neutral target molecule, originating from the large number of delocalized $\pi$ electrons, such maxima could be in fact indicative of the true physical response of the target to the incident charged projectile at a very low energy.
In both cases, in fact, qualitatively we see that the colliding electron has not enough kinetic energy to come very close to the target molecule, causing the short-range interaction terms (correlation and exchange) to be less important then the long-range interaction terms.
Thus, in order to provide an accurate physical description of the electron-C$_{24}$H$_{12}$ molecule collision, the asymptotic part of the long-range static interaction can be reasonably expressed as a sum of multipolar moments where the dipole term (the first one) is zero in the present case due to the inversion center that characterizes the present molecule at the equilibrium; furthermore, besides the 'pure' static interaction, the molecule is also perturbed by the presence of the incident electron which in fact induces an additional smaller dipole term, the effects of which is then considered by including our V$_{cp}$ potential in the ICS calculations.
In other words, since such a behaviour in the ICS for the E$_{1u}$ and E$_{2g}$ symmetries seems to be dramatically depleted when the long-range effects are not considered in the calculation (see figure \ref{fig_ch3.2_04}), such maxima tell us that the impinging electron induces an electric dipole moment, dependent on the molecular polarizability, which is strong enough to cause both the above ICS to have a pronounced peak close to threshold.

\begin {figure}[here]
\begin {center}
\includegraphics[width=\textwidth]{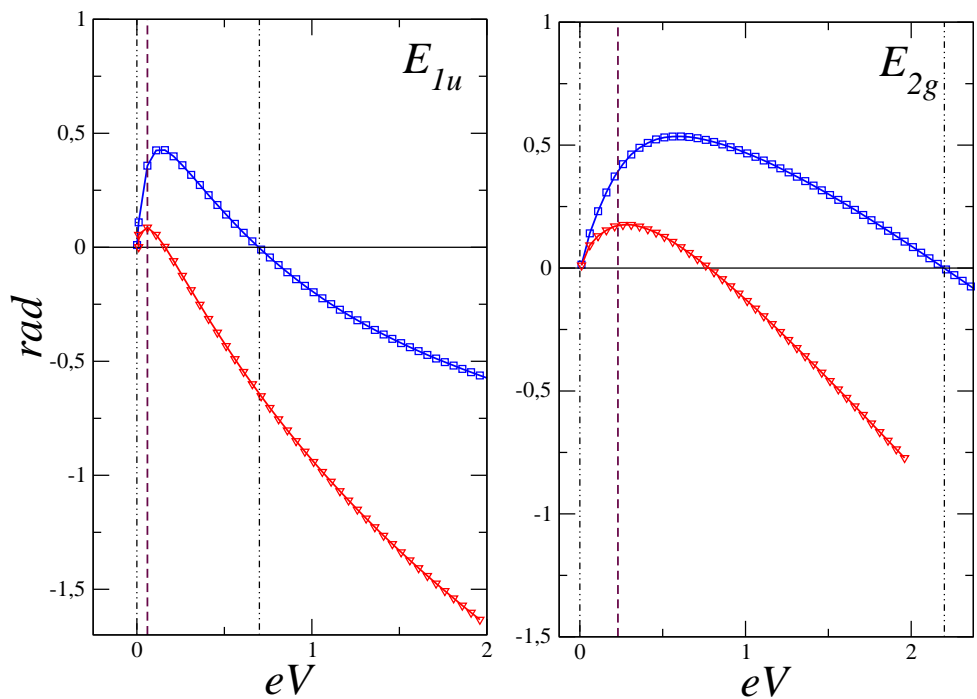}
\end {center}
\caption{\small{Low-energy (0-2 eV) eigenphase sum for the E$_{1u}$ and E$_{2g}$ symmetries.}} 
\label{fig_ch3.2_05}
\end {figure}

\noindent However, their not properly resonant character is confirmed by analysing the corresponding eigenphase sums that we depict in figure \ref{fig_ch3.2_05} as a function of collision energy.

\begin {figure}[here]
\begin {center}
\includegraphics[width=\textwidth]{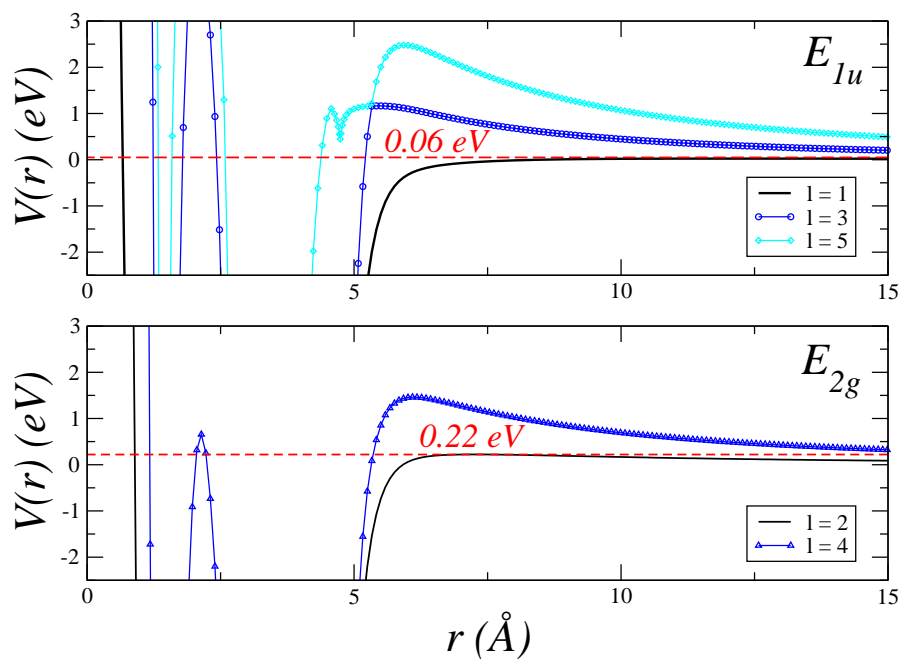}
\end {center}
\caption{\small{Computed lowest angular momenta diabatic potential curves for the E$_{1u}$ and E$_{2g}$ symmetries.}} 
\label{fig_ch3.2_06}
\end {figure}

\noindent As the panels corresponding to the E$_{1u}$ and E$_{2g}$ symmetries show, we do not have a $\pi$ jump around the energies corresponding to the maxima: for both of them, in fact, the rapidly rising cross sections moving away from threshold correspond in fact to a slow eigenphase rise.
The above low-energy features in both the partial ICS here under investigation are then followed by a Ramsauer-Townsend-type structures which occur in the energy window where the associated eigenphase sum crosses zero (for the E$_{2g}$ symmetry, to be precise, there is a little shift in energy of about 0.9 eV).
In order to glean further physical information about this couple of maxima, we have computed the corresponding diabatic potential curves that we report in the insets of figure \ref{fig_ch3.2_06}.

\noindent It is thus clear from them that the energy of each maximum corresponds to the top of the corresponding barrier for the lowest angular component in the computed diabatic potential curves: $l$ = 1 for the 0.06 eV peak in the E$_{1u}$ symmetry and $l$ = 2 for the 0.22 eV peak in the E$_{2g}$ symmetry, respectively.
Additionally, in both symmetries it is possible to recognize a rising eigenphase sum close to threshold and which has a maximum immediately after the top of each of the above potential barriers is reached, as shown by the vertical black dashed lines within the insets in figure \ref{fig_ch3.2_05}.
We can therefore conclude that the analysis of these two features leads us to say that they do not appear to be the result of some actual resonant structure since none of the required indicators does exist.
Qualitatively speaking, and using a semiclassical picture, we find it possible to speculate that these low energy rising eigenphase sums might be the quantum analog of the classical orbiting, where there is an increase of time delay that in fact takes place when the energy of the colliding electron equals the energy value associated to the maximum of its own centrifugal barrier.
This effect is reminiscent of that in a previous investigation \cite{gianturco3} that focused on the benzene molecule in its equilibrium geometry (which belongs to the same point group of the molecule here under investigation); in the present case, however, such non-resonant maxima are more pronounced and markedly stronger in intensity if compared with those that characterize the same partial cross section contributions for the C$_{6}$H$_{6}$ molecule.
Thus, according to the experiments \cite{abouaf}, we surmise that even if such maxima are not effectively describing an actual dynamical trapping of the electron, the markedly enhanced static polarizability for the coronene molecule could indeed make them participate in the formation of the parent anion at very low energy via such temporary states.

\subsubsection{A$_{1g}$ contribution threshold behaviour: zero energy resonance.}

\noindent The totalsymmetric partial contribution also deserves a detailed analysis: as we shall discuss in what follows, it plays in fact a key role in the formation of the resonant metastable anion at vanishing energies.

\begin {figure}[here]
\begin {center}
\includegraphics[scale=0.4]{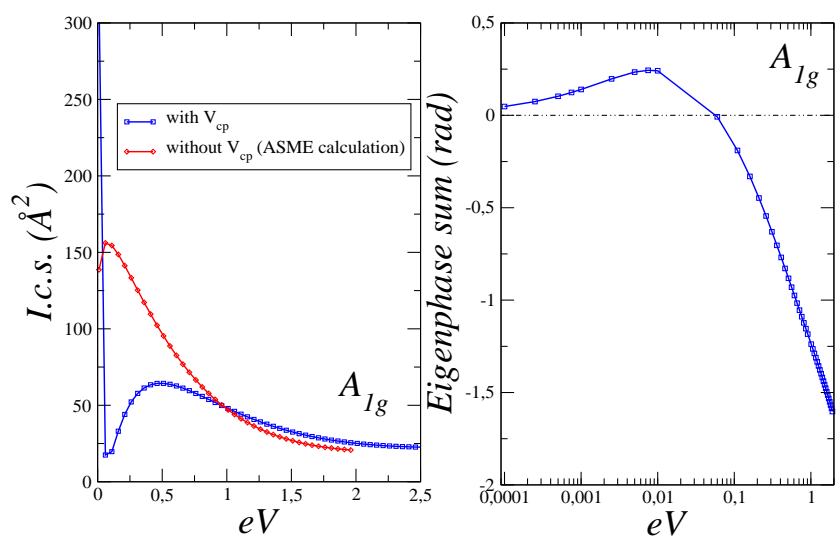}
\end {center}
\caption{\small{Left panel: behaviour of the A$_{1g}$ elastic (rotationally summed) ICS. The blue squares are referred to an ASMECP calculation; the red diamonds show the ICS behaviour when omitting the V$_{cp}$ interaction potential from the calculation (ASME calculation). Rigth panel: Computed ASMECP eigenphase sums for the same symmetry and for the same energy window.}} 
\label{fig_ch3.2_07}
\end {figure}

\noindent We show in figure \ref{fig_ch3.2_07} our computed A$_{1g}$ cross sections (left panel) and the corresponding eigenphase sum in the energy range between 10$^{-4}\:$eV and 2 eV.

\noindent The presence of a marked and very rapidly rising threshold value (for 10$^{-4}\:$eV our computed ICS value is 1091 \AA$^{2}$) in the totally symmetric component, followed by a Ramsauer-Townsend minimum located at 0.06 eV as the corresponding eigenphase sum moves down crossing the zero value, are features which demonstrate that the dominant contributing channel is the s-wave channel.
In particular, as for the case of electron-C$_{60}$ molecule low-energy scattering \cite{lucchese3}, we note a rising eigenphase sum at threshold which is clearly shown between 0.0001 and 0.01 eV in the right panel of figure \ref{fig_ch3.2_07}: such feature is indeed indicative of a virtual state \cite{newton}.

\begin {figure}[here]
\begin {center}
\includegraphics[scale=0.4]{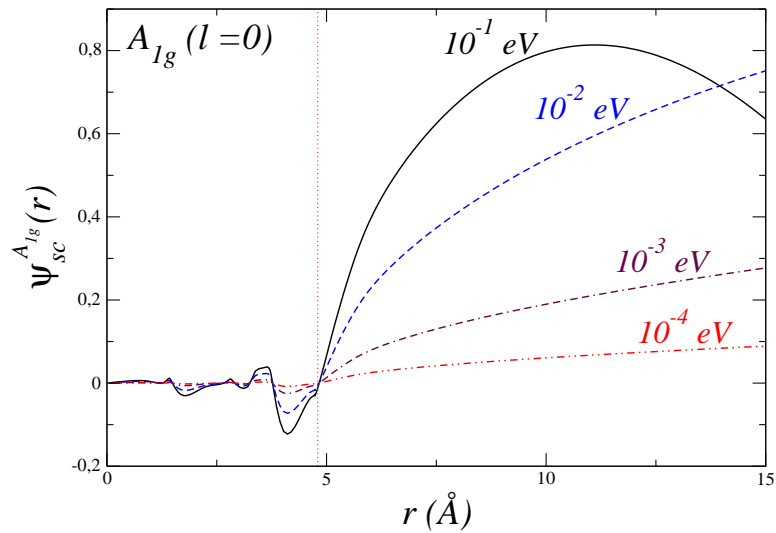}
\end {center}
\caption{\small{Behaviour of the A$_{1g}$ adiabatic scattering wavefunction for very low-energies as a function of the radial coordinate $r$ (0 $\leq$ $r$ $\leq$ 15 \AA).}} 
\label{fig_ch3.2_08}
\end {figure}

\noindent To better characterize the likely presence of a virtual state, we show in figure \ref{fig_ch3.2_08} the behaviour of the A$_{1g}$ adiabatic scattering wavefunction for very low-energies as a function of the radial coordinate $r$.

\noindent As that figure clearly shows, the adiabatic A$_{1g}$ resonant wavefunction becomes increasingly flattened, smoother and less prominent as the collision energy approaches zero.
The corresponding scattering length becomes negative and very large ($\alpha$ = -17.63) when computed at the lowest energy value (0.0001 eV) here considered.
When using a computationally more demanding partial wave expansion, i.e. varying $l_{max}$ from 60 up to $l_{max}$ = 80, we obtain $\alpha$ = -19.53.
According to the geometrical interpretation of $\alpha$, this strongly suggests the existence of a zero energy resonance ($\alpha\:\: \rightarrow \infty$ for an $s$-wave $virtual$ state) as the intermediate physical situation between an attractive potential which for vanishing energies is actually able to support an $s$-wave $bound$ state ($\alpha\:\:>\:\:$0), and an attractive potential which instead is not deep enough to produce an $s$-wave $bound$ state ($\alpha\:\:<\:\:$0) in the energy range close to zero.
The definitive confirmation that we are thus dealing with a zero energy resonance comes from the S matrix pole for the A$_{1g}$ scattering symmetry, that we locate at an energy of -0.037 eV on the unphysical Riemann sheet (associated to a pure imaginary negative linear moment of -0.05215$i$ a.u.).

\noindent These findings provide an intriguing and interesting explanation for the experimental scattering data reported by \cite{abouaf,denifl}, raising the possibility that the bound anion observed close to zero energy may have its origin in a virtual state scattering feature.
This argumentation is in fact only seemingly in contrast with the experimental findings, since in scattering of this nature the zero energy resonant negative ion does not form and indeed cannot be accessed, unless some changes in the nuclear geometry do occur.
In other words, we can surmise that the physical process leading to the zero-energy resonant anion is an $s$-wave dominated collision where the interaction, in the framework of the fixed nuclei approach, is not able to support a bound state.
However, the large polarizability of the target molecule allows us to suggest that there is time enough during that special complex formation for nuclear motion to take place in order to access the bound state, the latter being in fact located in a classically inaccessible regime.
In this sense, the virtual state scattering can thus act as a doorway state for the formation of negative ions like it does for C$_{60}$ electron scattering \cite{weber, lucchese3}).
Given the large number of closely spaced vibrational modes for the C$_{24}$H$_{12}$ molecule, we find it very likely that such a virtual state may decay via coupling with the dense manifold of excited roto-vibrational states in a sufficiently short time to compete with the autodetachment process, as indeed shown by the experiments \cite{abouaf}. 
The negative ion resulting by virtual state scattering is in fact a metastable species, even if for a small amount of energy, and thus its lifetime should be long enough so that the excess energy could be efficiently dissipated either by photon emission or via intramolecular vibrational redistribution (IVR).

\noindent Hence, the coronene molecule represents an example of medium size PAH species where virtual state scattering plays an important role for its parent negative ion formation.

\begin {figure}[here]
\begin {center}
\includegraphics[scale=0.4]{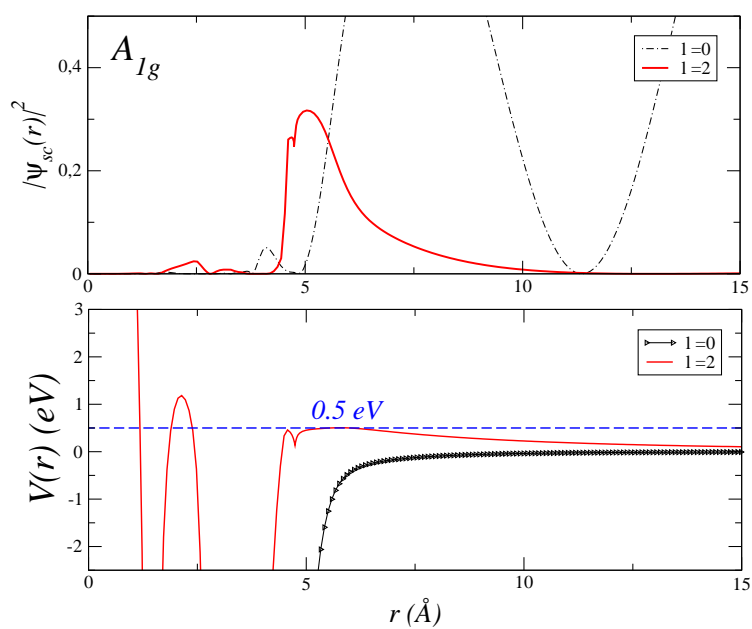}
\end {center}
\caption{\small{Upper panel: squared modulus of the lowest angular momentum contributions (dotted-dashed line: $l$ = 0; red straight line: $l$ = 2) of the totalsymmetric adiabatic scattering wavefunction when the incident energy is 0.5 eV $vs$ the radial coordinate $r$. Lower panel: diabatic potential curves that correspond to the above partial angular contributions computed at the same collision energy and depicted in the same radial domain.}} 
\label{fig_ch3.2_09}
\end {figure}

\noindent To conclude the present theoretical analysis of the threshold electron scattering behaviour for the coronene molecule at its equilibrium geometry, we report in figure \ref{fig_ch3.2_09} the computed squared modulus of the lowest angular contributions ($l$ = 0 and $l$ = 2) of the totalsymmetric adiabatic scattering wavefunction when the incident energy is 0.5 eV (upper panel), as well as the diabatic potential curves that correspond to the above partial angular contributions at the same energy. 

\noindent When the incident projectile has an energy of 0.5 eV, we can tentatively note that the $l = 2$ angular component (\footnote{a $d_{z^{2}}$-like angular wavefunction which belongs to the A$_{1g}$ symmetry; see the character table of the D$_{6h}$ point group of symmetry.}), included in the $A_{1g}$ scattering wavefunction, is further able to yield the same feature previously discussed for both the E$_{1u}$ and E$_{2g}$ symmetries, even if in the present case for an higher energy and with a comparatively reduced intensity.
The A$_{1g}$ ICS shows in fact a peak at $\sim$ 0.5 eV which is much broader when compared to those of E$_{1u}$ and E$_{2g}$ symmetry (see the top right panel in figure \ref{fig_ch3.2_03}).
Furthermore, such a collision energy corresponds to the top of the $l = 2$ centrifugal barrier (lower panel of figure \ref{fig_ch3.2_09}), and the squared modulus of the totalsymmetric scattering wavefunction, computed at 0.5 eV (see upper panel of figure \ref{fig_ch3.2_09}, red solid line), intriguingly exhibits just a small probability density mainly located at the outer edge of the region occupied by the target C$_{24}$H$_{12}$ molecule.
However, since in the present case this effect is not caused by the lowest angular momentum contribution ($l = 0$), we find such a probability density to be negligible so that the $l = 2$ partial wave contribution within the A$_{1g}$ symmetry should not be very effective for the low-energy trapping of the incoming electron.

\subsubsection{The low-energy shape resonances.}

\begin {figure}[here]
\begin {center}
\includegraphics[scale=0.25]{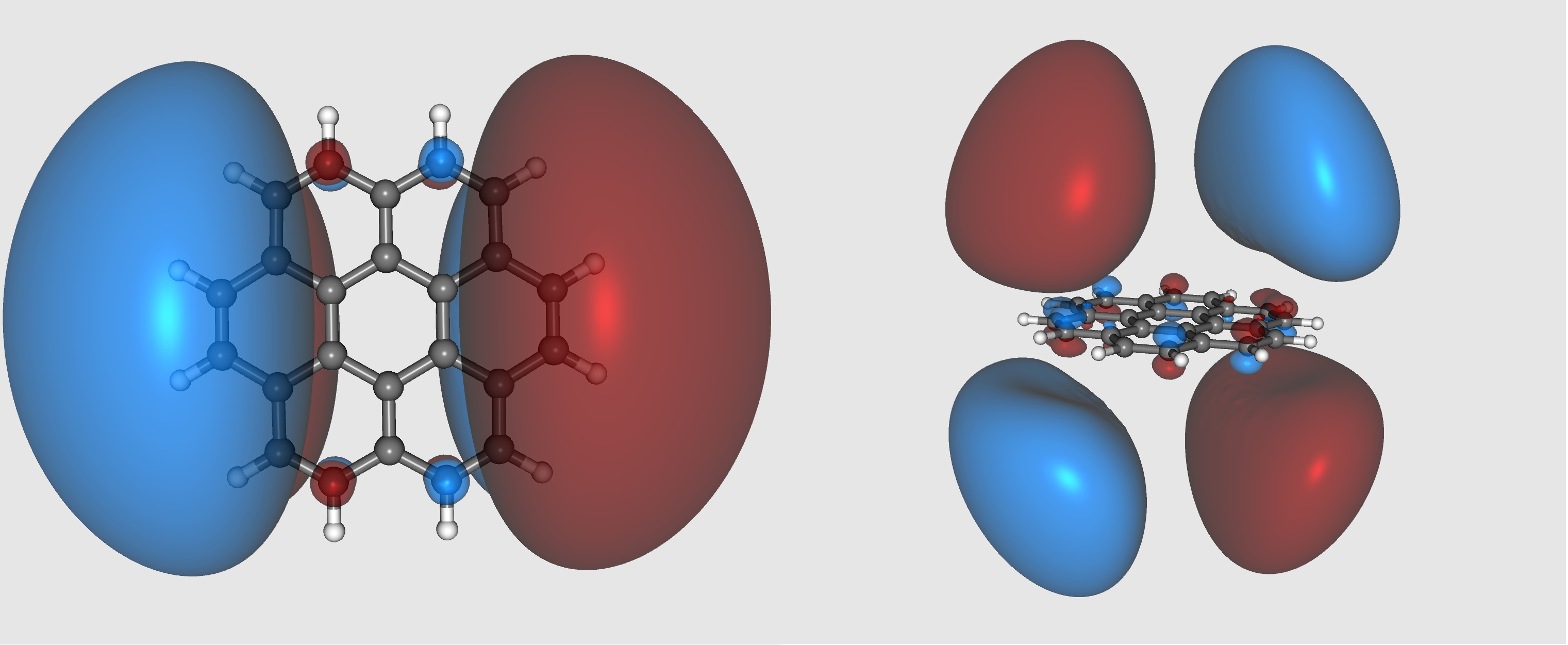}
\end {center}
\caption{\small{Spatial shape of the low-energy E$_{1g}$ (E = 1.277 eV) resonant wavefunction.}} 
\label{fig_ch3.2_10}
\end {figure}

\begin {figure}[here]
\begin {center}
\includegraphics[scale=0.4]{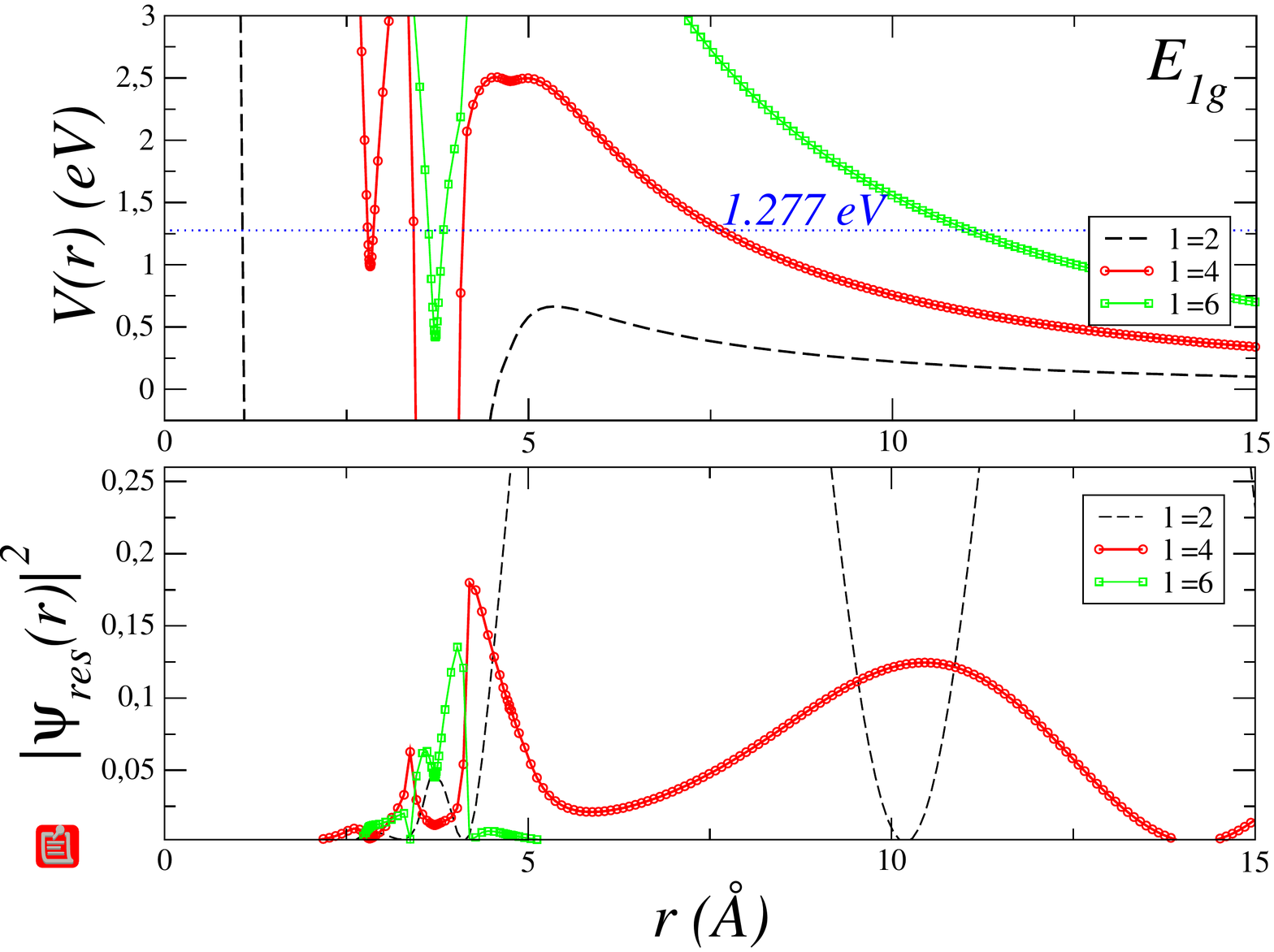}
\end {center}
\caption{\small{Upper panel: computed diabatic potential curves for the low-energy $E_{1g}$ (E = 1.277 eV) resonance. Lower panel: squared modulus for the adiabatic resonant wavefunction lowest angular contributions. Black-dashed line: $l$ = 2; red circles: $l$ = 4; green circles: $l$ = 6.}} 
\label{fig_ch3.2_11}
\end {figure}

\noindent Having discussed our findings for the marked but non-resonant features which characterize the E$_{1u}$, E$_{2g}$ and A$_{1g}$ scattering symmetries, we now pass to analyse more in detail the six one-particle shape resonances introduced in a previous section, starting from that one at the lowest energy.
To this purpose, we report in figure \ref{fig_ch3.2_10} the shape of the wavefunction corresponding to the very narrow E$_{1g}$ shape resonance that we locate at an incident energy of E = 1.277 eV.
Within the two panels of figure \ref{fig_ch3.2_11} we further show the associated diabatic potential curves for the lowest angular contributions ($l$ = 2, 4, 6; lower panel) involved in the dynamical trapping of the electron and (upper panel) the radial behaviour of the resonant wavefunction for each of the above angular contributions.

\noindent It is then clear from the two panels of figure \ref{fig_ch3.2_11} that in such a very narrow resonance the dominant dynamical trapping of the impinging electron arises from the $l$ = 4 centrifugal barrier, where the tunnelling would have to span a rather large radial region (4 $\leq$ $r$ $\leq$ 7.5 \AA).
Excluding for the moment the possible decay into a bound stable state, in most elastic one-electron resonances originated by an angular momentum barrier there are two possible mechanisms for their decay: the resonant state can either give rise to a direct decay since its probability density can tunnel backward through its own dominant centrifugal barrier, or it can decay due to a dynamical coupling with other partial waves having lower $l$ values.
In the present case, as shown pictorially in the lower panel of figure \ref{fig_ch3.2_11}, the probability densities associated with the lowest angular contributions ($l$ = 2 and $l$ = 4) can provide such a coupling, although a possible nonadiabatic coupling between the $l$ = 4 and $l$ = 6 partial resonant waves, occurring just behind the $l$ = 4 barrier, can play a role in preventing the rapid ejection of the extra electron, thus providing an additional qualitative explanation for the very long lifetime that we find to be associated to this resonance.
It is furthermore interesting to note that this resonance belongs to the same symmetry of the C$_{24}$H$_{12}$ HOMO-1 as well as in the LUMO (\footnote{where, making a comparison between our resonant 3D map and the 3D map for the E$_{1g}$ LUMO calculated by \cite{abouaf} at DFT level but with the same basis set, we can tentatively note some resemblance}); in a qualitative sense, the above E$_{1g}$ resonance should then be expected to be overlapped (i.e. interacting) with the lower-lying bound states which are more likely to participate in the formation of a stable bound anion.
In other words, according to our equilibrium geometry analysis, we surmise that such a very long-lived metastable anion could play a role in the formation of the negative parent species by dissipating the fairly small excess energy content via a radiationless intramolecular vibrational redistribution process.

\begin {figure}[here]
\begin {center}
\includegraphics[scale=0.6]{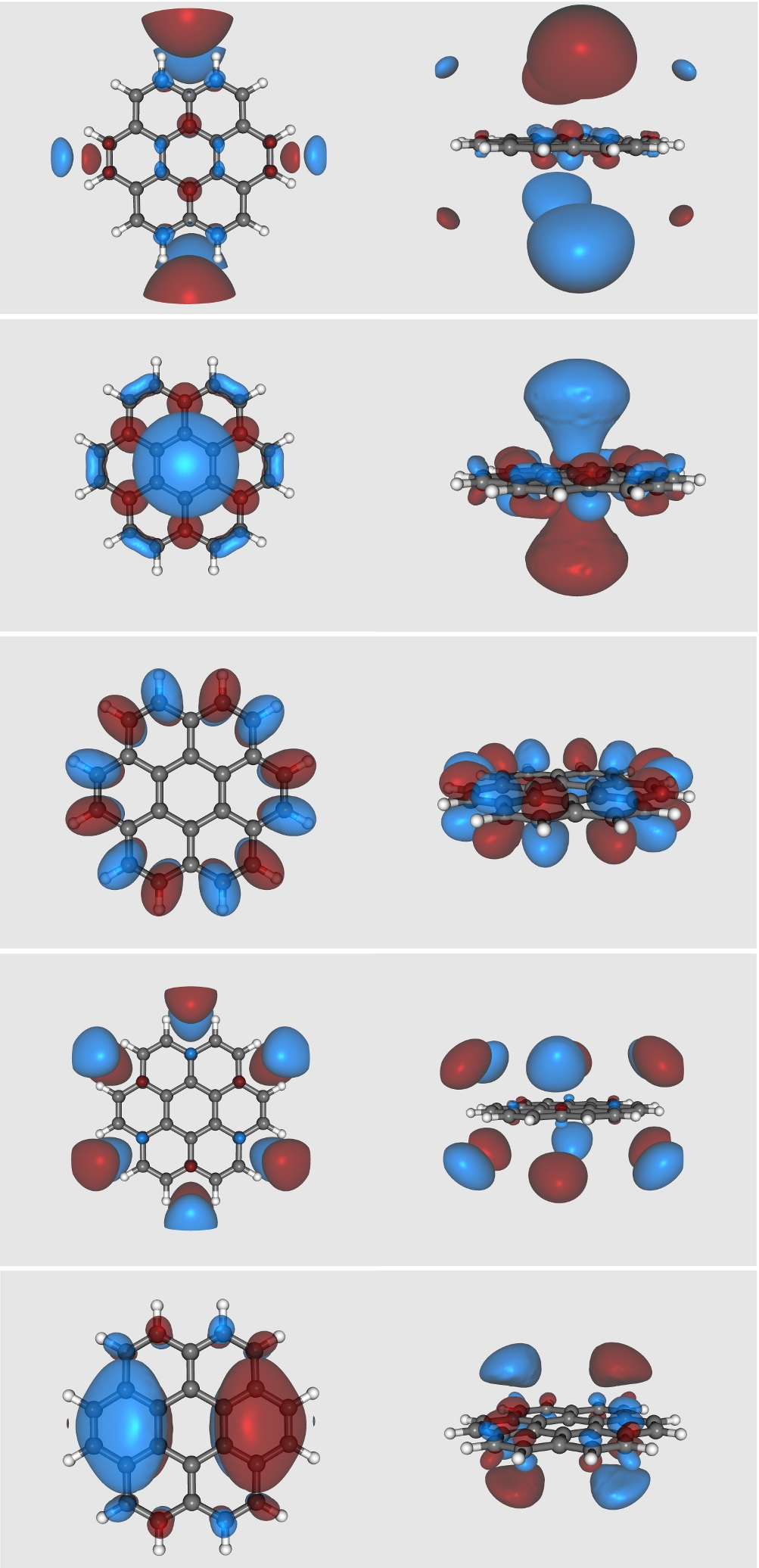}
\end {center}
\caption{\small{Spatial shape of the resonances located over 2 eV; see table 3.4. From top to bottom: $^{2}$E$_{2u}$, $^{2}$A$_{2u}$, $^{2}$A$_{1u}$, $^{2}$B$_{2g}$ and $^{2}$E$_{1g}$ }} 
\label{fig_ch3.2_12}
\end {figure}

\noindent In figure \ref{fig_ch3.2_12} we collect the real part of the wave functions for the remaining resonances, all of them occurring for energies which are greater than 2 eV; the above figure, from top to bottom, contains the panels ordered from the lower resonance energy (of the E$_{2u}$ resonance, E$_{r}$ = 2.26 eV) moving up to the higher one (E$_{1g}$, E$_{r}$ = 5.79 eV).

\noindent On the basis of these findings, to conclude this section, we tentatively try to make a comparison with the available fairly detailed experimental data \cite{denifl,abouaf}.
We can thus summarize the main points of contacts in the following way:

\noindent i) in the considered energy range, which is the one that is here astrophysically relevant, our calculations do not show the presence of any $\sigma^{*}$ resonance, as directly proved by the real 3D maps of the resonant wavefunctions that we collect and depict in figures \ref{fig_ch3.2_10} and \ref{fig_ch3.2_12};

\noindent ii) the resonance $^{2}$A$_{1u}$ located at E$_{r}$ = 3.62 eV, due to its relatively long lifetime ($\tau_{^{2}A_{2u}}$ $\sim$ 253 $\cdot 10^{-16}$ sec, see table 3.4), could be however involved in the excitation of some CH stretching modes: we can tentatively surmise it by looking at the spatial shape of the corresponding metastable wavefunction, depicted in the central inset of figure \ref{fig_ch3.2_12} since only in such a resonant state, in fact, the excess electronic wavefunction seems to efficienly 'cover' the periferal CH groups although located above and below the molecular plane.
Furthermore, we find it interesting that this resonant feature occurs at an energy which falls almost at the onset of the energy range where the experiments \cite{abouaf} show the $\nu_{CH}$ = 0.375 eV mode to be quantitatively excited;

\noindent iii) in general terms, the presence of several $\pi^{*}$  resonances up to about 7 eV for the coronene molecule is not surprising: besides the fact that the associated electron affinity is positive, \cite{chen,duncan}, the experiments indeed reveal that at incident energies higher or equal than 2.0 eV the variation of the relative intensities of different vibrational bands indicates the occurrence of several resonant excitation processes.
It would also be helpful \cite{abouaf} to investigate the coronene molecule by means of the electron transmission spectroscopy, since the findings of such experiments might enable us to have a better comparison with the present results, thus clarifying the role played by its metastable anions in the more general astrophysical context in which such electron-driven reactions have been studied.

\clearpage

\subsection{The scattered electron angular distributions at low energies}

\noindent The previous subsections have described in detail the behaviour of the elastic, partial and total, ICS for an energy range of the impinging electrons which is astrophysically relevant.
Now, the above results further suggest to be of interest knowing how the angular distributions would behave within the same range of energies, even if, due to the lack of specific laboratory data, we cannot make a direct comparison of our results with measured angular distributions for gas phase coronene molecule.
However, the present computational experiment already allows us to draw the following features from the computed behaviour that we report in figures \ref{fig_ch3.2_13} and \ref{fig_ch3.2_14}:

\begin {figure}[here]
\begin {center}
\includegraphics[scale=0.4]{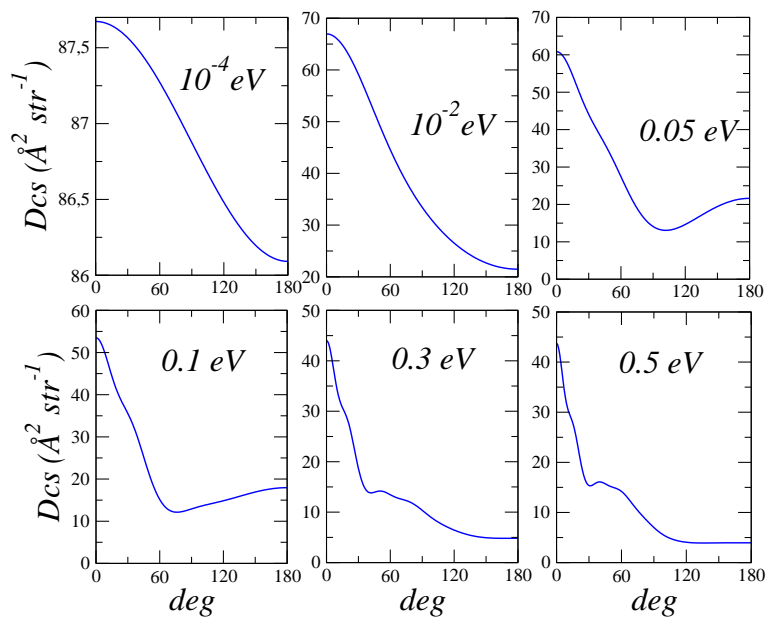}
\end {center}
\caption{\small{Computed differential elastic (rotationally summed) cross sections in the 'low-energy' domain. See main text for details and discussion.}} 
\label{fig_ch3.2_13}
\end {figure}

\begin {figure}[here]
\begin {center}
\includegraphics[scale=0.4]{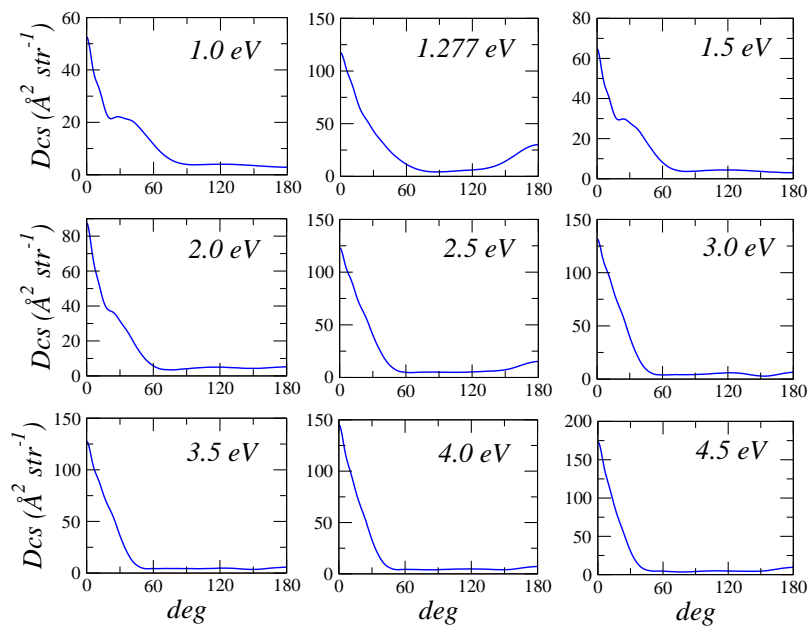}
\end {center}
\caption{\small{Computed differential elastic (rotationally summed) cross sections in the 'high-energy range'. See main text for details and discussion.}}
\label{fig_ch3.2_14}
\end {figure}

\noindent i) at the lowest collision energies considered here ($E\;=\;$10$^{-4}$ eV), the scattering is dominated by the dipole-induced interaction which, according to the results previously discussed, is however not able to support a bound state; as shown by figure \ref{fig_ch3.2_13}, the corresponding DCS is found to be very little angle-dependent and largely isotropic, thus confirming once again a behaviour chiefly due to ''s-wave'' scattering at those low energies;

\noindent ii) in the energy range between $10^{-2}$ eV and up to 0.5 eV (see figure \ref{fig_ch3.2_13}), with dominant contributions from E$_{1u}$ and E$_{2g}$ symmetries to the corresponding ICS (see figure \ref{fig_ch3.2_03}), our findings show a progressive increase of the forward peak, as well as a marked reduction of the backward scattering as indicated by all the panels of figure \ref{fig_ch3.2_13};

\noindent iii) as the collision energy moves across the narrow E$_{1g}$ ICS maximum associated to a strong resonance of that symmetry, the computed angular distribution at $E\;=\;$1.277 eV given by the top row of panels in figure \ref{fig_ch3.2_14} becomes much flatter (from $\theta\;=\;$60$^{0}$ up to $\theta\;=\;$120$^{0}$).
Such a feature occurs as soon as the energy leaves the very narrow resonance region but resumes again up to at least 1.5 eV;

\noindent iv) the data from figure \ref{fig_ch3.2_14} show that when the collision energy increases to higher values (which are still astrophysically relevant), we see that the DCS, as a function of the scattering angle $\theta$, shows an increase of the forward peaks at smaller angles while the backward scattering becomes largely negligible.

\noindent As a general trend in the energy range of interest, we thus find as dominant the presence of the forward scattering behaviour, a feature that involves a rather large geometrical cone ($\theta_{max}\;\sim\;$60$^{0}$).
Furthermore, the forward scattering appears to be dominant also when the collision energy matches the E$_{1g}$ resonance position (E = 1.277 eV), where using a qualitatively semiclassical picture, the 'orbiting' of a trapped electron in a shape resonance should distort the angular distribution generated by the induced dipole on the target molecule.

\subsection{Computed momentum-transfer cross sections and the rate coefficients}

\begin {figure}[here]
\begin {center}
\includegraphics[scale=0.45]{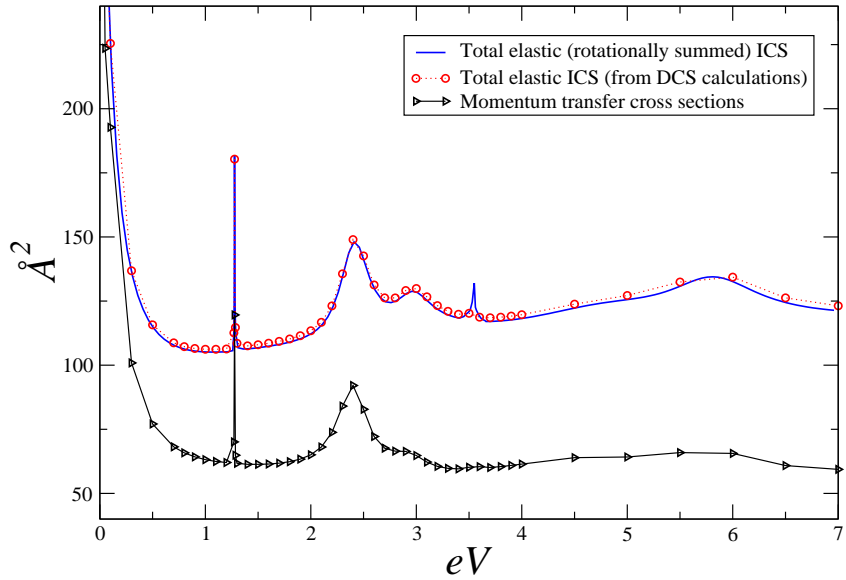}
\end {center}
\caption{\small{Comparison between computed elastic integral cross sections and momentum transfer cross sections. The blue straight line reports the ICS directly computed via K matrix elements; the red circles and the black triangles refer to the ICS and MTCS computed by numerical integration of the differential elastic (rotationally summed) cross sections. See text for details.}} 
\label{fig_ch3.2_15}
\end {figure}

\noindent In figure \ref{fig_ch3.2_15} we report again, as a remainder, the ground-state total elastic ICS (solid line) together with our computed momentum-transfer cross sections (MTCS; black right-oriented triangles), both as a function of the collision energy.
The red circles refer to the elastic ICS computed by numerical integration of the differential elastic cross sections $d\sigma$/$d\Omega$:

\begin{equation}
\label{sigma_int}
\sigma_{int} = 2\pi\; \int_{0}^{\pi} \frac{d\sigma}{d\Omega} (\hat k \cdot \hat r)\: sin\theta\:d\theta
\end{equation}

\noindent while the blue solid line reports the ICS that were calculated directly from the K matrices, i.e. without the recoupling operations necessary to express the DCS in terms of K matrices \cite{sanna}.
The overall trend for the MTCS is seen to follow the ICS profile but producing consistently smaller values, as expected from the behaviour of the DCS discussed before.
The rate coefficients can in turn be obtained by integration over the energy distribution of the scattered electrons present in the astronomical environment 

\begin{equation}
k(T_{e}) = \sqrt{\frac{8\:K\:T}{\pi\:\mu}} \frac{1}{(K\:T)^{2}} \int_{0}^{\infty} \sigma(E)\:E\:e^{\frac{-E}{K\:T}}\:dE
\end{equation}

\noindent as already discussed in our introduction.
They are computed using both the present ICS ($\sigma_{int}$) and MTCS ($\sigma_{m}$) values, and are summarized in table 3.5 as a function of the kinetic electronic temperature, $T_{e}$.
In that table the first column of data refers to a numerical quadrature using the integral cross sections of eq.\ref{sigma_int}, while the second column shows the data obtained from the momentum-transfer cross sections, that for elastic scattering are defined as

\begin{equation}
\label{sigma_mt}
\sigma_{m} = 2\pi \int_{0}^{\pi} (1-cos\theta) \: \frac{d\sigma}{d\Omega} \: sin\theta \: d\theta,
\end{equation}

\noindent $d\sigma$/$d\Omega$ being the elastic differential cross section (DCS).

\begin{table}
\begin{center}
\begin{tabular}{|l|l|l|}
\hline
$ \:\: T_{e} $ & $ \:\: k_{int}(T_{e})\;=\;\langle\sigma_{int}(E)\sqrt{E}\rangle $ & $ \:\: k_{m}(T_{e})\;=\;\langle\sigma_{m}(E)\sqrt{E}\rangle $ \\ 
$ \:\: (K)   $ & $ \:\: (cm^{3}\cdot sec^{-1})                                   $ & $ \:\: (cm^{3} \cdot sec^{-1})                              $ \\ \hline
\hline
\hline
10000  &   5.34 $\cdot 10^{-9}$   &   3.24 $\cdot 10^{-9}$  \\ 
\hline
5000   &   3.94 $\cdot 10^{-9}$   &   2.67 $\cdot 10^{-9}$  \\ 
\hline
2000   &   3.22 $\cdot 10^{-9}$   &   2.53 $\cdot 10^{-9}$  \\ 
\hline
1000   &   3.02 $\cdot 10^{-9}$   &   2.49 $\cdot 10^{-9}$  \\ 
\hline
500    &   2.88 $\cdot 10^{-9}$   &   2.25 $\cdot 10^{-9}$  \\ 
\hline
250    &   2.68 $\cdot 10^{-9}$   &   1.98 $\cdot 10^{-9}$  \\ 
\hline
150   &   2.43 $\cdot 10^{-9}$   &   1.83 $\cdot 10^{-9}$  \\ 
\hline
100   &   2.16 $\cdot 10^{-9}$   &   1.75 $\cdot 10^{-9}$  \\ 
\hline
50    &   1.76 $\cdot 10^{-9}$   &   1.62 $\cdot 10^{-9}$  \\ 
\hline
30    &   1.56 $\cdot 10^{-9}$   &   1.55 $\cdot 10^{-9}$  \\ 
\hline
\end{tabular}
\caption{\small{Computed electron-molecule rate coefficient using the cross sections from eq.\ref{sigma_int} (left column) and those from eq.\ref{sigma_mt} (right column).}}
\end{center}
\end{table}

\noindent An interesting comparison is provided by the Langevin rate coefficient, which is given by the following expression

\begin{equation}
\label{langevin}
k_{L} = 2\pi e \sqrt{\frac{\alpha}{\mu}},
\end{equation}

\noindent $\alpha$ and $\mu$ being the mean polarizability of the molecule entering the collision and the reduced collisional mass respectively.
After substituting $\alpha$ = 264.35 $bohr^{3}$, such an expression returns the value $k_{L}$ = 4.57$\cdot\:$10$^{-9}\:cm^{3}\:sec^{-1}$, which is not dependent on the temperature $T_{e}$.

\noindent We find that some comments should be now made on the above results:

\noindent (i) first of all, our computed rate coefficients (both $k_{int}(T_{e})$ and $k_{m}(T_{e})$) follow the same trend in the sense that they both decrease as the $T_{e}$ temperature moves down to the lowest value considered ($T_{e}$ = 30 K); (ii) they reach almost the same value ($\sim$ 1.55/1.56$\cdot$10$^{-9}cm^{3}sec^{-1}$) when $T_{e}$ = 30 K, according to the fact that both ICS and MTCS decrease close to each other in the very-low energy range.
(iii) The increasing of $T_{e}$ up to 10000 K causes both $k_{int}(T_{e})$ and $k_{m}(T_{e})$ to increase, although $k_{m}$(at $T_{e}$ = 10000 K) is lower than the Langevin $k_{L}$ rate coefficient value: the latter thus appears as an upper limit to the present modelling.
Hence, in this framework, the Langevin rate coefficient turns out not to be 'accurate' enough, expecially at low electronic temperatures ($T_{e}$ $\leq$ 150 K).

\noindent Since our DCS as well as our ICS are both rotationally summed, we cannot separate the rotationally elastic and inelastic contributions to the final rates.
Furthermore, within the FN approach we are neglecting the vibrational degrees of freedom; consequently, given the present collision energy range and the initial internal temperatures of the 'large' molecule of interest in the present study, (e.g. see also ref. \cite{gianturco4}), little vibrational excitation effects should be present as in fact confirmed by the results of \cite{abouaf}.
In other words, since the vibrationally inelastic processes require stronger penetration of the molecular volume by the impinging electron, they should correspond to different angular distributions for the scattered electrons: however, we expect that for the present case the latter would provide only small variat
ions.
We are in fact here interested in the global behaviour of the ICS and MTCS in providing a reasonable estimate for the TNI formation rate constants which could be thus viewed as restricted with respect to our estimates.
Our computed features for the scattered electrons would therefore lead us to suggest that gaseous coronene molecules are indeed relatively 'transparent' to free electrons in the considered energy range, a conclusion also confirmed by our computed DCS, which indicate negligible electron scattering flux for values of $\theta$ larger than 60$^{0}$.

\subsection{Present conclusions}
\label{conclusions}

\noindent In this work we have endeavoured in order to analyse in some detail the behaviour of a very typical polycyclic aromatic system which is considered to be part of the possible PAH species existing in different regions of the ISM like diffuse clouds \cite{draine4}, dense molecular clouds \cite{wakelam} and in circumstellar atmospheres \cite{cesarsky}.

\noindent In particular, having discussed in a previous section (\footnote{and discovered in our earlier published work \cite{carelli1, carelli2}}) that another possible candidate as a PAH precursor, the ortho-benzyne molecule, does indeed exhibit strong interaction with the electrons present in the astronomical environments under consideration, we have deemed to be now important to move to a much larger partner, the gas-phase coronene molecule, which has been often considered important in early PAH studies \cite{allamandola}, in order to try to understand its possible behaviour, at the molecular level, as a deflector of low-energy electrons present in those astrophysical contexts and therefore understand better its possible role as an energy absorber from the environmental electrons or within a low-temperature molecular plasma produced in those astronomical surroundings.
We note moreover that one important point that makes this comparison intriguing is provided by the electron affinity for such a couple of molecules; despite in fact their marked difference in size (thus in the number of normal modes) as well as the presence of a rather large permanent dipole that characterizes the o-benzyne (1.68 D, \cite{kraka}), the EA value is similar: about +0.5 eV for C$_{24}$H$_{12}$ (+0.54 $\pm$ 0.1 eV according to the kinetic measurements as provided by \cite{chen} and +0.47 $\pm$ 0.09 eV as measured by photodetachment experiments reported in \cite{duncan}) and +0.56 eV for o-C$_{6}$H$_{4}$ \cite{wenthold, leopold}, respectively.
This suggests that $thermodynamically$ both of them should be characterized by almost the same capability to form the low-lying parent not dehydrogenated metastable negative ion; according to this, we have consequently studied their kinetics in forming the metastable anions, as provided by our estimated corresponding rate coefficients.

\noindent The calculations of the present study are indeed the first example of such computational investigations on a 'large' (\footnote{where from the computational point of view, $large$ means $demanding$}) member of the PAHs family, one which reasonably can be considered to be present in several different astronomical environments (\cite{wakelam, bakes2}) where, together with larger PAHs, plays a role in contributing to both their chemical and physical properties (\cite{wakelam, malloci, cecchi, bakes2}).
Tackling such a large species with reliable ab initio methods is furthermore meant to provide a link with earlier studies on smaller members of that large family of molecules, and therefore to clarify to us the relative importance of at least two emblematic systems: benzyne and coronene; at the same time, one can also speculate that such an endeavour could be useful to better extrapolate which is the behaviour of polycyclic aromatic molecules interacting with low-energy electrons, moving from small-size PAHs (benzene, benzyne) toward large-size PAHs (with more than 30-40 C atoms): the coronene, in fact, as a medium-size PAH without permanent dipole, can be viewed as representing one possible ideal link between the previous two PAHs classes.

\noindent The present findings come from a realistic, quantum treatment of low-energy electron scattering and provide both integral and momentum-transfer cross-sections as well as angular distributions of the elastically scattered electrons over a broad range of collision energies, i.e. covering the range of the likely energetic of electrons in dense molecular cores and in protoplanetary atmospheres.

\noindent The detailed search of the possible formation of threshold, metastable anions for the coronene molecule was carried out and discussed under the ligth of the most recent available experimental data: our findings point toward virtual state scattering (zero energy resonance), occurring for vanishing energies, as the likely doorway to the production of the parent metastable not dehydrogenated negative ion.
This allows us to surmise that the nonadiabatic couplings of the trapped electron with the nuclear motion during the near zero-energy electron-molecule encounter could be particularly efficient to access the stable bound state.
Conversely to what is thought to occur for the benzene and naphtalene molecules, due to their negative EA, for the present molecule we are thus led to confirm an active involvement of the virtual state scattering in forming the corresponding resonant anion.
Thus, more in general, the coronene molecule would constitute a reliable first example where the virtual state scattering indeed plays a crucial role in the complex interstellar processes responsible for the production of stable negative ions, which currently are in turn understood to be foundamental both in modifying, even substantially, specific molecular abundances \cite{wakelam, lepp, bakes2} and even the global physical conditions \cite{cecchi,malloci} in several astrophysical environment.
In this framework, the coronene molecule can therefore be viewed as a relatively large polycyclic aromatic species that, on the basis of the results obtained in this study about its capability to form resonances in the 0-7 energy range at the equilibrium geometry, can form by different paths bound stable anions at low- and very low-energy in competition with autodetachement channels.
From the astrophysical point of view, this means that the C$_{24}$H$_{12}$ molecule, when considered either in a diffuse or in a dense and cold molecular cloud, should be able of soaking up a rather large amount of free electrons but only at selected energy windows and not over the broad spectrum of energy distributions of the environmental electrons.
This therefore confirms our previous conclusions about the small rate coefficient for the metastable anion formation and on its 'relatively transparency' under low- and very low-energy electron scattering.
Its marked features at vanishing energies, however, allows us to suggest that it could additionally strongly interact with different atomic Rydberg species, if present in the same environment, thus actively taking part in charge-transfer chemical processes at fairly low energies.

\noindent Furthermore, the present results provided by the angular redistributions of the incident electrons, definitely show that for all the contributing IRs of this molecule, the scattering is dominated essentially by forward ejection of the electrons and by negligible intensities in the angular regions beyond $\sim\:\theta$ = 60 $^{0}$.
Hence, the corresponding momentum-transfer cross sections, which are important indicators of electron deflection gas-phase molecules, turn out to be smaller than the corresponding integral cross sections calculated here for the same system with the same quantum method (as depicted in figure \ref{fig_ch3.2_15}).
In conclusion, the corresponding rates of electron-molecule interaction are seen to be smaller than those originated from the $\sigma_{int}$ cross sections and therefore, in a statistical sense, indicate the substantial inefficiency of the coronene molecule for deflecting electrons. 
We thus argue here that more in general the coronene molecule is likely to have a limited role in disturbing radiowave propagation through interstellar plasmas since it should not participate in producing conductivity inhomogeneity.

To conclude, we want to underline that all the present findings can only be had from a realistic treatment of the quantum collisions between coronene molecule and environmental electrons, as performed in the present study.

\clearpage

\section[The carbon nitride molecule, NC$_{2}$N]{The carbon nitride molecule, NC$_{2}$N}

\subsection[Introduction: role of metastable dicyanogen anions in the negative ion chemistry of Titan's upper atmosphere]{Introduction: role of metastable dicyanogen anions in the negative ion chemistry of Titan's upper atmosphere}

\noindent Titan has been the object of very substantial scrutiny since the arrival of Cassini-Huygens at the Saturn system in July 2004.
Titan has been found to be the only satellite of the Solar System with an extensive atmosphere largely composed of N$_{2}$, with abundant minor components like CH$_{4}$ and H$_{2}$ \cite{yelle}.
The presence of negative ions in its atmosphere had originally been considered only within the cosmic-ray-induced ionosphere which lies below 200 Km, since the pre-Voyager models \cite{capone, barucki} considered their formation, by three-body electron attachment in the nighttime ionosphere, strongly reduced in the daytime because of the large photodetachment rates \cite{yung}.

\noindent It therefore came as a surprise when the Electron Spectrometer (ELS), one of the sensor of the Cassini Plasma Spectrometer (CAPS), revealed the existence of numerous negative ions in Titan's upper atmosphere\cite{waite, coates}, and furthermore negative ions have been detected on all Titan encounters when the spacecraft altitude was low enough and pointing conditions were favourable \cite{coatesbis}.

\noindent In order to understand the presence of negative ions in Titan's atmosphere, the set up of an ionosphere chemistry model for the formation of lighter-mass negative ions was recently presented and discussed \cite{vuitton}.
The dominant presence of CN$^{-}$, followed by C$_{3}$N$^{-}$ and further down C$_{5}$N$^{-}$ had been the result of the CAPS-ELS measurements at the altitude of about 10$^{4}$ Km \cite{coatesbis}, where ions with a polyyne-like structure (C$_{4}$H$^{-}$ and C$_{6}$H$^{-}$) were also found to be fairly abundant.
The modeling included a description of the suprathermal electron production from solar flux photoionization that presented two broad peaks around 1 eV and between 4 and 6 eV, with the flux tapering down by several orders of magnitude beyond 10 eV \cite{vuitton}; it also included a description of the neutral constituents distribution, producing at the end an abundance of the same lighter-mass anions found by the observation \cite{coatesbis}.
It did conclude, however, by indicating the desperate need of knowing data on the radiative recombination of molecular radicals after the primary electron impact processes.

\noindent In the present study we shall therefore focus on one possible source for the formation of the most abundant anion observed in the Titan upper atmosphere, the CN$^{-}$ anionic species, i.e. on the role of electron resonant attachment to NC$_{2}$N and the possible pathways leading to the stable formation of both CN$^{-}$ and (NCCN)$^{-}$, another observed anion \cite{vuitton}.
The latter has also been observed as a trace consituent neutral species within the upper atmosphere of this large saturnian moon \cite{kunde, coustenis}, hence the existence of some amount of dicyanogen, which can be either depleted after electron attachment by the decay into CN$^{-}$ or survive as the full anionic (NCCN)$^{-}$, is an intriguing possibility that we shall explore in the present study.

\noindent The present analysis is also connected with our previous published work on a different anionic species of polyyne-like structure, the dicyanoacetylene NC$_{4}$N \cite{sebastianelli}, where our calculations demonstrated the likelyhood of stabilizing the above anion by either radiative emission or by internal vibrational rearrangement (IVR) paths, possible mechanisms after the formation of a series of transient negative ions (TNIs) intermediates.
It was found there that, besides the NC$_{4}$N$^{-}$ anion, other species like CN$^{-}$ and C$_{3}$N$^{-}$ (mentioned before as being the most abundant anions in Titan's upper ionosphere \cite{coustenis}) could also be produced via the above stabilization mechanisms and have indeed been observed in laboratory experiments \cite{graupner}.

\subsection[Electron scattering at the equilibrium geometry]{Electron scattering at the equilibrium geometry}

\begin {figure}[here]
\begin {center}
\includegraphics[scale=0.4]{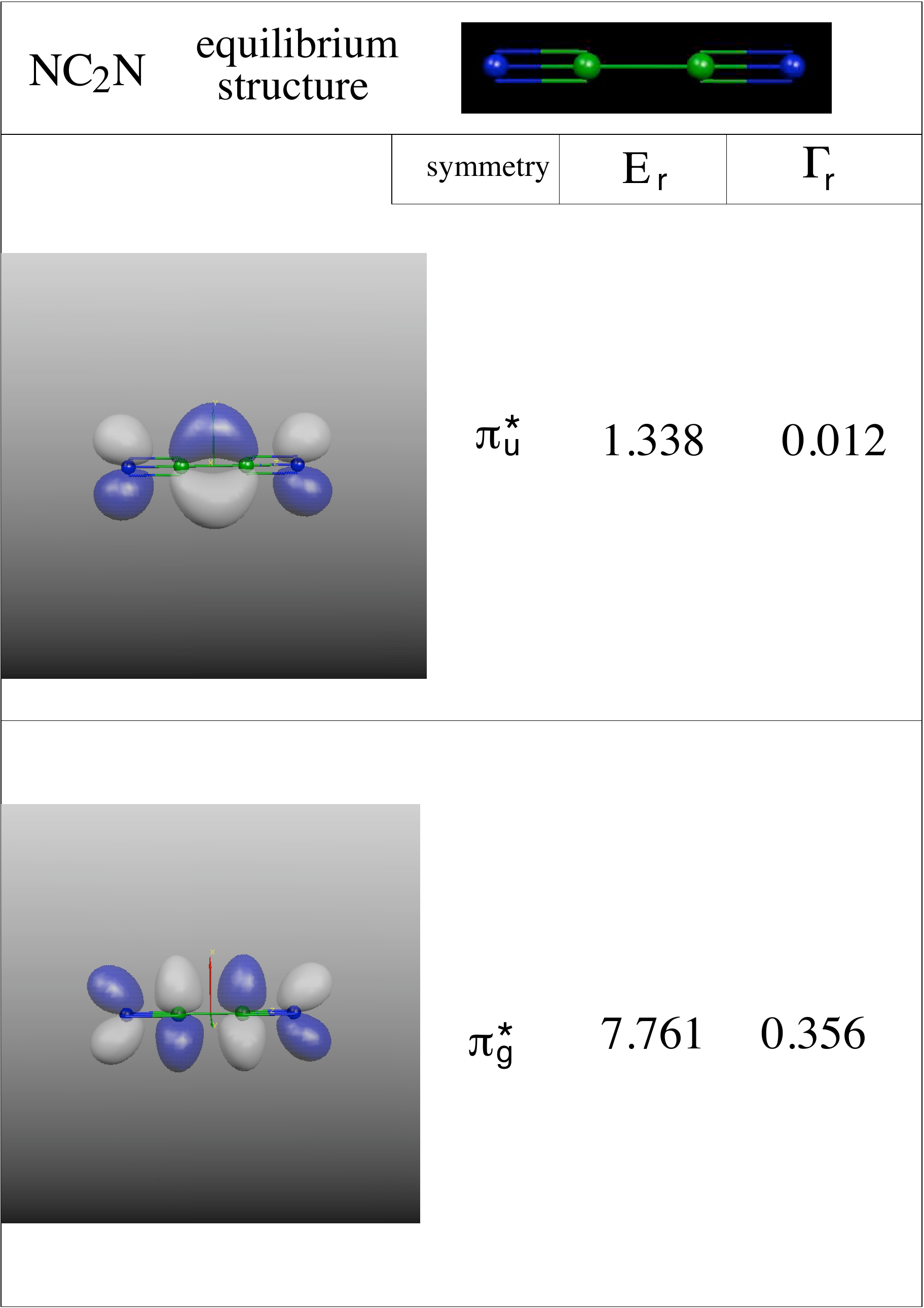}
\end {center}
\caption{\small{Spatial representation within the molecular space of the metastable excess electron for the two computed resonances of $\pi^{*}$ symmetry. The energy positions, $E_{R}$, and widths, $\Gamma_{R}$, are also reported.}}
\label{fig_ch3.3_01}
\end {figure}

\noindent The neutral molecule N-electron wavefunction, at its equilibrium geometry ($R_{NC}$ = 1.16 \AA\ and $R_{CC}$ = 1.39 \AA), was initially obtained at the Hartree-Fock (HF) level by using a 6-311++G(3df,3pd) basis set expansion within the Gaussian 03 suite of codes \cite{frisch}.
The computed polaryzabilities were $\alpha_{0}$ = 30.37 a.u., $\alpha_{2}$ = 19.59 a.u.: the experimental values are not available for comparison.
The employed range of integration for the scattering wavefunctions was extended up to 100 \AA, employing a multipolar expansion of the bound molecular orbitals up to L$_{max}$ = 60, as well as for the continuum electron, extending the computed multipoles of the e$^{-}$-molecule interaction up to $\lambda_{max}$ = 120.The total grid of integration within the ($r,\theta,\phi$) space is given by 1344 x 84 x 324 points.
The results for the two resonances of $\pi^{*}$ symmetry are shown in the panels of figure \ref{fig_ch3.3_01}.

\noindent We report in that figure the real components of the complex, continuum one-electron wavefunctions associated with the dynamical trapping (shape resonances) of the incoming electrons by the target molecule.
The formation of TNIs is seen to occur at two distinct energies, shown in the figure, and to have different widths: the lifetimes before detachment correspond to 2.7 x 10$^{-14}$s for the lower energy resonance and to 9.2 x 10$^{-16}$s for the resonance at higher energy.
According to what has been surmised \cite{vuitton} about the energetics of the electron flux in the Titan's ionosphere at 1015 Km, we see that the resonant state at lower energy falls within one of the maxima of electron flux of about 10$^{9}$cm$^{-2}$s$^{-1}$eV$^{-1}$, while the attachment at higher energy would occur within a flux range reduced by nearly two orders of magnitude.
The metastable anion of $\Pi_{u}$ symmetry also suggests that the extra density from the (N+1) electron is mostly located onto the C-C single bond and on the end N-atoms, as expected from chemical intuition.
The corresponding lowest unoccupied molecular orbital (LUMO) exhibits a very similar spatial shape, although located at a very different energy ($\sim$ 5.4 eV), and is also a $\pi_{u}^{*}$ orbital.
The metastable anion is therefore qualitatively described as binding the extra electron into that vacant orbital, although the more rigorous scattering calculations place the $\pi^{*}_{u}$ resonance at a much lower energy in the electronic continuum.
To help with a pictorical view of the quantum chemical calculations for the virtual orbitals, we report in figure 4.27 the spatial probabilities of the wavefunctions for the highest occupied molecular orbital (HOMO) and for the first three virtual orbitals of $\pi_{u}^{*}$, $\pi_{g}^{*}$ and $\sigma_{u}^{*}$ symmetries.

\begin {figure}[here]
\begin {center}
\includegraphics[scale=0.4]{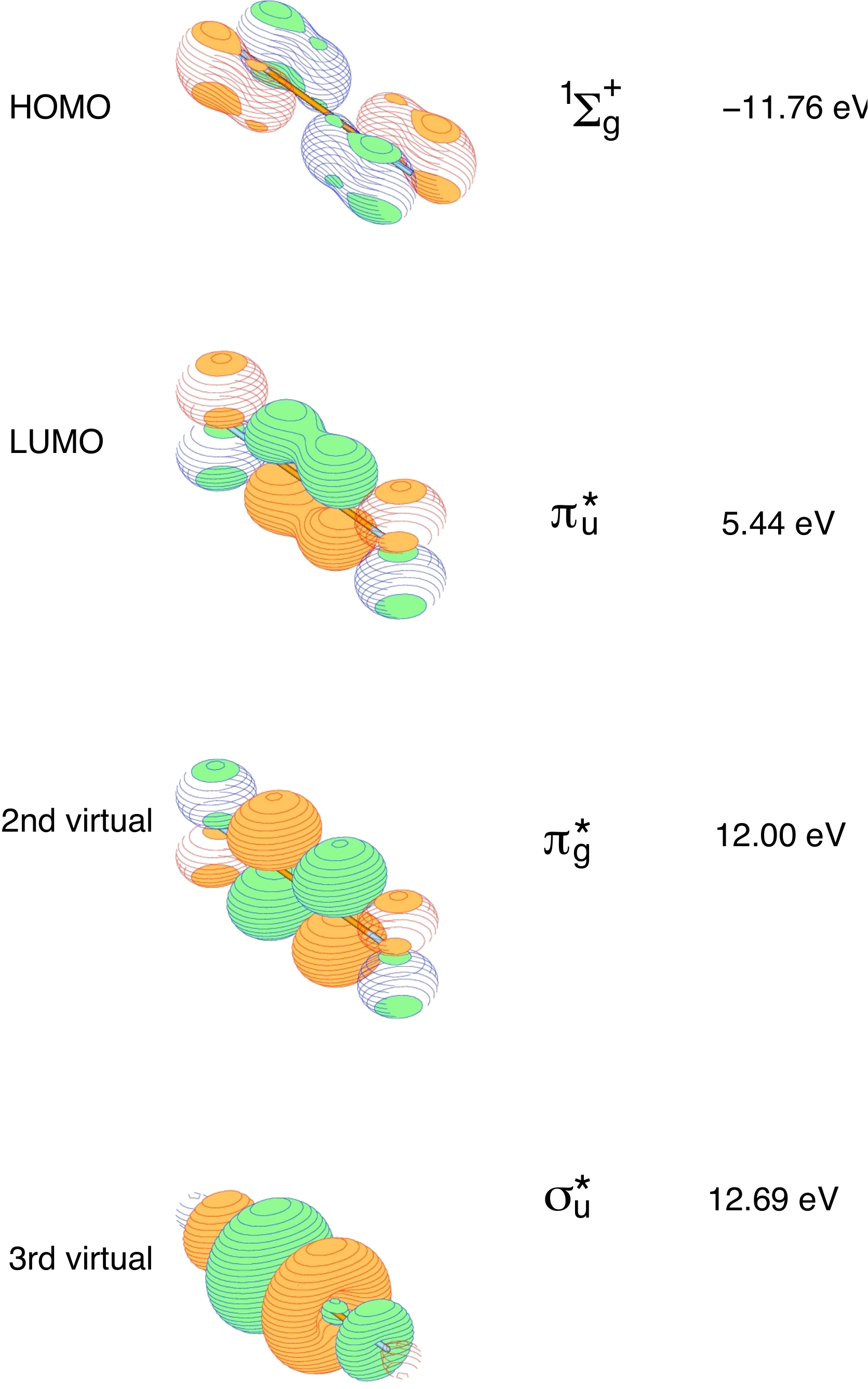}
\end {center}
\caption{\small{Computed molecular orbitals (MO) for the NC$_{2}$N at its equilibrium geometry. From top to bottom: highest occupied MO (HOMO), lowest unoccupied MO (LUMO), 2nd virtual MO (virtual02), 3rd virtual MO (virtual03).}}
\label{fig_ch3.3_02}
\end {figure}

\noindent We clearly see the strong spatial similarities between the resonant wavefunctions (real part) shown in figure \ref{fig_ch3.3_01} and the LUMO and virtual02 orbitals from quantum chemical calculations; the eigenvalues, however, are very different from the actual resonant positions obtained from our calculations.
All wavefunctions indeed show the charge accumulation along the C-C single bond for the extra electron attached to the target molecule in the $\pi^{*}_{u}$ orbital.

\begin {figure}[here]
\begin {center}
\includegraphics[scale=0.4]{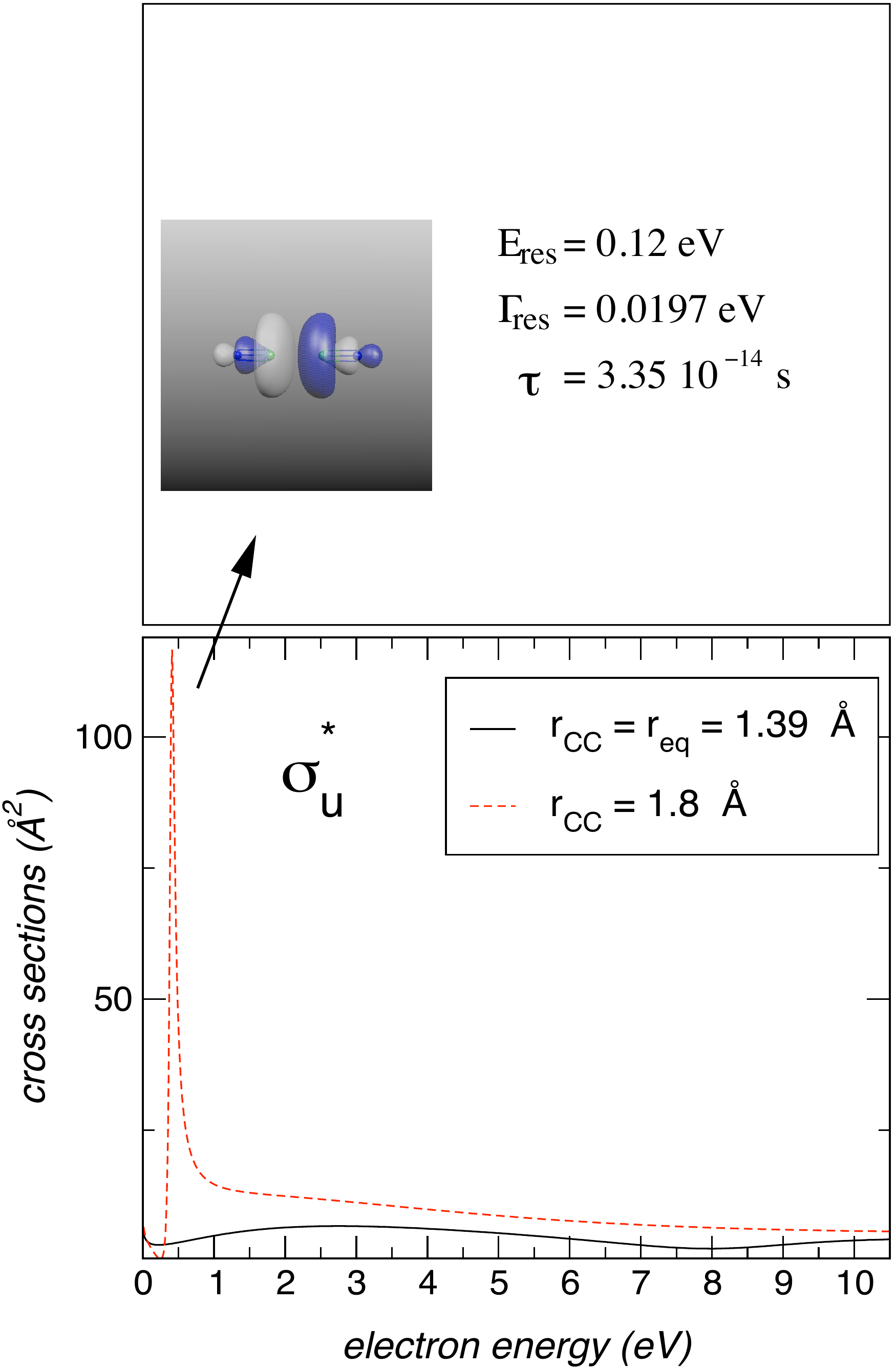}
\end {center}
\caption{\small{Computed resonance of $^{2} \Sigma_{u}$ symmetry obtained by stretching the single C-C bond. The upper panel reports the real part of the scattering wavefunction, while the lower panel show the location in energy of the strong resonant peak.}}
\label{fig_ch3.3_03}
\end {figure}

\noindent Another interesting result could be had by allowing the single C-C bond to stretch by about 30\%\ of its value up to 1.8 \AA: the scattering calculations now reveal the appearance of an additional $\sigma^{*}$ resonance very near threshold (at about 0.12 eV) and with a rather long lifetime of about 4 x 10$^{-14}$s: the data reported by figure \ref{fig_ch3.3_03} describe in more detail the behaviour of this resonance.

\noindent The spatial features of the resonant wavefunction associated with the $\sigma$-type excess electron clearly indicate that now the additional charge is chiefly localized onto the terminal C$\equiv$N groups of the linear system.
As one further stretches the C-C bond, however, we find that the resonance disappears into a bound state, i.e. the present resonance seems to correspond to a compound state of very-long lifetime and with very little residual energy that can rapidly dissipate through a linear IVR process and thus form a bound anion.
We can surmise, then, that it consitutes a possible candidate for the formation of a negative ion of the full molecule: (NCCN)$^{-}$, with a slightly stretched geometry.

\noindent It is also interesting to note here that the sequence of MOs from quantum chemical calculations (see the data in figure \ref{fig_ch3.3_02}) shows also the presence of a $\sigma^{*}_{u}$ MO after the two $\pi^{*}$ virtual orbitals, but with a very different energy collocation.

\subsection{Stretching and bending effects on the $\pi^{*}$ resonances.}

\noindent In order to follow more in detail the effects of bond deformations on the computed resonances, we carried out a series of calculations for the $\pi^{*}$ resonances depicted in figure \ref{fig_ch3.3_01}.
The first set of calculations involved the symmetric stretching of the C-C bond, leaving both C$\equiv$N bonds at their equilibrium values.

\begin {figure}[here]
\begin {center}
\includegraphics[scale=0.35]{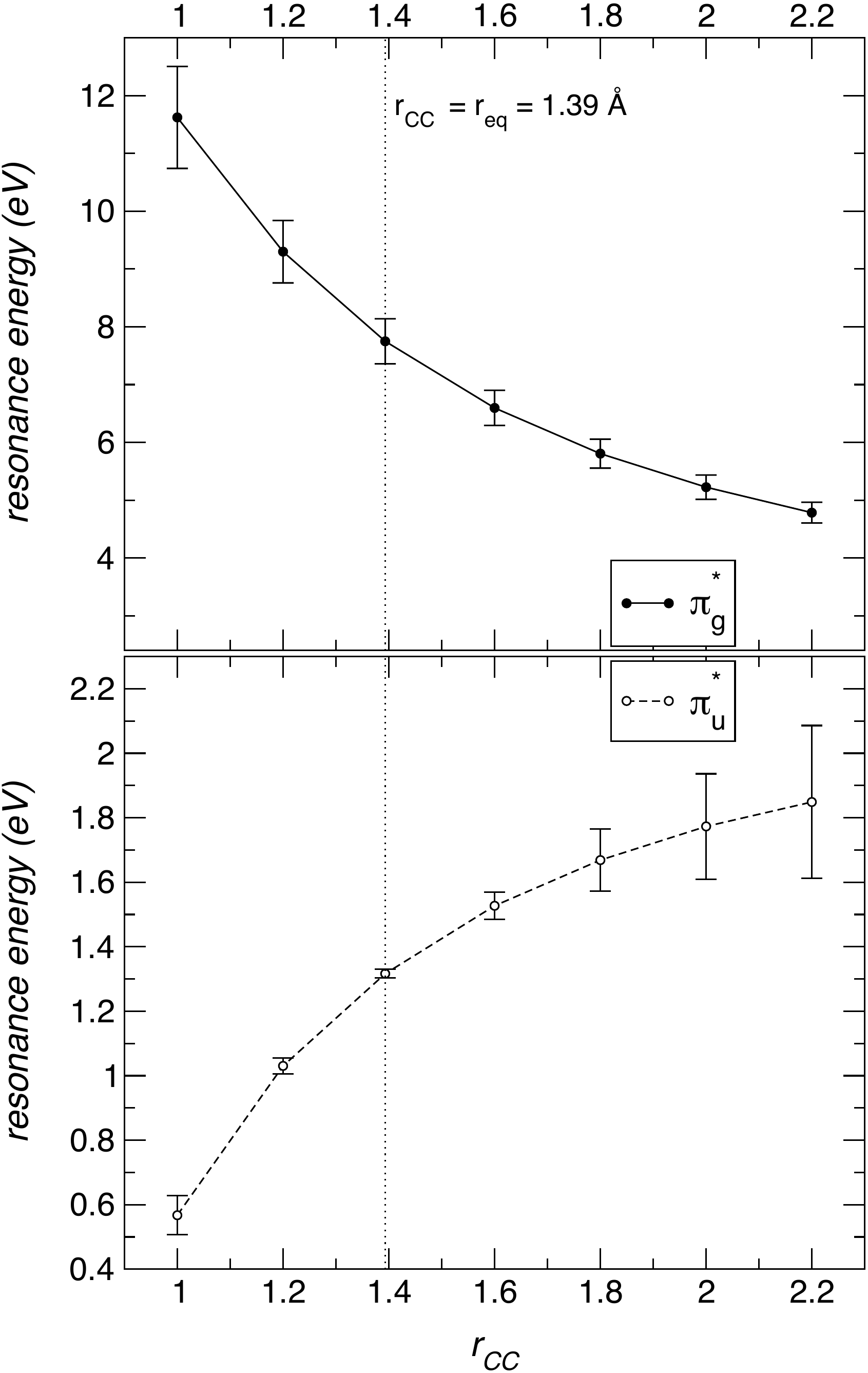}
\end {center}
\caption{\small{Computed behaviour of the resonance energy positions as a function of the stretching of the C-C bond (in \AA). The vertical dotted line shows the values at the equilibrium geometry. Solid line: $^{2}\Pi_{g}$ metastable anion; dashes: $^{2}\Pi_{u}$ metastable anion.}}
\label{fig_ch3.3_04}
\end {figure}

\noindent Figure \ref{fig_ch3.3_04} reports the findings for both resonances: we show there the resonance locations $E_{res}({\bf R})$ and indicate at each grid point the associated immaginary component, i.e. the width of each metastable state.

\noindent The calculations clearly indicate the dramatic difference of behaviour between the two resonances.
If one examines first the $\pi^{*}_{u}$ resonance, we see from figure \ref{fig_ch3.3_01} that the excess-electron wavefunction has no nodes across the bond in question and therefore it stands to reason that, in this case, the resonance gets destabilized as the nuclei move apart and its lifetime decreases.
In other words, any electron attachment into this TNI is seen to reduce its stabilization upon bond lengthening and to favour instead the electron detachment path: no stable anion can form by the single bond stretching, contrary to what occurred for the $\sigma^{*}$ resonance.

\noindent If we now turn to the $\pi^{*}_{g}$ resonance, the situation appears as entirely different: from figure \ref{fig_ch3.3_01} we see first that the excess-electron wavefunction exhibits a marked nodal plane across the C-C bond, a sign of strong antibonding character.
The stretching of the bond, therefore, now stabilizes the resonance and lengthens its lifetime, as clearly seen from the data on the top panel of figure \ref{fig_ch3.3_04}.
In other words, the formation of a $\pi^{*}_{g}$ TNI, although occurring at higher energies and therefore likely to be caused by the less intense component of the suprathermal electrons produced by solar flux photoionization \cite{vuitton}, is stabilizing the negative ion upon bond stretching rearrangements.

\begin {figure}[here]
\begin {center}
\includegraphics[scale=0.35]{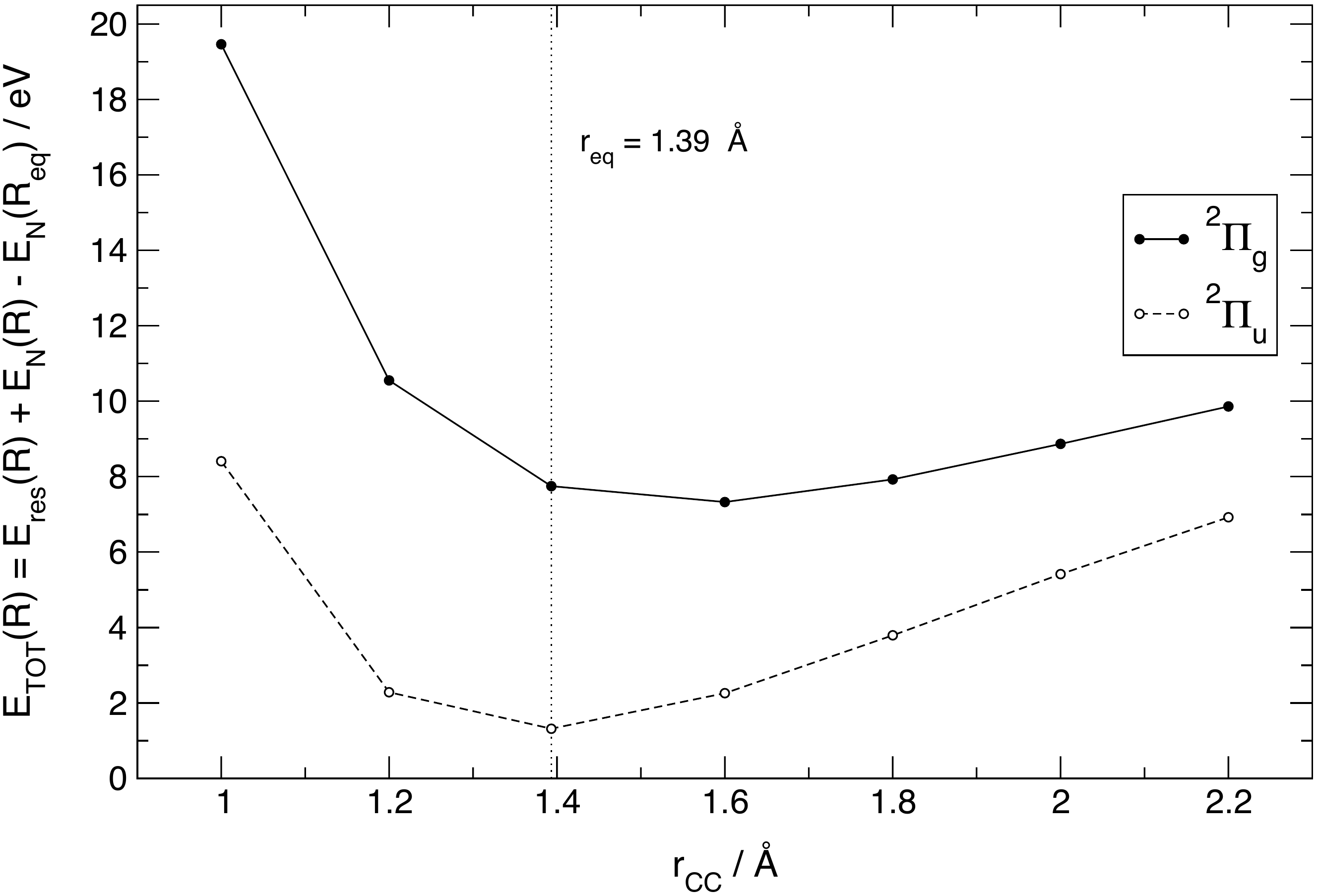}
\end {center}
\caption{\small{Computed potential energy curves of the two resonant states created in the $^{2}\Pi_{g}$ (solid line) and $^{2}\Pi_{u}$ (dashes) symmetries. The C-C bond distances are varied along the range shown. Vertical line: location of the equilibrium geometry of the neutral molecule.}}
\label{fig_ch3.3_05}
\end {figure}

\noindent This difference of behaviour can also be seen even more clearly from the potential energy curves of the resonant molecular states reported in figure \ref{fig_ch3.3_05}, obtained according to eq.\ref{Etotn+1}.

\noindent The data of figure \ref{fig_ch3.3_05} indicate that both resonances, as expected, do not lead to dissociative attachment (DA) processes since the considered cuts of the overall potential energy surfaces (PESs) do not lead to molecular dissociation.
On the other hand, the computed lifetimes of the associated resonances (see figure \ref{fig_ch3.3_04}) tell us that bond stretching stabilizes a negative ion for the $\pi^{*}_{g}$ resonance while leads to electron detachment for the $\pi^{*}_{u}$ resonance.
Thus, we could surmise that our calculations indicate already two possible ways to stable anionic formation after electron attachment: (i) via the $\sigma^{*}_{u}$ resonant intermediate near threshold and (ii) via the $\pi^{*}_{g}$ intermediate formation at higher energies.Both stabilization processes appear to possibly occur via IVR couplings for the initial excess energy of the electron.
It is also interesting to note that, when the resonant electron wavefunction is represented at the largest stretching value shown by figure \ref{fig_ch3.3_05}, it clearly exhibits a strong localization of the extra charge density over the two CN regions with no additional density along the bond.
Since the physical IVR process is really occurring on a multidimensional complex PES, then one could argue that more complicated molecular deformations after stretching could turn out to be dissociative deformations and therefore lead to final (CN)$^{-}$ detachment.

\noindent With regard to the stable molecular anion, earlier and highly accurate quantum chemical calculations \cite{nsangou} indicated that the most stable bound anionic species is the one given by the trans configuration of the N$\equiv$C-C$\equiv$N, with a C-C$\equiv$N angle distorted from 180$^{\circ}$ down to about 150$^{\circ}$, yielding a vertical electron affinity of 0.65 eV, not far from the experimental indication of 0.58 eV \cite{ng}.
In order to see the effects on the $\pi^{*}$ resonances of deforming the neutral molecule close to the ground state geometry of the bound anion suggested in ref \cite{nsangou}, we have carried out scattering calculations in which the angle mentioned before is distorted from the linear configuration ($\theta$ = 180$^{\circ}$) towards the value of the stable anion ($\theta$ = 154.7$^{\circ}$) for the trans configuration \cite{nsangou}; the bond distances were kept at $R_{CN}$ = 1.124 \AA\ and $R_{CC}$ = 1.393 \AA\ during the bending, as indicated by the quantum chemical study \cite{nsangou}.

\begin {figure}[here]
\begin {center}
\includegraphics[scale=0.35]{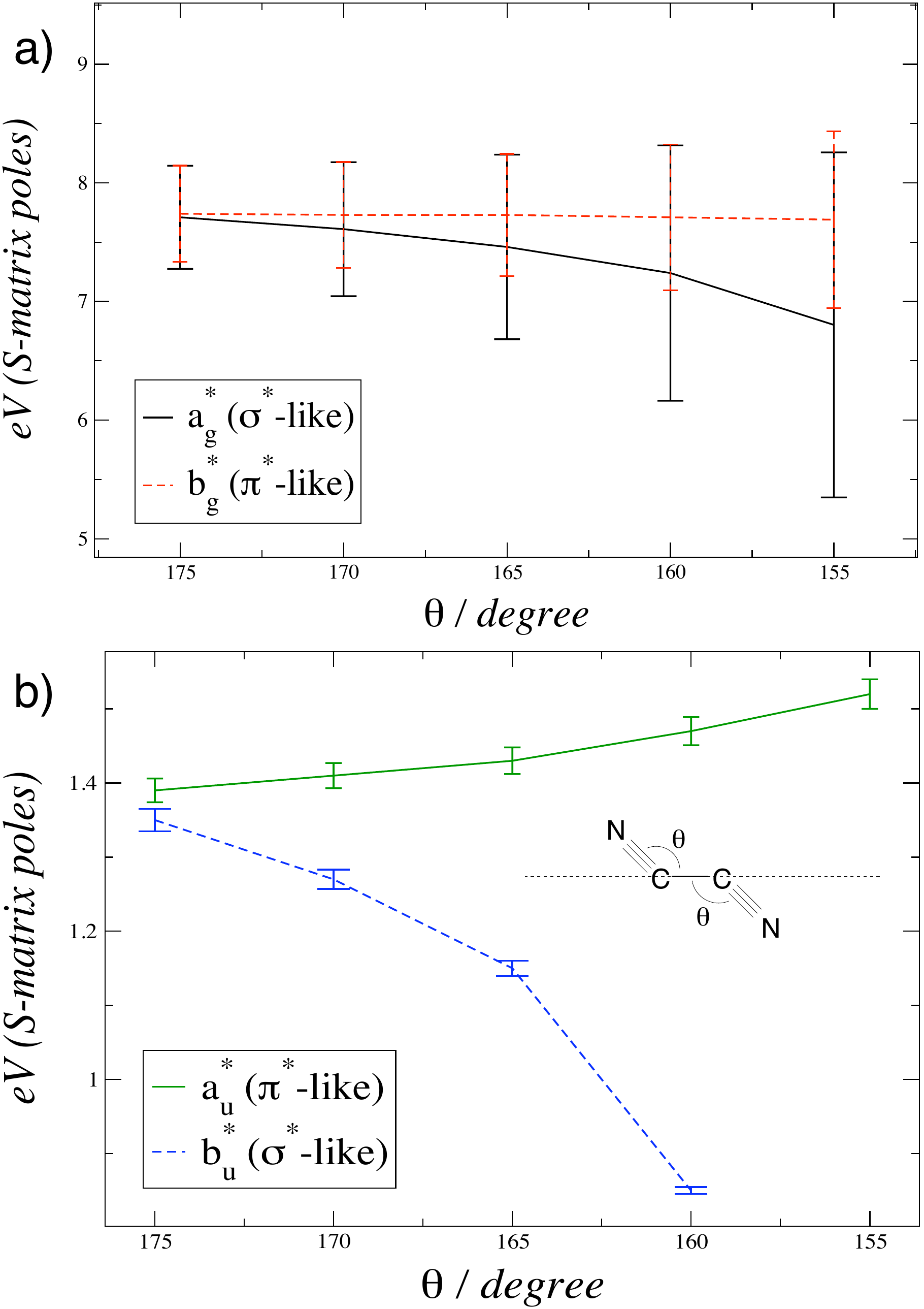}
\end {center}
\caption{\small{Computed resonance positions and widths (bars at the grid points) as a function of changing the angle $\theta$ of the trans configuration (in
sert). See main text for details. Upper panel: deformation effects on the $g$-type resonant components, $a^{*}_{g}$ and $b^{*}_{g}$, when the degeneracy is r
emoved. Lower panel: same evolution for the $u$-type resonance components.}}
\label{fig_ch3.3_06}
\end {figure}

\noindent The results of the calculations are shown in the two panels of figure \ref{fig_ch3.3_06} for both $\pi^{*}_{g}$ and $\pi^{*}_{u}$ resonances.

\noindent The data on the upper panel report the removal of the $\pi$-degeneracy as the symmetry lowers to the $D_{2h}$ symmetry: the $a^{*}_{g}$ resonance component acquires now the $\sigma^{*}$-like character existing in the linear configuration, while the $b^{*}_{g}$ component now follows the original $\pi^{*}$-like character of the linear geometry. 
The calculations confirm here what we had already surmised from the stretching effects seen in figure \ref{fig_ch3.3_02}, but now with an inverted behaviour: the higher resonance moves little in energy when the C-C bond does not change and only the orientation of the C$\equiv$N group changes, leading instead to a less stable compound state with strongly reduced lifetimes in both components.
In other words, we see that the bending deformation preferentially causes detachment of the resonant electron attached at the higher energy ($\sim$7.7 eV).

\begin {figure}[here]
\begin {center}
\includegraphics[scale=0.35]{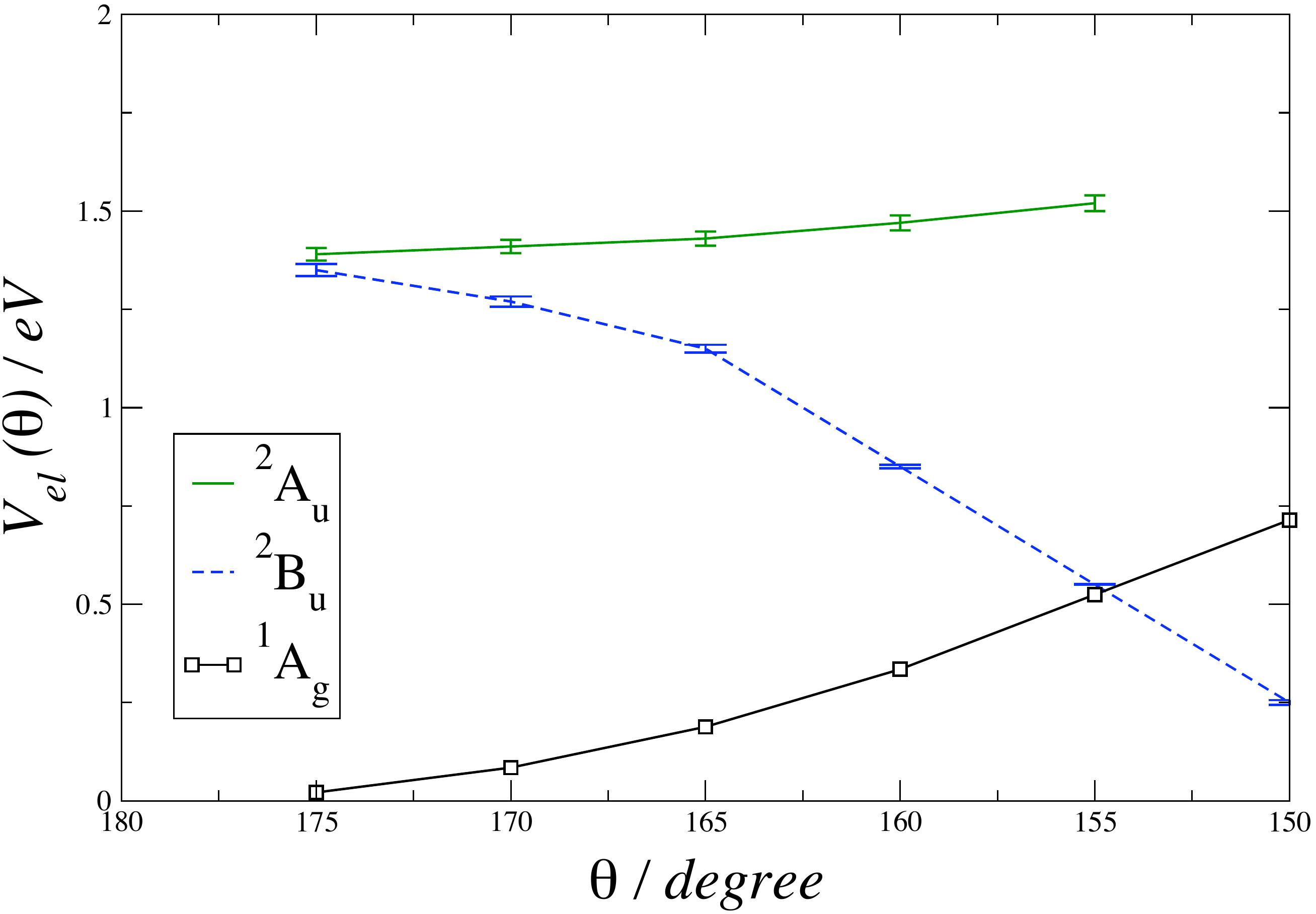}
\end {center}
\caption{\small{Computed electronic energies of the (N+1)-electron metastable states reported in fig. 4.31b ($A_{u}$, $B_{u}$) and of the (N)-electron neutral molecule ($A_{g}$) as a function of the deformation angle $\theta$. The reference energy is that of the linear neutral ($\theta$ = 180$^{\circ}$)}}
\label{fig_ch3.3_07}
\end {figure}

\noindent On the other hand, the data in the lower panel clearly show that the $\sigma^{*}$-like component of the lower resonance is dramatically affected by the bending motion, moves down to threshold energy and in our calculations becomes almost a bound state around $\theta$ = 160$^{\circ}$, i.e. not far the stable anion geometry ($\theta$ = 154.7$^{\circ}$) given by the quantum chemical calculations \cite{nsangou}.
This shows that electron attachment to the partner molecule, via the lower energy $\pi^{*}_{u}$ resonance of its equilibrium geometry, can evolve into a stabilized negative ion that undergoes efficient IVR coupling mechanism via the bending degree of freedom: this possibility could be examined more pictorically from the data reported in figure \ref{fig_ch3.3_07}.

\noindent We report in that figure the real energies of the (N+1)-electron metastable anions associated with the two resonances of figure \ref{fig_ch3.3_06}b.
We further show the energy dependence on the same angle $\theta$ for the case of the ground state neutral molecule.
In spite of the fairly simplified description of the quantum electronic state of the neutral molecule (given at the Unrestricted Hartree-Fock (UHF) level only), the general physical picture comes up rather clearly: as the system is deformed towards the optimal geometry of its stable anion \cite{nsangou} we see, in fact, that the $\sigma^{*}$-like resonant state moves down in energy and tends to become a bound state crossing the real energy curve of the corresponding neutral partner.
By extrapolating the scattering data to negative energy we see that stabilization of (NCCN)$^{-}$ starts around 155$^{\circ}$, i.e. very close to the structural value for the bound anion \cite{nsangou}.
If one further considers that the estimated adiabatic electron affinity is around 0.3 eV \cite{ng}, then we see that our extrapolation reaches that value around 150$^{\circ}$, a failry realistic estimate from scattering calculations of the bending angle of 154$^{\circ}$ of the stable anionic species given in \cite{nsangou}.

\begin {figure}[here]
\begin {center}
\includegraphics[scale=0.28]{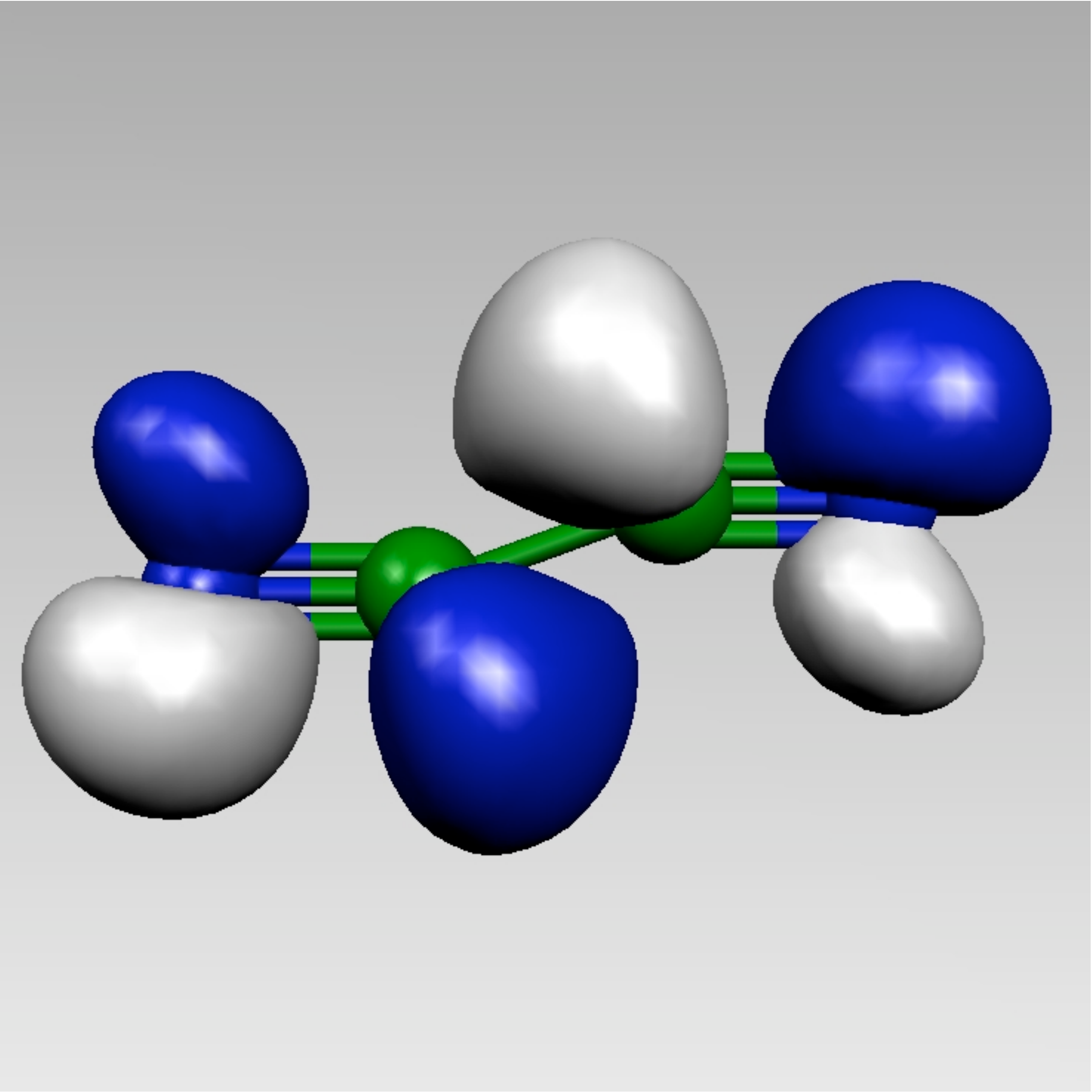}
\end {center}
\caption{\small{Computed real part of the wavefunction for the $b^{*}_{u}$ component of the resonant electron in $D_{2h}$ symmetry at the bending angle value of $\sim$160$^{\circ}$. See text for details.}}
\label{fig_ch3.3_08}
\end {figure}

\noindent A further way to qualitatively confirm the stabilization of the anionic molecule when following the bending path depicted in figure \ref{fig_ch3.3_06} is to look at the spatial features of the resonant excess electron at the last geometry of our scattering calculations, i.e. at the $\theta$ value of $\sim$160$^{\circ}$.
Those features are given by figure \ref{fig_ch3.3_08} and describe the $\sigma^{*}$-like component of the $u$-type resonant state.

\noindent We clearly see that the excess, resonant electron is distributed over all the atoms of the system, it has ''$\sigma_{u}$'' character with respect to the plane of symmetry in the bent molecule and it does not exhibit any ''bond rupture'' feature perpendicularly to the C-C bond.
In other words, we are producing a metastable precursor, at nearly zero energy, of the bent configuration of the (NCCN)$^{-}$ bound anionic species.

\noindent It is also interesting to note here that we classify under the 'IVR' heading a broad variety of processes where nonadiabtic coupling effects are induced by nuclear motion.
For instance, the use of Renner-Teller nonadiabatic effects to explain dissociative recombination processes as discussed in \cite{yungen} for the HCO$^{+}$ + e$^{-}$ system is another instance where our nuclear-induced nonadiabatic energy redistribution is seen to be at work.

\subsection{Present conclusions}
\label{conclusions}

The present work has investigated, at the nanoscopic level, the quantum dynamics of the stabilization paths which could lead to the formation of stable, bound anionic molecules of (NCCN) after the attachment of environment electrons.
The latter electrons span the energy range of the observed distributions of free electrons produced by the solar photons in the Titan's atmosphere \cite{vuitton} and the quantum description of the scattering processes between such electrons and the title molecule reveals that, at specific energies, the collision events cause formation of resonant compounds, i.e. of metastable anionic complexes temporarily bound within the continuum energy of the electrons.
In particular, our present calculations found that a $^{2}\Sigma_{u}$ resonance is formed at very low energy and with a very large lifetime, indicating its likely stabilization as a bound state by internal vibrational redistribution (IVR) processes of the small excess energy.

\noindent Additionally, we found two more resonant anionic states: a $^{2}\Pi_{u}$ state around 1.3 eV and a $^{2}\Pi_{g}$ one around 8 eV.
The former resonance exhibits a longer lifetime than the latter and, upon stretching the C-C bond, they follow different evolutions.
Thus, the resonance of $\pi^{*}_{g}$ character stabilizes into a very long-lived metastable state with strongly reduced residual energy of that excess electron, while the one of $\pi^{*}_{u}$ character moves up in energy and appears to follow preferentially an electron detachment path: its lifetime decreases upon bond stretching.
In other words, the stretching of the single C-C bond provides an efficient nuclear coupling of the excess electron motion only in the case of the $\pi^{*}_{g}$ resonant electron, where it can lead to the formation of a stable (NCCN)$^{-}$ molecule.

\noindent We have further investigated the effects of bending deformations on the above resonances and carried out calculations in which the C$\equiv$N groups are bent symmetrically from the linear configuration and move the molecule to a geometry suggested by earlier calculations \cite{nsangou} to be the ground state of its bound negative ion.

\noindent Our calculations find now that the reduction of the symmetry removes the degeneracy of the $\pi$-resonances and that each component follows a different path (see our data in figures \ref{fig_ch3.3_06}a, \ref{fig_ch3.3_06}b).
We see there that the $\sigma^{*}$-like component ($^{2}B_{u}$) of the $\pi^{*}_{u}$ resonant electron is now dramatically lowering its energy location and lengthening its lifetimes.
In comparison with the potential energy curve of the neutral molecule ($^{1}A_{g}$), our scattering calculations indeed suggest that the metastable potential energy component (real part) for the $^{2}B_{u}$ anion now crosses the neutral potential around a bent structure of 155$^{\circ}$ (see figure \ref{fig_ch3.3_07}), and therefore stabilizes the bound anionic configuration suggested by the earlier quantum chemical calculations.
In other words, our modelling of the dynamical evolution that follows electron attachment on a series of bent configurations for the title molecule indicates that the extra-electron of $b^{*}_{u}$ symmetry ($\sigma^{*}$-like wavefunction) couples very efficiently with that bending motion and can stabilize into a bound state of the anionic species.
These findings therefore confirm the $^{2}\Pi_{u}$ resonant complex to provide an efficient doorway state to the stabilization of the (NCCN)$^{-}$ anion in its $D_{2h}$ deformed configuration with respect to the linear neutral.
On the other hand, our calculations also suggest that metastable attachment of electrons to NCCN molecules does not lead to the formation of (CN)$^{-}$ fragments via shape-resonance mechanisms along simple pseudo-1D molecular deformations, but may require instead a more complicated set of nuclear motions.

\clearpage


\chapter*{Concluding remarks}
\addcontentsline{toc}{chapter}{Concluding remarks}
\chaptermark{Concluding remarks}
\label{conclusions}

\noindent The present thesis focuses on low-energy electron collisions with astrophysically relevant closed-shell molecules, therefore represents a theoretical/computational work which deals with an area placed at the boundary between (molecular) astrophysics, quantum collision thery, and of course theoretical chemistry.

\noindent The findings presented in this work demonstrate that taking in consideration resonant low-energy electron attachment processes associated not only to s-wave resonances, does provide many useful information.
From the astrophysical point of view, such a treatment reveals infact to be exhaustive in the sense that it enables to locate many more channels to form different metastable anions associated with a given neutral molecule.

\noindent In the framework of a rigorous ab initio approach, the ensuing characterization of the resonances by the 3D spatial maps describing the resonant wavefunctions enables to immediately visualize which are the likely reactive sites on the metastable anion, therefore allowing to also qualitatively surmise possible reaction paths involving that species.
In line with that, keeping in mind that the stable anion formation process by resonant electron collision is a unimolecular process, our pseudo-1D approach proves to be a powerful tool by which investigate the dynamical evolution of the resonant species; following the 'chemical intuition', when supported by the necessary decreasing of both the real and the imaginary (longer lifetimes) component of the resonance energy as function of the choosen molecular 1D deformation, the 3D maps for different molecular geometries could be also considered as indicative of an IVR process in the sense that the localization of the resonant electron on a specific fragment could suggest a non-ergodic IVR(\footnote{usually leading to fragmentation}), while, conversely, the 'broadening' of such spatial resonant electron density over a larger molecular region could suggest an actual ergodic multidimensional IVR.
Although we are not able to estimate the time requested by the IVR occurrence, we can however take advantage of our resonance lifetime estimates to qualitatively infer if there is time enough for the IVR to be effective.

\noindent Moreover, given a neutral molecule, which is astrophysically relevant, the qualitative comparison between the energies requested to the resonant species formation with the environmental conditions of the astrophysical region under investigation can then allow a deeper insigth into the feasibility of certain chemical networks; in this sense, one can further take advantage of the knowledge of the relevant ab initio integral, differential and momentum transfer cross sections to also estimate the metastable anion formation's rate coefficient, the latter being an important parameter needed by astrophysicists involved in modeling the chemical networks in the ISM.\\

The three molecular species that were computationally investigated about their behaviour under low-energy electron collisions, are the ortho-benzyne (o-C$_{6}$H$_{4}$), the coronene (C$_{24}$H$_{12}$), and the carbon nitride (NC$_{2}$N), respectively.
As already emphasized in the general introduction, due to their sizes, their peculiar structures, their chemical reactivity, their physical and chemical properties as well as according to astronomical observations, when available, the above three molecules belong to different astrophysical contexts in the sense that each of them is linked to an important example of a specific astrophysical puzzle that appear to currently constitute an intriguing scientific challenge.

\noindent The following important remarks can now de drawn from our present study:

\noindent i) \textbf{ortho-benzyne}

\noindent The computational investigation for this molecule focused on the possible role which could be played by the anionic ortho-benzyne as a very reactive intermediate in proto-planetary (PPN) atmospheres, where the physical conditions are suitable to justify its actual existence.
Due to the positive electron affinity that characterizes the parent neutral species, the o-C$_{6}$H$_{4}^{-}$ anion could be thus also viewed as a reasonable aromatic precursor of larger species belonging to the polycyclic aromatic hydrocarbons' family.

\noindent When considering the o-C$_{6}$H$_{4}$ as frozen at its equilibrium geometry, the results indicate that this molecule indeed exhibits four fairly narrow resonances below 10 eV, three of which can be classified as $\pi^*$ resonances while one of them is a low-energy $\sigma^*$ resonance exhibiting antibonding character across the triple bond.
When compared with the benzene molecule, it shows lower energy resonances: since the excess energy carried by the extra electron has to come from the environmental medium in such a stellar atmosphere, the attachment processes to benzene do require much higher electron temperatures than those needed for the formation of metastable anions of benzyne.
Furthermore, the (o-C$\mathrm{_{6}}$H$\mathrm{_{4}}^{-}$)$^{*}$ temporary negative ion is seen to have markedly longer lifetimes, thereby allowing more easily the dynamical coupling of the extra electron with the nuclear degrees of freedom which play a crucial role in the ensuing molecular break-up or also in possible radiationless non-dissociative stabilization paths.

\noindent Keeping in mind that the benzene has a rather large negative electron affinity (-1.12$\pm$0.003 eV, \cite{burrow}), it therefore appears as reasonable to argue about the possible surviving of the benzene as a neutral species against the more likely involvement of the longer lived negative ion o-C$_{6}$H$_{4}^{-}$ in several subsequent chemical reactions.
This provides a possible explanation about the presence of the former and, at the same time, the absence of the neutral counterpart \cite{widicus}. 

\noindent The pseudo-1D quantum dynamics, which was employed to study the possible dynamical evolution once the metastable anion is formed, revealed the stabilization of the initial metastable anion by small ring deformations which cannot lead to ring-breaking effects, and, even more interestingly, the formation of a bound aromatic anion by small direct stretching of the triple bond, therefore suggesting the possibility of reaching the bound (o-C$_{6}$H$_{4}$)$^{-}$ ground state by fast intramolecular vibrational redistribution processes.
The combination of the estimated rate coefficient value for the metastable ortho-benzyne anion with the possible fast radiationless stabilization of such a resonant anion, enabled us to suggest that the neutral o-C$_{6}$H$_{4}$ might actually play a role in the circumstellar chemistry of CRL-618 and more in general in the PPN chemistry, if the neutral parent molecule is actually produced in that environment.
In conclusion, in line with the above conjectures, which are in turn proven to be realistic by the findings of the present quantum scattering calculations, it was then possible to also tentatively indicate the reactions with H/H$^{+}$ and the following, albeit qualitative, chemical paths for an aromatic condensation reaction toward larger PAHs,

\begin{figure}[here]
\begin{center}
\includegraphics[scale=0.09]{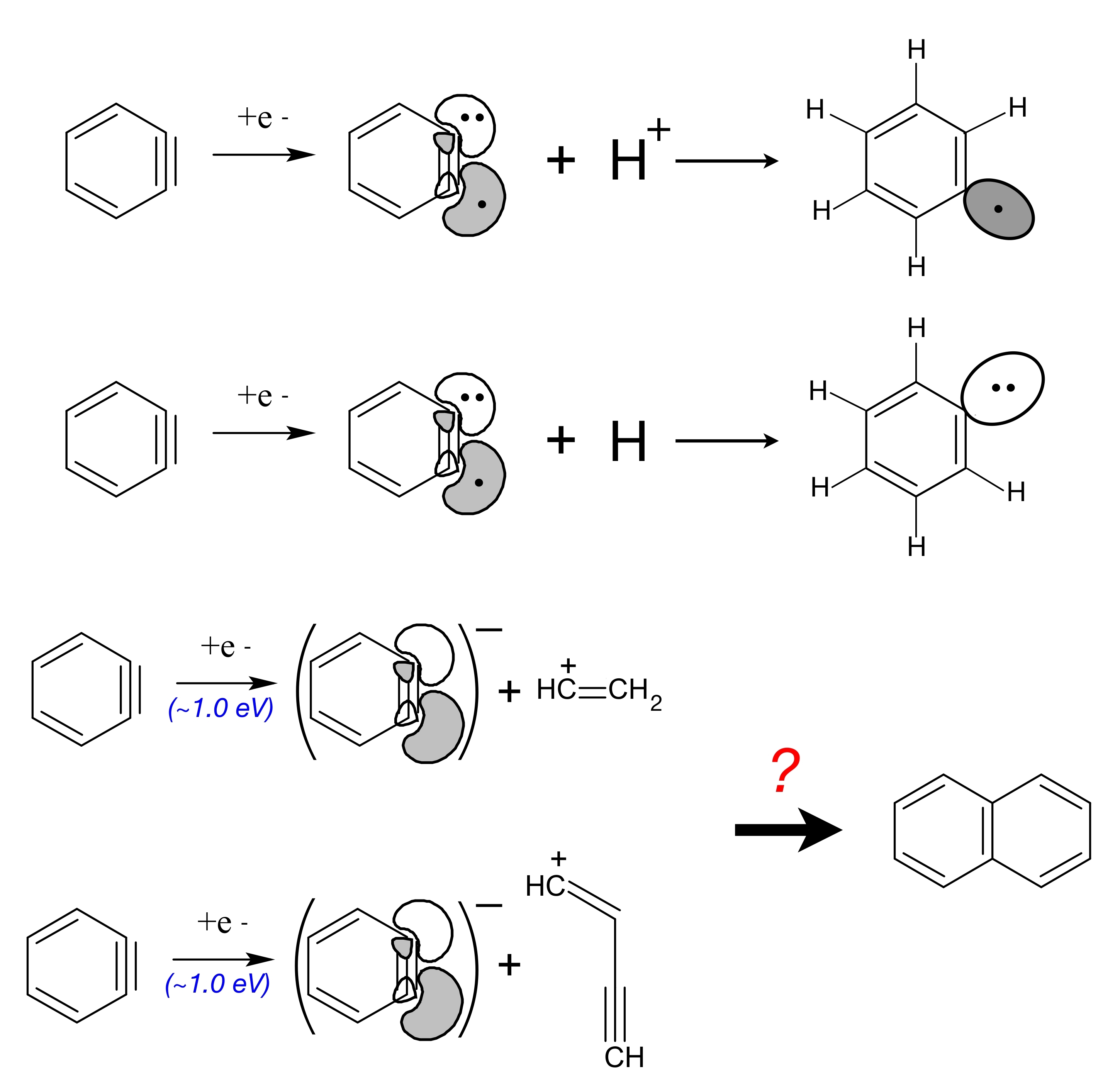}
\end{center}
\end{figure}

\noindent where the molecules chosen in the above figure indeed correspond to some of the most likely abundant carbonaceous species in the CRL-618 environment (\cite{woods2}).

\noindent ii) \textbf{coronene}

\noindent The present calculations for the coronene molecule, a larger partner when compared with the previous molecular system, are indeed the first example of a computational investigation of a medium-size member of the PAHs family, one which can be considered as likely to be present in several different astronomical environments \cite{wakelam, bakes} where, together with larger PAHs, plays a role in contributing to both their chemical and physical properties \cite{wakelam, malloci, cecchi, bakes}.
Tackling such a 'large' species with reliable ab initio methods has furthermore meant to provide a link with earlier studies on smaller members of that wide family of molecules, and therefore to clarify the relative importance of at least two emblematic systems: ortho-benzyne and coronene.
At the same time, one can also speculate that such an endeavour could be useful to better extrapolate the behaviour of polycyclic aromatic molecules interacting with low-energy electrons, moving from the small-size closed shell PAHs (benzene, benzyne) toward the large-size closed shell PAHs (with more than 25-20 C atoms): the coronene, in fact, as a medium-size PAH without permanent dipole, can be viewed as representing one possible ideal link between the previous two PAHs classes.
In this connection, we should note that one important point that makes intriguing the comparison between the coronene and the o-benzyne, is provided by the electron affinity for these molecules; despite their marked difference in size (thus in the number of normal modes), as well as the presence of a rather large permanent dipole that characterizes the o-benzyne (1.68 D, \cite{kraka}), the EA value is similar: about +0.5 eV for C$_{24}$H$_{12}$ (+0.54 $\pm$ 0.1 eV according to the kinetic measurements in \cite{chen} and +0.47 $\pm$ 0.09 eV as measured by photodetachment experiments reported in \cite{duncan}) and +0.56 eV for o-C$_{6}$H$_{4}$ \cite{wenthold, leopold}, respectively.

\noindent The detailed search of the possible formation of threshold, metastable anions for the coronene molecule was carried out and discussed under the ligth of the most recent available experimental data: our findings point to the virtual state scattering (zero energy resonance), occurring for vanishing energies, as the likely doorway to the production of the parent metastable not dehydrogenated negative ion, consequently allowing us to also surmise that an important role might be played by the nonadiabatic couplings with the nuclear motions during the zero-energy electron-molecule encounter in order to access the stable bound state.
Therefore we argue that the stable bound state, that experiments show to occur for an energy close to 0 eV, could be accessed via the zero energy resonance evolution due to the nuclear motions that increase the induced dipole moments interaction.

\noindent More in general, on the basis of the results obtained on its capability to form resonances in the 0-7 eV energy range at the equilibrium geometry, the coronene molecule could be viewed as a species that forms bound stable anions but almost only at very low-energy: according to experiments, in fact, at incident energies higher or equal to 2 eV, the present findings allow to surmise the occurrence of several resonant $excitation$ processes instead of actual bound stable negative ions.
From the astrophysical point of view, this means that the C$_{24}$H$_{12}$ molecule should not be able to strongly participate in soaking up a rather large amount of free electrons; on the other side we can speculate that it could strongly react with different atomic Rydberg species, if present in the same environment.

\noindent In order to try to understand its possible behaviour, at the molecular level, as a deflector of low-energy electrons present in those astrophysical contexts, and therefore understand better its possible role as an energy absorber from the environmental electrons or within a low-temperature molecular plasma produced in those astronomical surroundings, the elastic (rotationally summed) differential cross sections were investigated: the findings enable to look at this molecule as a relatively 'transparent' species for low-energy electrons, so that we can also conjecture that the C$_{24}$H$_{12}$ molecule is likely to have a limited role in disturbing radiowave propagation through interstellar plasmas since it should not participate in producing conductivity inhomogeneity.

\noindent iii) \textbf{dicyanogen}

\noindent This molecule is relevant for the chemical modeling of Titan's upper atmosphere, since according to the current reaction network \cite{vuitton}, such a species is supposed to efficiently yield, by dissociative electron attachment, the most abundant anion observed in that environment: the CN$^{-}$.
In this connection, following the interaction with low-energy environmental electrons by means of a pseudo-1D modelling of excess energy redistributions, we have analysed in some detail the possible dynamical paths which could lead, under the conditions observed in Titan's atmosphere, to the formation of both the above small anionic fragment and the stable (NCCN)$^{-}$.
The latter, in fact, represents another possible product that could be efficiently produced, once the primary electron attachment have taken place so that the corresponding metastable anion, (NC$_{2}$N$^{-}$)$^{*}$, is formed.

\noindent The present scattering calculations revealed that a $^{2}\Sigma_{u}$ resonance is formed at very low energy and with a very large lifetime, indicating its likely stabilization as a bound state by internal vibrational redistribution (IVR) processes of the small excess energy.
Additionally, we found two other resonant anionic states, a $^{2}\Pi_{u}$ state around 1.3 eV and a $^{2}\Pi_{g}$ one around 8 eV, respectively, which upon stretching the C-C bond follow different evolutions: the former, which at the equilibrium has a longer lifetime than the latter, moves up in energy with the C-C lengthening, and, as its lifetime decreases, it also appears to follow preferentially an electron detachment path.
The other one, conversely, appears to stabilize into a very long-lived metastable state characterized by a strongly reduced residual energy, so that in this case we are able to argue that the C-C stretching should provide an efficient nuclear coupling by means of which the formation of the stable (NCCN)$^{-}$ anion can occur.

\noindent The CN$^{-}$ fragment production is not however completely excluded: according to our pseudo-1D dynamical investigation, we can only argue that more complicated molecular deformations, following the C-C stretching, are likely to be involved in causing this fragmentation.

\noindent We have further investigated the effects of bending deformations on the above resonances and carried out calculations in which the C$\equiv$N groups are bent symmetrically from the linear configuration and move the molecule to a geometry suggested by earlier calculations \cite{nsangou} to be the ground state of its bound negative ion.

\noindent Our calculations find now that the reduction of the symmetry removes the degeneracy of the $\pi$-resonances and that each component follows a different path (see our data in figures 4.31a, 4.31b).
We see there that the $\sigma^{*}$-like component ($^{2}B_{u}$) of the $\pi^{*}_{u}$ resonant electron is now dramatically lowering its energy location and lengthening its lifetimes.
In comparison with the potential energy curve of the neutral molecule ($^{1}A_{g}$), our scattering calculations indeed suggest that the the real part of the metastable potential energy component for the $^{2}B_{u}$ anion now crosses the neutral potential around a bent structure of 155$^{\circ}$ (see figure 4.32), and therefore stabilizes the bound anionic configuration suggested by the earlier quantum chemical calculations \cite{nsangou}.

\noindent As a general conclusion from the astrophysical point of view, the above three in-depth studies thus provide important examples on the role of anionic species in different astronomical environment like a PPN atmosphere (where we surmise the involvement of o-benzyne anion in PAH synthesis), a dense interstellar cloud (where the coronene could represent a typical medium-size PAH) and a planetary ionosphere (the upper atmosphere of Titan where the anionic chemistry of nitrogenated polyynes as dicyanogen play a role) where, without any claim of completeness, each of the above three molecular systems investigated in this work represents in fact an emblematic example.

\clearpage

\appendix
\chapter{The elastic ICS and DCS}

\noindent As briefly introduced in section 2.4.8, in order to obtain an expression for the differential cross section one needs to introduce then a coordinate system $S'$ which is space-fixed (SF), the so-called LAB system, and another coordinate system $S$ which is fixed to the molecular highest symmetry axis and that is identified as body-fixed (BF).
The latter is obtained from $S'$ by a rotation through the appropriate Euler angles ($\alpha$, $\beta$, $\gamma$).
The previously discussed expression for the FN scattering amplitude (see eq. 2.66), through the symmetry adapted functions $X_{\ell h}^{p\mu}(\hat{\textbf{k}}_{0})$ and $X_{\ell h}^{p\mu}(\hat{\textbf{r}})$, contains angular factors like $Y_{\ell m}(\hat{\textbf{k}}_{0})$ and $Y_{\ell m}(\hat{\textbf{r}})$ that are in turn defined with respect the BF system, $S$ ($\hat{\textbf{k}}_{0}$ and $\hat{\textbf{r}}$ are defined with respect the S reference).
The corresponding expression in the SF frame is obviously given by 

\begin{equation}
Y_{\ell m}(\hat{\textbf{r}}') = \sum_{m'} D_{m m'}^{\ell}(\alpha, \beta, \gamma) Y_{\ell m'}(\hat{\textbf{r}})
\end{equation}

\noindent where $\hat{\textbf{r}}'$ is defined with respect to $S'$ and $\hat{\textbf{r}}$ with respect to $S$ and the coefficient $D_{m m'}^{\ell}(\alpha, \beta, \gamma)$ are the Wigner functions \cite{th-sakurai}.
So, to this purpose, one has to transfer the angular terms from the BF to the SF frame through the following relations

\begin{eqnarray}
X_{\ell' h'}^{p\mu}(\hat{\textbf{r}}') = \sum_{m'}\sum_{\lambda} b_{\ell' h' m'}^{p\mu}\mathcal{D}_{\lambda m'}^{\ell'}(\alpha \beta \gamma) Y_{\ell'}^{\lambda}(\hat{\textbf{r}})\\  
X_{\ell h}^{p\mu}(\hat{\textbf{k}}_{0}') = \sum_{m}\sum_{\lambda'} b_{\ell h m}^{p\mu}\mathcal{D}_{\lambda' m}^{\ell}(\alpha \beta \gamma) Y_{\ell}^{\lambda'}(\hat{\textbf{k}}_{0})
\end{eqnarray}

\noindent where, since in the BF frame the incoming electron points toward the molecular main symmetry axis which is in turn taken as the $z$ axis, in the last expression one should consider $\vartheta$=0 and $\lambda'$=0, therefore obtaining

\begin{equation}
X_{\ell h}^{p\mu}(\hat{\textbf{k}}_{0}') = \sum_{m} b_{\ell h m}^{p\mu}\mathcal{D}_{0 m}^{\ell}(\alpha \beta \gamma) \left( \frac{2\ell +1}{4\pi} \right)^{1/2}
\end{equation}

\noindent By substituting eqs. A.2 and A.4 in eq. 2.66, the SF scattering amplitude is given by

\begin{eqnarray}
\tilde{f}_{FN} (\hat{\textbf{k}_{0}}' \cdot \hat{\textbf{r}}'; \alpha \beta \gamma) = \sum_{\ell h m}\sum_{\ell' h' m'} \sum_{p\mu} \sum_{\lambda} \frac{\sqrt{\pi (2\ell +1)}}{ik_{0}} i^{\ell -\ell'} \\ \nonumber
b_{\ell' h' m'}^{p\mu} b_{\ell h m}^{p\mu} Y_{\ell'}^{\lambda}(\hat{\textbf{r}}) \mathcal{D}_{\lambda m'}^{\ell'}(\alpha \beta \gamma) \mathcal{D}_{0 m}^{\ell}(\alpha \beta \gamma) \left( \mathbb{S}_{\ell h,\ell' h'}^{p\mu} - \delta_{\ell \ell'} \delta_{hh'} \right)
\end{eqnarray}

\noindent which, remembering the definition of $\mathbb{T}$ matrix as $1 - \mathbb{S}$ becomes

\begin{eqnarray}
\tilde{f}_{FN} (\hat{\textbf{k}_{0}}' \cdot \hat{\textbf{r}}'; \alpha \beta \gamma) = \sum_{\ell h m}\sum_{\ell' h' m'} \sum_{p\mu} \sum_{\lambda} \frac{i\sqrt{\pi (2\ell +1)}}{k_{0}} i^{\ell -\ell'} \\ \nonumber
b_{\ell' h' m'}^{p\mu} b_{\ell h m}^{p\mu} Y_{\ell'}^{\lambda}(\hat{\textbf{r}}) \mathcal{D}_{\lambda m'}^{\ell'}(\alpha \beta \gamma) \mathcal{D}_{0 m}^{\ell}(\alpha \beta \gamma) \mathbb{T}_{\ell h,\ell' h'}^{p\mu}
\end{eqnarray}

\noindent The differential cross section (thus the angular distribution) for the scattering of an electron by randomly oriented molecules is then obtained by averaging the SF quantity $\left| \tilde{f}_{SF}(\hat{\textbf{k}}_{0}' \cdot \hat{\textbf{r}}';\alpha \beta \gamma) \right|^{2}$ over all the molecular orientations ($\alpha$, $\beta$, $\gamma$):

\begin{eqnarray}
\frac{d\sigma}{d\Omega} (\hat{\textbf{k}}_{0}' \cdot \hat{\textbf{r}}') = \frac{1}{8\pi^{2}} \int\;d\alpha d\beta d\gamma \sin\beta |\tilde{f}_{SF}(\hat{\textbf{k}}_{0}' \cdot \hat{\textbf{r}}';\alpha \beta \gamma)|^{2}.
\end{eqnarray}

\noindent It therefore follows that

\begin{eqnarray}
|\tilde{f}_{SF}(\hat{\textbf{k}}_{0}' \cdot \hat{\textbf{r}}';\;\alpha \beta \gamma)|^{2} = \sum_{\ell hm}\sum_{\bar{\ell} \bar{h} \bar{m}} \sum_{\ell' h' m'} \sum_{\bar{\ell}' \bar{h}' \bar{m}'} \sum_{p\mu} \sum_{\bar{p} \bar{\mu}} \sum_{\lambda,\bar{\lambda}} \frac{\pi}{k_{0}^2}\\ \nonumber
\sqrt{(2\ell+1)(2\bar{\ell}+1)} i^{\ell-\ell'-\bar{\ell}+\bar{\ell}'} b_{\ell' h' m'}^{p\mu} b_{\bar{\ell}' \bar{h}' \bar{m}'}^{\bar{p}\bar{\mu} *} b_{\ell h m}^{p\mu} b_{\bar{\ell} \bar{h} \bar{m}}^{\bar{p}\bar{\mu} *} Y_{\ell'}^{\lambda} Y_{\bar{\ell}'}^{\bar{\lambda}}\\ \nonumber
\mathcal{D}_{\lambda m'}^{\ell'} \mathcal{D}_{\bar{\lambda} \bar{m}'}^{\bar{\ell}' *} \mathcal{D}_{0 m}^{\ell} \mathcal{D}_{0 \bar{m}}^{\bar{\ell}*} \mathbb{T}_{\ell h,\ell' h'}^{p\mu} \mathbb{T}_{\bar{\ell} \bar{h},\bar{\ell}' \bar{h}'}^{\bar{p}\bar{\mu} *}
\end{eqnarray}

\noindent The last expression, when integrated as in eq. A.7, can be simplified by making use of the contraction theorem foe the spherical harmonics (see eq. 3.7.72 p.216 in \cite{sakuraibook})

\begin{eqnarray}
Y_{\ell'}^{\lambda}Y_{\bar{\ell}'}^{\bar{\lambda}*} = \sum_{L} \sum_{M} \sqrt{\frac{\left(2\ell'+1\right) \left(2\bar{\ell}'+1\right) \left(2L+1\right)}{4\pi}} \left( \begin{array}{ccc} \ell' & \bar{\ell}' & L \\ \lambda & -\bar{\lambda} & M \end{array} \right) \left(-1\right)^{\bar{\lambda}} \left( \begin{array}{ccc} \ell' & \bar{\ell}' & L \\ 0 & 0 & 0 \end{array} \right) Y_{L}^{M}\nonumber
\end{eqnarray}

\noindent as well as by using the following relations (see p.216 eq.3.7.69 in \cite{sakuraibook} and p.103 eqs. 3.116-3.118 in \cite{zarebook})

\begin{eqnarray}
\mathcal{D}_{m' m}^{\ell} \mathcal{D}_{\bar{m}' \bar{m}}^{\bar{\ell}} = \sum_{L=\left|\ell-\bar{\ell}\right|}^{\ell+\bar{\ell}} \sum_{M=-L}^{L} \sum_{M'=-L}^{L} (2L+1)\left( \begin{array}{ccc} \ell & \bar{\ell} & L \\ m & \bar{m} & M \end{array} \right) \left( \begin{array}{ccc} \ell & \bar{\ell} & L \\ m' & \bar{m}' & M' \end{array} \right) \mathcal{D}_{MM'}^{L*}\nonumber
\end{eqnarray}

\begin{eqnarray}
\int\; d\Omega\; \mathcal{D}_{\hat{m}' \hat{m}}^{\hat{\ell}} \mathcal{D}_{\bar{m}' \bar{m}}^{\bar{\ell}} \mathcal{D}_{m' m}^{\ell} = 8\pi^{2} \left( \begin{array}{ccc} \ell & \bar{\ell} & \hat{\ell} \\ m' & \bar{m}' & \hat{m}' \end{array} \right) \left( \begin{array}{ccc} \ell & \bar{\ell} & \hat{\ell} \\ m & \bar{m} & \hat{m} \end{array} \right)\nonumber
\end{eqnarray}

\noindent so that the angular integration of the four rotation matrices in eq. A.8 provides

\begin{eqnarray}
\int\; \sin\beta\; d\alpha d\beta d\gamma\; \mathcal{D}_{\lambda m'}^{\ell'} \mathcal{D}_{\bar{\lambda} \bar{m}'}^{\bar{\ell}' *} \textcolor{blue}{\mathcal{D}_{0 m}^{\ell} \mathcal{D}_{0 \bar{m}}^{\bar{\ell}*}} =
\end{eqnarray}

\begin{eqnarray}
\textcolor{blue}{=\sum_{L=|l - \bar{l}|}^{l+\bar{l}} \sum_{M=-L}^{+L}\left(2L+1\right)\left(-1\right)^{\bar{m}} \left( \begin{array}{ccc} \ell & \bar{\ell} & L \\ 0 & 0 & 0 \end{array} \right) \left( \begin{array}{ccc} \ell & \bar{\ell} & L \\ m & -\bar{m} & M \end{array} \right)} \int d\Omega\; \mathcal{D}_{\lambda m'}^{\ell'} \mathcal{D}_{\bar{\lambda} \bar{m}'}^{\bar{\ell}' *} \textcolor{blue}{\mathcal{D}_{0M}^{L*}} \nonumber
\end{eqnarray}

\begin{eqnarray}
=\sum_{L=|l - \bar{l}|}^{l+\bar{l}} \sum_{M=-L}^{+L}\left(2L+1\right)\left(-1\right)^{\bar{m}} \left( \begin{array}{ccc} \ell & \bar{\ell} & L \\ 0 & 0 & 0 \end{array} \right) \left( \begin{array}{ccc} \ell & \bar{\ell} & L \\ m & -\bar{m} & M \end{array} \right) \textcolor{blue}{\int d\Omega\; \mathcal{D}_{\lambda m'}^{\ell'} \mathcal{D}_{\bar{\lambda} \bar{m}'}^{\bar{\ell}' *} \mathcal{D}_{0M}^{L*}} \nonumber
\end{eqnarray}

\begin{eqnarray}
=\sum_{L=|l - \bar{l}|}^{l+\bar{l}} \sum_{M=-L}^{+L}\left(2L+1\right)\left(-1\right)^{\bar{m}} \left( \begin{array}{ccc} \ell & \bar{\ell} & L \\ 0 & 0 & 0 \end{array} \right) \left( \begin{array}{ccc} \ell & \bar{\ell} & L \\ m & -\bar{m} & M \end{array} \right)\nonumber
\end{eqnarray}

\begin{eqnarray}
\textcolor{blue}{8\pi^{2}\left( \begin{array}{ccc} \ell' & \bar{\ell}' & L \\ \lambda & -\bar{\lambda} & 0 \end{array} \right)\left(-1\right)^{\bar{\lambda}} \left( \begin{array}{ccc} \ell' & \bar{\ell}' & L \\ m' & -\bar{m}' & -M \end{array} \right) \left(-1\right)^{\bar{m}'+M}}. \nonumber 
\end{eqnarray}

\noindent Noting now the orthogonality between $3j$ coefficients as expressed by writing (see eq. 2.32 p.51 \cite{zarebook})

\begin{eqnarray}
\sum_{\lambda} \sum_{\bar{\lambda}} \left( \begin{array}{ccc} \ell' & \bar{\ell}' & L \\ \lambda & -\bar{\lambda} & M \end{array} \right) \left( \begin{array}{ccc} \ell' & \bar{\ell}' & L \\ \lambda & -\bar{\lambda} & 0 \end{array} \right) = \left(2L+1\right)^{-1} \delta_{M0}\nonumber
\end{eqnarray}

\noindent it is finally possible to write

\begin{eqnarray}
\frac{d\sigma}{d\Omega} (\hat{\textbf{k}}_{0}' \cdot \hat{\textbf{r}}') = \sum_{\mathcal{L}=0}^{+\infty} \mathcal{A}_{\mathcal{L}} \mathcal{P}_{\mathcal{L}}\left(\cos\vartheta \right),
\end{eqnarray}

\noindent where $\mathcal{P}_{\mathcal{L}}(\cos\vartheta)$ are Legendre polynomials and the coefficients $\mathcal{A}_{\mathcal{L}}$ are therefore found to be

\begin{eqnarray}
\mathcal{A}_{\mathcal{L}} = \frac{1}{4k_{0}^{2}} \left(2\mathcal{L}+1\right)^{2} \sum_{\ell hm}\sum_{\bar{\ell} \bar{h} \bar{m}} \sum_{\ell' h' m'} \sum_{\bar{\ell}' \bar{h}' \bar{m}'} \sum_{p\mu} \sum_{\bar{p} \bar{\mu}} \sqrt{\left(2\ell+1\right) \left(2\bar{\ell}+1\right) \left(2\ell'+1\right) \left(2\bar{\ell}'\right)}\nonumber
\end{eqnarray}

\begin{eqnarray}
i^{\ell + \bar{\ell} - \ell' -\bar{\ell}'} \left( \begin{array}{ccc} \ell & \bar{\ell} & \mathcal{L} \\ 0 & 0 & 0 \end{array} \right) \left( \begin{array}{ccc} \ell' & \bar{\ell}' & \mathcal{L} \\ 0 & 0 & 0 \end{array} \right)\left( \begin{array}{ccc} \ell & \bar{\ell} & \mathcal{L} \\ m & -\bar{m} & M \end{array} \right) \left( \begin{array}{ccc} \ell' & \bar{\ell}' & \mathcal{L} \\ m' & -\bar{m}' & -M \end{array} \right)  \nonumber
\end{eqnarray}

\begin{eqnarray}
\left(-1\right)^{\bar{m}+\bar{m}'} \left\{b_{\ell hm}^{p\mu} \mathbb{T}_{\ell h, \ell' h'}^{p\mu} b_{\ell' h'm'}^{p\mu} \right\} \left\{ b_{\bar{\ell} \bar{h} \bar{m}}^{\bar{p} \bar{\mu}*} \mathbb{T}_{\bar{\ell} \bar{h}, \bar{\ell}' \bar{h}'}^{\bar{p} \bar{\mu}*} b_{\bar{\ell}' \bar{h}' \bar{m}'}^{\bar{p} \bar{\mu}*} \right\}.
\end{eqnarray}

\noindent The $\textcolor{blue}{\textbf{total}}$ $\textcolor{blue}{\textbf{integral}}$ ($\textcolor{blue}{\sigma_{i}}$) and the $\textcolor{blue}{\textbf{momentum}}$ $\textcolor{blue}{\textbf{transfer}}$ ($\textcolor{blue}{\sigma_{m}}$) cross sections, (ICS and MTCS, respectively), follow from eqs. A.10 and A.11:

\begin{eqnarray}
\sigma_{i} = \int \frac{d\sigma}{d\Omega} (\hat{\textbf{k}}_{0}' \cdot \hat{\textbf{r}}') \sin\vartheta\; d\vartheta d\varphi = 4\pi \mathcal{A}_{0}
\end{eqnarray}

\begin{eqnarray}
\sigma_{m} = \int \frac{d\sigma}{d\Omega} (\hat{\textbf{k}}_{0}' \cdot \hat{\textbf{r}}') (1-\cos\vartheta) \sin\vartheta\; d\vartheta d\varphi = 4\pi\left(\mathcal{A}_{0} - \frac{1}{3}\mathcal{A}_{1}\right).
\end{eqnarray}

\noindent In conclusion, the expressions for $\mathcal{A}_{0}$ and $\mathcal{A}_{1}$ can be directly deduced from eq. A.11; as an example, $\mathcal{A}_{0}$ is given by

\begin{eqnarray}
\mathcal{A}_{0} = \frac{1}{4k_{0}^{2}} \sum_{\ell hm} \sum_{\ell' h'm'} \sum_{p\mu} \left| b_{\ell hm}^{p\mu} \mathbb{T}_{\ell h, \ell' h'}^{p\mu} b_{\ell' h'm'}^{p\mu} \right|^{2} (-1)^{m+m'}
\end{eqnarray}


\chapter{The rotational excitation cross sections}

\noindent In appendix A the way to obtain a \textbf{rotationally} \textbf{summed} \textbf{cross} \textbf{section} is provided.
In practice, however, it can be efficiently computed by sums over magnetic substates as briefly illustrated in the present appendix; see \cite{th-fag-physrep} for a full detailed treatment.

\noindent So far the participation of the rotational (and vibrational) molecular degrees of freedom were ignored during the scattering process.
In spite of the added complexity that such additional couplings introduce in the treatment of the collision, it is however still possible to extract some useful information by using the approximate treatment of the dynamics known as the adiabatic nuclei approximation (AN).

\noindent Briefly, the physics involved requires to consider the effect of the kinetic energy operator of nuclear motion as negligible with respect to that of the corresponding operator of the electronic motion.
As a consequence, the molecule is seen as having 'infinite' mass and to stand still during the scattering event, when the energy spacings of the molecular levels is treated as negligible with respect to the local kinetic energy of the impinging electron.
In other words, the AN approximation is a special case of the FN approximation; in fact, if the nuclei are allowed to be fixed (not necessarily in their equilibrium position), it is possible to neglect the coupling between the incoming particle's angular momenta and the rotational (and/or vibrational, in general) quantum number of the molecule.
It therefore follows that the electrons $adiabatically$ respond to the istantaneous position of the nuclei.
In the absence of long-range forces (non-polar molecular targets) this is indeed a very good approximation, but for polar molecules the coupling between the projectile and the nuclear motion must be taken into account: it can be relaxed in the core-region (in the inner-region) but in the outer region the nuclear Hamiltonian should be incorporated into the scattering equation.

\noindent Assuming the BO approximation is valid, the AN total scattering (electron + molecular target) wavefunction can be written as:

\begin{eqnarray}
\Psi(\textbf{r}_{e}, \textbf{r}, \textbf{R}_{N}) \cong \psi_{e}(\textbf{r}_{e}, \textbf{r}; \textbf{R}_{N}) \psi_{N}^{n}(\textbf{R}_{N})
\end{eqnarray}

\noindent where 'n' represents all the rotational (and vibrational) quantum numbers of the unperturbed nuclear wave function.
The adiabatic nuclei rotation (ANR) approximation for the scattering amplitude is obtained for the transition $n_{i}\rightarrow n_{f}$, following \cite{chase}, as

\begin{eqnarray}
\tilde{f}_{ANR}^{n_{i}\rightarrow n_{f}}(\textbf{k}_{0},\bar{\textbf{k}}) \approx \langle \psi_{N}^{n_{f}} \left|\tilde{f}_{FN}(\textbf{k}_{0}'\cdot \textbf{r}'; \textbf{R}_{N} | \alpha\beta\gamma) \right| \psi_{N}^{n_{i}}\rangle
\end{eqnarray}

\noindent where $n_i$ and $n_f$ specify only the rotational molecular quantum numbers and $\tilde{f}_{FN}(\textbf{k}_{0}'\cdot \textbf{r}'; \textbf{R}_{N} | \alpha\beta\gamma)$ is the FN elastic scattering amplitude in the SF frame (see sec. 2.4.8 and appendix A).
In this connection, one should note that the collision time $\tau_{coll}$ may be comparable to the vibrational period; however, the rotational period $\tau_{rot}$ ma still be such that $\tau_{coll}\ll\tau_{rot}$, where this condition is satisfied even at thermal energies since for low-lying rotational states the rotational spacing is of the order of 10$^{-2}\sim$10$^{-3}$ eV.
Full details about the specific expressions for asymmetric top, symmetric top and spherical top molecules may be found in \cite{th-fag-physrep}.

\noindent In the case of a symmetric top target molecule, the rotational wavefunction is given by the well-known formula

\begin{eqnarray}
\psi_{N}^{n_i} = \psi_{JKM}^{Symm}(\alpha\beta\gamma) = \left(\frac{2J+1}{8\pi^2}\right)^{2} \mathcal{D}_{KM}^{J*}(\alpha\beta\gamma)
\end{eqnarray}

\noindent where $M$ gives the projection of the total angular momentum $J$ along the 'z' axis within the SF frame, while $K$ gives its projection along the molecular figure axis that in turn corresponds with the 'z' axis in the BF frame; in this case, both $M$ and $K$ are good quantum numbers.

\noindent One therefore writes the explicit scattering amplitude as calculated from eq. B.2, $\tilde{f}(J_{i}K_{i}M_{i}\rightarrow J_{f}K_{f}M_{f})$ and obtain from it the differential cross sections summed over all final degenerate states $M_{f}$ and averaged over all initial degenerate magnetic substates $M_{i}$

\begin{eqnarray}
\frac{d\sigma}{d\Omega}(J_{i}K_{i}\rightarrow J_{f}K_{f}) = \frac{k_f}{k_i(=k_0)} \sum_{M_{i},M_{f}}\frac{1}{2J_{i}+1} \left| \tilde{f}(J_{i}K_{i}M_{i}\rightarrow J_{f}K_{f}M_{f}) \right|^{2}
\end{eqnarray}

\noindent where $k'^{2}$ = $k^{2} + 2(E_{J_{i}K_{i}} - E_{J_{f}K_{f}})$ and the formula above is appropriate for molecules like C$_{24}$H$_{12}$.

\noindent In the case of asymmetric top (like for the o-C$_{6}$H$_{4}$), $K$ is no longer a good quantum number and the rotational wavefunction needs to be expressed as a linear combination of asymmetric top functions:

\begin{eqnarray}
\psi_{J\tau M}^{Asym} = \sum_{K=-J}^{J} a_{\tau K}^{J} \psi_{JKM}^{Symm}(\alpha\beta\gamma)
\end{eqnarray}

\noindent where the coefficients $a_{\tau K}^{J}$ are determined by diagonalizing the asymmetric top rotational Hamiltonian and can be chosen so that the asymmetric top wavefunctions are orthonormal and $\tau$ is a pseudo quantum number which differentiate the different rotational levels with the same $J$.
The scattering amplitude can therefore be had once more from the B.2 formula and the corresponding cross section, summed over the final states $M_{f}$ and averaged over initial states $M_{i}$ is given by

\begin{eqnarray}
\frac{d\sigma}{d\Omega}(J_{i}\tau_{i}\rightarrow J_{f}\tau_{f}) = \frac{k_f}{k_i(=k_0)} \sum_{M_{i},M_{f}}\frac{1}{2J_{i}+1} \left| \tilde{f}(J_{i}\tau_{i}M_{i}\rightarrow J_{f}\tau_{f}M_{f}) \right|^{2}
\end{eqnarray}

\noindent where in the present case $k'^{2}$ = $k^{2} + 2(E_{J_{i}\tau_{i}} - E_{J_{f}\tau_{f}})$.

\noindent The corresponding rotational wavefunction for a spherical top molecule is the same as that for the symmetric top.
There is a further degeneracy in that the $(2J+1)^2$ states with the same $J$, but with different $M$ and $K$, have the same energy $E_J$, where

\begin{eqnarray}
E_{J} = BJ(J+1)
\end{eqnarray}

\noindent where $B$ is the molecular rotational constant (in unit of energy).
The corresponding differential cross section is therefore given by

\begin{eqnarray}
\frac{d\sigma}{d\Omega}(J_{i}\rightarrow J_{f}) = \frac{k_f}{k_i(=k_0)} \sum_{M_{i},M_{f}}\frac{1}{2J_{i}+1} \left| \tilde{f}(J_{i}K_{i}M_{i}\rightarrow J_{f}K_{f}M_{f}) \right|^{2}
\end{eqnarray}

\chapter{List of publications}

\noindent 1) Carelli F., Sebastianelli F., Baccarelli I., \& Gianturco F.A., 2008, Int. J. Mass Spec., \textbf{277}, 155-161, 'Following electron attachment to CS(X$^{1}\Sigma$): quantum scattering calculations of the lowest resonant state'.\\

\noindent 2) Carelli F., Sebastianelli F., Baccarelli I., \& Gianturco F.A., 2010, ApJ, \textbf{712}, 445-452, 'Electron-driven reactions in proto-planetary atmospheres: metastable anions of gaseous ortho-benzyne'.\\

\noindent 3) Carelli, F., Sebastianelli, F., Satta, M., \& Gianturco, F.A., 2011, MNRAS, \textbf{415}-1, 425-430, 'Gas-phase route to PAHs in proto-planetary atmospheres: role of stabilized ortho-benzyne anions'.\\

\noindent 4) Carelli, F., \& Gianturco, F.A., 2011, ApJ, \textbf{743}, 151-157, 'On the relative 'transparency' of gas-phase coronene molecules to low-energy electrons: effects on the interstellar medium'.\\

\noindent 5) Sebastianelli F., Carelli F., \& Gianturco F.A., 2011, J. Phys. Chem., accepted, 'Resonant paths to molecular fragmentation after electron attachment: quantum dynamics of CF$_{2}^{-}$ intermediates'.\\

\noindent 6) Sebastianelli F., Carelli F., \& Gianturco F.A., 2011, Chem. Phys., in press, 'Forming NCCN$^{-}$ by quantum scattering: a modeling for Titan's atmosphere'.\\

\noindent 7) Carelli F., \& Gianturco F.A., 2011, Teochem, accepted, 'Resonant dynamics of electron-driven reactions in the gas-phase: the coronene molecule as a prototype for planetary atmospheres and interstellar clouds'.\\

\noindent 8) Carelli F., \& Gianturco F.A., 2011, MNRAS, in preparation, 'PAH negative ion formations in interstellar cloud by low-energy electron attachment: a quantum study on gas-phase coronene'.\\

\clearpage




\begin{thebibliography}{999}
\addcontentsline{toc}{chapter}{Bibliography}
\bibliographystyle{plain}

\bibitem{intro-birtwistle} Birtwistle, D.T., \& Herzenberg, A., 1971, J. Phys. B: Atom. Molec. Phys., \textbf{4}, 53
\bibitem{intro-berman} Berman, M., Estrada, H., Cederbaum, L.S., \& Domcke, W., 1983, Phys. Rev. A., \textbf{28}, 1363
\bibitem{intro-fenzlaff} Fenzlaff, M., Gerhard, R., \& Illenberger, E., 1988, J. Chem. Phys., \textbf{88}, 149
\bibitem{intro-fenzlaff2} Fenzlaff, M., \& Illenberger, E., Chem. Phys., \textbf{136}, 443
\bibitem{intro-sakuraibook} Sakurai, J.J., 1985, 'Modern quantum collision', Addison-Wesley Publishing Company
\bibitem{intro-burkebook} Burke, P.G., 1977, 'Potential scattering in atomic physics', Plenum Press, New York
\bibitem{intro-joachainbook} Joachain, C.J., 1975, 'Quantum collision theory', North Holland, Amsterdam
\bibitem{intro-vuitton} Vuiton, V., Lavvas, P., Yelle, R.V., Galand, M., Wellbrock, A., Lewis, G.R., Coates, A.J., \& Wahlund, J.E., 2009, Planet. Space Sci., \textbf{57}, 1558
\bibitem{intro-dalgarno} Dalgarno, A., \& McCray, R.A., 1973, ApJ, \textbf{181}, 95
\bibitem{intro-sarre} Sarre, P.J., 1980, Journal de Chimie Physique, \textbf{77}, n$^{o}$7/8, 769
\bibitem{intro-herbst} Herbst, E., 1981, Nature, \textbf{289}, 656
\bibitem{intro-kawaguchi} Kawaguchi, K., Kasai, Y., Ishikawa, S., \& Kaifu, N., 1995, PASJ, \textbf{47}, 853
\bibitem{intro-mccarthy} McCarthy, M.C., Gottlieb, C.A., Gupta, H.C., \& Thaddeus, P., 2006, ApJ, \textbf{652}, L141
\bibitem{intro-gupta} Gupta, H., Brunken, S., Tamassia, F., Gottleib, C.A., McCarthy, M.C., \& Thaddeus, P., 2007, ApJ, \textbf{655}, L57
\bibitem{intro-thaddeus} Thaddeus, P., Gottlieb, C.A., Gupta, H., Brunken, S., McCarthy M.C., Agundez, M., Guelin, M., \& Cernicharo, J., 2008, ApJ, \textbf{677}, 1132
\bibitem{intro-cernicharo} Cernicharo, J., Guelin, M., Agundez, M., Kawaguchi, K., McCarthy, M., \& Thaddeus, P. 2007, A$\&$A, \textbf{437}, L67
\bibitem{intro-remijan} Remijan, A.J., Hollis, J.M., Lovas, F.J., Cordiner, M.A., Millar, T.J., Markwick-Kemper, A.J., \& Jewell, P.R., 2007, ApJ, \textbf{664}, L47
\bibitem{intro-brunken} Brunken, S., Gupta, H., Gottlieb, C.A., McCarthy, M.C., \& Thaddeus, P., 2007, ApJ, \textbf{664}, L43
\bibitem{intro-sakai} Sakai, N., Sakai, T., Osamura, Y., Yamamoto, S., 2007, ApJ, \textbf{667}, L65
\bibitem{intro-sakai2} Sakai, N., Sakai, T., Hirota, T., \& Yamamoto, S., 2008, ApJ, \textbf{672}, 371
\bibitem{intro-agundez} Agundez, M., Cernicharo, J., Guelin, M., Gerin, M., McCarthy, M.C., \& Thaddeus, P., 2008, A$\&$A, \textbf{478}, L19
\bibitem{intro-gupta2} Gupta, H., Gottleib, C.A., McCarthy, M.C., \& Thaddeus, P., 2009, ApJ, \textbf{691}, 1494
\bibitem{intro-cernicharo2} Cernicharo, J., Guelin, M., Agundez, M., McCarthy, M.C., \& Thaddeus, P., 2008, ApJ, \textbf{688}, L83
\bibitem{intro-agundez2} Agundez, M., Cernicharo, J., Guelin, M., Kahane, C., Roueff, E., Klos, J., Aoiz, F.J., \& Lique, F., 2010, A$\&$A, \textbf{517}, L2
\bibitem{intro-bakes} Bakes, E.L.O., \& Tielens, A.G.G.M., 1998, ApJ, \textbf{499}, 258
\bibitem{intro-wakelam} Wakelam, V., \& Herbst, E., 2008, ApJ, \textbf{680}, 371
\bibitem{intro-flower} Flower, D.R., Pineau des Forets, G., \& Walmsley, C.M., 2007, \textbf{474}, 923
\bibitem{intro-draine} Draine, B.T., 2003, ARA$\&$A, \textbf{41}, 241
\bibitem{intro-leger} Leger, A., D'Hendecourt, L., \& Defourneau, D., 1989, A$\&$A, \textbf{216}, 148
\bibitem{intro-leger2} Leger, A., \& D'Hendecourt, L., 1985, A$\&$A, \textbf{146}, 81
\bibitem{intro-allamandola} Allamandola, L.J., Tielens, A.G.G.M., \& Barker, J.R., 1985, ApJ, \textbf{290}, L25
\bibitem{intro-salama} Salama, F., Galazutdinov, G.A., Krelowski, J., Allamandola, L.J., \& Musaev, F.A., 1999, ApJ, \textbf{526}, 625
\bibitem{intro-cesarsky} Cesarsky, D., Lequeux, J., Abergel, A., et. al., 1996, A$\&$A, \textbf{315}, L305
\bibitem{intro-smith} Smith, J.D.T., Draine, B.T., Dole, D.A., et al., 2007, ApJ, \textbf{656}, 770
\bibitem{intro-mulas} Mulas, G., Malloci, G., Joblin, C., \& Toublanc, D., 2006, A$\&$A, \textbf{456}, 161
\bibitem{intro-tielensbook} Tielens, A.G.G.M., 2005, 'The physics and chemistry of the interstellar medium', Cambridge University Press
\bibitem{intro-allamandolarew} Allamandola, L.J., Tielens, A.G.G.M., \& Barker, J.R., 1989, ApJS, \textbf{71}, 733
\bibitem{intro-herbst2} Herbst, E., 1975, ApJ, \textbf{205}, 94
\bibitem{intro-herbst3} Herbst, E., 1979, J. Chem. Phys., \textbf{70}, 2201
\bibitem{intro-herbst4} Herbst, E., 1980, ApJ, \textbf{237}, 462
\bibitem{intro-herbst5} Herbst, E., 1980, ApJ, \textbf{241}, 197
\bibitem{intro-hotop} Hotop, H., \& Fabrikant, I.I., 2008, J. Chem. Phys., \textbf{128}, 124308
\bibitem{intro-flowerbook} Flower, D., 2007, 'Molecular collisions in the interstellar medium', Cambridge Astrophysics Series n.42
\bibitem{intro-troe} Troe, J., Miller, T.M., \& Viggiano, A.A., 2007, J. Chem. Phys., \textbf{127}, 244304
\bibitem{intro-robinson} Robinson, P.J., \& Holbrook, K.a., 1972, 'Unimolecular Reactions' Wiley, New York
\bibitem{intro-carelli0} Carelli, F., Sebastianelli, F., Baccarelli, I., \& Gianturco, F.A., 2008, Int. J. Mass Spec., \textbf{277}, 155
\bibitem{intro-carelli2} Carelli, F., Sebastianelli, F., Satta, M., Gianturco, F.A. 2011, MNRAS, \textbf{415}-1, 425
\bibitem{intro-carellicf2} Sebastianelli, F., Carelli. F., \& Gianturco, F.A., 2011, J. Phys. Chem., accepted
\bibitem{intro-romanini} Romanini, D., \& Lehmann, K.K., 1993, J. Chem. Phys., \textbf{98}, 6437
\bibitem{intro-marcus} Stuchebrukhov, A.A., \& Marcus, R.A., 1993, J. Chem. Phys., \textbf{98}-9, 6004
\bibitem{intro-bethardy} Bethardy, G.A., Wang, X., \& Perry, D.S., 1994, Can. J. Chem., \textbf{72}, 652
\bibitem{intro-gambogi} Gambogi, J.E., Timmermans, J.T., \& Lehmann, K.K., 1993, Chem. Phys., \textbf{99}, 9314
\bibitem{intro-christophe} Christophe, I., \& Wyatt, R.E., 1993, J. Chem. Phys., \textbf{99}-3, 2261
\bibitem{intro-petrie} Petrie, S., \& Herbst, E., 1997, ApJ, \textbf{491}, 210
\bibitem{intro-osamura} Herbst, E., \& Osamura, Y., 2008, ApJ, \textbf{679}, 1670
\bibitem{intro-taylorbook} Taylor, J.R., 1972, 'Scattering theory: the quantum theory of non-relativistic collisions', John Wiley $\&$ Sons, New York

\end{thebibliography}

\begin{thebibliography}{999}
\addcontentsline{toc}{chapter}{Bibliography}
\bibliographystyle{plain}

\bibitem{th-frisch} Frisch, M.J., et al., Gaussian 03, revision c.02 (2004) Gaussian Inc., Wallingford, CT
\bibitem{th-altman} Altman, S.L., \& Cracknell, A.P., 1965, Rev. Mod. Phys., \textbf{37}, 19
\bibitem{th-hara} Hara, S., 1967, J. Phys. Soc. Japan, \textbf{22}, 710
\bibitem{th-fag-VcorrDFT} Gianturco, F.A., \& Rodriguez-Ruiz, 1992, J. Mol. Struct. (Teochem), \textbf{260}, 99
\bibitem{th-lucchese} Lucchese, R.R., \& Gianturco, F.A., 1996, Int. Rev. Phys. Chem., \textbf{15}, 429
\bibitem{th-joachainbook} Joachain, C.J., 1975, 'Quantum collision theory', North Holland, Amsterdam
\bibitem{th-taylorbook} Taylor, J.R., 1972, 'Scattering theory: the quantum theory of nonrelativistic collisions', John Wiley \& Sons, New York
\bibitem{th-huo} Gianturco, F.A., Thompson, D.G., \& Jain, A., 1995, in 'Computational methods for electron-molecule collisions', Ed. by Gianturco, F.A., \& Huo, W.M., Plenum Press, New York
\bibitem{th-sakurai} Sakurai, J.J., 1985, 'Modern quantum mechanics', Addison-Wesley Publishing
\bibitem{th-fag-physrep} Gianturco, F.A., \& Jain, A., 1986, Phys. Rep., \textbf{143}, n$^{o}$6, 348-425
\bibitem{th-sanna} Sanna, N., \& Gianturco, F.A., 1998, Comp. Phys. Comm., \textbf{114}, 142
\end{thebibliography}

\begin{thebibliography}{999}
\addcontentsline{toc}{chapter}{Bibliography}
\bibliographystyle{plain}

\bibitem{altshuler} Altshuler, S. 1957, Phys. Rev., \textbf{107}, 114
\bibitem{wenthold} Wenthold, P.G., Squires, R.R., \& Lineberger, W.C., 1998, J. Am. Chem. Soc., \textbf{120}, 5279
\bibitem{fangtong} Zhang, F., Parker, D., Kim, Y.S., Kaiser, R.I., \& Mebel, A.M., 2011, ApJ, \textbf{728}, 141
\bibitem{burrow} Burrow, P.D., Michejda, J.A., \& Jordan, K.D., 1987, J. Chem. Phys., \textbf{86}, 9
\bibitem{tielensbook} Tielens, A.G.G.M., 2005, 'The Physics and Chemistry of the interstellar medium', Ed. Cambridge University Press
\bibitem{pardo} Pardo, J.R., Cernicharo, J., \& Goicoechea, J.R., 2007, ApJ, \textbf{661}, 250
\bibitem{kraka} Kraka, E., \& Cremer, D., 1993, Chem. Phys. Lett., \textbf{216}, 333 
\bibitem{mccarthy} McCarthy, M.C., Gottlieb, C.A., Gupta, H.C., \& Thaddeus, P., 2006, ApJ, \textbf{652}, L141
\bibitem{sakai} Sakai, N., Sakai, T., Osamura Y., Yamamoto, S., 2007, ApJ, \textbf{667}, L65 
\bibitem{lucchese} Lucchese, R.R., \& Gianturco, F.A., 1996, Int. Rev. Phys. Chem. \textbf{15}, 429 TH
\bibitem{tele} Telega, S., \& Gianturco, F.A., 2005, Eur. Phys. J. D,  \textbf{36}, 271 TH
\bibitem{telega} Telega, S., \& Gianturco, F.A., 2006, Eur. Phys. J. D,  \textbf{38}, 495 TH
\bibitem{carelli0} Carelli, F., Sebastianelli, F., Baccarelli, I., \& Gianturco, F.A., 2008, Int. J. Mass Spec., \textbf{277}, 155
\bibitem{frisch} Frisch, M.J., et al., Gaussian 03, revision c.02 (2004) Gaussian Inc., Wallingford, CT
\bibitem{groner} Groner, P., \& Kukolich, S.G., 2006, J. Mol. Struct. \textbf{780-1}, 178
\bibitem{epoly} Lucchese, R.R., Sanna, N., Natalense, A.P.P., \& Gianturco, F.A., ePolyScat (version E2) www.chem.tamu.edu/rgroup/lucchese/ePolyScat.E2.manual/manual.html available on request
\bibitem{nist} NIST Computational Chemistry Comparison and Benchmark Database
\bibitem{zhang} Zhang, X., \& Chen, P., 1992, J. Am. Chem. Soc. \textbf{114} 3147
\bibitem{allamandola} Allamandola, L.J., Tielens, A.G.G.M., \& Barker, J.R., 1989, ApJ, \textbf{290}, L25
\bibitem{allamandola2} Allamandola, L.J., Tielens, A.G.G.M., \& Barker, J.R., 1989, ApJS, \textbf{71}, 733
\bibitem{carelli1} Carelli, F., Sebastianelli, F., Baccarelli, I., \& Gianturco, F.A., 2010, ApJ, \textbf{712}, 445
\bibitem{cernicharo} Cernicharo, J., Heras, A.M., Tielens, A.G.G.M., Pardo, J.R., Herpin, F., Guelin, M., \& Waters, L.B.F.M., 2001, ApJ, \textbf{546}, L123
\bibitem{cherchneff} Cherchneff, I., Barker, J.R., \& Tielens, A.G.G.M., 1992, ApJ, \textbf{401}, 269
\bibitem{flowerbook} Flower, D., 2007, 'Molecular collisions in the interstellar medium', Cambridge Astrophysics Series n.42
\bibitem{frenklach} Frencklach, M., Clary, D.W., Yuan, T., Gardiner, W.C., \& Stein, S.E., 1986, Combustion Sci. Tech., \textbf{50}, 79
\bibitem{frenklach2} Frenklach, M., \& Feigelson, E.D., 1989, ApJ, \textbf{341}, 372
\bibitem{gaussian} Frisch, M.J. et al., 2004, GAUSSIAN 03, revision c.02 Gaussian Inc., Wallingford, CT
\bibitem{geballe} Geballe, T.R., \& Van der Veen, V.E.C.J., 1990, A$\&$A, \textbf{235}, L9
\bibitem{herbst} Herbst, E., 1981, Nature, \textbf{289}, L656
\bibitem{kwok} Kwok, S., 2007, 'The origin and evolution of planetary nebulae', Cambridge Astrophysics Series, vol.31
\bibitem{kwok2} Kwok, S., Nature, 2004, \textbf{430}, 985
\bibitem{mcmahon} McMahon, R.J., McCarthy, M.C., Gottlieb, C.A., Dudek, J.B., Stanton, J.F., \& Thaddeus, P., ApJ, 2003, \textbf{590}, L61
\bibitem{mebel} Mebel, A.M., Lin, M.C., Chakraborty, D., Park, J., Lin,  S.H., \& Lee, Y.T., 2001, J. Chem. Phys., \textbf{114}, 8421
\bibitem{meixner} Meixner, M., Campbell, M.T., Welch, W.J., \& Likkel, L., 1998, ApJ, \textbf{509}, 392
\bibitem{nash} Nash, J.J., \& Squires, R.R., 1996, J. Am. Chem. Soc., \textbf{118}, 11872
\bibitem{sanchez} Sanchez Contreras, C., Sahai, R., \& de Paz, G., 2002, ApJ, \textbf{578}, 269
\bibitem{sanchez2} Sanchez Contreras, C., \& Sahai, R., 2003, ApJ, \textbf{602}, 960
\bibitem{sanchez3} Sanchez Contreras, C., \& Sahai, R., 2004, ApJ, \textbf{617}, 1142
\bibitem{stein} Stein, S.E., \& Fahr, A., 1985, J.Phys. Chem., \textbf{89}, 3714
\bibitem{taylorbook} Taylor, J.R., 1972, 'Scattering theory: the quantum theory of non-relativistic collisions', John Wiley $\&$ Sons, New York
\bibitem{thanopulos} Thanopulos, I., Brumer, P., \& Shapiro, M., 2010, J. Chem. Phys., \textbf{133}, 154111
\bibitem{tielens} Tielens, A.G.G.M., 1993, ApJ, \textbf{271}, 702
\bibitem{valiev} Valiev, M. et al., 2010, Comput. Phys. Commun., \textbf{181}, 1477
\bibitem{widicus} Widicus Weaver, S.L., Remijan, A.J., McMahon, R.J., \& McCall, B.J., 2007, ApJ, \textbf{671}, L153
\bibitem{witterborn} Witterborn F.C., Sandford S.A., Bregman J.D., Allamandola, L.J., Cohen, M., Wooden, D.H., \& Graps, A.L., 1989, ApJ, \textbf{341}, 270
\bibitem{woods} Woods, P.M., Millar, T.J., Herbst, E., \& Zijlstra A.A., 2003, A$\&$A, \textbf{402}, 189
\bibitem{woods2} Woods, P.M., Millar, T.J., Zijlstra, A.A., \& Herbst, E., 2002, ApJ, \textbf{574}, L167
\bibitem{leger} Leger, A., \& Puget, J.L., 1984, A$\&$A, \textbf{137}, L5
\bibitem{giard} Giard, M., Lamarre, J.M., Pajot, E., \& Serra, G., 1994, A$\&$A, \textbf{286}, 203
\bibitem{flower2} Flower, D.R., Pineau des Forets, G., \& Walmsley, C.M., 2007, A$\&$A, \textbf{474}, 923
\bibitem{draine3} Draine, B.T., \& Sutin, B., 1987, ApJ, \textbf{320}, 803
\bibitem{weingartner} Weingartner, J.C., \& Draine, B.T., 2001, ApJ, \textbf{548}, 296
\bibitem{flower3} Flower, D.R., Pineau des Forets, G., Walmsley, C.M., 2006, A$\&$A, \textbf{449}, 621
\bibitem{cernicharo2} Cernicharo, J., Spielfiedel, A., Balanca, C., Dayou, F., Senent, M.-L., Feautrier, N., Faure, A., Cressiot-Vincent, L., Wiesenfeld, L., Pardo, J.R., 2011, A$\&$A, \textbf{531}, A103
\bibitem{lepp} Lepp, S. \& Dalgarno A. 1988, Apj, \textbf{324}, 553
\bibitem{lepp2} Lepp, S. \& Dalgarno A. 1988, Apj, \textbf{335}, 769
\bibitem{prasad} Prasad, S.S., \& Tarafdar, S.P., 1983, ApJ, \textbf{267}, 603
\bibitem{cravens} Cravens, T.E., \& Dalgarno, A., 1978, ApJ, \textbf{219}, 750
\bibitem{brunken} Brunken, S., Gupta, H., Gottlieb, C.A., McCarthy, M.C., \& Thaddeus, P., 2007, ApJ, \textbf{664}, L43
\bibitem{remijan} Remijan, A.J., Hollis, J.M., Lovas, F.J., Cordiner, M.A., Millar, T.J., Markwick-Kemper, A.J., \& Jewell, P.R., 2007, ApJ, \textbf{664}, L47
\bibitem{cernicharo3} Cernicharo, J., Guelin, M., Agundez, M., Kawaguchi, K., McCarthy, M., \& Thaddeus, P. 2007, A$\&$A, \textbf{437}, L67
\bibitem{taylort} Taylor, T.R., Xu, C., \& Neumark, D.M., 1998, J. Chem. Phys., \textbf{108}, n.24, 10018
\bibitem{modelli} Modelli, A., \& Mussoni, L., 2007, Chem. Phys., \textbf{332}, 367
\bibitem{mulas} Mulas, G., Malloci, G., Joblin, C., \& Toublanc, D., 2006, A$\&$A, \textbf{456}, 161
\bibitem{draine4} Draine, B.T., \& Li, A., 2007, ApJ, \textbf{657}, 810
\bibitem{boulanger} Boulanger, F., Falgarone, E., Puget, J.L., \& Helou, G., 1990, ApJ, \textbf{364}, 136
\bibitem{bernard} Bernard, J.P., Boulanger, F., \& Puget, J.L., 1993, A$\&$A, \textbf{277}, 609
\bibitem{rapacioli} Rapacioli, M., Joblin, C., \& Boissel, P., 2005, A$\&$A, \textbf{429}, 193
\bibitem{berne} Berne, O., et. al., 2007, A$\&$A, \textbf{469}, 575
\bibitem{velusamy} Velusamy, T., \& Langer, W.D., 2008, ApJ, \textbf{136}, 602
\bibitem{steglich} Steglich, M., Jager, C., \& Rouille, G., et. al., 2010, ApJ, \textbf{712}, L16
\bibitem{batesbook} Bates, D.R., \& Herbst, E., 1988, 'Rate coefficient in astrochemistry', ed. Millar, T.J., \& Williams, D.A., Kluwer
\bibitem{mulas2} Mulas, G., Malloci, G., Joblin, C., \& Cecchi-Pestellini, 2011, in 'PAH and the Universe', ed. Joblin, C., \& Tielens, A.G.G.M., EAS Publications Series, \textbf{46}, 327
\bibitem{peeters} Peeters, E., Hony, S., van Kerckhoven et. al., 2002, A$\&$A, \textbf{390}, 1089
\bibitem{rapacioli2} Rapacioli, M., Calvo, F., Joblin, C., Parneix, P., Toublanc, D., \& Spiegelman, F., 2006, A$\&$A, \textbf{460}, 519
\bibitem{lepage} Le Page, V., Snow, T.P., \& Bierbaum, V.M., 2001, ApJS, \textbf{132}, 233
\bibitem{abouaf} Abouaf, R., \& Diaz-Tendero, S. 2009, Chem. Phys. Phys. Chem., \textbf{11}, 5686
\bibitem{allamandola} Allamandola, L.J., Tielens A.G.G.M., \& Barker J.R. 1985, ApJ \textbf{290}, L25
\bibitem{bakes} Bakes, E.L.O., Bauschlicher, C.W., \& Tielens, A.G.G.M. 2004, in Astrophysics of Dust, ASP Conf. Ser., \textbf{309}, 731 
\bibitem{bakes2} Bakes, E.L.O., \& Tielens A.G.G.M., 1998, ApJ, \textbf{499}, 258
\bibitem{carelli2} Carelli, F., Sebastianelli, F., Satta, M., Gianturco, F.A. 2011, MNRAS, \textbf{415}-1, 425
\bibitem{cecchi} Cecchi-Pestellini, C., Malloci, G., Mulas, G., Joblin, C., \& Williams, D.A. 2008, A\&A, \textbf{486}, L25
\bibitem{cecchi2} Cecchi-Pestellini, C., \& Williams, D.A. 1998, MNRAS, \textbf{296}, 414
\bibitem{chen} Chen, G., Cooks, R.G., Corpuz, E. \& Scott L.T., 1996, J. Am. Soc. Mass. Spectrom 1996 \textbf{7}, 619
\bibitem{crawford} Crawford, M.K., Tielens, A.G.G.M., \& Allamandola, L.J. 1985, ApJ, \textbf{293}, L45
\bibitem{denifl} Denifl, S., Ptasinska, S., Sonnweber, B., Scheier, P., Liu, D., Hagelberg, F., Mach, J., Scott, L.T., \& Mark, T.D. 2005, J. Chem. Phys., \textbf{123}, 104308
\bibitem{draine} Draine, B.T. 2003, ARA\&A, \textbf{41}, 241
\bibitem{draine2} Draine, B.T. 2004, in Astrophysics of Dust, ASP Conf. Ser., \textbf{309}, 691
\bibitem{duncan} Duncan, M.A., Knight, A.M., Negishi, Y., Nagao, S., Nakamura, Y., Kato, A., Nakajima, A., \& Kaya, K. 1999, Chem. Phys. Lett. \textbf{309}, 49
\bibitem{gianturco3} Gianturco, F.A., \& Lucchese, R.R. 1998, J. Chem. Phys., \textbf{108}, n$^{o}$15, 6144 
\bibitem{gianturco4} Gianturco, F.A., Rodriguez-Ruiz, J.A., \& Sanna, N. 1995, Phys. Rev. A, \textbf{52} n$^{0}$2, 1257 
\bibitem{henning} Henning, T., \& Salama, F. 1998, Science, \textbf{282}, 2204
\bibitem{hudgins} Hudgins, D.M., \& Allamandola, L.J. 2004, in Astrophysics of Dust, ASP Conf. Ser., \textbf{309}, 665
\bibitem{iati} Iati', M.A., Saija, R., Borghese, F. et al. 2008, MNARAS, \textbf{384}, 591
\bibitem{joachain} e. g. see: Joachain. C.J., in 'Quantum collision theory', (North Holland, 1987)
\bibitem{khakoo} Khakoo, M.A., Ratliff, J.M., \& Trajmar, S. 1990, J. Phys. Chem., \textbf{93}, 8616
\bibitem{kraka} Kraka, E. \& Cremer, D. 1993, Chem. Phys. Lett., \textbf{216}, 333
\bibitem{li} Li, A., \& Greenberg, J.M. 1997, A\&A, \textbf{323}, 566
\bibitem{lepp} Lepp, S. \& Dalgarno A. 1988, Apj, \textbf{324}, 553
\bibitem{leopold} Leopold D.G., Miller, E.S. \& Lineberger W.C. 1986, J. Am. Chem. Soc., \textbf{108}, 1379
\bibitem{lucchese3} Lucchese, R.R., Gianturco, F.A., \& Sanna, N. 1999, Chem. Phys. Lett. \textbf{305}, 413
\bibitem{malloci} Malloci, G., Mulas, G., Cappellini, G., Fiorentini, V., \& Porceddu, I. 2005, A\&A, \textbf{432}, 585
\bibitem{newton} e.g. see: Newton, R.G. in 'Scattering theory of Waves and Particles', 2nd edition, Springer, New York, 1982 
\bibitem{sanna} Sanna, N., \& Gianturco, F.A., 1998, Comp. Phys. Comm., \textbf{114}, 142
\bibitem{song} Song, J.K., Lee, N.K., Kim, S.K. 2003, Angew. Chem., Int. Ed. \textbf{42}, 213
\bibitem{vidmar} Vidmar, R.J. 1990, IEEE Trans. Plasma Sci., \textbf{18}, 733
\bibitem{vijh} Vijh, U.P., Witt, A.N., \& Gordon, K.D. 2004, ApJ, \textbf{606}, L65
\bibitem{wakelam} Wakelam, V \& Herbst, E. 2008, Apj, \textbf{680}, 371
\bibitem{weber} Weber, J.M., \& Hotop, H. 1996, Z. Phys. D, \textbf{37}, 351
\bibitem{zwet} van der Zwet, G.P., \& Allamandola, L.J. 1985, A\&A, \textbf{146}, 76
\bibitem{cesarsky} Cesarsky, D., Lequeux, J., Abergel, A., et. al., 1996, A$\&$A, \textbf{315}, L305
\bibitem{yelle} Yelle, R.V., Borggren, N., De La Haye, V., Kasprzak, W.T., Niemann, H.B., Mueller-Wodarg, I.C.F., \& Waite, J.H., 2006, Icarus, \textbf{182}, 567
\bibitem{capone} Capone, L.A., Whitten, R.C., Dubach, J., Prasad, S.S., \& Huntress, W.T., 1976, Icarus, \textbf{28}, 367
\bibitem{barucki} Borucki, W.J., Levin, Z., Whitten, R.C., Keesee, R.G., Capone, L.A., Summers, A.L., Toon, O.B., \& Dubach, J., 1987, Icarus, \textbf{72}, 604
\bibitem{yung} Yung, Y.L., Allen, M., Pinto, J.P., 1984, Astrophys. J. Suppl. Ser., \textbf{55}, 465
\bibitem{waite} Waite, J.H., Young, D.T., Cravens, T.E., Coates, A.J., Crary, F.J., Magee, B., \& Westlake, J., 2007, Science, \textbf{316}, 870
\bibitem{coates} Coates, A.J., Crary, F.J., Lewis, G.R., Young, D.T., Waite. J.H., \& Sittler, E.C., 2007, Geophys. Res. Lett., \textbf{34}, L22103
\bibitem{coatesbis} Coates, A.J., Wellbrock, A., Lewis, G.R., Jones, G.H., Young, D.T., Crary, F.J., \& Waite, J.H., 2009, Planet Space Sci., \textbf{57}, 1573
\bibitem{vuitton} Vuitton, V., Lavvas, P., Yelle, R.V., Galand, M., Wellbrock, A., Lewis, G.R., Coates, A.J., \& Wahlund, J.E., 2009, Planetary and Space Science, \textbf{57}, 1558
\bibitem{kunde} Kunde, V.G., Aikin, A.V., Hauel, R.A., Jennings, D.E., Maguire, W.C., \& Sammelsan, R.E., 1981, Nature, \textbf{292}, 686
\bibitem{coustenis} Coustenis, A., Bezard, B., Gautier, D., Martin, A., \& Samuelson, R., 1991, Icarus, \textbf{89}, 152
\bibitem{sebastianelli} Sebastianelli, F., \& Gianturco, F.A., 2010, Eur. Phys. J. D, \textbf{59}, 389
\bibitem{graupner} Graupner, K., Field, T.A., \& Saunders, G.C., 2008, ApJ, \textbf{685}, L95
\bibitem{herbst5} Herbst, E., Osamura, Y., 2008, ApJ, \textbf{679}, 1670
\bibitem{nsangou} Nsangou, M., Senent, M.L., \& Hochlaf, M., 2008, Chem. Phys., \textbf{355}, 164
\bibitem{ng} Ng, L., Balaji, V., \& Jordan, K.D., 1983, Chem. Phys. Lett., \textbf{101}, 171
\bibitem{yungen} Jungen, Ch., \& Pratt, S.T., 2008, J. Chem. Phys., \textbf{129}, 164311
\bibitem{sakuraibook} Sakurai, J.J., 1985, 'Modern quantum mechanics', Addison-Wesley Publishing Company, USA
\bibitem{zarebook} Zare, R.N., 1986, 'Angular Momentum: understanding spatial aspects in chemistry and physics', John Wiley \& Sons, New York (USA)
\bibitem{chase} Chase, D.M., 1956, Phys. Rev., \textbf{104}, 838



\end{thebibliography}
\end{document}